\documentclass[12pt]{article}
\usepackage{amssymb,amsmath,amscd}
\usepackage[bbgreekl]{mathbbol}
\usepackage[mathscr]{eucal}
\usepackage[dvips]{color}
\usepackage{graphicx}
\usepackage{psfrag}
\DeclareFontFamily{U}{euc}{}
\DeclareFontShape{U}{euc}{m}{n}{<-6>eurm5<6-8>eurm7<8->eurm10}{}%
\DeclareSymbolFont{AMSc}{U}{euc}{m}{n} 
\DeclareMathSymbol{\uzeta}{\mathord}{AMSc}{"10} 
\DeclareMathSymbol{\uvarepsilon}{\mathord}{AMSc}{"22} 
\DeclareMathSymbol{\umu}{\mathord}{AMSc}{"16} 
\DeclareMathSymbol{\uphi}{\mathord}{AMSc}{"1E}

\usepackage{hyperref}
\usepackage[numbers,sort&compress]{natbib}
\usepackage{hypernat}
\hypersetup{linkbordercolor=.8 .8 1,
            citebordercolor=.7 1 .7}

\DeclareFontFamily{U}{rsf}{}
\DeclareFontShape{U}{rsf}{m}{n}{
  <5> <6> rsfs5 <7> <8> <9> rsfs7 <10-> rsfs10}{}
\DeclareMathAlphabet\Scr{U}{rsf}{m}{n}

\catcode`\@=11
\def\@citex[#1]#2{%
\if@filesw \immediate \write \@auxout {\string \citation {#2}}\fi
\@tempcntb\m@ne \let\@h@ld\relax \def\@citea{}%
\@cite{%
  \@for \@citeb:=#2\do {%
    \@ifundefined {b@\@citeb}%
      {\@h@ld\@citea\@tempcntb\m@ne{\bf ?}%
      \@warning {Citation `\@citeb ' on page \thepage \space undefined}}%
      {\@tempcnta\@tempcntb \advance\@tempcnta\@ne%
      \@tempcntb\number\csname b@\@citeb \endcsname \relax%
      \ifnum\@tempcnta=\@tempcntb 
        \ifx\@h@ld\relax%
          \edef \@h@ld{\@citea\csname b@\@citeb\endcsname}%
        \else%
          \edef\@h@ld{\ifmmode{-}\else--\fi\csname b@\@citeb\endcsname}%
        \fi%
      \else
        \@h@ld\@citea\csname b@\@citeb \endcsname%
        \let\@h@ld\relax%
      \fi}%
    \def\@citea{,\penalty\@highpenalty\,}%
  }\@h@ld
}{#1}}

\def\@citeb#1#2{{[#1]\if@tempswa , #2\fi}}
\def\@citeu#1#2{{$^{#1}$\if@tempswa , #2\fi }}
\def\@citep#1#2{{#1\if@tempswa , #2\fi}}

\def\bcites{         
        \catcode`\@=11
        \let\@cite=\@citeb
        \catcode`\@=12
}

\def\upcites{         
        \catcode`\@=11
        \let\@cite=\@citeu
        \catcode`\@=12
}

\def\plaincites{      
        \catcode`\@=11
        \let\@cite=\@citep
        \catcode`\@=12
}

\newcount\hour
\newcount\minute
\newtoks\amorpm
\hour=\time\divide\hour by 60
\minute=\time{\multiply\hour by 60 \global\advance\minute by-\hour}
\edef\standardtime{{\ifnum\hour<12 \global\amorpm={am}%
        \else\global\amorpm={pm}\advance\hour by-12 \fi
        \ifnum\hour=0 \hour=12 \fi
        \number\hour:\ifnum\minute<10 0\fi\number\minute\the\amorpm}}
\edef\militarytime{\number\hour:\ifnum\minute<10 0\fi\number\minute}

\def\draftlabel#1{{\@bsphack\if@filesw {\let\thepage\relax
   \xdef\@gtempa{\write\@auxout{\string
      \newlabel{#1}{{\@currentlabel}{\thepage}}}}}\@gtempa
   \if@nobreak \ifvmode\nobreak\fi\fi\fi\@esphack}
        \gdef\@eqnlabel{#1}}
\def\@eqnlabel{}
\def\@vacuum{}
\def\marginnote#1{}
\def\draftmarginnote#1{\marginpar{\raggedright\scriptsize\tt#1}}
\def\draft{
        \pagestyle{plain}
        \overfullrule=2pt
        \oddsidemargin -.5truein
        \def\@oddhead{\sl \phantom{\today\quad\militarytime} \hfil
        \smash{\Large\sl DRAFT} \hfil \today\quad\militarytime}
        \let\@evenhead\@oddhead
        \let\label=\draftlabel
        \let\marginnote=\draftmarginnote
        \def\ps@empty{\let\@mkboth\@gobbletwo
        \def\@oddfoot{\hfil \smash{\Large\sl DRAFT} \hfil}
        \let\@evenfoot\@oddhead}
        \def\@eqnnum{(\theequation)\rlap{\kern\marginparsep\tt\@eqnlabel}%
        \global\let\@eqnlabel\@vacuum}  }

\def\section{\@startsection {section}{1}{\z@}{3.ex plus 1ex minus
 .2ex}{2.ex plus .2ex}{\Large\bf}}
\def\subsection{\@startsection{subsection}{2}{\z@}{2.75ex plus 1ex minus
 .2ex}{1.5ex plus .2ex}{\large\bf}}        
\def\subsubsection{\@startsection{subsubsection}{2}{\z@}{2.75ex plus 1ex minus
 .2ex}{1.5ex plus .2ex}{\bf}}

\def\abstract{\if@twocolumn
\section*{Abstract}
\else 
\begin{center}
{\bf Abstract\vspace{-.5em}\vspace{0pt}}
\end{center}
\quotation
\fi}

\catcode`\@=12

\newcommand{\beq}{\begin{equation}}
\newcommand{\eeq}{\end{equation}}
\newcommand{\beqa}{\begin{eqnarray}}
\newcommand{\eeqa}{\end{eqnarray}}
\newcommand{\dd}{{\rm d}}

\newcommand{\Z}{{\bf Z}}

\newcommand{\R}{{\bf R}}

\newcommand{\C}{{\bf C}}
\newcommand{\CC}{{\bf C}}
\newcommand{\HH}{{\bf H}}
\newcommand{\PP}{{\bf P}}
\newcommand{\e}{\,{\rm e}}

\newcommand{\be}{\begin{equation}}
\newcommand{\ee}{\end{equation}}
\newcommand{\bea}{\begin{eqnarray}}
\newcommand{\eea}{\end{eqnarray}}

\def\to{\rightarrow}

\def\longlongrightarrow{\relbar\joinrel\relbar\joinrel\rightarrow}

\def\lae{\mathrel{\mathop{\smash{\lower .5 ex \hbox{$\stackrel<\sim$}}}}}
\def\lae{\mathrel{\mathop{\smash{\lower .5 ex \hbox{$\stackrel>\sim$}}}}}

\def\Tr{{\rm Tr}}
\def\l:{\mathopen{:}\,}
\def\r:{\,\mathclose{:}}

\catcode`\@=11
\def\theequation{\arabic{equation}}
\catcode`\@=12

\bcites

\catcode`\@=11
\def\theequation{\thesection.\arabic{equation}}
\@addtoreset{equation}{section}
\@addtoreset{footnote}{section}
\@addtoreset{footnote}{subsection}
\catcode`\@=12

\newcommand{\cQ}{{\mathcal Q}}
\newcommand{\cB}{{\mathcal B}}
\newcommand{\cC}{{\mathcal C}}
\newcommand{\bepsilon}{\overline{\epsilon}}

\newcommand{\bi}{{\overline{\imath}}}
\newcommand{\bj}{{\overline{\jmath}}}

\newcommand{\bz}{\overline{z}}
\newcommand{\bw}{\overline{w}}

\newcommand{\bareta}{\overline{\eta}}

\newcommand{\Ker}{{\rm Ker}}
\newcommand{\Coker}{{\rm Coker}}
\newcommand{\nn}{\nonumber}
\newcommand{\s}{\sigma}

\newcommand{\bartial}{\overline{\partial}}
\newcommand{\longto}{\longrightarrow}

\newcommand{\Hom}{{\rm Hom}}

\newcommand{\rank}{{\rm rank}}

\newcommand{\End}{{\rm End}}

\newcommand{\surface}{\Sigma}

\newcommand{\Fr}{{\Scr F}}
\newcommand{\Tc}{{\Scr T}}
\newcommand{\Hi}{{\rm H}}
\newcommand{\urT}{{\rm T}}

\newcommand{\skFr}{{\,\,\widehat{\!\!\Fr}_{\!*}}}
\newcommand{\Inv}{\mathscr{I}}
\newcommand{\Cl}{{\rm C}}
\newcommand{\J}{{\rm J}}
\newcommand{\Frst}{\Scr{F}_*}

\newcommand{\HRC}{\Hi_{\R}\otimes\C}
\newcommand{\rmA}{{\rm A}}
\newcommand{\K}{{\rm K}}
\newcommand{\rK}{\widetilde{\rm K}}
\newcommand{\KO}{{\rm KO}}
\newcommand{\KR}{{\rm KR}}

\newcommand{\GW}{{\rm GW}}
\newcommand{\Witt}{{\rm W}}

\begin{document}

\newcommand{\bX}{\overline{X}}
\newcommand{\bx}{\overline{x}}
\newcommand{\id}{{\rm id}}
\newcommand{\wA}{{\mathcal A}}
\newcommand{\xia}{\boldsymbol{\xi}}

\newcommand{\brane}{{\mathscr W}}
\newcommand{\wilson}{{\mathscr W}}
\newcommand{\cp}{{\mathscr V}}

\newcommand{\lpi}{\mbox{\large $\pi$}}
\newcommand{\Rp}{{\rm Re}}
\newcommand{\Ip}{{\rm Im}}

\newcommand{\wtOmega}{\widetilde{\Omega}}
\newcommand{\wtP}{\widetilde{P}}

\begin{titlepage}
\hfill \today

\begin{center}

\vskip 2.5 cm
{\large \bf On The Structure Of The Chan-Paton Factors
\\[0.2cm] For D-Branes In Type II Orientifolds}
\vskip 1 cm
{Dongfeng Gao $\!{}^{{}^{ u}}\,$\, and ~ Kentaro Hori $\!{}^{{}^{ v}}\,$}\\
\vskip 0.6cm
$\!{}^{{}^{ u}}\,${\sl Wuhan Institute of Physics and Mathematics, 
Chinese Academy of Sciences,\\ Wuhan 430071, China}

$\!{}^{{}^{ v}}\,${\sl IPMU, the University of Tokyo, Kashiwa, Japan}

\end{center}

\vskip 0.5 cm
\begin{abstract}
We determine the structure of the Chan-Paton factors of the open strings 
ending on space filling D-branes in Type II orientifolds.
Through the analysis, we obtain a rule concerning possible distribution 
of O-plane types.
The result is applied to classify the topology of D-branes in terms of 
Fredholm operators and K-theory, deriving a proposal made earlier
and extending it to more general cases.
It is also applied to
compactifications with ${\mathcal N}=1$ supersymmetry
in four-dimensions. We adapt and develop the language of category
in this context, and use it to describe some decay channels.
\end{abstract}

\end{titlepage}

\newpage

\pagestyle{empty}

{\footnotesize

\tableofcontents

}

\setcounter{page}{0}

\newpage

\pagestyle{plain}

\newcommand{\gE}{E}
\newcommand{\gT}{T}
\newcommand{\gA}{A}
\newcommand{\gF}{F}
\newcommand{\gU}{\mathbf{U}}
\newcommand{\indpU}{U}
\newcommand{\gc}{\mathbf{c}}
\newcommand{\indpc}{c}
\newcommand{\ugE}{\check{E}}
\newcommand{\ugT}{\check{T}}
\newcommand{\ugA}{\check{A}}
\newcommand{\ugF}{\check{F}}
\newcommand{\ugU}{\check{U}}
\newcommand{\ugM}{\check{M}}
\newcommand{\Pexp}{{\rm P\, exp}}
\newcommand{\tr}{{\rm tr}}
\newcommand{\sDelta}{\mit\Delta}
\newcommand{\upper}{\mathbb{H}^2}
\newcommand{\sA}{\boldsymbol{\alpha}}
\newcommand{\ev}{{\rm ev}}
\newcommand{\spst}{\uvarepsilon}
\newcommand{\bk}{[k]}
\newcommand{\abs}{{\!{}_{\rm ABS}}}
\newcommand{\bfQ}{{\bf Q}}
\newcommand{\Ptr}{\mathscr{P}}
\newcommand{\phxi}{\umu}
\newcommand{\Po}{{\bf P}}
\newcommand{\inv}{\boldsymbol{\tau}}
\newcommand{\invS}{\inv_{\rm S}}

\section{Introduction}
\label{sec:intro}

The Chan-Paton factors \cite{Chan-Paton,NSch,JSch,MS}
carry the gauge quantum numbers for Yang-Mills type fields on D-branes
and are important ingredients in any theory including open strings.
$N$ coincident BPS D-branes in Type II string theory
have $U(N)$ gauge symmetry on their worldvolume. 
In Type II orientifolds \cite{oriori,Horavaori,BSori,GP}, 
$N$ BPS D-branes on top of an orientifold plane
have either $O(N)$ or $USp(N)$ gauge symmetry
depending on the choice of orientifold action
on the Chan-Paton factor.
The choice is referred to as
{\it the type of O-plane} and is denoted by
O${}^-$ for $O(N)$ and O${}^+$ for $USp(N)$, reflecting the sign of
the tension of the plane.
In practice, one may be interested in D-branes which are not on top of 
the O-plane, and also, there can be several O-planes of different types in 
a given theory.
However, the orientifold projection condition of open string states
is known only in simple examples, and usually only in the
bosonic sector.
More fundamentally, a general condition
on allowed distributions of O-plane types for a given involution are not known.
In this paper, we approach these problems by studying the structure of
the Chan-Paton factor of the open string ending on space filling D-branes
(i.e. D9-branes).

One motivation of this work comes from classification of D-brane charges via
K-theory \cite{Atiyah}. For Type I, Type IIB
{\it resp}.\! Type IIA string theory on 
a spacetime $X$, D-brane charges take values in the group $\KO(X)$,
$\K(X)$ {\it resp}.\! $\K^{-1}(X)$ \cite{MM,WittenK}. 
This can be derived very naturally \cite{WittenK,Horava} from the study 
of topology of D9-brane configurations 
including tachyon condensation \cite{Senrev}.
For Type II orientifolds on a spacetime $X$ with an involution,
similar classification exists in terms of KR-theory \cite{KR}:
If the orientifold planes have O${}^-$-type and codimension $k$ 
(modulo $8$) or/and O${}^+$-type and codimension $(k\pm 4)$, 
D-brane charges take values in $\KR^{-k}(X)$ \cite{H,BGH}.
However, that is just a proposal based on consistency with T-duality.
Direct derivation from D9-brane configurations has been missing.
Furthermore, the proposal does not cover more general situations
such as coexistence of O${}^-$ and O${}^+$-planes of the same dimension
or of O-planes of the same type but with the dimensions differing by $4$.
One goal of the present paper is to fill the gap by finding
the structure of the D9-brane Chan-Paton factor in a general Type II 
orientifold.

Another motivation comes from four-dimensional ${\mathcal N}=1$ supersymmetric
compactifications with D-branes. Orientifold is an indispensable element in 
the tadpole cancellation \cite{PolCai}, which is required for models with 
non-zero gravitational coupling. We would like to have an approach to 
systematically construct and analyze such models.
One possible approach would be to realize D-branes as supersymmetric 
configurations on space filling D-branes in Type IIB string theory. 
At least before orientifolding, it comes with a useful mathematical language,
that of {\it D-brane category},
to describe an important part of the low energy
effective theory on D-branes, such as the low lying spectrum,
the tree level superpotential and the D-flatness condition \cite{Douglas}.
In order to adapt the approach to consistent string compactifications,
solid knowledge on the structure of the Chan-Paton factors for space filling
D-branes in Type IIB orientifold is required. 
Applying the understanding obtained in this paper and building 
on earlier steps taken in \cite{HW,Emanuel},
we develop the theory and put it into the right context.
We summarize the structure in the categorical language,
with the expectation that it can be used in non-geometric regimes
such as orbifolds and Gepner models, as well as in Type IIA models.

The primary tool of our study is the consistency condition
on the parity operator $\Po$,
with which we define the orientifold projection:
It must square to a gauge transformation $g$,
\beq
\Po^2\,=\, g,
\label{bid}
\eeq
on open strings stretched between all possible pairs of D-branes.
For the orientifold by an involution, which we consider in the present paper,
$g$ is the GSO operator $(-1)^F$ on the Neveu-Schwarz sector
and the identity on the Ramond sector.
This condition was employed by Gimon-Polchinski in \cite{GP}
to determine the gauge groups of D$5$ and D$1$-branes in Type I string theory.
Our work applies this method to study the local as well as global
properties of orientifold projection conditions 
in more general Type II orientifolds.
As important cases, we manage to find the structure of the Chan-Paton factors
for D-branes of all dimensions in Type I string theory,
extending earlier results by \cite{GP,WittenK,Oren,AST}.

In the next few paragraphs, we summarize the structure 
of the Chan-Paton factors which we find in this paper.

In an orientifold, we must consider an invariant configuration of D-branes.
Namely,
 a configuration ${\mathcal B}$ of D-branes and its orientifold image 
$\Ptr({\mathcal B})$ must be the ``same'' or, to be more precise, 
isomorphic. 
In fact, the isomorphism itself,
$\Ptr({\mathcal B})\cong {\mathcal B}$, 
carries an important information as that is used to 
define the orientifold projection of open string states.
We shall refer to it as the {\it orientifold isomorphism}, 
or the {\it o-isomorphism} for short, of the D-branes.
We would like to find the possible form of
parity transform, ${\mathcal B}\mapsto \Ptr({\mathcal B})$,
 and the condition on the isomorphisms
$\Ptr({\mathcal B})\cong{\mathcal B}$,
for configurations ${\mathcal B}$ of space-filling D-branes in the
Type II orientifold on $X$ with an involution $\inv:X\to X$.

A D9-anti-D9-brane configuration in Type IIB string theory on $X$
is determined by a choice of superconnection data \cite{Quillen}, 
i.e., a $\Z_2$-graded hermitian vector bundle $E$
on $X$ (the Chan-Paton bundle),
an even unitary connection $A$ of $E$ (the gauge field),
and an odd hermitian section $T$ of $\End(E)$ (the tachyon).
A configuration of non-BPS D9-branes in Type IIA is
determined by $(E,A,T)$ as above with a distinguished section $\xia$ of 
$\End(E)$ which is odd and obeys $\xia^2={\rm id}_E$, $[\xia,A]=0$ and 
$\{\xia,T\}=0$. 
The data $(E,A,T,\xia)$ can be obtained from 
an ungraded data $(\ugE,\ugA,\ugT)$, via $E=\ugE\oplus\ugE$
and
$$
A=\left(\begin{array}{cc}
\ugA&0\\
0&\ugA
\end{array}\right),
\quad
T=\left(\begin{array}{cc}
0&\ugT\\
\ugT&0
\end{array}\right),
\quad
\xia=\left(\begin{array}{cc}
0&-i\\
i&0
\end{array}\right).
$$

A parity exchanges the right and the left ends of the open string
and therefore must involve the transpose of the Chan-Paton factor.
Here, it is natural to use a $\Z_2$-graded version of the transpose, 
$f\mapsto f^T$, with the property $(fg)^T=(-1)^{|f|\cdot |g|}g^Tf^T$.
We shall find that the transform ${\mathcal B}\mapsto\Ptr({\mathcal B})$
is given by
\beqa
\Ptr(E)&=&\inv^*E^*\otimes {\mathcal L},\nn\\
\Ptr(A)&=&-\inv^*A^T+\sA,\\
\Ptr(T)&=&\spst\inv^*T^T.\nn
\eeqa
Here, $\spst$ is a phase, $i$ or $-i$,
that is associated with the parity action on the worldsheet fermion.
$({\mathcal L},\sA)$ is 
a hermitian line bundle with a unitary connection, which we call {\it twist}.
Invariance of the worldsheet action, which includes the B-field term
$\int_{\Sigma}x^*B$, requires the constraint 
$\dd\sA=\inv^*B+B$.

A natural candidate for
the o-isomorphism $\Ptr({\mathcal B})\cong {\mathcal B}$ is a
unitary map $\gU:\Ptr(E)\to E$ 
that transforms $\Ptr(A)$ and $\Ptr(T)$ back to $A$ and $T$,
\beq
\begin{array}{r}
\gU(-\inv^*A^T+\sA)\gU^{-1}+i^{-1}\gU\dd\gU^{-1}\,=\,A,\\[0.2cm]
(-1)^{|\gU|}\gU(\spst \inv^*T^T)\gU^{-1}\,=\,T.
\end{array}
\label{ITAcond}
\eeq
We have two maps, $\gU=\gU_{(i)}$ and $\gU_{(-i)}$, corresponding to the two
phases, $\spst=i$ and $-i$. This fact will be particularly 
important to define the orientifold projection condition in the Ramond sector.
The requirement (\ref{bid}) on
the parity operator $\Po$ defined via the o-isomorphism $\gU$ 
yields a condition of the form
\beq
\gU(\inv^*\gU^T)^{-1}\imath=\gc\cdot \sigma,
\label{IcondforU}
\eeq
for a section $\gc$ of $\inv^*{\mathcal L}\otimes {\mathcal L}^*$
which is parallel with respect to the connection $\inv^*\sA-\sA$.
Here, $\imath$ is the natural isomorphism $E\to E^{**}$
and $\sigma$ is the grading operator on $E$ which assigns
$1$ and $-1$ on even and odd elements.
For Type IIA, we need an additional condition
\beq
(-1)^{|\gU|}\gU\inv^*\xia^T\gU^{-1}=\phxi\xia,
\label{IxiU}
\eeq
where $\phxi$ is a phase $\pm i$ which is independent of $\spst$.

On the $\inv$-fixed point set, $\inv^*{\mathcal L}$ is canonically isomorphic
to ${\mathcal L}$ and $\inv^*\sA-\sA$ vanishes in the tangent direction. 
This means that the parallel section 
$\gc$ of $\inv^*{\mathcal L}\otimes {\mathcal L}^*$ can be regarded as 
a complex number at each O-plane. 
This number is related to the dimension and the type of the O-plane. 
Let $k$ be the codimension of the O-plane, which is even ({\it resp}. odd) 
for Type IIB ({\it resp}. IIA) orientifold, and put 
\beq
\bk:=\left\{\begin{array}{ll}
k ~~(\mbox{mod 8})&\mbox{for O$^-$}\\
k+4 ~~(\mbox{mod 8})&\mbox{for O$^+$}.
\end{array}\right.
\label{bkdef}
\eeq
Then,
\beq
\gc=\left\{\begin{array}{ll}
\spst^{\bk\over 2}&\mbox{$k$ even (IIB),}\\
\spst^{\bk-\spst\cdot\phxi\over 2}
&\mbox{$k$ odd (IIA).}
\end{array}\right.
\label{BaSt}
\eeq
When the twist $({\mathcal L},\sA)$ is trivial, $\gc$ is a constant
number over the entire spacetime and
all the O-planes must have the same $\bk$.
For example, O$p^-$ can coexist with O$(p\pm 4)^+$ but not
with O$p^+$ nor O$(p\pm 4)^-$.
For a non-trivial twist, the value of $\gc$ may differ from one O-plane
to another and such ``forbidden mixture'' becomes possible.

The local behaviour of the condition on the tachyon
near an O-plane can be written more explicitly.
We choose a trivialization of the Chan-Paton vector bundle
as well as the twist line bundle ${\mathcal L}$
in a neighborhood of a point of an O-plane.
For Type IIB, the condition reads
\beq
\begin{array}{ll}
\gT=\indpU\inv^*\gT^t\indpU^{-1},\,\,~\,\,\indpU\,\,{\rm even},\!\!&\left\{
\begin{array}{ll}
\indpU=\inv^*\indpU^t&\mbox{$\bk=0$~ (O9${}^-$/O5${}^+$/O1${}^-$)}\\[0.2cm]
\indpU=-\inv^*\indpU^t&\mbox{$\bk=4$~ (O9${}^+$/O5${}^-$/O1${}^+$)}
\end{array}\right.\\[0.6cm]
\gT=-\indpU\inv^*\gT^t\indpU^{-1},\,\,\,\indpU\,\,{\rm odd},\!\!\!\!&\left\{
\begin{array}{ll}
\indpU=\inv^*\indpU^t&\mbox{$\bk=2$~ (O7${}^-$/O3${}^+$)}\\[0.2cm]
\indpU=-\inv^*\indpU^t&\mbox{$\bk=6$~ (O7${}^+$/O3${}^-$)}
\end{array}\right.
\end{array}
\label{iiBcond}
\eeq
For Type IIA, we write the condition on the ungraded data $\ugT$:
\beq
\begin{array}{ll}
\ugT=\ugU\inv^*\ugT^t\ugU^{-1},\!&\left\{\begin{array}{ll}
\ugU=\inv^*\ugU^t&\mbox{$\bk=7$~ (O6${}^+$/O2${}^-$)}\\[0.2cm]
\ugU=-\inv^*\ugU^t&\mbox{$\bk=3$~ (O6${}^-$/O2${}^+$)}
\end{array}\right.\\[0.6cm]
\ugT=-\ugU\inv^*\ugT^t\ugU^{-1},\!&\left\{\begin{array}{ll}
\ugU=\inv^*\ugU^t&\mbox{$\bk=1$~ (O8${}^-$/O4${}^+$/O0${}^-$)}\\[0.2cm]
\ugU=-\inv^*\ugU^t&\mbox{$\bk=5$~ (O8${}^+$/O4${}^-$/O0${}^+$)}
\end{array}\right.
\end{array}
\label{iiAcond}
\eeq
$\indpU$ and $\ugU$ are determined from $\gU$, and
we use here the ordinary matrix transpose $f^t$ rather than 
the graded transpose $f^T$.

The structure (\ref{iiBcond})-(\ref{iiAcond}) can be described in terms of 
Fredholm operators of 
a Hilbert space with Clifford algebra action \cite{ABS,ASFred},
which are relevant in a formulation of K-theory $\KR^{-\bk}$.
In particular,  (\ref{iiBcond})-(\ref{iiAcond})
would follow from the proposal in \cite{H,BGH}
concerning the classification of D-brane charges in terms of the
KR-theory.
This is in fact the first way we obtained this structure
(summer, 1999).
In this paper, we directly derive this structure using a worldsheet
analysis, from which the proposal in \cite{H,BGH} follows.
In addition, we will also be able to find K-theory classification 
of D-brane charges 
in the case with a non-trivial twist
where the local behaviour changes from one O-plane to another.
We would also like to point out that the pattern
(\ref{iiBcond})-(\ref{iiAcond}) 
appears in the classification of random matrix ensembles 
or many body systems.
These eight cases plus two from Type IIB and Type IIA string theories
match the ``ten-fold way'' classification
based on symmetry properties \cite{Dyson,AltlandZirnbauer}.
The tachyon $T$ {\it resp}.\! $\ugT$ corresponds to the random matrix
(or the Hamiltonian) in a system with {\it resp}.\! without chirality,
and $U$ {\it resp}.\! $\ugU$
corresponds to the unitary matrix that enters into
either the time reversal or the charge conjugation symmetry.
In addition, we note that the structure for the case $[k]=2$
versus $6$ played an important r\^ole in \cite{CDE}.

The rest of the paper is organized as follows.

We make preliminary remarks in Section~\ref{sec:prelim}.
We fix our convention on the spin structure and parity action at the
boundary of the upper-half plane, introducing the phase $\spst=\mp i$.
We also describe the Chan-Paton structure of
space filling D-branes in Type II string theory
and duality in the category of graded vector spaces.

In Section~\ref{sec:main}, we explain all the structures 
summarized above except the formula (\ref{BaSt}).
We also discuss the orientifold projection in
the Ramond sector, which is defined using both of the two
o-isomorphisms, $\gU_{(i)}$ and $\gU_{(-i)}$.

In Section~\ref{sec:rbf}, we study orientifold action on systems with
boundary fermions.
This section provides the background for our treatment of
non-BPS D-branes. 
We also study the D9-brane 
configuration that represents D-branes on top of the O-plane,
and find an evidence of the formula (\ref{BaSt}).

Section~\ref{sec:direct} is the main section in which we determine the
structure of space filling D-branes in Type II orientifolds on the flat 
Minkowski space with a single O-plane.
We derive the formula (\ref{BaSt}) using the consistency condition 
(\ref{bid}). 

In Section~\ref{sec:TypeI}, we determine the tachyonic and massless
spectrum on D-branes in Type I string theory.

In Section~\ref{sec:example}, we illustrate 
how non-trivial twists give rise to
``forbidden mixture'' of O-plane types 
in explicit examples of toroidal and Calabi-Yau compactifications.
We classify the orientifold data $(\inv,B,{\mathcal L},\sA,\gc)$
and find agreement in well-studied examples as well as some new results.

In Section~\ref{sec:K}, we classify the topology of D9-brane configurations
in terms of Fredholm operators on a Hilbert space and/or K-theory. 
We see how it is organized in terms of the Clifford algebras.
This leads to the K-theory classification proposed in \cite{H,BGH}
for the cases with trivial twist.
We also introduce new K-theory in order to describe the cases with non-trivial
twists.

In Section~\ref{sec:SUSY}, we consider Type IIB orientifold on Calabi-Yau
manifolds with holomorphic involutions, with a focus on
${\mathcal N}=1$ spacetime supersymmetry.
We shall study the condition for the orientifold projection to be compatible
with ${\mathcal N}=1$ supersymmetry and find that
we can focus on D-branes with quasi-o-isomorphisms of a certain degree.
We also develop and/or adapt the language of category
and use it to describe some decay channels.

Throughout this paper, we set $\alpha'=1$.

\medskip

\section{Preliminaries}
\label{sec:prelim}

We make remarks on three independent subjects,
(i) Worldsheet spin structures and parity action on spinors,
(ii) Structure of the Chan-Paton factors on space filling D-branes in
Type II string theory, and
(iii) Transpose of linear maps between $\Z_2$-graded vector spaces.
The main purpose is to fix the convention and notation.

\subsection{Worldsheet Spin Structures}
\label{subsec:WSspin}

A Type II string theory is obtained by 
a chiral GSO projection ---
gauging the independent sign flips 
of left-handed and right-handed spinors on the worldsheet, 
$(-1)^{F_L}$ and $(-1)^{F_R}$.
This involves a sum over different spin structures.
In a flat cylinder region, the parallel transport
along the non-trivial circle is either the sign flip
(Neveu-Schwarz (NS) sector) or the identity (Ramond (R) sector)
for each of the two chiralities.
A closed string thus has four sectors in total, 
NS-NS, R-R, R-NS and NS-R.
If the worldsheet has a boundary, in order to specify a boundary condition,
we must choose an identification between left-handed and 
right-handed spinors at the boundary.
The choice is two-fold, related by a sign.
This is the boundary analog of the spin structure.
In a flat strip region, the identification at one boundary
is sent by the parallel transport to the one at the other boundary (R-sector) 
or to the one opposite to it (NS-sector).
Closed string states in the R-NS and NS-R sectors 
and open string states in the R-sector 
correspond to spacetime fermions,
while states in the other sectors correspond to spacetime bosons.

To each open string state corresponds a boundary vertex operator.
The boundary spin structure is continuous at the insertion point of
an NS-vertex operator, while it flips by a sign at a R-vertex operator.
\begin{figure}[htb]
\centerline{\includegraphics{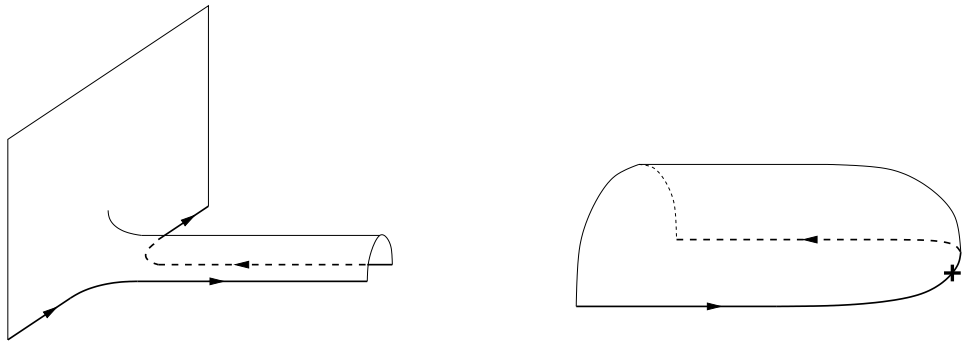}}
\caption{State-operator correspondence}
\label{fig:soc}
\end{figure}
This rule appears opposite to the one on the strip, but
that is because the worldsheets defining
the correspondence (Figure~\ref{fig:soc})
have the curvature ${1\over 4\pi}\int R\sqrt{g}\dd^2 \s={\mp}{1\over 2}$.
Likewise, to each closed string state corresponds a bulk vertex operator.
The holonomy around the insertion point is opposite to
the one along the closed string,
as the worldsheets defining the correspondence
(the doubles of the surfaces in Figure~\ref{fig:soc}) have the curvature
${1\over 4\pi}\int R\sqrt{g}\dd^2 \s={\mp}1$.
For example, the holonomy is trivial around the insertion point of
an NS-NS vertex operator.


We now introduce conventions 
concerning boundary spin structures
and parity transforms.
We consider the strip, with the space coordinate $\s^1=\s$
spanning the interval $-\pi\leq \s \leq 0$ and the time coordinate
$\s^0=t$.
Let us first look at the boundary on the right, $\s=0$.
The superpartner of the boundary value of the coordinate field, 
$x(t,0)$, is either of the following two
\beq
(\pm):~~\psi(t)=\psi_+(t,0)\pm\psi_-(t,0).
\label{cbcR}
\eeq
These two possibilities can be regarded as coming from the two
different choices of boundary spin structure.
More generally, the choice can be characterized in terms of 
the boundary condition for the ${\mathcal N}=(1,1)$ supercurrents,
$(\pm):G^1_+\pm G^1_-=0$.
Here
$G_{\pm}^1$ is the normal component of the supercurrent,
which reads as $G_{\pm}^1=\mp\psi_{\pm}\cdot(\partial_0\pm\partial_1)x$
for the sigma model.
Let us next look at the boundary on the left, $\s=-\pi$.
Again there are two possibilities for the superpartner of the
boundary value of $x(t,-\pi)$,
\beq
(\pm):~~\psi(t)=\psi_+(t,-\pi)\mp 
\psi_-(t,-\pi),
\label{cbcL}
\eeq
or more generally, two possibilities for the boundary condition on
the supercurrent $(\pm):G^1_+\mp G^1_-=0$.
In the NS-sector, the spin structures on the two boundaries are
$(++)$ or $(--)$, i.e., both $(+)$ or both $(-)$.
In the R-sector, they are $(-+)$ or $(+-)$.

Let us go from the Minkowski to the Euclidean strip
by the Wick rotation, $t\to -it_E$, 
and then to the upper-half plane,
$\Ip(z)\geq 0$, by
$$
z=\tau^1+i\tau^2=\e^{t_E-i\s}.
$$
The right and the left boundaries are mapped to real positive $z$
and real negative $z$.
The fermionic fields may be expressed as
$
\psi_-^{\rm strip}\left({\dd z\over z}\right)^{1\over 2}
=\psi_-^{\rm plane}(\dd z)^{1\over 2}
$
and
$
\psi_+^{\rm strip}\left({\dd \bz\over \bz}\right)^{1\over 2}
=\psi_+^{\rm plane}(\dd \bz)^{1\over 2},
$
from which we find the relation between the field components
\beq
\psi_{\pm}^{\rm strip}(t_E,\s)
=\e^{{1\over 2}(t_E\pm i\s)}
\psi_{\pm}^{\rm plane}(\tau^1,\tau^2).
\label{relcomp}
\eeq
In particular, 
$\psi_{\pm}^{\rm strip}=\e^{t_E/2}\psi_{\pm}^{\rm plane}$
at the right boundary and 
$\psi_{\pm}^{\rm strip}=\mp i\e^{t_E/2}\psi_{\pm}^{\rm plane}$
at the left boundary.
The superpartner of the coordinate field $x$ is
\beq
(\pm):~~\psi^{\rm plane}(\tau)
=\psi^{\rm plane}_+(\tau,0)\pm\psi^{\rm plane}_-(\tau,0)
\label{defpsi}
\eeq
at the boundary of the upper-half plane.

An orientifold is obtained by gauging 
a transformation of fields that involves a parity of the worldsheet.
Note that a parity swaps the chirality of spinors.
On the Minkowski strip, a parity acts on 
the fermions as
\beq
\Omega\,:\,\psi_{\pm}(t,\s)\,\longrightarrow\,
\mp\psi_{\mp}(t,-\pi-\s),
\label{defOmega}
\eeq
or as $(-1)^{F}\Omega$, $(-1)^{F_R}\Omega$ and $(-1)^{F_L}\Omega$
which have sign factor $\pm$, $+$ and $-$ instead of 
$\mp$ on the right hand side.
They obey
\beqa
&&\Omega^2=((-1)^F\Omega)^2=(-1)^F,
\label{omsq}\\
&&((-1)^{F_R}\Omega)^2=((-1)^{F_L}\Omega)^2={\rm id}.
\label{omtilsq}
\eeqa
$\Omega$ and $(-1)^F\Omega$ map the $(\pm)$ spin structure on one boundary 
to the $(\pm)$ on the other, and thus lift to transformations of NS sector.
On the other hand,
$(-1)^{F_R}\Omega$ and $(-1)^{F_L}\Omega$ map the $(\pm)$ on one to the
$(\mp)$ on the other, and lift to transformations of Ramond sector.

The action on the field components 
on the upper-half plane can be found from the relation (\ref{relcomp}).
$\Omega$ does
$\psi_{\pm}^{{\rm plane}}(\tau^1,\tau^2)
\to -i\psi_{\mp}^{{\rm plane}}(-\tau^1,\tau^2)$,
and the other three parities have phases
$+i$, $\pm i$ and $\mp i$ instead of $-i$.
The spin structure $(\pm)$ is invariant under
$\Omega$ and $(-1)^F\Omega$ while it is flipped under
$(-1)^{F_R}\Omega$ and $(-1)^{F_L}\Omega$.
The field (\ref{defpsi}) transforms as
\beq
\psi^{\rm plane}(\tau)\,\,\,\longrightarrow \,\,\,
\spst\psi^{\rm plane}(-\tau),\qquad
\label{Ompla}
\eeq
where
\vspace{-0.5cm}
\beq
\qquad\qquad\qquad
\spst\,=\,\mp i\quad\,
\mbox{for}\quad\,\left\{\begin{array}{rl}
\Omega:&\!\!\!(\pm)\to (\pm)\\
(-1)^F\Omega:&\!\!\!(\mp)\to (\mp)\\
(-1)^{F_R}\Omega:&\!\!\!(\pm)\to (\mp)\\
(-1)^{F_L}\Omega:&\!\!\!(\mp)\to (\pm).
\end{array}\right.
\label{defspst}
\eeq
This transformation rule may also be understood as follows.
Let $z$ be the complex coordinate as above.
The identification of the spinor bundles of the opposite chirality
is given either by $(+):\sqrt{\dd z}=\sqrt{\dd \bz}$
or $(-):\sqrt{\dd z}=-\sqrt{\dd \bz}$.
Under the parity $z\mapsto -\bz$, 
$(\dd z,\dd\bz)$ is mapped to $(-\dd\bz,-\dd z)$
and hence $(\sqrt{\dd z},\sqrt{\dd\bz})$ is mapped to
$(-i\sqrt{\dd \bz},-i\sqrt{\dd z})$
or to the other three combinations of the phases, corresponding to
$\Omega$ and the other three parities.
$\Omega$ preserves the identification 
$(\pm):\sqrt{\dd z}=\pm\sqrt{\dd \bz}$,
and maps $\sqrt{\dd \bz}$ to $\mp i\sqrt{\dd \bz}$.
This together with the consideration on the other parities
reproduces (\ref{Ompla}) with (\ref{defspst}).
This argument is useful in finding the parity transform of other fields.
For example, the spin $({3\over 2},-{1\over 2})$ super-ghost
system transforms as
\beq
\beta^{\rm plane}(\tau)\rightarrow \spst^3\beta^{\rm plane}(-\tau),
\quad\gamma^{\rm plane}(\tau)\rightarrow \spst^{-1}
\gamma^{\rm plane}(-\tau).
\label{betagammaP}
\eeq

For later use, let us record the $\Omega$ parity action on the Fourier
modes of the fermions for the open string stretched 
between D9-branes. The mode expansions are
\beq
\psi_{\pm}(t,\s)=\left\{\begin{array}{ll}
\sum_{r\in \Z+{1\over 2}}\psi_r(t)\e^{\mp i r\s}&(++)\\
\pm \sum_{r\in \Z+{1\over 2}}\psi_r(t)\e^{\mp i r\s}
&(--)\\
\sum_{n\in \Z}\psi_n(t)\e^{\mp i n\s}&(-+)\\
\pm \sum_{n\in \Z}\psi_n(t)\e^{\mp i n\s}&(+-)\\
\end{array}\right.
\label{99}
\eeq
and the parity action is
\beq
\Omega:
\left\{
\begin{array}{l}
\psi_r~~\mbox{in $(++)$}\longrightarrow
\e^{i\pi r}\psi_r~~\mbox{in $(++)$}\\
\psi_r~~\mbox{in $(--)$}\longrightarrow
-\e^{i\pi r}\psi_r~~\mbox{in $(--)$}\\
\psi_n~~\mbox{in $(-+)$}\longrightarrow
(-1)^n\psi_n~~\mbox{in $(+-)$}\\
\psi_n~~\mbox{in $(+-)$}\longrightarrow
-(-1)^n\psi_n~~\mbox{in $(-+)$}\\
\end{array}\right.
\label{Omps}
\eeq

We stress that the notations we have introduced, 
$(\pm)$, $\Omega$ and $\spst$, are simply to fix the convention 
in which we discuss parity transform on an open string or on a neighborhood
of boundary vertex operators.
They are by no means canonical, 
as the meaning can be easily changed, say, with a redefinition of the frame.

\subsection{D9-Branes In Type II String Theory}
\label{subsec:D9}

We describe the structure of the Chan-Paton factors on space filling D-branes 
in Type II string theory and write down the 
corresponding worldsheet boundary interaction.

\subsubsection*{\it Type IIB}

A D9-anti-D9-brane system in Type IIB string theory supports 
a $\Z_2$-graded vector bundle on $X$,
\beq
\gE=E^0\oplus E^1,
\label{IIBCP}
\eeq
with a hermitian inner product. $E^0$ and $E^1$ are the Chan-Paton bundles
of branes and antibranes respectively.
They are distinguished by the $\Z_2$-grading operator 
$$
\sigma=\left(\begin{array}{cc}
{\rm id}_{E^0}&0\\
0&-{\rm id}_{E^1}
\end{array}\right).
$$
The tachyon is an odd endomorphism of $\gE$, that is, 
a linear map $\gT:\gE\to \gE$ that exchanges $E^0$ and $E^1$.
It is assumed to be hermitian, $\gT^{\dag}=\gT$.
The gauge field $\gA$ is an even unitary connection of $\gE$,
$A=A^{\dag}$.
They can be written as
\beq
\gT=\left(\begin{array}{cc}
0&T_{01}\\
T_{10}&0
\end{array}\right),\qquad
\gA=\left(\begin{array}{cc}
A^0&0\\
0&A^1
\end{array}\right).
\eeq
The boundary interaction
corresponding to the configuration $(\gT,\gA)$ is
given by the path-ordered exponential,
\beq
\Pexp\left(-i\int_{t_i}^{t_f}{\mathcal A}_t\,\dd t\right)
\label{Wilson}
\eeq
with \cite{Hlin,KL,TTU}
\beq
{\mathcal A}_t=\dot{x}^{\mu}\gA_{\mu}
-{i\over 4}\psi^{\mu}\psi^{\nu}\gF_{\mu\nu}
+{i\over 2}\psi^{\mu}D_{\mu}\gT+{1\over 2}\gT^2.
\label{At}
\eeq
It depends on the the boundary value $x^{\mu}(t)$
of the sigma model field via $\gT=\gT(x(t))$ and
$\gA_{\mu}=\gA_{\mu}(x(t))$
as well as the boundary value $\psi^{\mu}(t)$
of a linear combination of the fermionic field
given by (\ref{cbcR}), (\ref{cbcL}) or (\ref{defpsi}).
$\gF_{\mu\nu}=\partial_{\mu}\gA_{\nu}-\partial_{\nu}\gA_{\mu}
+i[\gA_{\mu},\gA_{\nu}]$ 
is the field strength and $D_{\mu}\gT$ is the
covariant derivative $\partial_{\mu}\gT+i[\gA_{\mu},\gT]$.
The boundary interaction has ${\mathcal N}=1$ supersymmetry
\beq
\delta x^{\mu}=i\epsilon_1\psi^{\mu},
\quad
\delta \psi^{\mu}=-2\epsilon_1 \dot{x}^{\mu},
\label{N=1var}
\eeq
since it varies as
\beq
\delta{\mathcal A}_t~=~
-i{\mathcal D}_t\left(\epsilon_1 (T-\psi\cdot A)\right)+i\dot{\epsilon}_1T.
\label{deltaA1}
\eeq
For a closed boundary component of the worldsheet, $S^1\subset\partial\Sigma$,
with anti-periodic or periodic spin structure,
the interaction enters into the path-integral weight as
the trace or the supertrace factor,
\beq
\tr\left[\Pexp\left(-i\int_{S^1}{\mathcal A}_t\dd t\right)\right]
\qquad\mbox{or}\qquad
\tr\left[ \sigma \,\Pexp\left(-i\int_{S^1}{\mathcal A}_t\dd t\right)\right].
\eeq

\subsubsection*{\it Type IIA}

We refer the reader to Section~\ref{subsec:nonBPS} for the background of
what is said below.

To describe a system of non-BPS D9-branes in Type IIA string theory on $X$
we need to choose a hermitian vector bundle $\ugE$ on $X$ without 
$\Z_2$-grading.
This $\ugE$ is not exactly the Chan-Paton bundle but
``$1/\sqrt{2}$ of it''.
The corresponding worldsheet boundary has an extra degrees of
freedom and it is appropriate to
introduce the $\Z_2$-graded double of $\ugE$;
\beq
\gE=\ugE\oplus \ugE.
\label{IIACP}
\eeq
The boundary interaction is given again by (\ref{At})
with the condition that it commutes with the odd operator
\beq
\xia:=i\left(\begin{array}{cc}
0&-{\rm id}_{\ugE}\\
{\rm id}_{\ugE}&0
\end{array}\right).
\label{xidef}
\eeq
This means that the gauge field $\gA$ commutes
with $\xia$ and the tachyon $\gT$
{\it anticommutes} with $\xia$ since
$\xia$ is odd and hence anticommutes with $\psi^{\mu}$.
Namely, the tachyon and the gauge field can be written as
\beq
\gT=\left(\begin{array}{cc}
0&\ugT\\
\ugT&0
\end{array}\right),\qquad
\gA=\left(\begin{array}{cc}
\ugA&0\\
0&\ugA
\end{array}\right),
\label{IIA}
\eeq
for a unitary connection $\ugA$ of $\ugE$ and
a hermitian section $\ugT$ of $\End(\ugE)$. 
The system has ${\mathcal N}=1$ worldsheet supersymmetry (\ref{N=1var}).
For a boundary circle $S^1\subset \partial\Sigma$ 
with anti-periodic or periodic spin structure,
we have
\beq
{1\over \sqrt{2}}\,
\tr\left[\Pexp\left(-i\int_{S^1}{\mathcal A}_t\dd t\right)\right]
\qquad\mbox{or}\qquad
{\#\over \sqrt{2}}\,
\tr\left[ i\sigma
\xia \,\Pexp\left(-i\int_{S^1}{\mathcal A}_t\dd t\right)\right]
\label{IIAD9}
\eeq
where $\#$ is a certain phase.

Replacing a brane by its antibrane is done by $\sigma\to -\sigma$
with $\xia$ being fixed. Couplings to all the RR sector states change
by a sign since the second expression in (\ref{IIAD9}) does so.
By conjugation with $\xia$, this operation is equivalent to
keeping $\sigma$ and $\xia$ but doing $T\to -T$. That is,
$$
(\ugE,\ugA,\ugT)\,\,\longmapsto\,\,(\ugE,\ugA,-\ugT).
$$

\subsection{Linear Algebra --- Graded Duality}

We comment on duality operations in the category
of $\Z_2$-graded vector spaces,
in which a minus sign shows up when two odd elements 
exchange their positions. As usual, $|v|=0$ or $1$ (modulo 2) if
$v$ is even or odd with respect to a given grading.

First of all, the dual $V^*$ of a $\Z_2$-graded vector space $V$
has a natural $\Z_2$-grading --- even elements of $V^*$ are orthogonal to
odd elements of $V$, and vice versa.

For a linear map between graded vector spaces, $f:V\to W$, 
we define its {\it graded transpose} (or simply {\it transpose}), 
$f^T:W^*\to V^*$, by
\beq
\langle f^T(w^*),v\rangle=(-1)^{|f|\cdot |w^*|}\langle w^*,f(v)\rangle,
\label{grT}
\eeq
for $v\in V$ and $w^*\in W^*$. If $f$ is expressed as
\beq
f=\left(\begin{array}{cc}a&b\\
c&d
\end{array}\right)
\eeq
with respect to basis of $V$ and $W$ such that even elements precede 
odd elements, then, with respect to the dual basis,
the graded transpose is expressed as
\beq
f^T
=\left(\begin{array}{cc}
a^t&-c^t\\
b^t&d^t
\end{array}\right),
\label{matT}
\eeq
where $a^t, b^t,...$ are transpose matrices for $a,b,...$.
The graded transpose has the desired property under the composition of maps,
say $g:U\to V$ and $f:V\to W$,
\beq
(f\circ g)^T=(-1)^{|f|\cdot |g|}g^T\circ f^T.
\label{prgrT}
\eeq 
In particular, we have $(f^T)^{-1}=(-1)^{|f|}(f^{-1})^T$.
For an even map $f$, we may simply write $f^{-T}$ for the inverse transpose.

The definition (\ref{grT}) is by no means unique.
For example, we may take $\langle f^T(w^*),v\rangle
=(-1)^{|f|\cdot |v|}\langle w^*,f(v)\rangle$, which also satisfies the property 
(\ref{prgrT}). However, we must make some choice, and (\ref{grT})
is the choice we make throughout this paper.

There is a natural isomorphism $\iota$ from a vector space $V$ 
to its double dual $V^{**}$, 
\beq
\langle \iota(v),v^{*}\rangle=(-1)^{|v|\cdot |v^*|}\langle v^*,v\rangle,
\label{defi}
\eeq
for $v\in V$ and $v^*\in V^*$.
It is even with respect to the natural gradings.
The natural isomorphism for the dual, $\iota_{V^*}:V^*\to V^{***}$,
is equal to the inverse transpose of $\iota=\iota_V:V\to V^{**}$,
\beq
\iota_{V^*}=\iota_V^{-T}.
\label{psidual}
\eeq
For $f:V\to W$, its double transpose $f^{TT}:V^{**}\to W^{**}$ 
is conjugate to $f$,
\beq
\iota_{W}^{-1}f^{TT}\iota_V=f.
\label{TT}
\eeq

If the $\Z_2$-gradings are changed, the graded transpose of course changes.
Let us show the gradings explicitly as the subscript as
$f^T_{\sigma_V,\sigma_W}$ and $|f|_{\sigma_V,\sigma_W}$
for $f:V\to W$.
By definition, we have
$\langle f^T_{\sigma_V,\sigma_W}(w^*),v\rangle
=(-1)^{|f|_{\sigma_V,\sigma_W}\cdot |w^*|_{\sigma_{W^*}}}\langle w^*,f(v)\rangle$.
The change in the transpose is
\beqa
&f^T_{-\sigma_V,\sigma_W}=f^T_{\sigma_V,\sigma_W}\sigma_{W^*},
\nn\\
&
f^T_{\sigma_V,-\sigma_W}=-\sigma_{V^*}f^T_{\sigma_V,\sigma_W},
\label{grshift}
\\
&
f^T_{-\sigma_V,-\sigma_W}=\sigma_{V^*}f^T_{\sigma_V,\sigma_W}\sigma_{W^*}
=(-1)^{|f|_{\sigma_V,\sigma_W}}f^T_{\sigma_V,\sigma_W},
\nn
\eeqa
where $\sigma_{V^*}$ and $\sigma_{W^*}$ are the natural gradings of
$V^*$ and $W^*$ induced by $\sigma_V$ and $\sigma_W$.

\medskip

\section{D-Branes In Type II Orientifolds}
\label{sec:main}

In this section, we discuss how we want to define D-branes
in Type II orientifolds in terms of D9-brane configurations.

\subsection{Parity Actions On Boundary Interactions}
\label{subsec:paonbdry}

We now discuss the parity action on the boundary interaction
determined by the configuration $(\gT,\gA)$.
Let us first look at the parity action on the on-shell vertex operators 
for tachyons and massless vector bosons.
The corresponding states are
$$
k\cdot\psi_{-{1\over 2}}|k;i,j\rangle^{}_{{}_{(\pm\pm)}},\qquad
\left(\zeta\cdot\alpha_{-1}-\sqrt{2}\zeta\cdot\psi_{-{1\over 2}}
k\cdot\psi_{-{1\over 2}}\right)|k;i,j\rangle^{}_{{}_{(\pm\pm)}},
$$
where the subscript $(\pm\pm)$ shows the spin structure at the two 
boundaries.
The parity $\Omega$ transforms these states to
$$
\mp i k\cdot\psi_{-{1\over 2}}|k;j,i\rangle^{}_{{}_{(\pm\pm)}},
\qquad
-\left(\zeta\cdot\alpha_{-1}-\sqrt{2}\zeta\cdot\psi_{-{1\over 2}}
k\cdot\psi_{-{1\over 2}}\right)|k;j,i\rangle^{}_{{}_{(\pm\pm)}},
$$
possibly up to a linear transformation of Chan-Paton vectors.
The factor $\mp i$ in the transformation of tachyon
comes from the parity action on $\psi_{-{1\over 2}}^{\mu}$,
see (\ref{Omps}).
The Chan-Paton indices $i$ and $j$ are swapped because
the orientation of the string is reversed.
This would correspond to the transpose of the Chan-Paton factor.
However, since we are considering $\Z_2$-graded Chan-Paton vector spaces,
it would be more natural to take the graded transpose.
Thus, we find that the parity $\Omega$ transforms
the on-shell fluctuation of $\gT$ and $\gA$ as
\beq
\delta\gT\longrightarrow \mp i\delta\gT^T,\qquad
\delta\gA\longrightarrow -\delta\gA^T.
\label{onshinf}
\eeq
Note that this is compatible with the hermiticity of $\gT$
--- if $\delta\gT$ is odd and hermitian
then $\delta\gT^T$ is anti-hermitian,
as can be seen by the matrix representation
for the graded transpose, (\ref{matT}). Thus $\mp i\delta\gT^T$ is 
also hermitian.

Let us now look at the parity action on the boundary interaction
for {\it finite} $\gT$ and $\gA$.
For concreteness, we consider the upper-half plane, ${\rm Im}(z)\geq 0$,
on which the parity acts as $z\mapsto -\bz$.
The boundary interaction takes the form
\beq
W(\tau_f,\tau_i)
=\Pexp\left(-i\int_{\tau_i}^{\tau_f}
{\mathcal A}_{\tau}\,\dd \tau\right)
\label{WilsonE}
\eeq
with
\beq
i{\mathcal A}_{\tau}=i{\dd x^{\mu}\over \dd \tau}\gA_{\mu}
-{i\over 4}\psi^{\mu}\psi^{\nu}\gF_{\mu\nu}
+{i\over 2}\psi^{\mu}D_{\mu}\gT+{1\over 2}\gT^2,
\label{Atau}
\eeq
which is the Wick rotated version of (\ref{At}).
Note that $\psi^{\mu}$ is the $\mu$-th component of the fermion defined
in (\ref{defpsi}).
$W(\tau_f,\tau_i)$ is a linear map from $\gE$ at $x(\tau_i)$
to $\gE$ at $x(\tau_f)$. In the computation of correlation functions,
it is to be multiplied to the Chan-Paton factor
at $\tau=\tau_i$ and to be followed by the Chan-Paton factor 
at $\tau=\tau_f$. 
The parity reverses the orientation
of the boundary and acts on the Chan-Paton factors by graded transpose.
Thus, it acts on the boundary interaction as 
$W(\tau_f,\tau_i)\longrightarrow W(\tau_f,\tau_i)^T$
in addition to the action on the fields $x(\tau)$ and $\psi(\tau)$.
The parity transforms the relevant fields as follows 
(see (\ref{Ompla}) with (\ref{defspst})):
$$
x^{\mu}(\tau)\to x^{\mu}(-\tau),\qquad
\psi^{\mu}(\tau)\to \spst \psi^{\mu}(-\tau).
$$
The outcome is
$$
\Pexp\left(-i\int_{\tau_i}^{\tau_f}
{\mathcal A}_{\tau}\,\dd \tau\right)
~\longrightarrow~
\Pexp\left(-i\int_{-\tau_f}^{-\tau_i}
\widetilde{\mathcal A}_{\tau}\dd\tau\right)
$$
with
\beq
i\widetilde{\mathcal A}_{\tau}
=-i{\dd x^{\mu}\over \dd \tau}\gA^T_{\mu}
-{i\over 4}\spst^2\psi^{\mu}\psi^{\nu}\gF^T_{\mu\nu}
+{i\over 2}\spst\psi^{\mu}D_{\mu}\gT^T+{1\over 2}\left(\gT^2\right)^T.
\eeq
Note that we have a sign in the relation
$\left(\gT\circ \gT\right)^T=-\gT^T\circ \gT^T$, since the tachyon 
$\gT$ is odd.
Namely, $(\gT^2)^T=(\spst\gT^T)^2$.
We see that the effect of the parity action is
\beq
\gT\longrightarrow \spst\gT^T,\qquad
\gA\longrightarrow -\gA^T.
\label{offshfin}
\eeq
This is nothing but the off-shell and finite version of 
(\ref{onshinf}). Note that the Chan-Paton bundle $\gE$ has transformed
to its dual, $\gE^*$.

If the parity is combined with an involution $\inv:X\to X$
of the spacetime, then the transformation rule (\ref{offshfin}) 
is dressed by the pull back,
$\gT\to \spst\inv^*\gT^T$, $\gA\to-\inv^*\gA^T$, and
$\gE\mapsto \inv^*\gE^*$.
Furthermore, we may also combine it with a shift of the gauge field,
which is in fact enforced when there is a nonzero $B$-field.
The $B$-field enters into the (Euclidean) action as
\beq
S_B=-i\int_{\Sigma}x^*B+{i\over 4}\int_{\partial \Sigma}
B_{\mu\nu}(x)\psi^{\mu}\psi^{\nu}\dd \tau.
\eeq
Under the parity combined with
$x\mapsto \inv\circ x$ and 
$\psi\mapsto \inv_*\psi$,
this transforms to
\beqa
S_B&\mapsto&-i\int_{\Sigma}\Omega^*x^*\inv^*B
+{i\over 4}\spst^2\int_{\partial \Sigma}
[(\inv^*B)_{\mu\nu}(x)\psi^{\mu}\psi^{\nu}](-\tau) \dd \tau\nn\\
&=&i\int_{\Sigma}x^*(\inv^*B)-{i\over 4}\int_{\partial\Sigma}
[(\inv^*B)_{\mu\nu}\psi^{\mu}\psi^{\nu}](\tau)\dd\tau
\nn
\eeqa
where we have used the fact that $\Omega$ reverses the orientation
of the worldsheet.
Thus the change in the action is
\beq
\sDelta S_B
=i\int_{\Sigma}x^*(\inv^*B+B)
-{i\over 4}\int_{\partial\Sigma}
(\inv^*B+B)_{\mu\nu}\psi^{\mu}\psi^{\nu}\dd\tau.
\eeq
At this point, we recall the condition coming from the invariance of
the weight $\exp\left(i\int_{\surface}x^*B\right)$ 
under the parity $x\mapsto \inv\circ x\circ \Omega$,
for a {\it closed} worldsheet $\surface$ with an orientation reversing 
involution $\Omega:\surface\to\surface$.
The condition is
$\exp\left(-i\int_{\surface}x^*(\inv^*B+B)\right)=1$.
That is, $\inv^*B+B$ has value $2\pi$ times an integer
on any 2-cycle of $X$, i.e.,
\beq
[\inv^*B+B]\in H^2(X,2\pi \Z).
\label{integralB}
\eeq
This means that there is a complex line bundle ${\mathcal L}$
with a $U(1)$ connection $\sA$ such that
\beq
\dd \sA=\inv^*B+B,
\label{eqnsA}
\eeq
so that $-[\inv^*B+B]/2\pi$ is the first Chern class of ${\mathcal L}$.
Given the expression (\ref{eqnsA}), 
the change $\sDelta S_B$ can be written as a boundary term
which is equal to $i{\mathcal A}_{\tau}$ for
$(\gE,\gA,\gT)=({\mathcal L},\sA,0)$.
The net effect is therefore the shift of the gauge 
field by $\sA$.
We find that the parity transform is
\beq
\gT\longrightarrow \spst\inv^*\gT^T,\qquad
\gA\longrightarrow -\inv^*\gA^T+\sA.
\label{pagen}
\eeq
The Chan-Paton bundle $\gE$ is transformed
to $\inv^*\gE^*\otimes {\mathcal L}$.

Note that the line bundle with connection $({\mathcal L},\sA)$ 
obeying (\ref{eqnsA}) is not unique if $X$ is not simply connected --- 
shift of $\sA$ by a flat connection preserves the condition (\ref{eqnsA}).
Thus, we must make a choice of $({\mathcal L},\sA)$, and that is so even when
the B-field vanishes. 
This is an important part of the data of the orientifold,
which we call the {\it twist}.
As we will discuss in the next subsection,
there is a severe constraint on the twist
$({\mathcal L},\sA)$ in order to be able to impose a $\Z_2$ 
orientifold projection.
Note that the B-field gauge transformation,
$B\to B+\dd \Lambda$ and $\gA\to\gA+\Lambda$, shifts the twist
connection as $\sA\to \sA+\Lambda+\inv^*\Lambda$.
Here, $\Lambda$ is a connection of a $U(1)$ bundle and
in particular $\dd\Lambda$ must represent an element of
$H^2(X,2\pi \Z)$.

Let us discuss the parity mapping of open string states.
We consider the open string stretched between
two D-branes $\mathcal{B}_i$ ($i=1,2$) 
determined by the data $(\gE_i,\gA_i,\gT_i)$.
The wavefunctional for a string configuration $x:[0,1]\to X$
(here we suppress the fermionic configuration $\psi_{\pm}$
from the notation) is a linear map from
$E_1$ at $x(0)$ to $E_2$ at $x(1)$,
$$
\Phi[x]\in \Hom({E_1}_{x(0)},{E_2}_{x(1)}).
$$
Na\"ively, the parity image of this state is its transpose combined with
$x\mapsto \inv\Omega(x):=\inv\circ x\circ\Omega$ for
$\Omega(\s)=1-\s$,
$$
\Phi^T[\inv\Omega(x)]\in\Hom({E^*_2}_{\inv(x(0))},{E^*_1}_{\inv(x(1))})
=\Hom({\inv^*E^*_2}_{x(0)},{\inv^*E^*_1}_{x(1)}).
$$
However, what we want as the parity image must take value in
$$\Hom((\inv^*E^*_2\otimes {\mathcal L})_{x(0)},
(\inv^*E^*_1\otimes{\mathcal L})_{x(1)})
\cong
\Hom({\inv^*E^*_2}_{x(0)},{\inv^*E^*_1}_{x(1)})\otimes 
\Hom({\mathcal L}_{x(0)},{\mathcal L}_{x(1)}).
$$
Thus, we have to amend $\Phi^T[\inv\Omega(x)]$
by an element of $\Hom({\mathcal L}_{x(0)},{\mathcal L}_{x(1)})$.
One and only one natural candidate is the parallel transport along the
path $x([0,1])$ with respect to the connection $\sA$:
\beq
h_{\sA}[x]=\exp\left(-i\int_{x[0,1]}\sA\right):{\mathcal L}_{x(0)}
\longto {\mathcal L}_{x(1)}.
\eeq
We therefore define the parity mapping as
\beq
\Phi[x]~\longmapsto ~\Phi^T[\inv\Omega(x)]\otimes h_{\sA}[x].
\label{parmap}
\eeq

\subsection{The Orientifold Isomorphism}
\label{subsec:o-isom}

\newcommand{\Poi}{\Po}

We now discuss how to define D-branes in Type II orientifolds
in which a parity symmetry is gauged.
First of all, a D-brane must be invariant under the parity,
that is, the parity image must be physically equivalent
to the original brane. 
Second, it is not enough that the two are just physically equivalent, but
an explicit {\it isomorphism} must be specified.
This is needed in order to impose orientifold
projection that selects {\it invariant} open string states.
To see the necessity, let us take a D-brane ${\mathcal B}$ 
and denote its parity image by $\Ptr({\mathcal B})$. 
The parity maps the space of states
of the open string ending on $\mathcal{B}$, 
${\mathcal H}_{\mathcal{B},\mathcal{B}}$,
to the space of states of the open string ending on $\Ptr(\mathcal{B})$,
which is {\it another space} 
${\mathcal H}_{\Ptr(\mathcal{B}),\Ptr(\mathcal{B})}$.
For the orientifold projection, however, we need
a parity operator acting on the {\it same} space.
That would be provided by a map from 
${\mathcal H}_{\Ptr(\mathcal{B}),\Ptr(\mathcal{B})}$ back to 
${\mathcal H}_{\mathcal{B},\mathcal{B}}$, which can be defined if
 an isomorphism
\beq
\Ptr(\mathcal{B})\stackrel{\cong}{\longrightarrow} \mathcal{B}
\eeq
is specified.
We shall call it an {\it orientifold isomorphism} (or {\it o-isomorphism}
for short).
Thirdly, we would like to choose the o-isomorphism so that the parity operator 
acting on the open string states is an involution.
To be more precise, we would like the isomorphism to
respect the algebra of parity actions.
For example, if $\Po=\Po(\inv\Omega)$ denote the
parity operator corresponding to $\inv\Omega$,
we want it to respect the relation
 $(\inv\Omega)^2=(-1)^F$ that follows from (\ref{omsq}):
\beq
\Po^2\,=\,(-1)^F.
\label{basicreq}
\eeq
This yields an important constraint on the possible form of the
o-isomorphisms.

Let us consider D-branes determined by D9-brane configurations 
and their parity actions of the form (\ref{pagen}).
The typical case in which a brane $\mathcal{B}=(\gE,\gA,\gT)$
and its parity image
$\Ptr(\mathcal{B})
=(\inv^*\gE^*\otimes {\mathcal L},-\inv^*\gA^T+\sA,\spst \inv^*\gT^T)$
determine physically equivalent D-branes
is when there is a gauge transformation between them,
that is, a unitary map 
\beq
\gU:\inv^*\gE^*\otimes {\mathcal L}\longto E
\eeq
that transforms the boundary interaction $\Ptr(\mathcal{A})$
for the image brane $\Ptr(\mathcal{B})$
to the one $\mathcal{A}$ for the original brane $\mathcal{B}$:
$$
{\mathcal A}_{\tau}=
\gU(x)\Ptr({\mathcal A})_{\tau}\gU(x)^{-1}
+i^{-1}\gU(x){\dd\over \dd \tau}\gU(x)^{-1}.
$$
Namely,
\beqa
T&=&(-1)^{|\gU|}\gU(\spst \inv^*\gT^T)\gU^{-1},
\label{Tco}\\
A&=&\gU(-\inv^*\gA^T+\sA)\gU^{-1}+i^{-1}\gU\dd\gU^{-1}.
\label{Aco}
\eeqa
The sign factor $(-1)^{|\gU|}$ comes from the reordering
of $\gU(x)$ and $\psi^{\mu}$ 
in $\gU(x)\Ptr({\mathcal A})_{\tau}\gU(x)^{-1}$.
We see from (\ref{Tco}) that $\gU$ should depend on the phase $\spst=\mp i$.
If we want to be specific, we write $\gU=\gU_{(\spst)}$.
The relation of the form
$\gU_{(i)}\,\propto\, \gU_{(-i)}\circ \sigma^T$,
as well as $\gU_{(i)}\,\propto \,\xia\circ \gU_{(-i)}\circ\sigma^T$
for Type IIA,
is consistent with the conditions (\ref{Tco}) and (\ref{Aco}).

We would like to regard such $\gU$ as an orientifold isomorphism
with which we define the parity operator. 
Let us discuss the action of $\Po=\Po(\inv\Omega)$ on the NS sector.
We should take $\gU=\gU_{(-i)}$ or $\gU_{(i)}$ 
for the spin structure $(++)$ or $(--)$ respectively,
since $\spst=\mp i$ for $\Omega$ on $(\pm)$, see (\ref{defspst}).
We have already defined a map from ${\mathcal H}_{\mathcal{B},\mathcal{B}}$
to ${\mathcal H}_{\Ptr(\mathcal{B}),\Ptr(\mathcal{B})}$, 
as shown in
(\ref{parmap}). We want to compose it with a map
${\mathcal H}_{\Ptr(\mathcal{B}),\Ptr(\mathcal{B})}\to 
{\mathcal H}_{\mathcal{B},\mathcal{B}}$, 
which is obtained by composition with $\gU(x(1))$ and $\gU(x(0))^{-1}$.
The parity image of the state $\Phi$ is thus given by
$$
\Po(\Phi)[x]
=\gU(x(1))\circ \left(\Phi^T[\inv\Omega(x)]\otimes h_{\sA}[x]\right)
\circ \gU(x(0))^{-1}(-1)^{|\Phi||\gU|}
$$
If we introduce the evaluation map
$\ev_{\s}$ that associates to a string $x:[0,1]\to X$ 
the value at $\s$, $\ev_{\s}(x)=x(\s)$, 
we have a more concise expression
\beq
\Po(\Phi)= \ev_1^*\gU\circ ((\inv\Omega)^*\Phi^T\otimes h_{\sA})
\circ \ev_0^*\gU^{-1}(-1)^{|\Phi||\gU|}.
\label{Pdef}
\eeq

Let us compute the parity squared,
\beqa
\Po^2(\Phi)&=&\Po(\ev_1^*\gU\circ ((\inv\Omega)^*\Phi^T\otimes h_{\sA})
\circ \ev_0^*\gU^{-1}(-1)^{|\Phi||\gU|})
\nn\\
&=&\ev_1^*\gU\circ ((\inv\Omega)^*(\ev_1^*\gU\circ 
((\inv\Omega)^*\Phi^T\otimes h_{\sA})
\circ \ev_0^*\gU^{-1})^T\otimes h_{\sA})\circ \ev_0^*\gU^{-1}
\nn\\
&=&\ev_1^*(\gU\circ\inv^*(\gU^{T})^{-1})\circ
((\inv\Omega)^{2*}\Phi^{TT}\otimes (\inv\Omega)^*h_{\sA}^T\otimes h_{\sA})
\circ \ev_0^*(\inv^*\gU^T\circ \gU^{-1}).
\nn
\eeqa
Here, we used the relation 
$(\inv\Omega)^*\ev_1^*\gU=\ev_0^*\inv^*\gU$, etc, that results from
$(\ev_1\circ \inv\Omega)(x)
=\inv(x(0))=(\inv\circ \ev_0)(x)$, etc.
We have also used the identities that involve the graded transpose,
$(\gU\Phi^T \gU^{-1})^T=(-1)^{|\gU|}(\gU^{-1})^T\Phi^{TT}\gU^T
=(\gU^T)^{-1}\Phi^{TT}\gU^T$.
We may further use the identity $\Phi^{TT}=\iota\circ \Phi\circ \iota^{-1}$
 from (\ref{TT}).
Note that $h_{\sA}^T[\inv\Omega(x)]$ can be regarded as the parallel transport
of $\inv^*{\mathcal L}^*$ along the path $x[0,1]$
with respect to the connection $-\inv^*\sA$,
$$
(\inv\Omega)^*h^T_{\sA}=h_{-\inv^*\sA}.
$$
Collecting all, we obtain the expression for the parity squared
\beq
\Po^2(\Phi)=\ev_1^*(\gU\inv^*(\gU^T)^{-1}\iota)\circ 
((\inv\Omega)^{2*}\Phi \otimes h_{-\inv^*\sA+\sA})
\circ \ev_0^*(\gU\inv^*(\gU^T)^{-1}\iota)^{-1}.
\label{P2Phi}
\eeq

Let us now impose the basic requirement (\ref{basicreq}): 
$\Po^2=(-1)^F$.
Note that $(-1)^F$ can be realized on the wavefunctional $\Phi[x]$
by the action of 
$(\inv\Omega)^{2}$ on $x$ combined with the conjugation by
the $\Z_2$-grading operator $\sigma$ on the Chan-Paton factor.
Therefore we would like (\ref{P2Phi}) to be equal to 
$\sigma\circ (\inv\Omega)^{2*}\Phi\circ \sigma^{-1}$.
This would be the case if
\beq
\gU\inv^*(\gU^T)^{-1}\iota=\sigma\otimes \gc
\label{condforU}
\eeq
where $\gc$ is a ``scalar'' such that
\beq
\gc(x(1))\cdot h_{-\inv^*\sA+\sA}[x]\cdot \gc(x(0))^{-1}=1
\label{condforc}
\eeq
for any open string configuration $x:[0,1]\to X$.
Note that it may depend on the phase $\spst$,
$\gc=\gc_{(\spst)}$, corresponding to $\gU=\gU_{(\spst)}$.
Equation (\ref{condforU}) is the condition for $\gU$ to be an o-isomorphism.

It follows from the definition of $\gU$ as a map 
$\inv^*\gE^*\otimes {\mathcal L}\to E$ that
the ``scalar'' $\gc$ in (\ref{condforU}) 
can be regarded as a section of the line bundle
$(\inv^*{\mathcal L}^*\otimes {\mathcal L})^{-1}$.
Then, the condition (\ref{condforc}) means that $\gc^{-1}$
is a globally defined parallel section of 
$\inv^*{\mathcal L}^*\otimes {\mathcal L}$ with respect to the connection
$-\inv^*\sA+\sA$.
This in particular means that {\it the connection $-\inv^*\sA+\sA$
must be flat and have trivial holonomy along any loop.}
This provides a severe constraint on the choice of $({\mathcal L},\sA)$.

The parallel section $\gc$ must be common for all D-branes in the theory.
To see this, note that a formula like (\ref{P2Phi}) holds also for
a wavefunction $\Phi$ of the open string stretched between
different D-branes, $\mathcal{B}_1$ and $\mathcal{B}_2$, 
\beq
\Po^2(\Phi)=\ev_1^*(\gU_2\inv^*(\gU_2^T)^{-1}\iota_2)\circ 
((\inv\Omega)^{2*}\Phi\otimes h_{-\inv^*\sA+\sA})
\circ \ev_0^*(\gU_1\inv^*(\gU_1^T)^{-1}\iota_1)^{-1}.
\label{P2Phi2}
\eeq
It then follows from the requirement
$\Po^2(\Phi)=(-1)^F\Phi$ that the parallel section $\gc$ for $\mathcal{B}_1$
must be the same as the one for $\mathcal{B}_2$.

\subsubsection*{\it Type IIA Case}

As stated in Section~\ref{subsec:D9}, 
non-BPS D9-branes in Type IIA string theory supports
a Chan-Paton bundle $E$ with
a special structure $\xia:E\to E$, see (\ref{IIACP}) and (\ref{xidef}). 
The tachyon and the gauge field obey the constraint
$\{T,\xia\}=0$ and $[A,\xia]=0$.
In fact, as will be explained in Section~\ref{subsec:nonBPS},
{\it all} states must obey such a constraint, i.e.,
\beq
\xia\circ \Phi=(-1)^{|\Phi|}\Phi\circ \xia.
\label{IIAcons}
\eeq
The parity operation (\ref{pagen}) with (\ref{parmap}) preserves this
structure --- we have $\inv^*\xia^T$ on the parity image 
$\inv^*E^*\otimes {\mathcal L}$. 
We require that the o-isomorphism $\gU$ maps it back to $\xia$,
namely, 
\beq
(-1)^{|\gU|}\gU\inv^*\xia^T \gU^{-1}=\phxi\xia,
\label{xicond2}
\eeq
for some proportionality constant $\phxi$ which must be
$i$ or $-i$ by the hermiticity of $\xia$. 
The two choices of $\phxi$ are related by the exchange
$\gU\longleftrightarrow\xia\circ \gU$ because 
$(-1)^{|\xia \gU|}=-(-1)^{|\gU|}$.
Note that the conditions (\ref{Tco}) and (\ref{Aco}) are maintained
by this exchange, thanks to $\xia T \xia^{-1}=-T$ and $\xia A\xia^{-1}=A$.
More generally, $\gU$ and $\xia\circ \gU$ give rise to the same parity
operator on open string states that
obey the constraint (\ref{IIAcons}).
Therefore $\phxi=i$ and $\phxi=-i$ are physically equivalent
although we need to make a choice once and for all.

\subsubsection*{\it Isomorphisms}

Let $(\cB_1,\gU_1)$ and $(\cB_2,\gU_2)$ be D-branes with o-isomorphisms,
where ${\mathcal B}_i=(E_i,A_i,T_i)$ for $i=1,2$.
We would like to discuss the condition for the two to be physically 
equivalent as D-branes in the orientifold?
Suppose there is an even and unitary bundle map $f:E_1\to E_2$ 
that sends $(A_1,T_1)$ to $(A_2,T_2)$ in the obvious sense
(and $\xia$ of ${\mathcal B}_1$ to $\xia$ of ${\mathcal B}_2$ for
Type IIA).
What should $f$ do on the o-isomorphisms?
We require that, given a third brane,
$({\mathcal B}_3,\gU_3)$, the parity operator
$\Po:{\mathcal H}_{{\mathcal B}_1,{\mathcal B}_3}\to
{\mathcal H}_{{\mathcal B}_3,{\mathcal B}_1}$ 
is equal to 
$\Po:{\mathcal H}_{{\mathcal B}_2,{\mathcal B}_3}\to
{\mathcal H}_{{\mathcal B}_3,{\mathcal B}_2}$
under the natural relations
between the domains and the targets which are determined by $f$.
That is, for any state $\Phi\in {\mathcal H}_{{\mathcal B}_1,{\mathcal B}_3}$,
we require
$\ev_1^*f\circ \Po(\Phi)=\Po(\Phi\circ \ev_0^*f^{-1})$
at ${\mathcal H}_{{\mathcal B}_3,{\mathcal B}_2}$.
A direct computation shows that this condition is
\beq
f\circ \gU_1=\gU_2\circ \inv^*(f^{-1})^T.
\label{isomDO}
\eeq
We shall call such an $f$ an {\it isomorphism} from 
$(\cB_1,\gU_1)$ to $(\cB_2,\gU_2)$.

\subsection{The Type Of O-Planes}
\label{subsec:O-plane}

We learned that we must choose a twist, i.e., 
a hermitian line bundle ${\mathcal L}$ with a unitary connection $\sA$
such that (i) the curvature equals $\inv^*B+B$
and (ii) the connection $-\inv^*\sA+\sA$ of 
$\inv^*{\mathcal L}^*\otimes {\mathcal L}$, which is flat by (i), 
has trivial holonomy along closed loops.
Furthermore, we also found that we must specify an o-isomorphism,
i.e., an isomorphism 
$\gU:\inv^*E^*\otimes {\mathcal L}\to E$ which obeys
$\gU\inv^*(\gU^T)^{-1}\iota=\sigma\otimes \gc$. Here, $\gc$ is
a parallel section, common for all D-branes, of the line bundle 
$(\inv^*{\mathcal L}^*\otimes {\mathcal L})^{-1}$ with respect to the 
flat connection $\inv^*\sA-\sA$. 
Note that the identity $\gU^{TT}\sim \gU$ (\ref{TT})
yields the constraint on it,
\beq
\inv^*\gc\cdot \gc=(-1)^{|\gU|}.
\label{csq}
\eeq

One very important fact is that $\inv^*{\mathcal L}$ is canonically
isomorphic to ${\mathcal L}$ over the fixed point set $X^{\inv}$
of the involution.
To see this, we first recall that the total space of the pull back
$\inv^*{\mathcal L}$ is defined as the subspace of
$X\times {\mathcal L}$ consisting of points
$(x,v)$ such that $v$ is in the fibre of
$\inv(x)$, $v\in {\mathcal L}\bigl|_{\inv(x)}$.
Then the canonical isomorphism is given by
\beq
(x,v)\in \inv^*{\mathcal L}\bigr|_x~\longleftrightarrow~
v\in {\mathcal L}\bigr|_x,
\qquad \forall x\in X^{\inv}.
\label{canoiso}
\eeq
In other words, 
$\inv^*{\mathcal L}^*\otimes  {\mathcal L}$ is canonically trivial 
over $X^{\inv}$.
Also, the connection $-\inv^*\sA+\sA$ is canonically flat
when restricted to $X^{\inv}$.
In particular, the parallel section $\gc$ of 
$(\inv^*{\mathcal L}^*\otimes {\mathcal L})^{-1}$ can be defined 
on $X^{\inv}$ as a locally constant function with values in
complex numbers. The possible values are constrained by 
$\gc^2=(-1)^{|\gU|}$ from (\ref{csq}) --- 
$\pm 1$ if $\gU$ is even and $\pm i$ if $\gU$ is odd.
We claim that it is the sign of this value that 
determines the type of the O-plane.

To be precise, 
we claim that the value of $\gc$ at an O-plane
is related to its type and dimension
by (\ref{BaSt}), which we repeat here:
\beq
\gc=\left\{\begin{array}{ll}
\pm \spst^{k\over 2}&\mbox{at O$(9-k)^{\mp}$-plane (Type IIB)},\\
\pm \spst^{k-\spst\cdot\phxi\over 2}&\mbox{at O$(9-k)^{\mp}$-plane (Type IIA)}.
\end{array}
\right.
\label{basic2}
\eeq
In particular, we have
\beq
(-1)^{|\gU|}=
\left\{\begin{array}{ll}
(-1)^{k\over 2}&\mbox{(Type IIB)},\\
(-1)^{k-\spst\cdot\phxi\over 2}&\mbox{(Type IIA)}.
\end{array}
\right.
\label{statU}
\eeq
Due to its local nature, it is enough to prove 
the formula (\ref{basic2}) in the simplest case of
orientifold of the Minkowski space
with a single flat O$(9-k)$-plane. 
This will be done in Section~\ref{sec:direct}.

The value of $\gc$ can be different at different O-plane components
if the twist $({\mathcal L},\sA)$ is non-trivial.
This leads, via the formula (\ref{basic2}), to mixture of O-plane types.
One of the first examples of mixed type O-planes in the literature
is Type IIA orientifold with one O$8^-$ and one O$8^+$ at antipodal points
of a circle.
This theory is T-dual to Type IIB orientifold on the dual circle with
a half-period shift \cite{DP,NVS}.
Back in the Type IIA orientifold, {\it the half-period shift occurs on the 
Wilson line}, and that is nothing but a non-trivial twist $\sA$.
This example is in fact how we discovered that mixed O-plane types
can be made possible by non-trivial twists.
We shall describe more examples in Section~\ref{sec:example}, 
including the details of O$8^-$-O$8^+$.

In what follows, we shall call $\gc$ the {\it crosscap section}.

\subsubsection*{\it The Four Cases}

According to (\ref{statU}), the statistcs of the o-isomorphism $\gU$ 
is determined by the codimension of the O-plane modulo $4$.
This in particular means that the components of the fixed point set $X^{\inv}$ 
must have the same codimensions modulo $4$.
This is guaranteed when the involution $\inv:X\to X$ has a lift 
$\invS$ to an action on Majorana spinors on $X$.
To see this, suppose there is a codimension $k$ O-plane
and let us choose local coordinates so that $\inv$ acts as the sign flip of
$x^1,...,x^k$. At this O-plane, the lift $\invS$ is realized by the
multiplication by $\pm \Gamma^1\cdots\Gamma^k$ if $k$ is even,
and by $\pm i \Gamma_{11}\Gamma^1\cdots\Gamma^k$ if $k$ is odd.
Here $\Gamma^i$'s are the Gamma matrices, $\{\Gamma^{\mu},\Gamma^{\nu}\}=
-2\eta^{\mu\nu}$ (with the $(-+\cdots +)$ convention for $\eta_{\mu\nu}$),
and $\Gamma_{11}=\Gamma^0\Gamma^1\cdots\Gamma^9$.
Then its square is
\beq
\invS^2=\left\{
\begin{array}{ll}
(\pm\Gamma^1\cdots\Gamma^k)^2=(-1)^{k\over 2}
&\mbox{$k$ even}\\
(\pm i\Gamma_{11}\Gamma^1\cdots\Gamma^k)^2
=(-1)^{k+1\over 2}
&\mbox{$k$ odd}.
\end{array}\right.
\label{sqti}
\eeq
Since $\invS^2$ is either $1$ or $-1$ globally, 
the codimension $k$ modulo 4 must be common to all O-planes.

Let us classify the possibilities into four cases
\beqa
({\rm B}_{\pm})&&\!\!\!\mbox{$\inv$ is orientation preserving and 
$\invS^2=\pm {\rm id}$},\nn\\
({\rm A}_{\pm})&&\!\!\!\mbox{$\inv$ is orientation reversing and 
$\invS^2=\pm {\rm id}$}.\nn
\eeqa
$({\rm B}_{\pm})$ and $({\rm A}_{\pm})$ are for Type IIB and
Type IIA orientifolds respectively.
O-planes that can appear are determined by (\ref{sqti}), i.e.
\beqa
({\rm B}_+)&&\mbox{O9/O5/O1}\nn\\
({\rm B}_-)&&\mbox{O7/O3}\nn\\
({\rm A}_+)&&\mbox{O6/O2}\nn\\
({\rm A}_-)&&\mbox{O8/O4/O0}\nn
\eeqa
The statistics of the o-isomorphisms is
\beq
\begin{array}{cl}
({\rm B}_{\pm})&(-1)^{|\gU_{(-i)}|}=(-1)^{|\gU_{(i)}|}=\pm 1,\\[0.2cm]
({\rm A}_{\pm})&(-1)^{|\gU_{(-i)}|}=-(-1)^{|\gU_{(i)}|}=\pm i\cdot \phxi.
\end{array}
\label{statU4}
\eeq
This allows us to generalize the structure of the D9-brane Chan-Paton factor
to the case when the involution $\inv:X\to X$ is
fixed point free --- the case without O-plane.

\subsection{Ramond Sector}
\label{subsec:Ramond}

\newcommand{\wtPo}{\widetilde{\Po}}

Let us discuss the parity action on the Ramond sector. 
For orientifold projection, we need to use
the operator corresponding to the parity,
$(-1)^{F_R}\inv\Omega$ or $(-1)^{F_L}\inv\Omega$,
that lifts to an action on spinors in this sector, 
i.e., preserves each of the $(-+)$ and $(+-)$ spin structures of the strip.
Let us discuss the action of 
$\inv\widetilde{\Omega}=(-1)^{F_R}\inv\Omega$ 
which has $\spst=\mp i$ for $(\pm)\to (\mp)$, see (\ref{defspst}).
The corresponding operator $\wtPo=\Po(\inv\wtOmega)$
in the $(+-)$ sector is defined by
\beq
\wtPo(\Phi)=\ev_1^*\gU_{(-i)}\circ 
((\inv\widetilde{\Omega})^*\Phi^T\otimes h_{\sA})
\circ \ev_0^*\gU_{(i)}^{-1}(-1)^{|\gU_{(i)}||(\inv\widetilde{\Omega})^*\Phi|}.
\label{deftPR}
\eeq
The definition in the $(-+)$ sector is the same except that 
$\gU_{(i)}$ and $\gU_{(-i)}$ must be interchanged.
As we will see in Section~\ref{subsec:Ra}, 
the operator $(\inv\widetilde{\Omega})^*$ can be odd
in the Ramond sector --- it is odd for Type IIA and even for Type IIB
---
and that is why the sign factor is written as
$(-1)^{|\gU_{(i)}||(\inv\widetilde{\Omega})^*\Phi|}$ rather than
$(-1)^{|\gU_{(i)}||\Phi|}$.
In order for the total parity to be even, the statistics 
of $\gU_{(i)}$ and $\gU_{(-i)}$
must be equal for Type IIB and opposite for Type IIA.
A natural relation between $\gU_{(i)}$ and $\gU_{(-i)}$ is then
\beqa
\mbox{IIB}:&&\gU_{(i)}\,=\,\kappa\,\gU_{(-i)}\circ\sigma^T,\label{Bmp}\\
\mbox{IIA}:&&\gU_{(i)}\,=\,\phxi\kappa\,\xia \circ \gU_{(-i)}\circ\sigma^T,
\label{Amp}
\eeqa
for some constant phase $\kappa$.
Recall that such relations are consistent with the conditions 
(\ref{Tco}) and (\ref{Aco}).

In the Type IIA case, there is an additional reason that
the relation must be (\ref{Amp}) rather than (\ref{Bmp}).
The constraint (\ref{IIAcons}), which we require also in the Ramond sector,
is preserved under the parity (\ref{deftPR}) 
only if $\gU_{(i)}$ and $\gU_{(-i)}$ obey (\ref{xicond2}) with 
the same phase $\phxi$.
This is the case if the relation is 
$\gU_{(i)}\propto \xia \gU_{(-i)}\sigma^T$ but not if 
$\gU_{(i)}\propto \gU_{(-i)}\sigma^T$.
That $\gU_{(i)}$ and $\gU_{(-i)}$ have opposite statistics 
in Type IIA orientifolds is also suggested
in the formula (\ref{statU}) (or (\ref{statU4})).
We must also make sure that 
the operator $\wtPo$ is independent of the choice of $\phxi$
($i$ or $-i$).
We noted earlier that $\phxi\to -\phxi$ is implemented by $\gU\to \xia\gU$, but
we have not fixed the proportionality constant.
In order for the parity $\wtPo$ to be the same, we need
$\gU_{(\pm i)}\to \pm\xia\gU_{(\pm i)}$ up to an overall constant.
We have placed $\phxi$ in the relation (\ref{Amp}) so that
we do not need to change $\kappa$ as $\phxi\to -\phxi$.

We would like the operator $\wtPo=\Po(\inv\wtOmega)$
to obey the same algebraic relation as $\inv\wtOmega$ (c.f. (\ref{omtilsq})),
\beq
\wtPo^2\,=\,{\rm id}.
\eeq
Let us compute the left hand side:
\beqa
\lefteqn{\wtPo^2(\Phi)}\nn\\
&=&\ev_1^*(\gU_{(-i)}\inv^*(\gU_{(i)}^T)^{-1})\circ 
((\inv\widetilde{\Omega})^{*2}\Phi^{TT}\otimes h_{-\inv^*\sA+\sA})
\circ \ev_0^*(\gU_{(i)}\inv^*(\gU_{(-i)}^T)^{-1})^{-1}\nn\\
&&~~~~~~~~~~~~~~~~~~~~~~~~~~~~~~~~~~~~~~~~~~~~~~~~~~~~~~~~~~~~~~~~~~
\times(-1)^{|(\inv\widetilde{\Omega})^*|\cdot |(\inv\widetilde{\Omega})^*\Phi|}
\nn\\
&=&\left\{\begin{array}{ll}
\kappa^{-2}(-1)^{|\gU_{(-i)}|}
(\inv\widetilde{\Omega})^{*2}\Phi&\mbox{(IIB)}\\
\kappa^{-2}\phxi^{-1}(-1)^{|\gU_{(-i)}|}
\xia^{-1}\circ (\inv\widetilde{\Omega})^{*2}\Phi\circ \xia (-1)^{|\Phi|}
&\mbox{(IIA)}\\
\end{array}\right.
\nn\\
&=&(\inv\widetilde{\Omega})^{*2}\Phi\times
\left\{\begin{array}{ll}
\kappa^{-2}(-1)^{|\gU|}&\mbox{(IIB)}\\
\kappa^{-2}\phxi^{-1}(-1)^{|\gU_{(-i)}|}&\mbox{(IIA).}
\end{array}\right.
\label{wtP2}
\eeqa
In the second equality, we used the relation (\ref{Bmp})-(\ref{Amp}) and
(\ref{xicond2}), along with (\ref{condforU}) and (\ref{condforc}).
In the third equality, we used (\ref{IIAcons}).
The consistency condition $\wtPo^2={\rm id}$
determines the phase $\kappa$ up to a sign.

Alternatively, we may define $\wtPo$ as
the composition $(-1)^{F_R}\circ \Po$ or equivalently as 
$\Po\circ (-1)^{F_L}$.
Note that each of $\Po$, $(-1)^{F_R}$ and $(-1)^{F_L}$
exchanges $(+-)$ and $(-+)$ but a product of two of them preserves them.
The operator $\Po:(+-)\to (-+)$ is defined by the same expression
as (\ref{deftPR}) except that $\inv\wtOmega$ is replaced by $\inv\Omega$,
and similarly for $\Po:(-+)\to (+-)$.
The square $\Po^2:(+-)\to (+-)$ is
\beqa
\Po^2(\Phi)&=&\ev_1^*(\gU_{(i)}\inv^*(\gU_{(i)}^T)^{-1}\iota)\circ 
((\inv\Omega)^{2*}\Phi\otimes h_{-\inv^*\sA+\sA})
\circ \ev_0^*(\gU_{(-i)}\inv^*(\gU_{(-i)}^T)^{-1}\iota)^{-1}\nn\\
&=&\sigma\circ (\inv\Omega)^{2*}\Phi\circ \sigma^{-1}\times
\gc_{(i)}\gc_{(-i)}^{-1}.
\nn
\eeqa
The ratio $\gc_{(i)}/\gc_{(-i)}$ can be found from (\ref{Bmp})
and (\ref{Amp}) along with (\ref{xicond2}):
\beq
\gc_{(i)}\gc_{(-i)}^{-1}=\left\{\begin{array}{ll}
(-1)^{|\gU|}&\mbox{(IIB)}\\
\phxi(-1)^{|\gU_{(-i)}|}&\mbox{(IIA)}.
\end{array}\right.
\label{cci}
\eeq
If we use (\ref{statU4}) it may further be evaluated to 
be $\pm 1$ for $({\rm B}_{\pm})$ and
$\mp i$ for $({\rm A}_{\pm})$.
As we will see in Section~\ref{subsec:Ra}, 
$\sigma\circ(\inv\Omega)^{2*}\Phi\circ \sigma^{-1}$ is not always
equal to $(-1)^F\Phi$ but only up to a certain phase
which exactly cancels this phase,
so that the relation $\Po^2(\Phi)=(-1)^F\Phi$ holds.
Let us next discuss the definition of the 
operators $(-1)^{F_R}$ and $(-1)^{F_L}$.
Note that they transform
the field (\ref{defpsi}) as $(-1)^{F_R}:\psi^{\mu}\to \psi^{\mu}$ 
and $(-1)^{F_L}:\psi^{\mu}\to-\psi^{\mu}$.
In order to be a symmetry of the boundary interaction,
which includes ${i\over 2}\psi^{\mu}D_{\mu}T$,
$(-1)^{F_R}$ {\it resp}.\! $(-1)^{F_L}$
 may be defined to act on the Chan-Paton factor
as the identity {\it resp}.\! the $\Z_2$-grading.
Now that we have $\Po$, $(-1)^{F_R}$ and $(-1)^{F_L}$,
the operator $\wtPo$ can be defined as, say, $(-1)^{F_R}\circ \Po$.
The consistency condition $\wtPo^2={\rm id}$
follows from $\Po^2=(-1)^F$ provided that the relation
$(-1)^{F_R}\circ \Po=\Po\circ (-1)^{F_L}$ holds. However,
the last relation is not automatic but imposes a constraint 
on the square of the relative phase $\kappa$.
As it must be the case, that constraint agrees with the one from
$\wtPo^2={\rm id}$ via (\ref{wtP2}).

Repeating the same analysis for the open string between different branes,
say ${\mathcal B}_1$ and ${\mathcal B}_2$, we find that the product
 $\kappa_1\kappa_2$ obeys the same condition as
the squares, $\kappa_1^2$ and $\kappa_2^2$. This
in particular means $\kappa_1=\kappa_2$:
{\it All D-branes must have the same value of $\kappa$.}
Since $\kappa^2$ is determined, we only have to choose the sign of (common)
$\kappa$. To be precise, the sign of $\kappa$ 
for a fixed definition of $(\inv\wtOmega)^*$. 
We next argue that this is related to the ``orientation of the orientifold''.

\subsubsection*{\it The Phase $\kappa$ And The Orientation Of Orientifold}


The key is the open-closed channel duality.
Let $|C_*\rangle$ denote the crosscap state for a parity operator $*$,
and $|B\rangle$ be the boundary state of an invariant brane.
Then we have
\beq
\begin{array}{ccl}\Tr^{}_{\rm NS}\Po q_o^{H_o}&\propto&
\langle C_{\Po}|q_c^{H_c}|B\rangle^{}_{{}_{\rm NSNS}},
\\[0.2cm]
\Tr_{\rm R}^{}\wtPo q_o^{H_o}&\propto&
\langle C_{\wtPo}|q_c^{H_c}|B\rangle^{}_{{}_{\rm RR}},
\end{array}
\label{ocd}
\eeq
where
$H_o$ and $q_o$ ({\it resp}.\! $H_c$ and $q_c$) are 
the Hamiltonian and the modular parameter
in the open string ({\it resp}.\! closed string) channel.

Now, the effect of the sign flip $\kappa\to -\kappa$
is to flip the sign of the parity operator $\wtPo$
in the Ramond sector while keeping the operator $\Po$
in the Neveu-Schwarz sector.
By the channel duality (\ref{ocd}), this is to change the sign of
either $|B\rangle^{}_{{}_{\rm RR}}$ or
$|C_{\wtPo}\rangle$ while keeping 
$|B\rangle^{}_{{}_{\rm NSNS}}$ and $|C_{\Po}\rangle$ untouched.
Reversing the sign of the RR-part of the boundary state
while keeping the NSNS part is nothing but 
 replacing the brane by its antibrane,
i.e., reversing the oriention of the brane.
However, the sign flip of $\kappa$ has no such effect.
Therefore, we must conclude that the sign flip of $\kappa$
corresponds to the sign flip of the RR-part $|C_{\wtPo}\rangle$
of the crosscap state.
This may be regarded as the orientation reversal of the orientifold
--- 
it is nothing but the orientation reversal of the O-planes when 
they exist.
In this sense, the phase $\kappa$ can be interpreted as the parameter for
the orientation of the orientifold.

As a side remark, we note that our formulation can directly derive
the consequence of the channel duality (\ref{ocd})
that the brane orientation reversal
flips the sign of the parity operator $\wtPo$
in the Ramond sector
while keeping $\Po$ in the NS sector.
In Type IIB string theory, the orientation reversal of a brane
is simply to flip the $\Z_2$-grading, $\sigma\to \overline{\sigma}=-\sigma$.
That changes the graded transpose according to (\ref{grshift}).
In order to maintain the condition (\ref{Tco}) we need to change the 
o-isomorphism as $\gU\to \overline{\gU}=\gU\sigma^T$ 
up to a multiplicative constant.
This constant must be opposite between $\spst=i$ and $-i$,
say, 
\beq
\overline{\gU}_{(i)}=\gU_{(i)}\sigma^T\quad\mbox{and}\quad
\overline{\gU}_{(-i)}= -\gU_{(-i)}\sigma^T,
\label{antiU}
\eeq
in order for the brane and its antibrane to have the same 
$\kappa$,
i.e., for $\gU_{(i)}=\kappa \gU_{(-i)}\sigma^T$ and 
$\overline{\gU}_{(i)}=\kappa\overline{\gU}_{(-i)}\overline{\sigma}^T$
to hold at the same time.
Under the grading flip with $\gU_{(\pm i)}\to 
\overline{\gU}_{(\pm i)}=\pm \gU_{(\pm i)}\sigma^T$,
the parity operator $\Po$ in the NS sector remains the same
but the operator $\wtPo$ in the Ramond sector is reversed.
In Type IIA string theory, the brane orientation reversal
is done by $(\sigma,\xia)\to (-\sigma,\xia)$ and it again leads to
the transformation of the o-isomorphism as
$\gU_{(\pm i)}\to \pm \gU_{(\pm i)}\sigma^T$.
We find the same effect.

\subsubsection*{\it Worldsheet Supersymmetry}

There is an ${\mathcal N}=1$ supersymmetry in Ramond sector.
The expression for the supercharge is found by the Noether procedure
--- on a $(-+)$ sector state $\Phi$, it acts as
\beqa
\bfQ_1\Phi&=&\int_0^1\dd\s\left(\psi\cdot (p+B\cdot x')
+\widetilde{\psi}\cdot x'\right)
\Phi
\nn\\
&&+\,\ev_1^*(\psi\cdot A-T)\circ\Phi
-(-1)^{|\Phi|}\Phi\circ\ev_0^*(\psi\cdot A+iT),
\label{Q1}
\eeqa
where $p$ is the conjugate momentum for $x$,
and we used the notation
$\psi=\psi_++\psi_-$, $\widetilde{\psi}=g\cdot(\psi_+-\psi_-)$
and $x'=\partial_{\s}x$.
The appearance of $B$ and $A$ is due to the relation
between the time derivative of $x$ and the conjugate momentum $p$,
$$
g\cdot\dot{x}(\s)=p(\s)+B\cdot x'(\s)+A\delta(\s-1)\circ-\circ A\delta(\s).
$$
The appearance of $T$ is due to the $i\dot{\epsilon}_1T$ term in
$\delta {\mathcal A}_t$, see (\ref{deltaA1}).
We have $-iT$ inside $\ev_0^*$, unlike $T$ inside $\ev_1^*$,
because the time runs in the opposite direction on the left boundary
---
the precise phase realtion, $-i$ versus $1$, 
can be found by the relation (\ref{relcomp})
between the field variables on the strip and those on the upper-half plane.
It is important for the hermiticity
of $\bfQ_1$ that $T$ is multiplied by $i$ in the last term, since
$\Phi\mapsto (-1)^{|\Phi|}\Phi\circ T$ is an anti-hermitian
operator if $T$ is hermitian and odd. 
It is straightforward to see the supersymmetry relation
\beq
(\bfQ_1)^2=2H,
\eeq
where the Hamiltonian $H$ includes the action on the Chan-Paton factor 
$\Phi\mapsto \ev_1^*({\mathcal A}_t)\circ \Phi
-\Phi\circ\ev_0^*({\mathcal A}_{t})$.

The parity operator $\wtPo=\Po(\inv\wtOmega)$ 
commutes with the supercharge,
\beq
\bfQ_1\, \wtPo=\wtPo\,\bfQ_1.
\label{Q1P}
\eeq
To prove this relation, all of (\ref{Tco}), (\ref{Aco}) 
as well as (\ref{eqnsA}) are required.
For example, $\bfQ_1$
acts on $h_{\sA}$ that enters in the definition of $\wtPo$,
and produces the factor
\beqa
\bfQ_1\left(-i\int_0^1\dd \s\,
{\dd x^{\mu}\over\dd\s}\sA_{\mu}(x)\right)\!\!
&=&\!\!-\int_0^1\dd\s\,\left(
{\dd\psi^{\mu}\over \dd\s}\sA_{\mu}(x)
+{\dd x^{\mu}\over\dd\s}\psi^{\nu}\partial_{\nu}\sA_{\mu}(x)\right)
\nn\\
&=&-\ev_1^*(\psi\cdot\sA)+\ev_0^*(\psi\cdot \sA)
-\int_0^1\dd\s\,\psi\cdot(\inv^*B+B)\cdot x',
\nn
\eeqa
where (\ref{eqnsA}) is used in the last equality. 
The terms $\ev^*(\psi\cdot \sA)$ contribute
in cancellation of the terms from the $\bfQ_1$ action of $\gU$,
via the relation (\ref{Aco}). The term involving $(\inv^*B+B)$
cancels with the B-field term in the expression (\ref{Q1}) for $\bfQ_1$.
The complete proof of (\ref{Q1P}) is left as an exercise to the reader.

\subsection{The Data --- Summary}
\label{subsec:summary}

Through the analysis, we have identified the data to specify the orientifold 
itself and D-branes in it. It can be summarized as follows.

We consider the Type II orientifold on a ten dimensional spin manifold
$X$ by an involution $\inv:X\to X$ with a lift
$\invS$ to an action on Majorana spinors.
It is classified into four cases, $({\rm B}_{\pm})$ and $({\rm A}_{\pm})$,
depending on whether $\inv$ is orientation preserving or reversing
and whether $\invS^2$ is $1$ or $-1$.
To specify the theory, we need to choose additional data ---
the B-field $B$,
a hermitian line bundle ${\mathcal L}$ over $X$ (the twist bundle), 
a unitary connecton $\sA$ of ${\mathcal L}$ (the twist connection),
and a section $\indpc$ of $\inv^*{\mathcal L}\otimes {\mathcal L}^*$
(the crosscap section).
These are required to obey the constraints:

(i)~ $\dd \sA=B+\inv^*B$,

\vspace{-0.2cm}

(ii)~ the connection $\inv^*\sA-\sA$  
of $\inv^*{\mathcal L}\otimes {\mathcal L}^*$
has trivial holonomy along any loop,

\vspace{-0.2cm}

(iii)~ $\indpc$ is a parallel section with respect to that connection
such that $\inv^*\indpc\cdot \indpc=1$.

\noindent
On a $\inv$-fixed locus, the section $\indpc$ can be regarded as
a number, $+1$ or $-1$, which determines the type of the O-plane:
\beq
\begin{array}{|c|cccccccccc|}
\hline
\mbox{O-plane}
&{\rm O9}^{\mp}&{\rm O8}^{\mp}&{\rm O7}^{\mp}&{\rm O6}^{\mp}&{\rm O5}^{\mp}
&{\rm O4}^{\mp}&{\rm O3}^{\mp}&{\rm O2}^{\mp}&{\rm O1}^{\mp}&{\rm O0}^{\mp}\\
\hline
\indpc&\pm 1&\pm 1&\pm 1&\mp 1&\mp 1&\mp 1&\mp 1&\pm 1
&\pm 1&\pm 1\\
\hline
\end{array}
\label{Otype}
\eeq

\noindent
The data for D-branes depend on the cases:

\noindent
$({\rm B}_{\pm})$: a $\Z_2$-graded hermitian vector bundle $E$ 
with an even unitary connection
$A$ and an odd hermitian endomorphism $T$, and a unitary
isomorphism $\indpU:\inv^*E^*\otimes {\mathcal L}\to E$ such that
\beqa
&&U\,=\,\indpc\cdot\inv^*U^t,\quad
(-1)^{|\indpU|}=\pm 1,\nn\\[0.1cm]
&&
A\,=\, \indpU(-\inv^*A^t+\sA)\indpU^{-1}+i^{-1}\indpU\dd \indpU^{-1},~
\label{indpcondIIB}\\[0.1cm]
&&T\,=\, \pm\,\indpU\inv^*T^t\indpU^{-1}.\nn
\eeqa
$({\rm A}_{\pm})$: an ungraded hermitian vector bundle $\ugE$ 
with a unitary connection
$\ugA$ and a hermitian endomorphism $\ugT$, and a unitary
isomorphism $\ugU:\inv^*\ugE^*\otimes {\mathcal L}\to \ugE$
such that
\beqa
&&\ugU\,=\,\indpc\cdot \inv^*\ugU^t,\nn\\[0.1cm]
&&\ugA\,=\,\ugU(-\inv^*\ugA^t+\sA)\ugU^{-1}+i^{-1}\ugU\dd \ugU^{-1},~
\label{indpcondIIA}\\[0.1cm]
&&\ugT\,=\,\pm\, \ugU\inv^*\ugT^t\ugU^{-1}.
\nn
\eeqa


The section $\indpc$ is independent of the phase $\spst$ 
and is related to the one we introduced earlier by 
$\gc= \indpc$, 
$\spst\indpc$,
$\spst^{7-\spst\cdot\phxi\over 2}\indpc$ and
$\spst^{1-\spst\cdot\phxi\over 2}\indpc$, for
$({\rm B}_+)$, $({\rm B}_-)$, $({\rm A}_+)$ and $({\rm A}_-)$ respectively.
The D-brane o-isomorphism $\gU$ is obtained from $\indpU$ or $\ugU$
via
\beqa
({\rm B}_+):&&
\gU_{(-i)}=\left(\begin{array}{cc}
U_{00}&0\\
0&i U_{11}
\end{array}\right),\quad
\gU_{(i)}=\kappa\left(\begin{array}{cc}
U_{00}&0\\
0&-i U_{11}
\end{array}\right),
\nn\\
({\rm B}_-):&&
\gU_{(-i)}=\left(\begin{array}{cc}
0&iU_{01}\\
U_{10}&0
\end{array}\right),\quad
\gU_{(i)}=\kappa\left(\begin{array}{cc}
0&-iU_{01}\\
U_{10}&0
\end{array}\right),
\nn\\[-0.3cm]
\label{relnU}\\[-0.3cm]
({\rm A}_+):&&
\gU_{(-i)}=\left(\begin{array}{cc}
\ugU&0\\
0&i\ugU
\end{array}\right),\quad
\gU_{(i)}= 
\kappa\left(\begin{array}{cc}
0& i\ugU\\
\ugU&0
\end{array}\right),
\nn\\
({\rm A}_-):&&
\gU_{(-i)}=\left(\begin{array}{cc}
0&i\ugU\\
\ugU&0
\end{array}\right),\quad
\gU_{(i)} 
=-\kappa\left(\begin{array}{cc}
\ugU&0\\
0&i \ugU
\end{array}\right).
\nn
\eeqa
$U_{ij}$ in Case $({\rm B}_{\pm})$ are the blocks of $U$ with respect to
the decomposition $E=E^0\oplus E^1$.
The expressions for $\gU$ in Case $({\rm A}_{\pm})$
are for the choice $\phxi=-i$. The expressions for the other choice
$\phxi=+i$ can be obtained by
the replacement $\gU_{(\pm i)}\to \pm\xia\gU_{(\pm i)}$.
The conditions (\ref{iiBcond})-(\ref{iiAcond}) in the introduction section
are written in terms of $\indpU$ and $\ugU$ in this summary.
Note that there is an ambiguity in $\indpU$: it could be replaced by
$\sigma\indpU$ in which case the equation involving the tachyon
 has an extra sign, 
i.e. we have $T=\mp \indpU\inv^*T^t\indpU^{-1}$ for $({\rm B}_{\pm})$, 
while the section $\indpc$ becomes $\pm\indpc$ for $({\rm B}_{\pm})$.
Of course, this is simply a matter of convention and has no physical effect.
In an announcement \cite{KITPtalk} of the present work, 
we reported the result partly in this different convention.

\subsubsection*{\it Gauge Transformations}

As discussed earlier, the B-field gauge transformation,
$B\to B+\dd\Lambda$ and $A\to A+ \Lambda$, shifts the twist connection
$\sA$ by $\Lambda+\inv^*\Lambda$.
Accordingly, the Chan-Paton bundle $E$ and the twist bundle ${\mathcal L}$ 
are mapped to $E\otimes L$ and ${\mathcal L}\otimes L\otimes \inv^*L$ 
respectively, where $L$ is the hermitian line bundle which has 
$\Lambda$ as a unitary connection.
The o-isomorphism is now from
$\inv^*(E\otimes L)^*\otimes({\mathcal L}\otimes L\otimes\inv^*L)
\cong\inv^*E^*\otimes {\mathcal L}\otimes L$ to
 $E\otimes L$ and can be taken as $\gU\otimes {\rm id}_L^{}$,
which we write simply as $\gU$. 
In particular the crosscap section $\indpc$ 
does not change under this transformation.
We may also consider ordinary gauge transformation of the twist connection,
$i\sA\to i\sA+\lambda^{-1}\dd \lambda$,
for a $U(1)$-valued function $\lambda$.
The simplest way to maintain the relation (\ref{Aco}) is to transform
the o-isomorphism as $\gU\to \lambda \gU$.
The section $\indpc$ is then transformed to
$\lambda\cdot\inv^*\lambda^{-1}\cdot \indpc$.
Note that $\lambda\cdot\inv^*\lambda^{-1}=1$ at $\inv$-fixed points,
in accordance with the fact that the O-plane type cannot change under
 gauge transformations.
To summarize, we have found gauge transformations 
which map the orientifold data
$(B,{\mathcal L},\sA, \indpc)$ to
\beq
(B+\dd\Lambda,\,{\mathcal L}\otimes L\otimes \inv^*L,\,
\sA+\Lambda+\inv^*\Lambda,\,\indpc)
\quad\mbox{and}\quad
(B,\,{\mathcal L},\,\sA-i\lambda^{-1}\dd\lambda,
\,\lambda\!\cdot\!\inv^*\lambda^{-1}\!\cdot\indpc),
\label{gtra}
\eeq
and the D-brane data $(E,A,T,\gU)$ to
\beq
(E\otimes L,\, A+\Lambda,\,T,\,\gU)\quad\mbox{and}\quad
(E,\,A,\,T,\,\lambda\cdot \gU).
\label{gtrab}
\eeq

Classification of orientifold data has been discussed in 
\cite{BrSt} and \cite{DFM}.
It would be interesting to find relation of the present result to these
works.



\medskip

\section{Boundary Fermions And Parity Actions}
\label{sec:rbf}

In this section,
we study parity operation on worldsheet theory with
boundary fermions, which has boundary action of the form
\beq
S_{\rm bdry}=\int_{\partial\Sigma}\dd t
\left\{{i\over 4}\sum_{i=1}^s
\xi_i{\dd\over\dd t}\xi_i+\cdots\,\right\},
\label{Sbdry}
\eeq
where the ellipses are interaction terms of $\xi_i$'s and 
the boundary values of the bulk fields.
This study serves as a preparation for 
a part of the direct CFT determination of the structure of 
the D9-brane Chan-Paton factor to be done in next section.
It also provides a background for
the treatment of non-BPS D-branes in Type II string theory,
such as D9-branes in Type IIA,
which is employed in the present paper.
At the end of this section, we study a particular configuration
on unstable D9-branes that represents
BPS D-branes on top of an orientifold plane.

Boundary fermion realization of Chan-Paton factors was first studied in 
\cite{MSbf}.
Parity action on boundary fermion system was discussed in \cite{CCHbf}.

\subsection{Open String States}\label{subsec:spectrum}

Let us consider an open string with $s$ fermions
on the right boundary and $r$ fermions on the left boundary.
The boundary action takes the form
\beq
S_{\rm bdry}
=\int \dd t\left.\left\{{i\over 4}\sum_{i=1}^s
\xi^{(R)}_i{\dd\over \dd t}\xi^{(R)}_i
+\cdots\right\}\right|_{\rm right}
+\int \dd t\left.\left\{-{i\over 4}\sum_{i'=1}^r
\xi^{(L)}_{i'}{\dd\over \dd t}\xi^{(L)}_{i'}
+\cdots\right\}\right|_{\rm left}.
\label{Sbdryop}
\eeq
The sign of the kinetic terms for $\xi^{(L)}_{i'}$'s is opposite
to the standard one
since the natural orientation of the left boundary is 
pointing toward the ``past''.
The canonical anticommutation relations are
\beq
\{\xi^{(R)}_i,\xi^{(R)}_j\}=2\delta_{i,j},\qquad
\{\xi^{(L)}_{i'},\xi^{(L)}_{j'}\}=-2\delta_{i',j'},\qquad
\{\xi^{(R)}_i,\xi^{(L)}_{j'}\}=0,
\label{CCR}
\eeq
and the hermiticity is
\beq
\xi^{(R)\dag}_i=\xi^{(R)}_i,\qquad
\xi^{(L)\dag}_{i'}=-\xi^{(L)}_{i'}.
\label{Herm}
\eeq
The numbers $r$ and $s$ must be related to the boundary conditions on 
the bulk fields, say $\beta$ (right) and $\alpha$ (left),
so that the space of open string states 
has a GSO operator --- a $\Z_2$-grading operator $(-1)^F$
that anticommutes with $\xi^{(R)}_i$,
$\xi^{(L)}_{i'}$ as well as the bulk fermions.
When $r+s$ is even, 
the boundary fermion algebra (\ref{CCR}) has a 
unique $\Z_2$-graded irreducible representation.
Then, the boundary conditions $(\alpha,\beta)$
must be such that the bulk fields have their own quantization
with a $\Z_2$-graded space of states, so that
the space of total open string states 
is the (graded) tensor product of the boundary fermion factor
and the bulk factor
\beq
{\mathcal H}^{\rm tot}_{(r,\alpha),(s,\beta)}={\mathcal H}^{\rm b.f.}_{r,s}
\otimes {\mathcal H}^{\rm bulk}_{\alpha,\beta}.
\label{Hfactor}
\eeq
When $r+s$ is odd, the algebra (\ref{CCR}) has two distinct irreducible 
representations and neither of them is $\Z_2$-graded.
In this case, the boundary conditions
$(\alpha,\beta)$ must be such that the bulk fields have
an unpaired fermionic mode, so that the combined bulk-boundary system
has a unique irreducible representation with a GSO operator $(-1)^F$.
The present discussion follows \cite{WittenK}
where the open string stretched between a BPS D9-brane and
a non-BPS D0-brane in Type IIB or Type I string theory is found
to have odd number of bulk fermion zero modes.
This was recognized as a problem against
 natural quantization with GSO projection and,
as a solution,
it was proposed to place a single fermion on the D0 boundary.
We shall say more on this momentarily.

Before giving a more detailed description of the space of states,
let us introduce some notations on representations of the Clifford algebra
\beq
\{\xi_i,\xi_j\}=2\delta_{i,j},\qquad
i,j=1,\ldots,s,
\label{Clifford}
\eeq
with even $s$.
We take complex combinations of the generators,
$\eta_i={1\over 2}(\xi_{2i-1}+i\xi_{2i})$ and
$\bareta_i={1\over 2}(\xi_{2i-1}-i\xi_{2i})$
($i=1,...,{s\over 2}$),
which obey the relations
$$
\{\eta_i,\bareta_j\}=\delta_{i,j},\qquad
\{\eta_i,\eta_j\}=\{\bareta_i,\bareta_j\}=0.
$$
An irreducible representation is 
build on a vector $|0\rangle$ annihilated by all
$\eta_i$'s, and is spanned by 
$|0\rangle$, $\bareta_i|0\rangle$,
$\bareta_i\bareta_j|0\rangle$, \ldots , 
$\bareta_1\cdots\bareta_{s\over 2}|0\rangle$.
It has a $\Z_2$-grading. For example,
even multiples of $\bareta_i$'s on $|0\rangle$
are even and odd multiples are odd.
The hermitian inner product such that 
the above $2^{s\over 2}$ vectors form an orthonormal basis has the property
$\eta_i^{\dag}=\bareta_i$ (equivalently $\xi_j^{\dag}=\xi_j$).
We shall denote this graded irreducible representation 
with inner product by $V_{s}$.

\subsubsection*{(i) $r$ and $s$ even}

The boundary fermion factor in (\ref{Hfactor}) is naturally 
of the Chan-Paton form, that is, the space 
of linear maps
\beq
{\mathcal H}^{\rm b.f.}_{r,s}=\Hom^{}_{\C}(V_r,V_s).
\label{CPeveneven}
\eeq
On this space, $\xi^{(R)}_i$'s and $\xi^{(L)}_{i'}$'s act as
\beq
\xi^{(R)}_i\phi=\xi_i\circ \phi,\qquad
\xi^{(L)}_{i'}\phi=(-1)^{\phi}\phi\circ \xi_{i'}.
\label{defXiLR}
\eeq 
It is straightforward to check the anticommutation relation (\ref{CCR}),
as well as the hermiticity (\ref{Herm})
with respect to the inner product 
$(\phi_1,\phi_2)=\tr^{}_{V_r}(\phi_1^{\dag}\phi_2)
=\tr^{}_{V_s}(\phi_2\phi_1^{\dag})$.
The space $\Hom^{}_{\C}(V_r,V_s)$ has a natural $\Z_2$-grading
induced by those of $V_r$ and $V_s$.

\subsubsection*{(ii) $r$ and $s$ odd}

The boundary fermion factor in (\ref{Hfactor}),
though $\Z_2$-graded by itself, 
does not have the structure of the space of linear maps
between $\Z_2$-graded vector spaces.
Such a Chan-Paton form would be advantageous though, for example,
in the consideration of product of open string states.
For this purpose, we introduce auxiliary
boundary fermions, one at each boundary ---
 $\xia^{(R)}$ on the right
and $\xia^{(L)}$ on the left,
with the action
\beq
S_{\rm aux}=\int_{-\infty}^{\infty}\dd t\, {i\over 4}\xia^{(R)}
{\dd\over\dd t}\xia^{(R)}
-\int_{-\infty}^{\infty}\dd t\, {i\over 4}\xia^{(L)}
{\dd\over\dd t}\xia^{(L)}.
\eeq
The space of states of the extended system is of the Chan-Paton type, 
$\Hom^{}_{\C}(V_{r+1},V_{s+1})$.
It can alternatively be defined as the (graded) tensor product
of the original space and the space from
the auxiliary fermions.
The second factor is a 2-dimensional space consisting 
of one even and one odd states.
We suppose that the even state $|0\rangle_{{}_{\rm aux}}$
satisfies the continuity condition 
$\xia^{(R)}|0\rangle_{{}_{\rm aux}}=
\xia^{(L)}|0\rangle_{{}_{\rm aux}}$.
To get back the original space,
we select only the states of the form
$\phi'\otimes |0\rangle_{{}_{\rm aux}}$.
This is equivalent to imposing the projection condition
\beq
\xia^{(R)}\xia^{(L)}=1.
\label{oddoddproj}
\eeq
The boundary fermion sector can thus be realized as
\beq
{\mathcal H}^{\rm b.f.}_{r,s}=\Hom^{}_{\C}(V_{r+1},V_{s+1})
\Bigr|_{\xia^{(R)}\xia^{(L)}=1}.
\label{CPoddodd}
\eeq
Note that the $\xia^{(R)}\xia^{(L)}=1$ condition amounts for 
$\phi\in \Hom^{}_{\C}(V_{r+1},V_{s+1})$ to
\beq
\xia\circ \phi=(-1)^{\phi}\phi\circ\xia,
\label{originIIAcons}
\eeq
where $\xia$ acts on $V_{r+1}$ and $V_{s+1}$ as the ``last'' Clifford 
generator.
We see that this is compatible with the
product of open string states.

\subsubsection*{(iii) $r$ even and $s$ odd ({\it resp}.
$r$ odd and $s$ even)}

In this case, as remarked above,
the boundary conditions $(\alpha,\beta)$ must be such that the bulk fields 
have an unpaired fermionic mode.
We introduce a pair of auxiliary fermions,
$\xia$ and $\xia'$,
which shall be included into the boundary and the bulk sectors respectively.
We choose the sign of the kinetic term of $\xia$ 
to be the same as the one for the odd boundary fermions,
i.e. positive if $s$ is odd ({\it resp}. negative if $r$ is odd),
and opposite to the one for $\xia'$.
Then the extended boundary fermion system
has a $\Z_2$-grading and is of the Chan-Paton form 
\beq
{\mathcal H}^{\rm b.f.+aux}_{r,s}=\Hom^{}_{\C}(V_r,V_{s+1})
\quad ({\it resp.}\,\,\, \Hom^{}_{\C}(V_{r+1},V_s)).
\eeq
Likewise the extended bulk system also has a $\Z_2$-grading.
To remove the extra degrees of freedom coming from the auxiliary fermions,
we impose the projection condition
$\xia\xia'=1$.
The space of states is therefore
\beq
{\mathcal H}_{(r,\alpha),(s,\beta)}^{\rm tot}=
{\mathcal H}^{\rm b.f.+aux}_{r,s}
\otimes {\mathcal H}^{\rm bulk+aux}_{\alpha,\beta}\Bigr|_{\xia\xia'=1}.
\label{extdecompo}
\eeq

\subsection{Non-BPS D-Branes In Type II Superstrings}
\label{subsec:nonBPS}

It is a good point to provide a background for 
the description of non-BPS D-branes used in this paper.

Type IIA (IIB) string theory has non-BPS D$p$-branes 
in addition to BPS D$q$-branes, where
$p$ is odd (even) and $q$ is even (odd).
As mentioned above, it was found in \cite{WittenK}
that a natural quantization of open strings with GSO projection 
is possible by placing an odd number of fermions along the boundary 
for non-BPS D-branes.
More generally, we may also have an additional 
$\Z_2$-graded vector space $V'$ along such a boundary. 
If we allow this, we may assume that the number of boundary fermions is one,  
since the additional, even number of boundary fermions may be included in $V'$.
The single boundary fermion $\xi_1$ and $V'$ may have interaction
among themselves and with the boundary values of the bulk fields.

In order to present the total degrees of freedom at the boundary,
i.e., $\xi_1$ and $V'$, in the more standard Chan-Paton form,
we introduce a single auxiliary boundary fermion $\xia$.
Then, we have a $\Z_2$-graded Chan-Paton vector space $V$ which has twice as
many dimensions as $V'$.
Of course, we have to make sure that the introduction of $\xia$ does not
change the content of the theory.
One necessary condition is that $\xia$ has only the kinetic term and has no 
interaction with other fields.
On the open string states, we need to impose an appropriate
projection condition, as we discuss below.

On a boundary circle $C$ without insertion of open string state,
the inclusion of auxiliary fermion has the effect of multiplying ``1'',
provided that the normalization is done correctly.
When the boundary circle $C$ has the anti-periodic spin structure,
we use
\beq
{1\over \sqrt{2}}\int {\mathcal D}\xia\,
\exp\left(i\int_{C}{i\over 4}\xia{\dd\over \dd\tau}\xia\,\dd\tau\right)=1.
\label{APid}
\eeq
The factor of ${1\over \sqrt{2}}$ is required for the reason explained in
\cite{WittenK}.
For the periodic spin structure, we use
\beq
{\#\over\sqrt{2}}\int {\mathcal D}\xia\,\xia(\tau_0)
\exp\left(i\int_{C}{i\over 4}\xia{\dd\over \dd\tau}\xia\,\dd\tau\right)=1,
\label{Pid}
\eeq
for some phase $\#$
where $\tau_0$ is an arbitrary point of the circle.
If we perform the path-integral of $\xia$ together 
with $\xi_1$ and include $V'$ as well,
we have the trace (for antiperiodic circle) or 
the supertrace (for periodic circle) over the $\Z_2$-graded Chan-Paton space
$V$ of a Wilson line $\Pexp\left(-i\int_C{\mathcal A}\right)$.
This results in the factors of the form (\ref{IIAD9}).
Note that ${\mathcal A}$ commutes with the auxiliary fermion $\xia$
since the original boundary interaction does not involve $\xia$.
In particular, if ${\mathcal A}$ is written in the form (\ref{At}),
then the tachyon $T$ anticommutes with $\xia$ and the gauge field $A$
commutes with $\xia$.
That is, they are of the form (\ref{IIA}).

Next, let us consider a boundary circle $C$ with insertion of open string 
states, assuming that all the segments correspond to non-BPS D-branes. 
Thus, all the open string states are of the type {\bf (ii)},
and the auxiliary fermion $\xia$ goes around the circle $C$.
As discussed in the previous subsection, we require
that the open string states in the extended system
obey the constraint $\xia^{(R)}\xia^{(L)}=1$.
This means that it is of the factorized form 
$\Phi\otimes |0\rangle^{}_{{}_{\rm aux}}$ where
$\Phi$ is the state of the original system and
$|0\rangle^{}_{{}_{\rm aux}}$ is the state of the auxiliary fermion system
obeying 
$\xia^{(R)}|0\rangle^{}_{{}_{\rm aux}}=\xia^{(L)}|0\rangle^{}_{{}_{\rm aux}}$.
Then, the auxiliary path-integral factors out and,
with the correct normalization, gives us
$1$ again, leaving us with the system before the extension.
Thus, the extension by auxiliary fermion again has no effect as long as
the constraint $\xia^{(R)}\xia^{(L)}=1$ is satisfied on the open string states.

\begin{figure}[htb]
\psfrag{odd}{non-BPS}
\psfrag{even}{BPS}
\centerline{\includegraphics{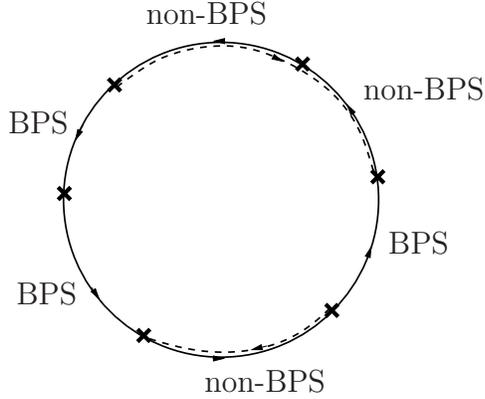}}
\caption{Representation of an amplitude in the extended system}
\label{fig:disk-ext}
\end{figure}

Finally, let us consider the boundary circle $C$ where some segments 
correspond to BPS D-branes while others correspond to non-BPS D-branes.
In this case, open string states of the type {\bf (iii)} must be present.
Following the previous subsection,
along the segment corresponding to non-BPS D-branes,
we introduce additional auxiliary fermion $\xia'$ with
kinetic term of the opposite sign  compared to the one for $\xia$.
The resulting amplitude in the extended system is of the form as depicted
in Figure~\ref{fig:disk-ext} for the case of a disc.
The dashed lines correspond to the $\xia'$-lines.
With the requirement of $\xia\xia'=1$ at {\bf (iii)}
and $\xia^{(R)}\xia^{(L)}=\xia^{\prime (R)}\xia^{\prime (L)}=1$ at {\bf (ii)},
the $\xia$-$\xia'$ path-integrals factor out to give ``1''
and we get back the amplitude of the original system.

The resulting description is similar to the one for
Type IIB and Type I D0-branes given by A. Sen
\cite{SenD0inI,SenD0,SenNonSUSY,Senrev}.
If we set $V'=\C$ and turn off boundary interaction,
our construction has the same Chan-Paton structure and the open string
interaction rule as the ones given by Sen,
which were first proposed in \cite{SenD0} and later rationalized 
in \cite{SenNonSUSY,Senrev} using the $(-1)^{{\bf F}_L}$ orbifold.
Here $(-1)^{{\bf F}_L}$ is the mod 2 number of spacetime fermions from
the left movers, and the corresponding orbifold maps Type IIA to Type IIB and
vice versa. 
However, unlike in the Green-Schwarz formalism,
this orbifold is not natural in the NSR formalism we are working with. 
We have given a natural derivation of the structure within
the NSR formalism.

\subsection{Parity Action}
\label{subsec:bfPa}

We now study parity actions.
First, we would like to find parity transformations of 
boundary fermions that leave invariant the boundary action
which, on the boundary of the (Euclidean) upper-half plane, takes the form
\beq
iS_{{\rm bdry}}=\int_{\partial\Sigma}\dd\tau
\left\{{i\over 4}\sum_{i=1}^s\xi_i{\dd\over\dd\tau}\xi_i+\cdots\right\}.
\label{Sbk}
\eeq
We assume a linear transformation which then must be of the form
\beq
\xi_i(\tau)\longrightarrow \spst\sum_{i=1}^s O_{ij}\xi_j(-\tau),
\label{Odef}
\eeq
where $\spst$ is the phase (\ref{defspst})
that appears in the transformation
$\psi(\tau)\to \spst\inv_*\psi(-\tau)$.
This is so that the orientifold projection condition
on GSO even expressions $(\psi)^n(\xi)^m$ ($n+m$ even)
does not depend on 
the choice of boundary spin structure.
Invariance of the kinetic term in (\ref{Sbk}) requires
that $O_{ij}$ is an orthogonal matrix,
$O^tO={\bf 1}_s$.
The algebra of the parity transformations, $(\inv\Omega)^2=(-1)^F$,
$((-1)^{F_R}\inv\Omega)^2={\rm id}$ etc,
requires that it is involutive, $O^2={\bf 1}_s$.
It follows from unitarity of the parity operator
on open strings that 
$O$ must be real, $O^*=O$ (see below).
In particular,
with a real orthogonal transformation of $\xi_i$'s, 
the matrix $O$ can be made into the diagonal form
\beq
O={\rm diag}(\underbrace{1,1,\ldots, 1}_{s_+},
\underbrace{-1,-1,\ldots,-1}_{s_-}).
\label{rprO}
\eeq
Additional condition on $O$ may come from the invariance of the
interaction terms, ``$+\cdots$'' in (\ref{Sbk}).

We study the parity action on the open string states,
in the set-up of Section~\ref{subsec:spectrum}.
For concreteness, we consider $\Po=\Po(\inv\Omega)$
for $\Omega$ as in (\ref{defOmega}). 
We assume for simplicity that the boundary conditions for the bulk fields, 
$\alpha$ and $\beta$, are also invariant under the parity.
We would like to find an even operator
\beq
\Po:{\mathcal H}^{\rm tot}_{(r,\alpha),(s,\beta)}
\longrightarrow {\mathcal H}^{\rm tot}_{(s,\beta),(r,\alpha)},
\label{PonH}
\eeq
corresponding to the parity $\inv\Omega$
that transforms the boundary fermions as
\beqa
&&\xi_i^{(R)}\longrightarrow
\Po\xi_i^{(R)}\Po^{-1}=\spst\sum_{j=1}^sO_{ij}\xi_j^{(L)},
\label{PRL}
\\
&&\xi_{i'}^{(L)}\longrightarrow
\Po\xi_{i'}^{(L)}\Po^{-1}=\spst'\sum_{j'=1}^rO'_{i'j'}\xi_{j'}^{(R)}.
\label{PLR}
\eeqa
Here $\spst$ and $\spst'$ are $\mp i$ depending on
the spin structures of the right and the left boundaries of the
domain strip. 
In view of the hermiticity (\ref{Herm}) of $\xi^{(R)}_i$ and 
$\xi^{(L)}_i$, 
unitarity of the operator $\Po$ indeed
requires that the matrices $O_{ij}$ and $O'_{i'j'}$ must be real.
In Case {\bf (ii)} and {\bf (iii)}, we also need to specify the transformation
of the auxiliary fermions.
We look for a parity operator that preserves the factorized form
of the space of states, (\ref{Hfactor}) or (\ref{extdecompo}),
\beq
\Po(\phi\otimes\psi)=
(-1)^{|\phi||\Po_{\rm bulk}|}
\Po_{\rm CP}(\phi)\otimes \Po_{\rm bulk}(\psi).
\eeq
We shall determine the Chan-Paton part $\Po_{\rm CP}$ below.

\subsubsection*{(i) $r$ and $s$ even}

We look for an operator
$\Po_{\rm CP}:\Hom(V_r,V_s)\to \Hom(V_s,V_r)$ of the from
\beq
\Po_{\rm CP}(\phi)~=~ (-1)^{|\phi||\gU|}
\gU'\circ\phi^T\circ \gU^{-1},
\label{PCP}
\eeq
for linear maps $\gU:V_s^*\to V_s$ and $\gU':V_r^*\to V_r$.
As $\xi_i^{(L)}$ and $\xi_i^{(R)}$ are given by (\ref{defXiLR}), 
the transformation (\ref{PRL}) is realized if
\beq
\gU \xi_i^T \gU^{-1}
=(-1)^{|\gU|}\spst \sum_{j=1}^sO_{ij}\xi_j.
\label{confU}
\eeq
Since $V_s$ is an irreducible representation of the Clifford algebra,
this condition uniquely fixes $\gU$ up to a scalar multiplication.
The same can be said on $\gU'$.
We would like $\gU$ and $\gU'$ to play the r\^ole of
the o-isomorphisms of the boundary fermion systems.

Let us explicitly construct such $\gU$ in the case where $O_{ij}=\delta_{i,j}$.
The equation (\ref{confU}) reads
for $\eta_i$ and $\bareta_i$ ($i=1,\ldots, {s\over 2}$) as
\beq
\gU\eta_i^T\gU^{-1}=(-1)^{|\gU|}\spst\eta_i,\qquad
\gU\bareta_i^T\gU^{-1}=(-1)^{|\gU|}\spst\bareta_i.
\label{etacond}
\eeq
From the first set of equations, we find
$\eta_i^T\gU^{-1}|0\rangle={\rm const}\times \gU^{-1}\eta_i|0\rangle=0$,
which means that $\gU^{-1}|0\rangle$ is proportional to
$\langle 0|\eta_1\cdots\eta_{s\over 2}$. 
Using the rest of the conditions, we find
\beq
\gU^{-1}\bareta_{i_1}\cdots\bareta_{i_a}|0\rangle
=\spst^{-a} 
\langle 0|\eta_1\cdots\eta_{s\over 2}\bareta_{i_1}\cdots\bareta_{i_a},
\label{slnforU}
\eeq
for $1\leq i_1<\cdots <i_a\leq \mbox{${s\over 2}$}$.
This is the solution to (\ref{eqnforU}). 
We shall denote it by $\gU_{s}$.
Note that the right hand side of (\ref{slnforU}) 
changes by a factor of $(-1)^a$ if we switch the sign of $\spst$.
Therefore $\gU_s$'s for the opposite phase $\spst$ 
are related by $\gU_{s(i)}=\sigma\circ \gU_{s(-i)}$.
Note that
\beq
(-1)^{|\gU_s|}=(-1)^{s\over 2}.
\eeq
Let us see if $\gU=\gU_{s}$ satisfies the condition (\ref{condforU})
for an o-isomorphism.
Let us compute the pairing $\langle (\gU^T)^{-1}\iota v,w\rangle$
for $v=\bareta_{i_1}\cdots\bareta_{i_a}|0\rangle$ and
$w=\bareta_{j_1}\cdots\bareta_{j_b}|0\rangle$:
\beqa
\langle (\gU^T)^{-1}\iota v,w\rangle&=&
(-1)^{|\gU|}\langle (\gU^{-1})^T\iota v,w\rangle
\nn\\
&=&(-1)^{|\gU|}(-1)^{a|\gU|}
\langle \iota v,\gU^{-1}w\rangle
\nn\\
&=&(-1)^{|\gU|}(-1)^{a|\gU|}(-1)^a\langle \gU^{-1}w,v\rangle
\nn\\
&=&(-1)^{|\gU|+a|\gU|+a}\spst^{-b}\langle 0|\eta_1\cdots\eta_{s\over 2}
\bareta_{j_1}\cdots\bareta_{j_b}
\bareta_{i_1}\cdots\bareta_{i_a}|0\rangle
\nn
\eeqa
On the other hand,  we have
\beqa
\langle \gU^{-1}\sigma v,w\rangle&=&
(-1)^a\langle \gU^{-1}v,w\rangle
\nn\\
&=&
(-1)^a\spst^{-a}\langle 0|\eta_1\cdots\eta_{s\over 2}
\bareta_{i_1}\cdots\bareta_{i_a}
\bareta_{j_1}\cdots\bareta_{j_b}|0\rangle
\nn\\
&=&(-1)^{a+ab}\spst^{-a}\langle 0|\eta_1\cdots\eta_{s\over 2}
\bareta_{j_1}\cdots\bareta_{j_b}
\bareta_{i_1}\cdots\bareta_{i_a}|0\rangle
\nn
\eeqa
Note that these are non-zero only when $a+b={s\over 2}$.
Using this fact and also the relation $|\gU|\equiv {s\over 2}$ (mod 2), 
we find
$$
\langle (\gU^T)^{-1}\iota v,w\rangle
=(-1)^{{s\over 2}+a}\spst^{-b+a}\langle \gU^{-1}\sigma v,w\rangle
=\spst^{s\over 2}\langle \gU^{-1}\sigma v,w\rangle.
$$
This proves that (\ref{condforU}) is indeed satisfied,
\beq
\gU_s(\gU_s^T)^{-1}\imath=
\spst^{s\over 2}\sigma.
\label{Uscon}
\eeq

The construction can be done in the same way in the general case (\ref{rprO}).
When the multiplicity $s_+$ of eigenvalue $1$ is even
(and hence so is $s_-$), we can find complex combinations of $\xi_i$'s,
$\eta_i$ and $\bareta_i$ ($i=1,\ldots, {s\over 2}$),
such that $\gU\eta_i^T \gU^{-1}$ is proportional to
$\eta_i$ for all $i$. 
Then we find that $\gU^{-1}|0\rangle$ is proportional to
$\langle 0|\eta_1\cdots\eta_{s\over 2}$.
When $s_+$ is odd (and hence so is $s_-$), one can find complex combinations
such that $\gU\eta_i^T \gU^{-1}$ is proportional to
$\eta_i$ for all but one $i$, say $i=1$, where it is proportional to 
$\bareta_1$. We then find that $\gU^{-1}|0\rangle$ is proportional to
$\langle 0|\eta_2\cdots\eta_{s\over 2}$.
The $\gU^{-1}$ transform of other vectors can be obtained by using 
(\ref{confU}).
We can show
\beq
\begin{array}{rcl}
(-1)^{|\gU|}\!\!&=&\!\!(-1)^{s_+-s_-\over 2},\\[0.2cm]
\gU(\gU^T)^{-1}\imath \!\!&=&\!\!\spst^{s_+-s_-\over 2}\sigma.
\end{array}
\label{UUTs}
\eeq

\subsubsection*{(ii) $r$ and $s$ odd}

Let us first discuss the parity transformation of the auxiliary fermions,
$\xia^{(R)}$ and $\xia^{(L)}$.
We require that the $\xia^{(R)}\xia^{(L)}=1$ condition is maintained
so that the parity preserves the from $\phi\otimes |0\rangle_{{}_{\rm aux}}$
of vectors. We must also require the condition $\Po^2=(-1)^F$ and
that preserves the hermiticity $\xia^{(R)\dag}=\xia^{(R)}$
and $\xia^{(L)\dag}=-\xia^{(L)}$.
This is satisfied if the transformation is
\beq
\xia^{(R)}\longrightarrow \phxi\xia^{(L)},
\qquad
\xia^{(L)}\longrightarrow \phxi\xia^{(R)},
\label{say1}
\eeq
where $\phxi$ is $i$ or $-i$ and is independent 
of the phase $\spst$.
Note that the reasoning for the correlation of parity transformation
of the physical boundary fermions $\xi_i$'s
with the phase $\spst$ does not apply to to auxiliary fermions. 
Also, the $\xia^{(R)}\xia^{(L)}=1$ condition would 
be violated in the Ramond-sector if we insisted such a correlation.

If the parity on the Chan-Paton factor is written as (\ref{PCP}), then 
$\gU=\gU_{(\spst)}$
must satisfy (\ref{confU}) as well as
\beq
\gU\xia^T\gU^{-1}=(-1)^{|\gU|}\phxi\xia
\label{UxU1}
\eeq
for both $\spst=-i$ and $+i$.
If we put $\xi_{s+1}=\xia$ as the last member of the extended
Clifford algebra, we have 
\beq
\gU_{(\spst)}\xi_i^T\gU_{(\spst)}^{-1}
=(-1)^{|\gU_{(\spst)}|}\spst\sum_{j=1}^{s+1}O^{(\spst)}_{ij}\xi_j
\label{UxU2}
\eeq
with
$$
O^{(\mp i)}=\left(\begin{array}{c|c}
O&{\bf 0}\\
\hline
{\bf 0}^t&\pm i\phxi
\end{array}\right).
$$
The solutions for $\gU_{(i)}$ and $\gU_{(-i)}$ must be related by
$$
\gU_{(i)}=\xia\circ\sigma\circ \gU_{(-i)},
$$
up to scalar multiplication. In particular, they have opposite statistics.
Using (\ref{UUTs}) we find
\beqa
\gU_{(\mp i)}(\gU_{(\mp i)}^T)^{-1}\imath
&=&(\mp i)^{s_+-s_-\pm i\phxi\over 2}\sigma,
\label{UUTi}
\\
(-1)^{|\gU_{(\mp i)}|}&=&(-1)^{s_+-s_-\pm i\phxi\over 2}.
\label{grUpm}
\eeqa

The present discussion provides the background for the structure
of Type IIA D9-branes described in Section~\ref{subsec:o-isom}
and \ref{subsec:Ramond}.
For example, the condition (\ref{UxU1}) is nothing but (\ref{xicond2}),
and the relation between $\gU_{(i)}$ and $\gU_{(-i)}$ shown above
yields (\ref{Amp}).

\subsubsection*{(iii) $r$ even and $s$ odd
({\it resp}. $r$ odd and $s$ even)}

The parity transform of $\xia$ is already determined in {\bf (ii)},
and the transformation of $\xia'$ is uniquely fixed by the requirement that
the projection condition $\xia\xia'=1$ is invariant:
\beq
\xia\longrightarrow \phxi\xia\quad
\mbox{and}\quad
\xia'\longrightarrow -\phxi\xia'.
\eeq
The parity on the Chan-Paton factor,
$\Po_{\rm CP}:\Hom^{}_{\C}(V_r,V_{s+1})
\to \Hom^{}_{\C}(V_{s+1},V_r)$ 
({\it resp}. $\Po_{\rm CP}:\Hom^{}_{\C}(V_{r+1},V_s)
\to \Hom^{}_{\C}(V_s,V_{r+1})$),
is defined as in (\ref{PCP})
where $\gU'$ and $\gU$ are as in {\bf (i)} and {\bf (ii)}
({\it resp}. {\bf (ii)} and {\bf (i)}).

\subsection{D-Branes On Top Of The O-Plane}
\label{subsec:ABS}

We study the orientifold action on a particular class of D-branes 
--- BPS D-branes on top of the orientifold plane.

Let us consider the BPS D$(9-k)$-brane in the Minkowski spacetime
at $x^1=\cdots =x^k=0$.
Its realization as a D9-brane configuration is well-known
and is referred to as the ``Atiyah-Bott-Shapiro (ABS) construction''
in string theory literature.
It can be described as a system of $k$ boundary fermions, 
$\xi_1,\ldots,\xi_k$, with the action
\beq
S_{\rm bdry}=\int_{\partial\Sigma}\dd t \left\{\, 
{i\over 4}\sum_{i=1}^k\xi_i{\dd\over \dd t}\xi_i
-{i\over 2}\sum_{i=1}^k\psi^i\xi_i
-{1\over 2}\sum_{i=1}^k(x^i)^2
\,\right\}.
\label{ABSbf}
\eeq
After quantizing $\xi_i$'s (together with the auxiliary fermion $\xia$
for Type IIA ($k$ odd)), the boundary interaction takes the form (\ref{At})
where the gauge field is trivial, $A=0$,
and the tachyon has the profile
\beq
T(x)=\vec{x}\cdot\vec{\xi}=\sum_{i=1}^kx^i\xi_i.
\label{ABS}
\eeq
It is represented on the trivial vector bundle with the fibre
$$
V=\left\{\begin{array}{ll}
V_k&\mbox{IIB ($k$ even)},
\\
V_{k+1}&\mbox{IIA ($k$ odd)}.
\end{array}\right.
$$

Now we put this configuration on top of the O$(9-k)$-plane
of the involution $\inv$ that acts by the sign flip of 
$x^1,\ldots, x^k$.
Since the fermions $\psi^i(\tau)$ transform under $\inv\Omega$
into $-\spst\psi^i(-\tau)$ for
 $i=1,\ldots,k$, the invariance of the boundary interaction 
$\int_{\partial\Sigma}\dd\tau\sum_{i=1}^s\psi^i(\tau)\xi_i(\tau)$
requires that
the parity transform of the boundary fermions is
\beq
\xi_i(\tau)\,\longrightarrow \,\spst\,\xi_i(-\tau).
\label{ptxk}
\eeq
Then the condition for the o-isomorphism (\ref{confU}) reads
\beq
\gU\,\xi_i^T\,\gU^{-1}=(-1)^{|\gU|}\spst\,\xi_i,\qquad i=1,\ldots, k.
\label{eqnforU}
\eeq
This also follows from the condition (\ref{Tco})
of the orientifold invariance of the tachyon profile,
which reads for (\ref{ABS}) as
$$
\sum_{i=1}^kx^i\xi_i=(-1)^{|\gU|}\gU\left(\spst\sum_{i=1}^k(-x^i)\xi_i^T\right)
\gU^{-1}.
$$
For $k$ odd (Type IIA), we also need
$$
(-1)^{|\gU|}\gU\xia^T\gU^{-1}=\phxi\xia.
$$
The solutions are obtained in the previous subsection and are found to satisfy
the o-isomorphism condition (\ref{condforU}).
Let us write them down together with the values of the crosscap section
$\gc$. 
For $k$ even (Type IIB),
\beq
\gU=\gU_{k},\,\,\quad
\gc=\spst^{k\over 2}.
\label{cforiib}
\eeq
For $k$ odd (Type IIA),
\beq
~~\left\{\begin{array}{lll}
\gU=\gU_{k+1},&
\gc=\spst^{k+1\over 2}&\mbox{~for~ $\spst=\phxi$},\\
\gU=\xia\circ\gU_{k+1},~&
\gc=\spst^{k-1\over 2}&\mbox{~for~ $\spst=-\phxi$}.
\end{array}\right.
\label{cforiia}
\eeq

Let us next put $N$ BPS D-branes on top of the O-plane.
An open string state has a general form $|\psi,ij\rangle$ where 
$\psi$ is the state of the conformal field theory and $i,j=1,\ldots, N$
are the Chan-Paton indices. The parity acts on the states as
\beq
\Po:~|\psi,ij\rangle\longmapsto \sum_{i',j'=1}^N
\gamma_{j'j}|\inv\Omega(\psi),j'i'\rangle\gamma^{-1}_{ii'}.
\label{Pcft}
\eeq
In terms of the D9-brane configuration, the $N$ D-branes
can be realized on the Chan-Paton space
$V=V_\abs\otimes \C^N$ with the grading
$\sigma=\sigma_\abs\otimes {\bf 1}_N$ and
the tachyon profile
$T(x)=T_\abs(x)\otimes {\bf 1}_N$.
Here we put the subscript ``ABS'' for all
the quantities found above for a single D-brane.
The parity action (\ref{Pcft}) corresponds to the choice
\beq
\gU=\gU_\abs\otimes\gamma
\label{NABS}
\eeq
of an o-isomorphism on the D9-branes. For this $\gU$ we have
\beq
\gU(\gU^T)^{-1}\imath=\gU_\abs(\gU^T_\abs)^{-1}\imath_\abs\otimes 
\gamma(\gamma^t)^{-1}
=\gc_\abs\sigma_\abs\otimes \gamma(\gamma^t)^{-1}.
\label{relUabsU}
\eeq
Compare the number $\gc_\abs$, given in (\ref{cforiib}) and (\ref{cforiia}),
and the number $\gc$ in the formula (\ref{BaSt}).
We see that they agree for the O${}^-$-type
and is opposite in sign for the O${}^+$-type.
Therefore, the formula (\ref{BaSt}) leads to the condition
\beq
\gamma(\gamma^t)^{-1}=\pm {\bf 1}_N\qquad
\mbox{for O${}^{\mp}$-type}.
\label{reqorcons}
\eeq
Namely, $\gamma$ is symmetric for O${}^-$ and
antisymmetric for O${}^+$.
By a suitable basis change of $\C^N$, which does $\gamma\to M\gamma M^t$,
we can set $\gamma={\bf 1}_N$ for O${}^-$
and $\gamma={\bf J}_N$ for O${}^+$, where
$$
{\bf J}_N=
\left(\begin{array}{cc}
0&-{\bf 1}_{N/2}\\
{\bf 1}_{N/2}&0
\end{array}\right).
$$
This means that the gauge group on
the D-branes is $O(N)$ for O${}^-$
and $USp(N)$ for O${}^+$. 
We have seen that the formula (\ref{BaSt}) leads to this standard
fact on BPS D-branes on top of the O-plane.

Turning around the logic, let us require that
the $N$ BPS D-branes on top of the O-plane have gauge group
$O(N)$ for O${}^-$ and $USp(N)$ for O${}^+$,
namely, that $\gamma$ must obey the equation (\ref{reqorcons}).
Via the relation (\ref{relUabsU}), this means that the number $\gc$ 
is given by (\ref{BaSt}) for this particular configuration.
As remarked in Section~\ref{subsec:o-isom},
the section $\gc$ is common for all D-branes in the theory.
We therefore conclude that the formula (\ref{BaSt}) must
hold in {\it any} D9-brane configuration.

This may be regarded as the first though indirect derivation of 
the formula (\ref{BaSt}).
In the next section, we will give a direct derivation of the formula
by studying D9-branes without excitation.

\medskip

\section{The Structure Of D9-Brane Chan-Paton Factor}
\label{sec:direct}

\newcommand{\Star}{\mbox{\Large $\star$}}

In this section, we study the consistency condition of
the o-isomorphism
for the D9-branes in the Type II orientifold on $\R^{10}$
associated with the involution
$$
\inv:(x^0,x^1,\ldots,x^k,x^{k+1},\ldots, x^9)\,\,\longmapsto\,\,
(x^0,-x^1,\ldots,-x^k,x^{k+1},\ldots, x^9)
$$
which has a single O$(9-k)$-plane. In particular, we will prove the formula
(\ref{BaSt}) that relates the value of the crosscap section $\gc$
at the O-plane to its type and dimension.

The main ground of study will be the open strings stretched between
the D$(9-k)$-branes at the O$(9-k)$-plane and the D9-branes.
The idea is to use the basic requirement $\Po^2=(-1)^F$ on these sectors
together with the knowledge on the Chan-Paton factor for the D$(9-k)$-branes.
However, it is extremely subtle to define the parity on such boundary changing
sectors, especially with the freedom of choosing a phase factor.
Facing this problem, we follow Gimon-Polchinski \cite{GP}
and employ the operator product rule
\beq
\Po(\Phi_2\cdot \Phi_1)=(-1)^{|\Phi_1|\cdot |\Phi_2|}
\Po(\Phi_1)\cdot \Po(\Phi_2)
\label{prodrule}
\eeq
which holds if at least one of $\Phi_1$ and $\Phi_2$ is in the NS sector.
(In this section, 
we denote open string states and the corresponding vertex operators
using the same symbols. We shall also abbreviate $(-1)^{|\Phi|}$
as $(-1)^{\Phi}$ when there is no danger of confusion.)
\begin{figure}[htb]
\psfrag{phi}{\small $\Phi$}
\psfrag{Pphi}{\small $\Po(\Phi)$}
\psfrag{P2phi}{\small $\Po^2(\Phi)$}
\psfrag{to}{$\longrightarrow$}
\psfrag{x}{$\times \,\,(-1)^{\Phi}$}
\psfrag{P}{\small $\Po$}
\psfrag{9-k}{\small D${}_{9-k}$}
\psfrag{9}{\small D${}_9$}
\centerline{\includegraphics{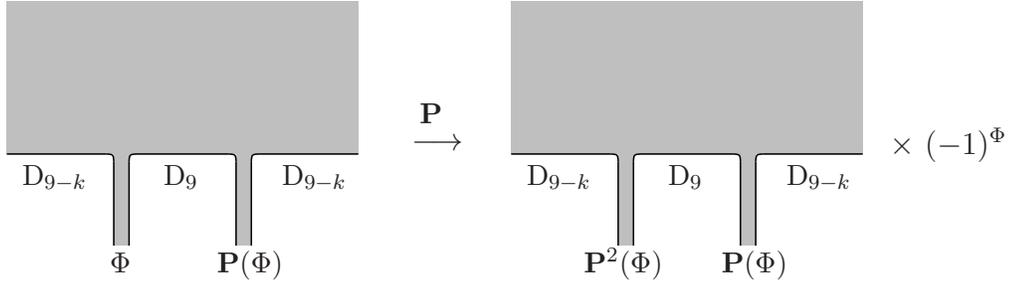}}
\caption{The product $\Po(\Phi)\cdot \Phi$ and its parity transform.}
\label{fig:OPE}
\end{figure}
Applying this in the set-up of Figure~\ref{fig:OPE},
we find another form of the basic requirement
\beq
\Po(\Po(\Phi)\cdot \Phi)=(-1)^{\Phi}\Po(\Phi)\cdot \Po^2(\Phi)
=\Po(\Phi)\cdot \Phi.
\label{basicid}
\eeq
If $\Phi$ is in the D$(9-k)$-D9 sector, then the product 
$\Po(\Phi)\cdot \Phi$
is in the D$(9-k)$-D$(9-k)$ sector in which we know very well about the parity
operator.
The equation (\ref{basicid}) will give us a strong constraint on the structure
of Chan-Paton factor of the D9-branes.

A simple reinterpretation of this analysis
will also determine the structure of Chan-Paton factor for
D-branes of all dimensions in Type I string theory.
In addition, in Section~\ref{subsec:Ra}
we shall use the same type of argument to 
show the relation $\Po^2=(-1)^F$ in the Ramond sector for the D$9$-D$9$ string,
as promised in Section~\ref{subsec:Ramond}.

For the most part, where we discuss the product NS $\times$ NS $\to$ NS, 
we shall use vertex operators in the $0$-picture \cite{FMS} 
so that we can ignore the ghost sector ---
all the states and vertex operators that appear in the discussion will be
the ones from the ``matter'' sector, i.e., the $c=15$ superconformal sigma model
on the ten dimensional Minkowski space.
In Section~\ref{subsec:Ra}, 
we consider the product of Ramond vertex operators 
and the ghost sector needs to be included
in the discussion.

\subsection{D$p$-D$q$ Strings}

We first record the mode expansions of the fermions
and the parity action on the modes, for the open string stretched between
a flat D$q$-brane in $\R^{10}$ and a flat D$p$-brane inside it 
($p\leq q$).
The boundary condition of the field $x^{\mu}$ at the two ends of the string
is, depending of the direction $\mu$,
NN, ND (or DN), or DD, where ``N'' and ``D'' stand for
Neumann and Dirichlet respectively.
The mode expansion of fermions is, for the NS-sector of the type 
$(++)$,
\beq
\psi^{\mu}_{\pm}(t,\s)=
\left\{\begin{array}{ll}
\sum_{r\in \Z+{1\over 2}}\psi^{\mu}_r(t)\e^{\mp ir\s}&{\rm NN},\\
\pm\sum_{n\in \Z}\psi^{\mu}_n(t)\e^{\mp in\s}&{\rm ND},\\
\sum_{n\in\Z}\psi^{\mu}_n(t)\e^{\mp in\s}&{\rm DN},\\
\pm \sum_{r\in \Z+{1\over 2}}\psi^{\mu}_r(t)\e^{\mp ir\s}&{\rm DD}.
\end{array}\right.
\label{fexpa}
\eeq
The mode expansion for the other pairs of spin structures is
easy to obtain from this by noting that 
the replacement $(+)\to (-)$ can be implemented by
N $\to$ D and D $\to$ N.
For example, the expansion of an NN direction 
in the R-sector of the type $(-+)$
is the same as the one for a DN direction in (\ref{fexpa}).

The space of states has a degeneracy due to the
fermionic zero modes that obey the Clifford algebra relations,
$\{\psi_0^{\mu},\psi_0^{\nu}\}=\eta^{\mu\nu}$.
The space of states is thus in a spinor representation of $SO(|p-q|)$
in the NS-sector, and of $SO(p,1)\times SO(9-q)$ in the R-sector.
When $|p-q|$ is even, there is a $\Z_2$-grading operator $(-1)^F$ with which
we can define the GSO projection.
When $|p-q|$ is odd, there is no $\Z_2$-grading 
as there are odd number of zero modes. 
This requires us to
have an odd number of fermions from boundaries, as suggested in
\cite{WittenK}, so that the total space of states has a $\Z_2$-grading.

Let us now look at the action of the parity on the mode that appear
in (\ref{fexpa}).
The $\Omega$ parity,
$\psi_{\pm}(t,\s)\to\mp\psi_{\mp}(t,-\s-\pi)$, transforms the modes 
in the NS-sector of the type $(++)$ as
\beqa
&\mbox{NN$\to$ NN}:\psi^{\mu}_r\to \e^{\pi i r}\psi^{\mu}_r,\qquad
\label{NN}
\\
&\mbox{ND $\to$ DN}:
\psi_n^{\mu}\to -(-1)^n\psi_n^{\mu},\qquad
\mbox{DN $\to$ ND}:
\psi_n^{\mu}\to (-1)^n\psi_n^{\mu},\label{NDDN}\\
&\mbox{DD$\to$ DD}:\psi^{\mu}_r\to -\e^{\pi i r}\psi^{\mu}_r.
\label{DD}
\eeqa

Let us describe the mode expansions and parity transform
of fields $\psi_{\pm}^{\mu}$ for $\mu=1,\ldots, k$
in the set-up of Figure~\ref{fig:OPE} such that the state $\Phi$ is in the
NS-sector.
Let $\psi^{\rm plane}_{\pm}$ be the components 
on the upper-half plane of one of these fields.
For the $(+)$ spin structure at the boundary ${\rm Im}(z)=0$,
they obey the boundary condition
$\psi^{\rm plane}_+=\psi^{\rm plane}_-$ on the segment $0<z<1$
and the opposite boundary condition 
$\psi^{\rm plane}_+=-\psi^{\rm plane}_-$
on the other parts $z<0$ and $z>1$.
\begin{figure}[htb]
\psfrag{0}{$0$}
\psfrag{1}{$1$}
\psfrag{p1}{$\psi_n'$}
\psfrag{p2}{$\psi_n''$}
\psfrag{p3}{$\psi_r$}
\psfrag{D8}{D${}_{9-k}$}
\psfrag{D9}{D${}_9$}
\centerline{\includegraphics{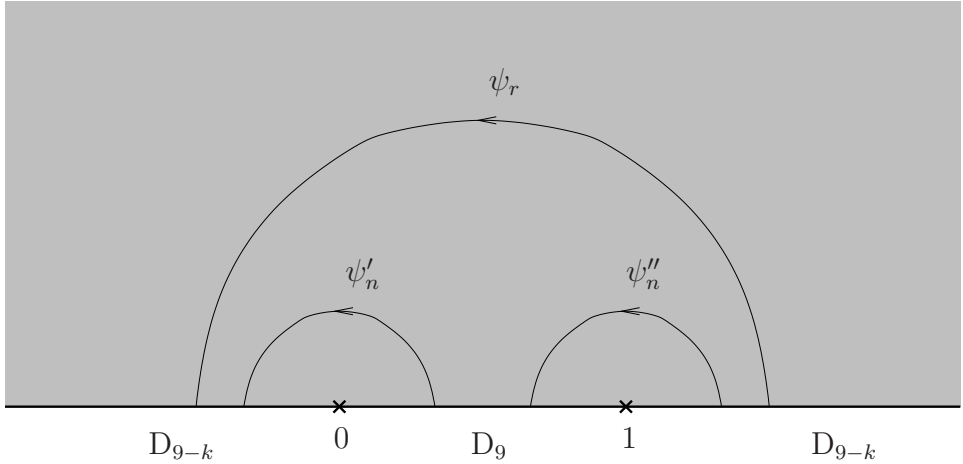}}
\caption{The mode expansions}
\label{fig:modes}
\end{figure}
We consider three kinds of mode expansion as depicted in
Figure~\ref{fig:modes},
\beq
\begin{array}{l}
\displaystyle 
\psi^{\rm plane}_-=-\sum_{r\in \Z+{1\over 2}}\psi_r
(z-\mbox{${1\over 2}$})^{-r-{1\over 2}}
~=\,\sum_{n\in\Z}\psi_n'z^{-n-{1\over 2}}
~=-\sum_{n\in\Z}\psi_n''(z-1)^{-n-{1\over 2}},
\\
\displaystyle 
\psi^{\rm plane}_+=~~\,\sum_{r\in \Z+{1\over 2}}\psi_r
(\bz-\mbox{${1\over 2}$})^{-r-{1\over 2}}
~=\,\sum_{n\in\Z}\psi_n'\bz^{-n-{1\over 2}}
~=~~\,\sum_{n\in\Z}\psi_n''(\bz-1)^{-n-{1\over 2}},
\end{array}
\label{exp3}
\eeq
where the square roots are defined so that $\sqrt{a}$ is real positive for
real positive $a$.
Let us look at the action of the parity $z\to 1-\bz$ that exchanges
$z=0$ and $z=1$.
The $\inv\Omega$ parity transforms the fields as
$\psi_{\pm}^{\rm plane}(z,\bz)
\to -(-i)\psi_{\mp}^{\rm plane}(1-\bz,1-z)$.
The first minus sign comes from the involution $\inv$
and the phase $(-i)$ is from the definition of $\Omega$,
see (\ref{Ompla}).
The transformation of the modes are
\beqa
&&\psi_r\longrightarrow \e^{\pi i r}\psi_r,\label{DD1}\\
&&\psi_n'\longrightarrow -(-1)^n\psi_n'',\label{DNND1}\\
&&\psi_n''\longrightarrow (-1)^n\psi_n'.\label{NDDN1}
\eeqa
For the other spin structure $(-)$ of the boundary,
there is an extra minus sign for the expansion of
$\psi_-^{\rm plane}$ in (\ref{exp3})
and on the hand sides in all of (\ref{DD1}) (\ref{DNND1}) and (\ref{NDDN1}).

\subsection{$k$ Even (Type IIB)}
\label{subsec:keven}

We work with the formulation in which
an open string state takes the form
\beq
\Phi=\phi\otimes \psi,
\label{vo}
\eeq
where $\phi$ is from the Chan-Paton factor and $\psi$ is a state
in the bulk sector.
The product is given by
\beq
(\phi_1\otimes \psi_1)\cdot (\phi_2\otimes \psi_2)
=(-1)^{\psi_1\phi_2}(\phi_1\cdot\phi_2)\otimes (\psi_1\cdot \psi_2),
\eeq
and the parity takes the form
\beq
\Po(\phi\otimes \psi)=\Po_{\rm CP}(\phi)\otimes \Po_{\rm bulk}(\psi)
(-1)^{\phi \Po_{\rm bulk}}.
\eeq
Let $\Phi$ be in the NS sector of the D$(9-k)$-D9 string
and let us take the product with its parity image $\Po(\Phi)$,
\beq
\Po(\Phi)\cdot \Phi=(-1)^{\phi \psi}(\Po_{\rm CP}(\phi)\cdot\phi)
\otimes (\Po_{\rm bulk}(\psi)\cdot \psi).
\label{PVV}
\eeq
It is in the NS sector of the D$(9-k)$-D$(9-k)$ string.
The bulk parity $\Po_{\rm bulk}$ is even in this sector
and hence the parity image of $\Po(\Phi)\cdot \Phi$ is
\beq
\Po(\Po(\Phi)\cdot \Phi)=
(-1)^{\phi \psi}\Po_{\rm CP}(\Po_{\rm CP}(\phi)\cdot \phi)
\otimes \Po_{\rm bulk}(\Po_{\rm bulk}(\psi)\cdot \psi).
\label{PPVV}
\eeq
In view of the requirement (\ref{basicid}), we would like to compare
the expressions (\ref{PVV}) and (\ref{PPVV}).
To simplify the Chan-Paton factor of (\ref{PPVV}), 
we note the following property of the
parity operators of the form (\ref{PCP}):
\beq
\Po_{\rm CP}^{a,c}(\phi_2\cdot \phi_1)
=(-1)^{\phi_1\phi_2+\phi_1\Po_{\rm CP}^{b,c}}
\Po_{\rm CP}^{a,b}(\phi_1)\cdot \Po_{\rm CP}^{b,c}(\phi_2).
\label{PCPpr}
\eeq
Here $\phi_1$ and $\phi_2$ are the Chan-Paton factors for the
$a$-$b$ and $b$-$c$ strings and we put the superscript as
$\Po_{\rm CP}^{a,b}$ in order to show the domain sector
of the parity.
The property (\ref{PCPpr}) can be proved by a straightforward computation:
\beqa
{\rm LHS}&=&\gU_a(\phi_2\cdot\phi_1)^T\gU_c^{-1}(-1)^{\gU_c(\phi_1+\phi_2)}
\nn\\
&=&(-1)^{\phi_1\phi_2}\gU_a\phi_1^T\gU_b^{-1}\cdot \gU_b\phi_2^T\gU_c^{-1}
(-1)^{\gU_c(\phi_1+\phi_2)}
\nn\\
&=&(-1)^{\phi_1\phi_2}\Po_{\rm CP}^{a,b}(\phi_1)(-1)^{\phi_1 \gU_b}
\cdot \Po_{\rm CP}^{b,c}(\phi_2)(-1)^{\phi_2 \gU_c}
(-1)^{\gU_c(\phi_1+\phi_2)}
\nn\\
&=&(-1)^{\phi_1\phi_2}\Po_{\rm CP}^{a,b}(\phi_1)\cdot 
\Po_{\rm CP}^{b,c}(\phi_2)
(-1)^{\phi_1(\gU_b+\gU_c)}~=~{\rm RHS}.
\nn
\eeqa
Using (\ref{PCPpr}) we find that the Chan-Paton factor of (\ref{PPVV})
is given by
\beq
\Po_{\rm CP}(\Po_{\rm CP}(\phi)\cdot \phi)=(-1)^{\phi}\Po_{\rm CP}(\phi)
\cdot \Po_{\rm CP}^2(\phi).
\eeq
The requirement (\ref{basicid}) on (\ref{PVV}) and (\ref{PPVV})
yields the condition that 
\beq
\Po_{\rm bulk}(\Po_{\rm bulk}(\psi)\cdot \psi)=\Star
 \, \Po_{\rm bulk}(\psi)\cdot \psi,
\label{star}
\eeq
for some scalar $\Star\in \C$, and that
\beq
\Po_{\rm CP}^2(\phi)=\Star^{-1}(-1)^\phi \phi.
\label{coCP}
\eeq
The latter gives a strong constraint on the structure of
the Chan-Paton factor for the D9-brane.
For this purpose, it is important to find
the scalar $\Star$ in the relation (\ref{star}).

We first consider the case $k=2$. We are interested 
in the NS sector of the D7-D9 and D9-D7 strings.
Due to the fermionic zero modes in the DN/ND directions,
$\psi_0^1$ and $\psi_0^2$, the ground states
are two-fold degenerate. Let $|\!\uparrow\rangle$
and $|\!\downarrow\rangle$ be the ground states which
are characterized by
$$
\left(\psi_0^1+i\psi_0^2\right)|\!\downarrow\rangle=0\quad
\mbox{and}\quad
|\!\uparrow\rangle=\left(\psi_0^1-i\psi_0^2\right)|\!\downarrow\rangle.
$$
With respect to the $U(1)$ symmetry of rotations in the $\mu=1,2$ directions,
the two states have different charges --- they differ
by the charge of $\left(\psi_0^1-i\psi_0^2\right)$, 
which we normalize to be $1$.
In addition, the spectrum must be symmetric under the charge conjugation.
Thus, the ground states $|\!\uparrow\rangle$ 
and $|\!\downarrow\rangle$ have charges ${1\over 2}$
and $-{1\over 2}$ respectively.
Note that the parity action $\inv\Omega$ commutes with the
rotational symmetry and hence conserves the charges.
We thus find that 
$\Po_{\rm bulk}|\alpha\rangle_{{}_{79}}\propto |\alpha\rangle_{{}_{97}}$,
for both
$\alpha=\,\uparrow$ and $\downarrow$.
In particular the product
$\Po_{\rm bulk}|\alpha\rangle_{{}_{79}}
\cdot |\alpha\rangle_{{}_{79}}$ 
is proportional to
$|\alpha\rangle_{{}_{97}}\cdot |\alpha\rangle_{{}_{79}}$ 
and hence has charge 
${\epsilon_{\alpha}\over 2}+{\epsilon_{\alpha}\over 2}=\epsilon_{\alpha}$, 
where we introduced
$\epsilon_{\uparrow}=1$ and $\epsilon_{\downarrow}=-1$.
The leading term in the operator product expansion is
the primary state in that charge sector,
\beq
\Po_{\rm bulk}|\alpha\rangle_{{}_{79}}
\cdot |\alpha\rangle_{{}_{79}}~\sim~ 
(\psi_{-{1\over 2}}^1-i\epsilon_{\alpha}
\psi^2_{-{1\over 2}})|0\rangle_{{}_{77}}.
\label{77}
\eeq
Since the right hand side is odd, 
we find that the parity $\Po_{\rm bulk}$ is odd
in the D7-D9 sector.
Let us look at the parity action on the state (\ref{77}).
We refer to (\ref{DD1}) for the $\inv\Omega$ parity
transform of the modes $\psi^{\mu}_{-{1\over 2}}$ with $\mu=1,2$.
In the NS sector of the type $(++)$, it is
$\psi_{-{1\over 2}}^{\mu}\to \e^{-{\pi i\over 2}}\psi_{-{1\over 2}}^{\mu}
=-i\psi^{\mu}_{-{1\over 2}}$.
In the NS sector of the type $(--)$ we have the opposite sign.
Therefore,
$\Po_{\rm bulk}$ acts
on $(\psi_{-{1\over 2}}^1-i\epsilon_{\alpha}
\psi^2_{-{1\over 2}})|0\rangle_{{}_{77}}$ as multiplication by
$\mp i$ on the $(\pm\pm)$ sector, i.e., by $\spst$ as defined in 
(\ref{defspst}).
We thus found
$$
\Po_{\rm bulk}(\Po_{\rm bulk}|\alpha\rangle_{{}_{79}}
\cdot |\alpha\rangle_{{}_{79}})
=\spst \Po_{\rm bulk}|\alpha\rangle_{{}_{79}}
\cdot |\alpha\rangle_{{}_{79}}.
$$
This shows that the scalar in (\ref{star}) is $\Star=\spst$.

Generalization to all even $k$ is straightforward.
We look at the D$(9-k)$-D9 ground state of the form 
$|\vec{\alpha}\rangle^{}_{{}_{(9-k)9}}
=|\alpha_1...\alpha_{k\over 2}\rangle_{{}_{(9-k)9}}$ 
which is annihilated by
$(\psi_0^{2j-1}-i\epsilon_{\alpha_j}\psi_0^{2j})$ for 
all $j=1,\ldots,{k-1\over 2}$.
The operator product with its parity image has the leading term
\beq
\Po_{\rm bulk}|\vec{\alpha}\rangle^{}_{\!{}_{(9-k)9}}\cdot
|\vec{\alpha}\rangle^{}_{\!{}_{(9-k)9}}
~\sim~(\psi^1_{-{1\over 2}}-i\epsilon_{\alpha_1}\psi^2_{-{1\over 2}})\cdots 
(\psi^{k-1}_{-{1\over 2}}-i\epsilon_{\alpha_{k\over 2}}\psi^k_{-{1\over 2}})
|0\rangle^{}_{\!{}_{(9-k)(9-k)}}.
\label{Pvvk}
\eeq
The parity $\Po_{\rm bulk}$ acts on this state
as multiplication by $\spst\cdots\spst$ (${k\over 2}$ times), which shows that
the scalar in (\ref{star}) is $\Star=\spst^{k\over 2}$.
Also, we find $(-1)^{\Po_{\rm bulk}}=(-1)^{k\over 2}$ 
in the D$(9-k)$-D9 sector.

Let us now shift our attention to the Chan-Paton factor.
We know it for the BPS
D$(9-k)$-branes on top of the O-plane (see Section~\ref{subsec:ABS}):
The Chan-Paton vector space is a purely even space $\C^N$
and the o-isomorphism is given by a
matrix $\gamma$ such that
$$
\gamma^t\gamma^{-1}=\pm 1\qquad\mbox{for O${}^{\mp}$-type}.
$$
As the D9-branes, we take the conformally invariant boundary condition
for which the gauge and the tachyon fields have trivial profile.
Let $V$ be the Chan-Paton vector space and $\gU:V^*\to V$ be the o-isomorphism.
The Chan-Paton part of the parity operator is then defined by
\beq
\Po_{\rm CP}(\phi)=\gamma \circ \phi^T \circ \gU^{-1}(-1)^{\phi \gU}
\eeq
for $\phi\in \Hom^{}_{\C}(\C^N,V)$.
As $\gamma$ is even,
the statistics of $\gU$ coincides with that of $\Po_{\rm CP}$,
which in turn is equal to that of $\Po_{\rm bulk}$
since the total parity $\Po=\Po_{\rm CP}\otimes \Po_{\rm bulk}$ must be even.
Therefore we have
$$
(-1)^{\gU}=(-1)^{\Po_{\rm CP}}=(-1)^{\Po_{\rm bulk}}=(-1)^{k\over 2}.
$$
The square of $\Po_{\rm CP}$ is given by
\beq
\Po_{\rm CP}^2(\phi)=(-1)^{\Po_{\rm CP}} \left(\gU(\gU^T)^{-1}\imath\right) 
\circ \phi \circ \left(\gamma^t\gamma^{-1}\right).
\eeq
Since we had found $\Star=\spst^{k\over 2}$, the constraint
(\ref{coCP}) leads to
\beq
\gU(\gU^T)^{-1}\imath=\pm \spst^{k\over 2}\sigma\qquad
\mbox{for O${}^{\mp}$-type}.
\label{evenkeqn}
\eeq
This derives the formula (\ref{BaSt}) for
the D$9$-branes in the Type IIB orientifold.

\subsection{$k$ Odd (Type IIA)}
\label{subsec:kodd}

We next consider the case where $k$ is odd.
There is an odd number of fermionic zero modes
in the D$(9-k)$-D9 and D9-D$(9-k)$ strings, and
we need to place an odd number of fermions at the boundary for the D9-branes
so that we have a $\Z_2$-grading in the total sector.

\subsubsection{O8${}^-$}

We begin with the O8${}^-$ case ($k=1$).
Let us see if a single boundary fermion $\xi_1$ can do the job, 
i.e., if we can find a parity transformation of
$\xi_1$ so that we have an even parity operator $\Po$ on the D8-D9 and D9-D8 
sectors satisfying the requirement $\Po^2=(-1)^F$.
We first work with the formulation that does not introduce
auxiliary fermions of Section~\ref{sec:rbf}.
To simplify the story, we start with a single D9-brane with $\xi_1$.
We may also focus on a single D8-brane since 
$\gamma={\bf 1}$ for the D8-branes on top of O8${}^-$.
Thus, we may assume that the open string has a trivial Chan-Paton factor.

In the NS-sector of each of the D8-D9 and D9-D8 strings,
there is a single fermionic zero mode from $\psi^1_{\pm}$.
In the set up of Figure~\ref{fig:OPE} and Figure~\ref{fig:modes}
(with D${}_{9-k}$ $=$ D${}_8$)
it is $\psi_0'$ on D8-D9 and $\psi_0''$ on D9-D8
in the expansion (\ref{exp3}) of $\psi^{{\rm plane}}_{\pm}$. 
Together with the fermion $\xi_1$ which
runs along the boundary of the D9-brane condition,
they obey the algebra
\beqa
\{\psi_0',\psi'_0\}=1,\quad (\xi_1)^2=1,\quad \{\psi_0',\xi_1\}=0&&
\mbox{on D8-D9},\nn\\
\{\psi_0'',\psi_0''\}=1,\quad (\xi_1)^2=-1,\quad \{\psi_0'',\xi_1\}=0&&
\mbox{on D9-D8}.\nn
\eeqa
$\psi_0'$ and $\xi_1$ are both hermitian in the D8-D9 string sector
while $\psi_0''$ is hermitian and $\xi_1$ is
antihermitian in the D9-D8 string sector.
Due to these modes, the ground states are two-fold degenerate in each sector, 
one even and one odd.
Let $|\pm \rangle^{}_{{}_{89}}$ and 
$|\pm \rangle^{}_{{}_{98}}$
be the D8-D9 and D9-D8 ground states
characterized by
\beqa
&&\left(\sqrt{2}\psi_0'\pm i\xi_1\right)|\pm\rangle^{}_{{}_{89}}=0,
\label{condpm89}\\
&&\left(\sqrt{2}\psi_0''\mp \xi_1\right)|\pm\rangle^{}_{{}_{98}}=0.
\label{condpm98}
\eeqa
Of course, being ground states, they are annihilated by all
positive frequency modes including
$\psi'_n$ or $\psi''_n$ with $n\geq 1$ (for $|\pm\rangle^{}_{{}_{89}}$
or $|\pm\rangle^{}_{{}_{98}}$).
We shall establish the following operator product rule,
\beqa
&&|+\rangle_{{}_{98}}\cdot |+\rangle_{{}_{89}}~\sim~
|-\rangle_{{}_{98}}\cdot |-\rangle_{{}_{89}}~\sim~
|0\rangle^{}_{{}_{88}},\label{OPE+}\\
&&|-\rangle_{{}_{98}}\cdot |+\rangle_{{}_{89}}~\sim~
|+\rangle_{{}_{98}}\cdot |-\rangle_{{}_{89}}~\sim~
\psi_{-{1\over 2}}|0\rangle^{}_{{}_{88}}.\label{OPE-}
\eeqa
This implies, and in fact is equivalent to, the statement
concerning the relation between the gradings in
the D8-D9 and D9-D8 sectors: 
$|+\rangle^{}_{{}_{89}}$ and $|+\rangle^{}_{{}_{98}}$ 
are both even (or both odd). In other words,
\beq
(-1)^F_{{}_{89}}=\sqrt{2}i\xi_1\psi'_0(-1)^{F_{\rm nz}}_{{}_{89}}
\quad\mbox{and}\quad
(-1)^F_{{}_{98}}=-\sqrt{2}\xi_1\psi''_0(-1)^{F_{\rm nz}}_{{}_{98}}
\label{relgr}
\eeq
(or the simultaneous sign flip), where $(-1)^{F_{\rm nz}}_{{}_{89}}$
and  $(-1)^{F_{\rm nz}}_{{}_{98}}$
are the canonical $\Z_2$-gradings for
the non-zero modes.

For the proof of (\ref{OPE+}) and (\ref{OPE-}), we may focus on the
$c={1\over 2}$ conformal field theory of a single Majorana fermion
(i.e. $\psi_{\pm}=\psi_{\pm}^1$) with a single boundary fermion ($\xi_1$)
on the segment $0<z<1$.
We first consider a conformal mapping of the upper-half plane
\beq
w={z\over 1-z},
\label{coma}
\eeq
that maps $z=0,1$ to $w=0,\infty$.
We consider the mode expansion of $\psi_{\pm}$ with respect to
$w$ (and $\bw$)
$$
\psi_-=-\sum_{n\in\Z}\psi_nw^{-n}\sqrt{\dd w\over w},
\qquad
\psi_+=\sum_{n\in\Z}\psi_n\bw^{-n}\sqrt{\dd \bw\over \bw},
$$
and compare them with
$$
\psi_-=\psi_-^{\rm plane}\sqrt{\dd z},\qquad
\psi_+=\psi_+^{\rm plane}\sqrt{\dd \bz},
$$
where $\psi_{\mp}^{\rm plane}$ are given by (\ref{exp3}).
To make it more precise, we have 
$\sqrt{\dd w}=\sqrt{\dd\bw}$ and $\sqrt{\dd z}=\sqrt{\dd\bz}$
at the boundary with the $(+)$ spin structure. Also 
the square roots that appear in the expansions are
defined so that $\sqrt{a}$ is real positive for real positive $a$.
Then we find the following relations among the modes
\beqa
\psi_n'&=&~~~~~~\psi_n~~~~~+~a_1\,\psi_{n+1}\,\,\,+~a_2\,\psi_{n+2}\,\,
+~\cdots,
\label{relpsi1}\\
\psi_n''&=&\!\!\!-i(-1)^n\psi_{-n}+~b_1\psi_{-n-1}\,+~b_2\psi_{-n-2}\,
\label{relpsi2}
+~\cdots.
\eeqa
Note that the state $|+\rangle^{}_{{}_{89}}$ is annihilated by
$(\sqrt{2}\psi_0+i\xi_1)$ and $\psi_n$ for all $n\geq 1$.
Taking the complex conjugation, we have
\beq
{}_{{}_{89}}\langle +|\left(\sqrt{2}\psi_0-i\xi_1\right)=0,
\qquad
{}_{{}_{89}}\langle +|\psi_{-n}=0\quad(\forall n\geq 1),
\eeq
where we have used the fact that $\psi_0$ and $\xi_1$ are both hermitian
and that $\psi_n^{\dag}=\psi_{-n}$.
These can be regarded as conditions on a state inserted at $w=\infty$.
When mapped back to the $z$-plane, using (\ref{relpsi2}),
these become the following conditions at $z=1$:
\beq
\left(\sqrt{2}\psi_0''-\xi_1\right)|?\rangle^{}_{{}_{98}}=0,
\qquad
\psi_n''|?\rangle^{}_{{}_{98}}=0 \quad(\forall n\geq 1).
\eeq
These are nothing but the conditions that define the state
$|+\rangle^{}_{{}_{98}}$.
The fact that ${}_{{}_{89}}\langle +|+\rangle^{}_{{}_{89}}$ is nonzero
means that the two point function 
of $|+\rangle^{}_{{}_{89}}$ and $|+\rangle^{}_{{}_{98}}$
on the upper-half plane, or equivalently on the disc, is non-zero.
This proves that their operator product starts with $|0\rangle^{}_{{}_{88}}$.
The same argument holds for
the product of $|-\rangle^{}_{{}_{89}}$ and $|-\rangle^{}_{{}_{98}}$.
Thus we have established (\ref{OPE+}).
This also shows that $|+\rangle^{}_{{}_{89}}$ and 
$|+\rangle^{}_{{}_{98}}$ must be both even or both odd, establishing
(\ref{relgr}) up to a simultaneous sign flip.
In particular, the product of 
$|+\rangle^{}_{{}_{89}}$ and $|-\rangle^{}_{{}_{98}}$ must be
odd. 
Their operator product must start with
$\psi_{-{1\over 2}}|0\rangle^{}_{{}_{88}}$, which is the unique odd
primary state in this sector of the $c={1\over 2}$ boundary
conformal field theory.
This is also supported by the fact that
${}_{{}_{89}}\langle -|\psi_0|+\rangle^{}_{{}_{89}}$ is non-zero.
This establishes (\ref{OPE-}).

Let us now discuss the parity operator.
Recall the transformation of the modes $\psi_n'$ and $\psi_n''$
for the spin structure $(+)$ from (\ref{DNND1}) and (\ref{NDDN1}):
$$
\psi'_n\to -(-1)^n\psi''_n,\qquad
\psi''_n\to (-1)^n\psi'_n.
$$
As for $\xi_1$, 
there are two possibilities for the parity transform:
$\xi_1\to +i\xi_1$ or $\xi_1\to -i\xi_1$.
The factor of $i$ is required since $\xi_1$ is hermitian at D8-D9 
and antihermitian at D9-D8.
If we use $\xi_1\to -i\xi_1$, the condition for 
$|\pm\rangle^{}_{{}_{89}}$ in (\ref{condpm89})
is mapped to the condition for 
$|\pm\rangle^{}_{{}_{98}}$ in (\ref{condpm98}).
That is, we find $\Po|\pm\rangle^{}_{{}_{89}}\propto
|\pm\rangle^{}_{{}_{98}}$, which means that $\Po$ is even.
Applying this to the operator product rule (\ref{OPE+}), we find
$\Po|\pm\rangle^{}_{{}_{89}}\cdot |\pm\rangle^{}_{{}_{89}}\sim 
|0\rangle^{}_{{}_{88}}$.
Since $\Po|0\rangle^{}_{{}_{88}}
=|0\rangle^{}_{{}_{88}}$, we see that
$$
\Po(\Po|\pm\rangle^{}_{{}_{89}}\cdot |\pm\rangle^{}_{{}_{89}})
=\Po|\pm\rangle^{}_{{}_{89}}\cdot |\pm\rangle^{}_{{}_{89}}.
$$
Thus, we successfully find an even parity operator satisfying
(\ref{basicid}) and hence $\Po^2=(-1)^F$.
If we use $\xi_1\to +i\xi_1$ instead, the condition for 
$|\pm\rangle^{}_{{}_{89}}$ 
is mapped to the condition for 
$|\mp\rangle^{}_{{}_{98}}$.
That is, $\Po|\pm\rangle^{}_{{}_{89}}\propto
|\mp\rangle^{}_{{}_{98}}$, and hence $\Po$ is odd.
Applying this to (\ref{OPE-}), we find
$\Po|\pm\rangle^{}_{{}_{89}}\cdot |\pm\rangle^{}_{{}_{89}}\sim 
\psi_{-{1\over 2}}|0\rangle^{}_{{}_{88}}$.
Since $\Po\psi_{-{1\over 2}}|0\rangle^{}_{{}_{88}}
=-i\psi_{-{1\over 2}}|0\rangle^{}_{{}_{88}}$, we see that
(\ref{basicid}) fails but instead we have
$$
\Po(\Po|\pm\rangle^{}_{{}_{89}}\cdot |\pm\rangle^{}_{{}_{89}})
=-i \Po|\pm\rangle^{}_{{}_{89}}\cdot |\pm\rangle^{}_{{}_{89}}.
$$
For the spin structure $(-)$, $\xi_1\to +i\xi_1$ yields
an even parity operator $\Po$ which satisfies (\ref{basicid})
while
$\xi_1\to -i\xi_1$ yields an odd parity $\Po$ such that
(\ref{basicid}) fails but 
$
\Po(\Po|\pm\rangle^{}_{{}_{89}}\cdot |\pm\rangle^{}_{{}_{89}})
=+i \Po|\pm\rangle^{}_{{}_{89}}\cdot |\pm\rangle^{}_{{}_{89}}
$
holds.

To summarize: {\it For O8${}^-$, a single D9-brane
with a single boundary fermion $\xi_1$ is admissible, as long as
the parity transform is $\xi_1\to \spst\xi_1$.
}

Let us discuss the Chan-Paton structure in the formalism that
includes auxiliary fermions.
The D8-D9 or D9-D8 string with a single boundary fermion $\xi_1$
on D9 is of the type {\bf (iii)} in the terminology of Section~\ref{sec:rbf}.
Thus we introduce a pair of auxiliary fermions,
$\xia$ and $\xia'$,
and impose the projection condition 
$\xia\xia'=1$.
These two fermions transform oppositely under the parity:
if $\xia\to\phxi\xia$
then $\xia'\to -\phxi\xia'$,
where $\phxi$ is fixed ($i$ or $-i$)
independently of the boundary spin structure.
$\xia$ is included in the Chan-Paton factor
while $\xia'$ is quantized together with the bulk modes.
Thus, the Chan-Paton factor consists of
$\xi_1$ and $\xia$ that transform under the parity as
$$
\xi_1\to \spst\xi_1,\quad
\xia\to \phxi\xia.
$$
The extended Chan-Paton vector space is therefore 
the 2-dimensional space $V_2$
of the irreducible representation of the Clifford algebra
generated by $\xi_1$ and $\xia$.
When $\spst=\phxi$, the two transform
in the same way, and therefore the o-isomorphism is $\gU=\gU_{2(\phxi)}$.
Here $\gU_{s(\mp i)}$ is the isomorphism introduced in 
Section~\ref{subsec:bfPa}. 
When $\spst=-\phxi$, they transform oppositely
and hence we take $\gU=\xia\gU_{2(-\phxi)}$ 
as the o-isomorphism.
Note that they obey
\beq
\gU(\gU^T)^{-1}\imath
=\left\{\begin{array}{lll}
\phxi\sigma&\mbox{for $\gU=\gU_{2(\phxi)}$}& (\spst=\phxi),\\
\sigma&\mbox{for $\gU=\xia\gU_{2(-\phxi)}$}&(\spst=-\phxi).
\end{array}\right.
\label{O8pr}
\eeq
This shows the formula (\ref{BaSt}) for the O$8^-$ case ($[k]=1$).

\subsubsection{The General Case}

It is straightforward to extend the above analysis for the general 
(odd) codimension $k$ and the type of the O-plane.

Let us first study the parity operator
on the open strings stretched between a single D$(9-k)$-brane 
and a single D9-brane equipped with a single boundary fermion $\xi_1$.
On the D$(9-k)$-D9 and D9-D$(9-k)$ strings, there are $k$ fermionic zero 
modes, $\psi_0^1,\ldots, \psi_0^k$. As the basis of the ground states,
we take
$$
|\vec{\alpha},\pm\rangle^{}_{{}_{(9-k)9}}:=
|\alpha_1...\alpha_{k-1\over 2},\pm\rangle^{}_{{}_{(9-k)9}},
\qquad
|\vec{\alpha},\pm\rangle^{}_{{}_{9(9-k)}}:=
|\alpha_1...\alpha_{k-1\over 2},\pm\rangle^{}_{{}_{9(9-k)}}.
$$
These are annihilated by
$(\psi_0^{2j-1}-i\epsilon_{\alpha_j}\psi_0^{2j})$ for
$j=1,\ldots,{k-1\over 2}$, where $\epsilon_{\uparrow}=1$ and 
$\epsilon_{\downarrow}=-1$,
and satisfy the conditions of the form (\ref{condpm89}) and
(\ref{condpm98}), in which $\psi_0'$ and $\psi_0''$ are the zero modes of
$\psi_{\pm}^k$.
Their operator product expansions are of the form
\beq
\begin{array}{l}
|\vec{\alpha},\pm\rangle^{}_{{}_{9(9-k)}}
\cdot |\vec{\alpha},\pm\rangle^{}_{{}_{(9-k)9}}
\sim
(\psi^1_{-{1\over 2}}-i\epsilon_{\alpha_1}\psi^2_{-{1\over 2}})\cdots 
(\psi^{k-2}_{-{1\over 2}}
-i\epsilon_{\alpha_{k-1\over 2}}\psi^{k-1}_{-{1\over 2}})
|0\rangle^{}_{{}_{(9-k)(9-k)}},
\\
|\vec{\alpha},\mp\rangle^{}_{{}_{9(9-k)}}
\cdot |\vec{\alpha},\pm\rangle^{}_{{}_{(9-k)9}}
\sim
(\psi^1_{-{1\over 2}}-i\epsilon_{\alpha_1}\psi^2_{-{1\over 2}})\cdots 
(\psi^{k-2}_{-{1\over 2}}
-i\epsilon_{\alpha_{k-1\over 2}}\psi^{k-1}_{-{1\over 2}})
\psi_{-{1\over 2}}^k|0\rangle^{}_{{}_{(9-k)(9-k)}}.
\end{array}
\label{OPE2}
\eeq
If we choose $\xi_1\to \spst\xi_1$ as the parity transform, then
the corresponding parity operator $\Po_1$ maps
$|\vec{\alpha},\pm\rangle^{}_{{}_{(9-k)9}}$ to
$|\vec{\alpha},\pm\rangle^{}_{{}_{9(9-k)}}$.
In view of the operator product rule (\ref{OPE2}), we see that
$(-1)^{\Po_1}=(-1)^{k-1\over 2}$
and that
\beq
\Po_1(\Po_1|\vec{\alpha},\pm\rangle^{}_{{}_{(9-k)9}}
\cdot |\vec{\alpha},\pm\rangle^{}_{{}_{(9-k)9}})
= \spst^{k- 1\over 2}\Po_1|\vec{\alpha},\pm\rangle^{}_{{}_{(9-k)9}}
\cdot |\vec{\alpha},\pm\rangle^{}_{{}_{(9-k)9}}.
\label{PPaa}
\eeq
For the other choice, $\xi_1\to -\spst\xi_1$,
the corresponding operator $\Po_1$ maps 
$|\vec{\alpha},\pm\rangle^{}_{{}_{(9-k)9}}$
to $|\vec{\alpha},\mp\rangle^{}_{{}_{9(9-k)}}$.
This together with (\ref{OPE2}) yields
$(-1)^{\Po_1}=(-1)^{k+1\over 2}$
and a formula of the form (\ref{PPaa}) in which
$\spst^{k-1\over 2}$ is replaced by $\spst^{k+1\over 2}$.

In order to find an even parity operator $\Po$ satisfying (\ref{basicid}),
we include the Chan-Paton factor into the discussion.
We consider $N$ D$(9-k)$-branes
and place a graded vector space $V'$ on the D9-brane boundary.
A D$(9-k)$-D9 string state is of the form 
$\phi'\otimes \psi$ where $\phi'$ is a map $\C^N\to V'$ and
$\psi$ is a state like $|\vec{\alpha},\pm\rangle^{}_{{}_{(9-k)9}}$.
We consider the parity of the form
$\Po(\phi'\otimes \psi)=(-1)^{\phi' \Po_1}\Po'_{\rm CP}(\phi')
\otimes \Po_1(\psi)$,
in which $\Po_{\rm CP}'$ is given by
$\Po'_{\rm CP}(\phi')=\gamma\phi^{\prime T}\gU^{\prime -1}(-1)^{\phi' \gU'}$
for the o-isomorphism $\gamma$ of the $N$ D$(9-k)$-brane
and $\gU':V^{\prime *}\to V'$.
If we choose $\xi_1\to \spst\xi_1$ as the parity transform,
the condition for the Chan-Paton part is
$(-1)^{\Po'_{\rm CP}}=(-1)^{k- 1\over 2}$
and $\Po_{\rm CP}^{\prime 2}(\phi')
=\spst^{-{k-1\over 2}}(-1)^{\phi'}\phi'$.
This is achieved when $(-1)^{\gU'}=(-1)^{k-1\over 2}$ and
\beq
\gU'(\gU^{\prime T})^{-1}\imath'
=\pm\spst^{k-1\over 2}\sigma'\qquad
\mbox{for O${}^{\mp}$-type}.
\label{upr}
\eeq
For the other choice $\xi_1\to -\spst\xi_1$, 
$\gU'$ has the opposite statistics, $(-1)^{\gU'}=(-1)^{k+1\over 2}$,
and satisfies (\ref{upr}) in which
the phase $\spst^{k-1\over 2}$
is replace by $\spst^{k+1\over 2}$.

Let us now include the auxiliary fermions, $\xia$
and $\xia'$.
The extended Chan-Paton vector space $V$ is the tensor product of 
$V'$ and the 2-dimensional space $V_2$ coming from ($\xi_1$, 
$\xia$), and the o-isomorphism $\gU$ is 
the tensor product of $\gU'$ and the one for
($\xi_1$, $\xia$). If we take
$\xi_1\to\spst\xi_1$, then the latter factor is identical to the
one chosen in the case of O$8^-$, i.e.
$\gU_{2(\phxi)}$ for $\spst=\phxi$ and 
$\xia\gU_{2(-\phxi)}$ for $\spst=-\phxi$.
Then, $\gU(\gU^T)^{-1}\imath$ is the product of
(\ref{upr}) and (\ref{O8pr}),
\beq
\gU(\gU^T)^{-1}\imath=\left\{\begin{array}{ll}
\pm\spst^{k+1\over 2}\sigma,&~\spst=\phxi,\\
\pm\spst^{k-1\over 2}\sigma,&~\spst=-\phxi,
\end{array}\right.\qquad
\mbox{for O${}^{\mp}$-type}.
\label{genupr}
\eeq
This proves the formula (\ref{BaSt}) for the D9-branes
in the Type IIA orientifold.

\subsubsection*{\it A Family Of Solutions}

Let us record a solution to the condition in the theory with
O$(9-k)^-$-plane for arbitrary (even or odd) $k$.
It is to have $k$ boundary fermions,
$\xi_1,\ldots,\xi_k$, that transform under the parity as
\beq
\xi_j(\tau)
\,\,\longrightarrow\,\,\spst\xi_j(-\tau),\quad j=1,\ldots, k.
\label{solK}
\eeq
Indeed, for even $k$, 
this leads to the o-isomorphism $\gU_k$ that satisfies the condition,
i.e., (\ref{evenkeqn}) with the plus sign.
The condition for odd $k$, (\ref{upr}) with the plus sign,
is solved by $\gU'=\gU_{k-1}$, which is
associated with $k-1$ boundary fermions that transforms as 
$\xi_j(\tau)\to \spst\xi_j(-\tau)$.
Including the single boundary fermion $\xi_1(\tau)$, 
we have (\ref{solK}).
Note that the solution (\ref{solK}) can be obtained from
the ABS configuration for the D-brane on top of the O-plane, 
(\ref{ABSbf}) with (\ref{ptxk}), by turning off the tachyon.

\subsection{Ramond Sector}\label{subsec:Ra}

As another application of the analysis developed in 
Sections~\ref{subsec:keven} and \ref{subsec:kodd},
we study the parity operator on the Ramond-sector of the D9-D9 string,
both in Type IIB ($k$ even) and Type IIA ($k$ odd).
We shall show, as promised in Section~\ref{subsec:Ramond},
that the relation $\Po^2=(-1)^F$ does indeed hold under the 
formula (\ref{BaSt}) of the Chan-Paton factor.

There are two novelties in this discussion. One is that the operator
product rule (\ref{prodrule}) is modified by a sign if both
of the two states are in the Ramond sector, i.e., 
spacetime fermions,
$\Po(\Psi_2\cdot\Psi_1)=-(-1)^{|\Psi_1|\cdot|\Psi_2|}\Po(\Psi_1)
\cdot \Po(\Psi_2)$.
In particular, the basic requirement takes the form
\beq
\Po(\Po(\Psi)\cdot\Psi)=-\Po(\Psi)\cdot \Psi.
\label{extram}
\eeq
Second, vertex operators in the Ramond sector must be
in half-integer pictures and thus the ghost sector cannot be
ignored. We shall consider the product of two Ramond vertex operators in
the $(-{1\over 2})$-picture that results in an NS vertex operator
in the $(-1)$-picture.

We recall that the mode expansions of the fermions
$\psi_{\pm}$ in the D9-D9 string
are given in (\ref{99}) and that the $\inv\Omega$
parity action on the modes
is
\beq
\inv\Omega:
\left\{
\begin{array}{l}
\psi_r~~\mbox{in $(++)$}\longrightarrow
\e^{i\pi r}\inv_*\psi_r~~\mbox{in $(++)$}\\
\psi_r~~\mbox{in $(--)$}\longrightarrow
-\e^{i\pi r}\inv_*\psi_r~~\mbox{in $(--)$}\\
\psi_n~~\mbox{in $(-+)$}\longrightarrow
(-1)^n\inv_*\psi_n~~\mbox{in $(+-)$}\\
\psi_n~~\mbox{in $(+-)$}\longrightarrow
-(-1)^n\inv_*\psi_n~~\mbox{in $(-+)$}\\
\end{array}\right.
\label{99Ptr}
\eeq
where
\beq
(\inv_*\psi)^{\mu}=\left\{\begin{array}{ll}
-\psi^{\mu}&\mu=1,\ldots,k\\
+\psi^{\mu}&\mu\ne 1,\ldots,k.
\end{array}\right.
\eeq
We shall sometimes write $-i\psi_r^{10}$ for $\psi_r^0$. 
We find zero modes $\psi^{\mu}_0$ in the Ramond-sector, 
i.e., in the $(+-)$ or $(-+)$ sector.
There are massless spacetime fermions, labeled by a quintuplet
$\vec{\alpha}=\alpha_1\cdots\alpha_5$
of ups and downs, $\alpha_j=\uparrow,\downarrow$, which satisfy
\beq
\left(\,\psi_0^{2j-1}-i\epsilon_{\alpha_j}\psi_0^{2j}\,\right)
|\vec{\alpha}\rangle^{}_{{}_{(\pm\mp)}}=0,\qquad
j=1,\ldots,5.
\eeq
We have chosen them to be in the $(-{1\over 2})$-picture,
annihilated by
$\beta_n$ and $\gamma_{n+1}$ for all $n\geq 0$.
When $k$ is even, these defining conditions are invariant under the parity.
Thus $\Po_{\rm bulk}$ maps $|\vec{\alpha}\rangle^{}_{{}_{(+-)}}$ 
to $|\vec{\alpha}\rangle^{}_{{}_{(-+)}}$ up to a constant, and vice versa.
When $k$ is odd, the condition changes at $j={k+1\over 2}$.
That is, the arrow $\alpha_j$ flips at $j={k+1\over 2}$
and remains the same for the other $j$'s.
For example, for $k=3$, we have
\beq
\Po_{\rm bulk}
|\uparrow\uparrow\uparrow\uparrow\uparrow\rangle^{}_{{}_{(\pm\mp)}}
\propto |\uparrow\downarrow\uparrow\uparrow\uparrow\rangle^{}_{{}_{(\mp\pm)}}.
\label{Pupupdndn}
\eeq
We therefore find that
\beq
(-1)^{\Po_{\rm bulk}}=(-1)^k.
\eeq
For even $k$, the operator product of a ground state and its parity image
is
\beq
\Po_{\rm bulk}|\vec{\alpha}\rangle^{}_{{}_{(+-)}}\cdot
|\vec{\alpha}\rangle^{}_{{}_{(+-)}}
~\sim~
\left(\psi^1_{-{1\over 2}}-i\epsilon_{\alpha_1}\psi^2_{-{1\over 2}}\right)
\cdots
\left(\psi^9_{-{1\over 2}}-i\epsilon_{\alpha_5}\psi^{10}_{-{1\over 2}}\right)
|-1\rangle^{}_{{}_{(++)}},
\label{prodRR}
\eeq
where $|-1\rangle^{}_{{}_{{(++)}}}$ is the vacuum in the $(-1)$-picture,
annihilated by $\beta_r$ and $\gamma_r$ for all $r\geq {1\over 2}$.
Using $\psi_{-{1\over 2}}\to 
\e^{-{\pi i\over 2}}\inv_*\psi_{-{1\over 2}}$
from (\ref{99Ptr}), we see that the parity acts on this state
by multiplication by 
$i^{k\over 2}\cdot (-i)^{5-{k\over 2}}\cdot (-i)=-(-1)^{k\over 2}$.
The last factor of $(-i)$ is from the transformation of the
vertex operator $\delta(\gamma)$ corresponding to $|-1\rangle^{}_{{}_{(++)}}$,
see (\ref{betagammaP}).
For odd $k$, 
a ground state and its parity image have opposite arrows at
$j={k+1\over 2}$, and therefore
their product misses the factor
$\left(\,\psi^k_{-{1\over 2}}
-i\epsilon_{\alpha_{k+1\over 2}}\psi^{k+1}_{-{1\over 2}}\right)$
compared to (\ref{prodRR}).
Hence the parity action on the product is multiplication by
$i^{k-1\over 2}\cdot (-i)^{5-{k+1\over 2}}\cdot (-i)
=i(-1)^{k+1\over 2}$.
Thus we found 
\beq
\Po_{\rm bulk}(\Po_{\rm bulk}(\psi)\cdot \psi)=
\Po_{\rm bulk}(\psi)\cdot \psi
\times\left\{\begin{array}{ll}
-(-1)^{k\over 2}&\mbox{$k$ even}\\
i(-1)^{k+1\over 2}&\mbox{$k$ odd},
\end{array}\right.
\label{Starex}
\eeq
for a bulk state $\psi$ in the $(+-)$ sector.

On the other hand, we have
$\Po_{\rm CP}^2(\phi)=(-1)^{\Po_{\rm CP}}
\gc_{(i)}\gc_{(-i)}^{-1}(-1)^{\phi}\phi$
on the Chan-Paton factor in the $(+-)$ sector.
Using the formula (\ref{cci})
in which we insert (\ref{statU}) for the value of $(-1)^{|\gU_{(\spst)}|}$, 
or directly using (\ref{BaSt}),
we find $\gc_{(i)}\gc_{(-i)}^{-1}=(-1)^{k\over 2}$ for $k$ even 
and $-i(-1)^{k+1\over 2}$ for $k$ odd.
Applying the identity (\ref{PCPpr}), we find
\beq
\Po_{\rm CP}(\Po_{\rm CP}(\phi)\cdot\phi)
=\Po_{\rm CP}(\phi)\cdot\phi\times\left\{\begin{array}{ll}
(-1)^{k\over 2}&\mbox{$k$ even}\\
i(-1)^{k+1\over 2}&\mbox{$k$ odd}
\end{array}\right.
\label{PPCP99}
\eeq

Note that the formula (\ref{PPVV}) holds in the present case since
$\Po_{\rm bulk}$ in the NS sector of the D9-D9 string is even.
Inserting (\ref{Starex}) and (\ref{PPCP99}) into that formula,
we see that the required relation (\ref{extram}) holds.
Thus, we have proved the promised relation $\Po^2=(-1)^F$ in the Ramond-sector.
We remark that {\it we needed to use the ten-dimensionalilty of the spacetime}
in this discussion.

\medskip

\section{D-Branes In Type I String Theory}
\label{sec:TypeI}

In this section, we analyze the massless and tachyonic spectrum 
on D-branes of various dimensions in Type I string theory.
We use two descriptions --- conformal field theory with
the standard D-brane boundary condition on the one hand,
and the D9-brane configurations with nontrivial tachyon profiles on the other.
The former is a direct application of the result of the previous section.
We will encounter an ambiguity in the parity action in the Ramond sector,
which may be fixed by an input from spacetime physics.
In the second approach, we find no ambiguity and 
can honestly derive the spectrum.

\subsection{CFT Analysis Of The Spectrum}
\label{subsec:CFTa}

What is done in the previous section can be interpreted as the study of
the Chan-Paton factor of D-branes of all dimensions
in Type I string theory. In particular, 
as the o-isomorphisms for Type I D$p$-brane,
we can use the ones obtained for the D9-branes in the presence of O$p^-$-plane,
which may be taken as follows:
\beq
\begin{array}{|c|c|}
\hline
p&\gU_{(\mp i)}
\\\hline
9,1&
\left(\begin{array}{cc}
{\bf 1}_{N^0}&0\\
0&\pm i{\bf 1}_{N^1}
\end{array}\right)
\\\hline
7&
\left(\begin{array}{cc}
0&\pm i{\bf 1}_{N}\\
{\bf 1}_{N}&0
\end{array}\right)
\\\hline
5&
\left(\begin{array}{cc}
{\bf J}_{N^0}&0\\
0&\pm i{\bf J}_{N^1}
\end{array}\right)
\\\hline
3&
\left(\begin{array}{cc}
0&\mp i{\bf 1}_{N}\\
{\bf 1}_{N}&0
\end{array}\right)
\\\hline
\end{array}
\quad
\begin{array}{|c|cc|}
\hline
p&\gU_{(-i)}&\gU_{(i)}\\\hline
8,0&
\left(\begin{array}{cc}
0&i{\bf 1}_{N}\\
{\bf 1}_{N}&0
\end{array}\right)&
\left(\begin{array}{cc}
{\bf 1}_{N}&0\\
0&i{\bf 1}_{N}
\end{array}\right)\\\hline
6&
\left(\begin{array}{cc}
{\bf J}_{N}&0\\
0&i{\bf J}_{N}
\end{array}\right)&
\left(\begin{array}{cc}
0&i{\bf J}_{N}\\
{\bf J}_{N}&0
\end{array}\right)\\\hline
4&
\left(\begin{array}{cc}
0&i{\bf J}_{N}\\
{\bf J}_{N}&0
\end{array}\right)&
\left(\begin{array}{cc}
{\bf J}_{N}&0\\
0&i{\bf J}_{N}
\end{array}\right)\\\hline
2&
\left(\begin{array}{cc}
{\bf 1}_{N}&0\\
0&i{\bf 1}_{N}
\end{array}\right)&
\left(\begin{array}{cc}
0&i{\bf 1}_{N}\\
{\bf 1}_{N}&0
\end{array}\right)\\\hline
\end{array}
\label{U+U-}
\eeq
where we have chosen $\phxi=-i$, see (\ref{relnU}). 
The relative phase between $\gU_{(i)}$ and $\gU_{(-i)}$ has been
chosen arbitrarily.

This enters into the parity operator
with which we define the orientifold projection of the degrees of freedom on
the D$p$-brane worldvolume.
Let us look at the tachyons and massless particles
from the $p$-$p$ strings.
The parity, $\Po=\Po(\Omega)$ in the NS-sector and
$\wtPo=(-1)^{F_R}\Po(\Omega)$ in the R-sector,
acts on the relevant states as follows
\beqa
k_t\cdot\psi|k_t\rangle^{}_{{}_{(++)}}&\longmapsto& 
-ik_t\cdot\psi|k_t\rangle^{}_{{}_{(++)}},\nn\\
(\zeta\cdot\alpha_{-1}+\cdots)|k_b\rangle^{}_{{}_{(++)}}
&\longmapsto&
 -(-1)^{|\zeta|}(\zeta\cdot\alpha_{-1}+\cdots)|k_b\rangle^{}_{{}_{(++)}},
\label{PAonCFT}\\
|k_f,\vec{\alpha}\rangle^{}_{{}_{(-+)}}&\longmapsto&
z_{\vec{\alpha}}\, |k_f,\vec{\alpha}'\rangle^{}_{{}_{(-+)}},
\nn
\eeqa
where $k_t^2=1$, $k_b^2=k_f^2=0$ and $\zeta\cdot k_b=0$. In the second line, 
$(-1)^{|\zeta|}=+1$ {\it resp}.\! $-1$
if $\zeta$ is tangent {\it resp}.\! transverse to the brane.
The important information in the Ramond sector analysis is
the transformation of the fermionic zero modes
which is, in the $(-+)$ sector,
\beq
\psi_0^{\mu}\,\longrightarrow \, \left\{\begin{array}{ll}
\psi_0^{\mu}&\mbox{if $x^{\mu}$ is tangent to the brane,}\\
-\psi_0^{\mu}&\mbox{if $x^{\mu}$ is normal to the brane}.
\end{array}\right.
\label{PAonfz}
\eeq
$\vec{\alpha}$ and $\vec{\alpha}'$ in (\ref{PAonCFT})
are the labels of the spin
which are equal for odd $p$ and different for even $p$, as in 
(\ref{Pupupdndn}).
$z_{\vec{\alpha}}$ is some phase.
For odd $p$, it is of the form 
$$
z_{\vec{\alpha}}=\e^{i\theta}\chi_n(\vec{\alpha}),
$$
where $\e^{i\theta}$ is an $\vec{\alpha}$-independent phase
and $\chi_n(\vec{\alpha})=\pm 1$ is the chirality
of $\vec{\alpha}$ in the directions normal to the brane
--- it comes from the minus sign in (\ref{PAonfz}).
(In the $(+-)$ sector, the transformation is opposite to
(\ref{PAonfz})
and we have the chirality in the tangent directions. However, after GSO
projection, that is equal to the chirality in the normal directions.)
The action (\ref{PAonCFT}) 
is to be combined with the action on the Chan-Paton factor
\beq
\phi~\longmapsto ~ \left\{\begin{array}{ll}
\gU_{(-i)}\circ\phi^T\circ\gU_{(-i)}^{-1}
(-1)^{|\gU_{(-i)}||\phi|}&\mbox{(NS)},\\[0.2cm]
\gU_{(i)}\circ\phi^T\circ\gU_{(-i)}^{-1}(-1)^{|\gU_{(-i)}||\phi|}&\mbox{(R)}.
\end{array}\right.
\eeq

The spectrum analysis in the NS-sector (spacetime bosons)
is straightforward, and only the result will be presented.
The analysis in the R-sector (spacetime fermions) is more interesting.
The phase $z_{\vec{\alpha}}$ must be determined in order to 
specify the orientifold projection. 
The consistency condition $\wtPo^2={\rm id}$ fixes it
only up to a sign. 
As far as the spectrum analysis is concerned,
this sign is irrelevant for even $p$ cases, since the 
orientifold projection simply relates
the Chan-Paton factors multiplying the two different vectors,
$|\vec{\alpha}\rangle$ and $|\vec{\alpha}'\rangle$.
The sign turns out to be irrelevant also in the cases $p=7,3$.
The sign does affect the spectrum for $p=9,5,1$ and thus we need to know it
for an honest analysis.
At this moment, we do not know how to determine it purely within 
the conformal field theory --- the analysis as in
the previous section is not sufficient.
Facing this problem, for now, we resort for
help to the information of spacetime physics,
in particular spacetime supersymmetry.
In the next subsection, we will see that the sign can be determined
by our formulation based on tachyon configurations on D9-branes.

For $p=9,5,1$, the o-isomorphisms $\gU_{(\pm i)}$ 
are even and hence
the even part (D$p$-branes) and odd part ($\overline{{\rm D}p}$-branes)
are individually invariant under the orientifold.
The projection conditions for fermions in these two sectors
are opposite due to the difference between
$\gU_{(i)}$ and $\gU_{(-i)}$ ---
if one is symmetric then the other is antisymmetric ---
as it must be the case by the open-closed channel duality (see
 Section~\ref{subsec:Ramond}).
The sign of $\e^{i\theta}$ determines which is which.
Note that the distinction between branes and antibranes are up to us,
and we take the convention that the D9-brane preserves the same
supersymmetry as the O9-plane, and that
the D9-D5 and D9-D1 strings have massless fermions of positive chirality
in $5+1$ and $1+1$ dimensions respectively.
Then, information about spacetime supersymmetry can tell us which fermions 
are supposed to survive the orientifold projection:
(i) D9-D9 spectrum must include superpartners of the gauge bosons,
(ii) D5-D5 spectrum must include a $(1,0)$ gauge multiplet,
and the gaugino in it must have negative chirality so that
D9-D5 string yields a massless $(1,0)$ hypermultiplet \cite{WsmallI}, and
(iii) on D1-brane the spacetime supersymmetry has positive chirality
and hence the superpartner of the massless scalar {\it resp}.\! vector
must have negative {\it resp}.\! positive chirality \cite{PolWi}.
These requirements fix the sign as $\e^{i\theta}=-1$ for $p=9,1$ 
and $\e^{i\theta}=1$ for $p=5$ under
the relative phase given in (\ref{U+U-}) in which we have
$\gU_{(i)}=\sigma\gU_{(-i)}$.
The result is summarized in the table below.\\[0.1cm]
$$
\begin{array}{|c||c|c|c|c|}
\hline
p&\mbox{gauge group}&\mbox{tachyon}&\mbox{massless scalar}&
\mbox{massless fermion}\\\hline\hline
9&O(N^0)\times O(N^1)&{\bf bi}&{\rm none}&({\bf A},{\bf 1})_+,
({\bf 1},{\bf S})_+,{\bf bi}_-\\\hline
8&O(N)&{\bf A}&{\bf S}&{\bf A}, {\bf S}\\\hline
7&U(N)&{\bf A}&{\bf adj}&{\bf adj}\\\hline
6&USp(N)&{\bf A}&{\bf A}&{\bf A},{\bf S}\\\hline
5&USp(N^0)\times USp(N^1)&{\bf bi}&({\bf A},{\bf 1}), ({\bf 1},{\bf A})&
\begin{array}{c}
({\bf A},{\bf 1})_+^+,\\[-0.1cm]
({\bf S},{\bf 1})_-^-,
\end{array}\!\!\!\!
\begin{array}{c}
({\bf 1},{\bf S})_+^+,\\[-0.1cm]
({\bf 1},{\bf A})_-^-,
\end{array}
{\bf bi}_+^-,{\bf bi}_-^+\\\hline
4&USp(N)&{\bf S}&{\bf A}&{\bf A},{\bf S}\\\hline
3&U(N)&{\bf S}&{\bf adj}&{\bf adj}\\\hline
2&O(N)&{\bf S}&{\bf S}&{\bf A},{\bf S}\\\hline
1&O(N^0)\times O(N^1)&{\bf bi}&({\bf S},{\bf 1}),({\bf 1},{\bf S})&
\begin{array}{c}
({\bf A},{\bf 1})_+^+,\\[-0.1cm]
({\bf S},{\bf 1})_-^-,
\end{array}\!\!\!\!
\begin{array}{c}
({\bf 1},{\bf S})_+^+,\\[-0.1cm]
({\bf 1},{\bf A})_-^-,
\end{array}
{\bf bi}_+^-,{\bf bi}_-^+\\\hline
0&O(N)&{\bf A}&{\bf S}&{\bf A}, {\bf S}\\\hline
\end{array}
$$
\\[0.1cm]
`{\bf bi}', `{\bf adj}', `{\bf 1}', `{\bf S}' and `{\bf A}' stand for
the bifundamental, adjoint, singlet, symmetric tensor and antisymmetric tensor
representations respectively. Note that ${\bf adj}={\bf A}$ for $O(n)$
and ${\bf adj}\cong{\bf S}$ for $USp(n)$. 
The massless scalar is tensored with a normal vector to the brane.
The massless fermion is a worldvolume spinor tensored with a spinor
of the normal bundle.
For $p=9,5,1$, there is a restriction
on the chirality in the tangent {\it resp}.\! normal directions,
shown by the subscript {\it resp}.\! superscript.
For $p=7,3$, the massless
fermion in the adjoint representation of $U(N)$ 
can have all four chirality pairs.

Parts of the result had been obtained earlier.
Gimon-Polchinski \cite{GP} and Witten \cite{WittenK}
applied the same method to find the projection condition on the
bosonic sector in $p$ odd cases.
References \cite{Oren} and \cite{AST} proposed/obtained 
the list of tachyons and massless bosons using different methods.
Ref. \cite{Sugi} determined the orientifold projection of
massless fermions from the $\overline{D9}$-$\overline{D9}$ string
 using the channel duality argument.

\subsection{Spectrum Via D9-Brane Configurations}

Let us realize the same D-branes as D9-anti-D9-brane systems
and study the orientifold projection of massless fermions,
in particular for the cases $p=9,5,1$.

\subsubsection{The Configurations}

A D$p$-brane in Type IIB string theory is provided by the 
tachyon configuration (\ref{ABS}),
$$
T(x)=\sum_{i=1}^kx^i\xi_i,
$$
for $k=9-p$, represented on a graded vector space $V$.
To have it as a configuration in Type I string theory, we need an 
o-isomorphism $\gU:V^*\to V$ which is even and has $\gc=1$.
One way to achieve this is to introduce $k$ additional boundary fermions
$\xi_{k+1},\ldots,\xi_{2k}$ and take
\beq
V=V_{2k}\quad\mbox{and}\quad
O\,=\,\left(\begin{array}{cc}
-{\bf 1}_k&\\
&{\bf 1}_k
\end{array}\right),
\label{soln2k}
\eeq
where $O$ is the matrix (\ref{Odef}) that specifies
the parity transformation.
The first $k$ eigenvalues are chosen to be $-1$ for the invariance of
the boundary interaction $\int\dd\tau\sum_{i=1}^k\psi^i\xi_i$, 
or equivalently, for the o-isomorphism condition 
$\gU(\spst T(x)^T)\gU^{-1}=T(x)$; 
we have the opposite sign compared to (\ref{eqnforU}) 
since the Type I involution is $x^i\to x^i$ rather than $x^i\to -x^i$. 
Then, having eigenvalue $+1$ with multiplicity $k$ guarantees
 $(-1)^{|\gU|}=1$ and $\gc=1$, see (\ref{UUTs}).

The boundary theory flows in the infra-red limit to the D$p$-brane
boundary condition for $x^{\mu}$ and $\psi^{\mu}_{\pm}$,
and only $\xi_{k+1},\ldots,\xi_{2k}$ remain as the boundary degrees of freedom.
They transform as $\xi_i(\tau)\to \spst\xi_i(-\tau)$
under the parity and hence are nothing but 
(the Type I versions of) the consistent boundary degrees 
of freedom recorded in (\ref{solK}).
These boundary fermions are represented on $V_{\rm IR}=V_k$ for even $k$ and,
together with an auxiliary fermion $\xia$, on $V_{\rm IR}=V_{k+1}$ for odd $k$.

Of course, (\ref{soln2k}) is not the only solution and is not even the
minimal one except for low values of $k$.
In the notation of Table (\ref{U+U-}), it corresponds to
$N=2^{k-1\over 2}$ for odd $k$ (i.e., even $p$),
$N=2^{{k\over 2}-1}$ for $k=2,6$ ($p=7,3$)
and $N^0=N^1=2^{{k\over 2}-1}$ for $k=4,8$ ($p=5,1$).
We can of course construct other cases including the minimal ones.
To be specific, however, we shall discuss the orientifold projection
for the solution (\ref{soln2k}).

\subsubsection{The Massless Fermions}

Next, we write down the wavefunctions for the supersymmetric ground states
in the Ramond sector.
We employ the zero mode approximation which is sufficient for the purpose of
 finding the parity action.
We denote by $S$ the spinor representation of the 
algebra $\{\psi_0^{\mu},\psi_0^{\nu}\}=\eta^{\mu\nu}$
of the fermionic zero modes
 $\psi_0^{\mu}$ ($\mu=0,1,\ldots, 9$).
We first consider the wavefunction for
the $p$-$p$ string.
We represent it as $\Psi_{pp}=\phi\otimes {\rm s}$ where 
$\phi$ and ${\rm s}$ take values in $\Hom(V,V)$ and $S$ respectively.
The supercharge (\ref{Q1}) acts on this state as
\beq
{\bf Q}_1(\phi\otimes {\rm s})=
-i\sum_{\mu=0}^9\psi_0^{\mu}{\partial\over \partial x^{\mu}}
(\phi\otimes {\rm s})
-(T\circ\phi)\otimes {\rm s}-(-1)^{|\phi|}i(\phi\circ T)\otimes {\rm s}.
\eeq
A general solution to the supersymmetry condition
${\bf Q}_1\Psi_{pp}=0$ is
\beq
\Psi_{pp}
=\e^{-\sum_{i=1}^k(x^i)^2}\prod_{j=1}^k\Bigl(1+(i-1)\xi_j\psi_0^j\Bigr)
\tilde{\phi}(\xi_{k+1},\ldots,\xi_{2k})
\otimes {\rm s}(x)
\eeq
where $\tilde{\phi}(\xi_{k+1},\ldots,\xi_{2k})$ is a sum of products of
$\xi_{k+1},\ldots,\xi_{2k}$ only, and ${\rm s}(x)$
solves the $5+1$ dimensional 
massless Dirac equation $\sum_j\psi_0^j\partial_js(x)=0$ (where
$j$ runs over $0,k+1,\ldots,9$).
This corresponds to the state $\tilde{\phi}\otimes |{\rm s}\rangle$
in the CFT description, where $\tilde{\phi}$ is regarded as
an element of $\Hom(V_{\rm IR},V_{\rm IR})$
with the constraint of the graded commutativity with $\xia$ in the
odd $k$ case, and $|{\rm s}\rangle$ is the state corresponding to
the solution $s(x)$ of the Dirac equation.

For our purpose, we also need to 
know the wavefunction for the Ramond ground states of the
$9$-$p$ string, for $p=5,1$ (i.e., $k=4,8$). 
Assuming that the number of D9-branes is 1, 
the wavefunction $\Psi_{9p}$ takes values in $V\otimes S$. The supersymmetry
condition reads ${\bf Q}_1\Psi_{9p}
=-i\sum_{\mu=0}^9\psi_0^i{\partial\over \partial x^i}\Psi_{9p}
-T \Psi_{9p}=0$ and a general solution is of the form
\beq
\Psi_{9p}=\e^{-{1\over \sqrt{2}}\sum_{i=1}^k(x^i)^2}
\prod_{j=1}^k\left(1+i\sqrt{2}\xi_j\psi_0^j\right)v\otimes s(x)
\label{9pgs}
\eeq
where $s(x)$ solves the $5+1$ dimensional Dirac equation.
Quantization of the $2k+10$ fermions $\xi_i$ and $\psi_0^{\mu}$
can be grouped into the $k$ pairs, $(\xi_i,\psi_0^i)$ for $i=1,\ldots, k$,
and the remaining ten.
The factor $(1+i\sqrt{2}\xi_i\psi_0^i)$ acts as the projection
operator into one out of two states in the $(\xi_i,\psi_0^i)$ system.
Thus, the dimension of the space of solutions is $2^5$, which is exactly 
what we expect in the CFT description.

\newcommand{\chir}{\overline{\Gamma}}

Let us fix our convention of the GSO projection $(-1)^F=1$.
We define the $\Z_2$-grading on the vector space $V$ by
\beq
\sigma=i^k \xi_1\cdots\xi_{2k}
\eeq
so that the GSO operator is given by
$(-1)^F:=\sigma\otimes \chir_{9+1}$,
where the second factor is the ten-dimensional
chirality $\chir_{9+1}:=2^5\psi_0^0\psi_0^1\cdots\psi_0^9$.
Let us look at the GSO projection of the $9$-$p$ string.
Using $(i\sqrt{2}\xi_j\psi_0^j)^2=1$
we find, for the state given by (\ref{9pgs}),
$(-1)^F\Psi_{9p}=(-1)^{k(k-1)\over 2}\xi_{k+1}\cdots\xi_{2k}\cdot
2^{5-{k\over 2}}\psi_0^0\psi_0^{k+1}\cdots\psi_0^9
\Psi_{9p}.$
In the cases of our interest, $k=4,8$ (i.e., $p=5,1$),
this may be written as
\beq
(-1)^F\Psi_{9p}=\xi_{k+1}\cdots\xi_{2k}\chir_{p+1}\Psi_{9p},
\eeq
where $\chir_{p+1}:=2^{p+1\over 2}\psi_0^0\psi_0^{k+1}\cdots\psi_0^9$
is the chirality in $p+1$ dimensions.
Thus, the convention taken in Section~\ref{subsec:CFTa} (that
D9-D5 and D9-D1 strings yield massless fermions of
positive chirality in $5+1$ and $1+1$ dimensions)
corresponds to the choice
\beq
\sigma_{\rm IR}=\xi_{k+1}\cdots\xi_{2k}
\label{sigIR}
\eeq
for the $\Z_2$-grading in the infra-red 
Chan-Paton vector space $V_{\rm IR}=V_k$ ($k=4,8$).

\subsubsection{Orientifold Projection}

Now we look at the orientifold projection of
massless fermions $\Psi=\Psi_{pp}$ for $p=5$ and $1$.
For the factorized expression of the state,
$\Psi=\phi\otimes {\rm s}$,
we have $\wtPo(\Psi)=\wtPo_{\rm CP}(\phi)\otimes
\wtPo_{\rm 99}({\rm s})$ where, in the $(-+)$ sector 
$$
\wtPo_{\rm CP}(\phi)=\gU_{(i)}\circ 
\phi^T\circ \gU_{(-i)}^{-1}
$$
and $\wtPo_{99}$ is an action on $S$ associated with
the parity $\wtOmega_{99}$ on the worldsheet fermions
with Neumann boundary condition at both boundaries.
The reason we put the subscript ``99'' is to distinguish it from 
the parity action on the fermions with
D$p$-brane boundary condition at both boundaries.
Note that the action on the modes is 
$\wtOmega_{99}:\psi_0^{\mu}\to\psi_0^{\mu}$
for $\mu=0,1,\ldots,9$, in the $(-+)$-sector. 
Hence $\wtPo_{99}$ is equal to
the identity up to a phase, $\wtPo_{99}({\rm s})
=\e^{i\delta}{\rm s}$.
Let us compute $\wtPo(\Psi)$.
Using $\gU_{(i)}=\kappa\gU_{(-i)}\sigma^T=\kappa\sigma\gU_{(-i)}$, we find
\beqa
\lefteqn{
\gU_{(i)}\left[\prod_{j=1}^k\Bigl(1+(i-1)\xi_j\psi_0^j\Bigr)\tilde{\phi}
\right]^T\!\!\gU_{(-i)}^{-1}}\nn\\
&=&\kappa\sigma\prod_{j=1}^k\left[i^2\e^{\pi i\over 4}\sqrt{2}\xi_j\psi_0^j
\Bigl(1-(i-1)\xi_j\psi_0^j\Bigr)\right]\gU_{(-i)}\tilde{\phi}^T\gU_{(-i)}^{-1}
\nn\\
&=&\kappa \cdot i^{3k}\e^{{\pi i\over 4}k}\xi_{k+1}\cdots\xi_{2k}\cdot
2^{k\over 2}\psi_0^1\cdots\psi_0^k\prod_{j=1}^k\Bigl(1-(i-1)\xi_j\psi_0^j\Bigr)
\gU_{(-i)}\tilde{\phi}^T\gU_{(-i)}^{-1}\nn\\
&=&\kappa\cdot\e^{{\pi i\over 4}k}\prod_{j=1}^k\Bigl(1+(i-1)\xi_j\psi_0^j\Bigr)
\xi_{k+1}\cdots\xi_{2k}\cdot
2^{k\over 2}\psi_0^1\cdots\psi_0^k
\gU_{(-i)}\tilde{\phi}^T\gU_{(-i)}^{-1}\nn\\
&=&
\kappa\cdot\e^{{\pi i\over 4}k}\prod_{j=1}^k\Bigl(1+(i-1)\xi_j\psi_0^j\Bigr)
\gU_{{\rm IR}(i)}\tilde{\phi}^T\gU_{{\rm IR}(-i)}^{-1}\cdot
2^{k\over 2}\psi_0^1\cdots\psi_0^k.\nn
\eeqa
Here $\gU_{{\rm IR}(\pm i)}$ are the $\xi_{k+1},\ldots,\xi_{2k}$ parts of 
$\gU_{(\pm i)}$,
and we decided to take them to obey the relation
$\gU_{{\rm IR}(i)}=\xi_{k+1}\cdots\xi_{2k}\gU_{{\rm IR}(-i)}$ which means
\beq
\gU_{{\rm IR}(i)}=\sigma_{\rm IR}\gU_{{\rm IR}(-i)}
\label{IRrelU}
\eeq
in view of (\ref{sigIR}). 
Thus, $\Psi\mapsto \wtPo(\Psi)$ corresponds in the CFT description
to
\beq
\tilde{\phi}\otimes |{\rm s}\rangle\,\,\longmapsto\,\,
\kappa\e^{{\pi i\over 4}k}
\gU_{{\rm IR}(i)}\tilde{\phi}^T\gU_{{\rm IR}(-i)}^{-1}\otimes 
\chi_n({\rm s})\e^{i\delta}|{\rm s}\rangle,
\eeq
where $\chi_n({\rm s})$ is the chirality of ${\rm s}$ in the normal
directions,
$\chi_n({\rm s})|{\rm s}\rangle=2^{k\over 2}\psi_0^1\cdots\psi_0^k
|{\rm s}\rangle$.
We read from this that the phase $\e^{i\theta}$ is given by
$$
\e^{i\theta}=\kappa\e^{{\pi i\over 4}k+i\delta}.
$$
Since all branes have the same value of $\kappa$,
we see that $\e^{i\theta}$ for $k=4$ is opposite to
the one for $k=0,8$.
Orientifold projection with
these values of $\e^{i\theta}$ and with the relation (\ref{IRrelU})
is exactly what we have seen to be consistent with spacetime supersymmetry.
Thus, we obtained the ``correct'' spectrum of massless spacetime fermions
without any input from spacetime physics.
This computation exhibits the power of our formulation.

\medskip

\section{Twists --- Illustration By Examples}
\label{sec:example}

We illustrate our general considerations on the orientifold data,
in particular the relation between the twisting and mixed type O-planes,
in explicit examples of toriodal and Calabi-Yau compactifications.
We classify the orientifold data $(\inv,B,{\mathcal L},\sA,c)$ on tori 
and discuss T-duality relations.
It matches with the known results
in the well-studied examples of $S^1$ and $T^2$
and also leads to new results for higher dimensional tori.
For orientifolds of Calabi-Yau manifolds by holomorphic involutions,
we find a convenient way to read off the type of O-planes 
using holomorphy.

\subsection{Circle}

As the first example, let us consider Type II orientifolds
on $S^1\times \R^9$. 
We parametrize the circle by a coordinate $x$ 
with periodicity $x\equiv x+1$.
There are three inequivalent involutions:
(i) the identity, $x\mapsto x$, (ii) the half-period shift, 
$x\mapsto x+{1\over 2}$, and (iii) the inversion, $x\mapsto -x$.
(i) and (ii) are orientation preserving and are for Type IIB orientifolds
while (iii) is orientation reversing and is for Type IIA orientifolds.
In (i) the whole spacetime is the fixed point
set (O9-plane), and (ii) is fixed point free (no O-plane).
(iii) has two fixed point sets,
one at $x=0$ and another at $x={1\over 2}$ (two O8-planes).

Let us classify possible choices of the data $(B,{\mathcal L},\sA,c)$.
Note that we may assume that the B-field is zero,
$B=0$, since any flat B-field on $S^1\times \R^9$
is exact and can be gauged away. 
Likewise, we may assume that ${\mathcal L}$ 
is the trivial line bundle.
Since $B=0$, the twist connection $\sA$ must be flat, and we write
$\sA=\alpha\dd x$ for a real parameter $\alpha$, with the
gauge equivalence relation $\alpha\sim \alpha+2\pi$. 
The $\Lambda$ gauge transformation that preserves $B=0$
must also be flat, $\Lambda=\lambda\dd x$, and it acts on
the twist connection as $\alpha\to\alpha+2\lambda$ for (i) and (ii) and
trivially, $\alpha\to \alpha$, for (iii).

\noindent
(i) The identity.\\
We can turn off the twist connection $\sA$
by the gauge transformation $\Lambda=-{1\over 2}\alpha\dd x$.
The crosscap section $c$ is thus a constant function that squares to $1$.
Thus there are two cases: $c=1$ and $c=-1$,
giving rise to the O9${}^-$ and O9${}^+$ planes respectively.

\noindent
(ii) The half-period shift.\\
Again $\sA$ can be turned off
and there are two cases: $c=1$ and $c=-1$. But
these two cases are equivalent since they are related 
by a combination of the $\Lambda$ and $\lambda$ gauge transformations 
(\ref{gtra}) --- take $\Lambda=-\pi\dd x$ and $\lambda=\e^{2\pi i x}$.

\noindent
(iii) The inversion.\\
The holonomy of $\sA-\inv^*\sA$ along the circle is
$\e^{2i\alpha}$ and it must be trivial, $\e^{2i\alpha}=1$.
Up to gauge equivalence
we find two possibilities:
$\alpha=0$ or $\pi$ (mod $2\pi$).
The crosscap section $c$ is given by
\beq
c(x)=\exp\left(-i\int_0^x\left(\inv^*\sA-\sA\right)\right)\cdot c(0)
=\e^{2i\alpha x}c(0).
\label{pc1}
\eeq
This expression is in reference to
the frame of $\inv^*{\mathcal L}\otimes {\mathcal L}^{-1}$
that comes from the original trivialization of
${\mathcal L}$, and this frame 
matches with the canonical trivialization of 
$\inv^*{\mathcal L}\otimes {\mathcal L}^{-1}$ at the two fixed points.
Therefore the values of $c(0)$ and $c({1\over 2})$ according to (\ref{pc1})
directly show the type of the O8-planes.
For $\alpha=0$, we have $c(0)=c({1\over 2})=1$ 
(both O8${}^-$) or $c(0)=c({1\over 2})=-1$ (both O8${}^+$).
For $\alpha=\pi$, we have $c(0)=-c({1\over 2})=1$
(O8${}^-$ at $x=0$ and O8${}^+$ at $x={1\over 2}$)
or $c(0)=-c({1\over 2})=-1$ 
(O8${}^+$ at $x=0$ and O8${}^-$ at $x={1\over 2}$).
The last two cases are obviously equivalent as
they are related by a diffeomorphism,
$x\mapsto x+{1\over 2}$.

Let us look at T-duality relation among the above orientifolds. 
We first recall that the T-dual of a circle $S^1$ is 
its dual circle $\widetilde{S}^1=H^1(S^1,U(1))$ which parametrizes a flat
$U(1)$ bundle on $S^1$. The orientifold action on a $U(1)$ gauge field,
$A\to -\inv^*A+\sA$, reads for the flat field $A=a\dd x$
as $a\to a+\alpha$ (with $\alpha=0$ or $\pi$) for (iii) 
and $a\to -a$ for (i) and (ii). 
Thus we find
that (i) is T-dual to (iii) with $\alpha=0$
while (ii) is T-dual to (iii) with $\alpha=\pi$.
This is consistent with the known facts:
Type I string theory on a circle ((i) with $c=1$) is T-dual to
Type IIA orientifold on the dual circle with two O8${}^-$ planes,
while Type IIB orientifold on a circle by a half-period shift
is T-dual to Type IIA orientifold with
O8${}^-$ and O8${}^+$ \cite{NVS,DP}.
Note that the gauge field shifts as $A\mapsto A-\pi \dd x$
under the $(\Lambda,\lambda)=(-\pi \dd x,\e^{2\pi i x})$ transformation 
that relates the two cases in (ii),
see (\ref{gtrab}). This corresponds under T-duality to the fact that
the two cases in (iii) with $\alpha=\pi$
are related by the shift of the coordinate $x\mapsto x+{1\over 2}$.

Let us find some simple D9-brane configurations, say,
in Type IIA orientifolds of (iii).
The condition for $(\ugA,\ugT,\ugU)$ reads
\beqa
&&\ugU(x)\,=\,c(x)\cdot \ugU(-x)^t,
\nn\\
&&\ugA_x(x)=\,\ugU(x)(\ugA_x(-x)^t+\alpha\,)\ugU(x)^{-1}
+i^{-1}\ugU(x){\dd\over \dd x}\ugU(x)^{-1},
\nn\\
&&\ugT(x)\,=\,-\ugU(x)\ugT(-x)^t\ugU(x)^{-1}.
\nn
\eeqa

\noindent
\underline{O8$^-$ and O8$^-$}\, ($\alpha=0$, $c=1$)\\
There is a rank one solution
\beq
\ugU= 1,\quad
\ugA_x=a,\quad
\ugT= f(x),
\label{soln--}
\eeq
where $f(x)$ is any odd and periodic function.
For a generic $f(x)$, this corresponds to D8 and anti-D8 branes at points of
the circle --- at the zeroes of $f(x)$ with positive and negative slopes
respectively. (There is a single D8 at an O$8^-$
and a single anti-D8 at the other O$8^-$. This is inconsistent
 \cite{triples,HIS}
for a reason that cannot be detected from the open string tree-level
analysis.) 
Note that the tachyon must vanish if we insist it to be a constant.
But non-zero constant tachyon is allowed if take the sum of the two,
$$
\ugU=\left(\begin{array}{cc}
1&0\\
0&1
\end{array}\right),\quad
\ugA_x=\left(\begin{array}{cc}
a_1&0\\
0&a_2
\end{array}\right),\quad
\ugT=t\left(\begin{array}{cc}
0&-i\\
i&0
\end{array}\right).
$$
In the T-dual picture, Type I on $\widetilde{S}^1$,
the solution (\ref{soln--}) with $\ugT=0$ 
corresponds to a single non-BPS D8-brane
at $a\in \widetilde{S}^1$. Absence and presence of constant tachyon
 for the rank one and two cases correspond to the fact that
the non-BPS D8-brane in Type I is stable but its charge is conserved only
modulo 2 \cite{SenD0inI,WittenK}.

\noindent
\underline{O8$^+$ and O8$^+$}\, ($\alpha=0$, $c=-1$)\\
We see that $\ugU(0)$ and $\ugU({1\over 2})$ are both antisymmetric
and hence there is no rank one solution.
There are solutions for even ranks. For example, a rank two solution is,
\beq
\ugU=\left(\begin{array}{cc}
0&-1\\
1&0\end{array}\right),\quad
\ugA_x=\left(\begin{array}{cc}
a&0\\
0&a\end{array}\right),\quad
\ugT=\left(\begin{array}{cc}
t_1&t_2\\
t^*_2&-t_1
\end{array}\right).
\label{soln++}
\eeq
This corresponds to the non-BPS D8-brane in the T-dual theory,
the USp-version of Type I \cite{Sugi}.

\begin{figure}[htb]
\psfrag{o8minus}{O8$^-$}
\psfrag{o8plus}{O8$^+$}
\psfrag{2D9}{2D9's}
\psfrag{D8}{D8}
\centerline{\includegraphics{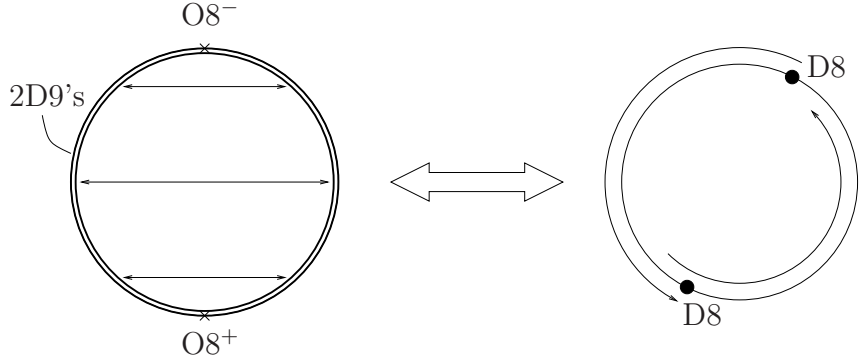}}
\caption{T-duality between Type IIA (left) and Type IIB (right)
orientifolds} 
\label{fig:Tduality}
\end{figure}

\noindent
\underline{O8$^-$ and O8$^+$}\, ($\alpha=\pi$, $c(x)=\e^{2\pi i x}$)\\
Note that $\ugU(0)$ is symmetric but $\ugU({1\over 2})$ is antisymmetric,
and again there is no rank one solution.
There is a rank two solution
\beq
\ugU=\left(\begin{array}{cc}
0&\e^{2\pi i x}\\
1&0
\end{array}\right),\quad
\ugA_x=\left(\begin{array}{cc}
a&0\\
0&a+\pi\end{array}\right),\quad
\ugT=t\left(\begin{array}{cc}
1&0\\
0&-1\end{array}\right).
\label{soln+-}
\eeq
In the T-dual Type IIB orientifold on $\widetilde{S}^1$
by the half-period shift,
this corresponds to two non-BPS D8-branes at the opposite points of
$\widetilde{S}^1$, one at $a$ and the other at $a+\pi$. 
See Figure~\ref{fig:Tduality}.
Existence of non-zero constant tachyon corresponds to the instability
of the non-BPS D8-branes in Type IIB string theory.

\subsection{Two-Torus}

Let us next study Type II orientifolds on $T^2\times \R^8$.
We use the coordinates $(x,y)$ of $T^2$ which have
periodicity $(x,y)\equiv (x+1,y)\equiv (x,y+1)$.
Up to diffeomorphisms, there are six distinct involutions:
(i) the identity, $(x,y)\mapsto (x,y)$, 
(ii) the half-period shift in one direction, $(x,y)\mapsto (x+{1\over 2},y)$,
(iii) the reflection of the type $(x,y)\mapsto (-x,y)$,
(iv) the reflection plus shift, $(x,y)\mapsto (-x,y+{1\over 2})$,
(v) the reflection of the type $(x,y)\mapsto (y,x)$, and
(vi) the inversion, $(x,y)\mapsto (-x,-y)$.
The properties of these involutions are summarized in the table:
$$
\begin{array}{|c|cccccc|}
\hline
\mbox{involution}&\mbox{(i)}&\mbox{(ii)}&\mbox{(iii)}&\mbox{(iv)}&
\mbox{(v)}&\mbox{(vi)}\\
\hline
\mbox{Type}&\mbox{B}&\mbox{B}&\mbox{A}&\mbox{A}&\mbox{A}&
\mbox{B}\\
\mbox{O-plane}&~\mbox{O9}~&\mbox{none}&\mbox{$2\times$O8}&\mbox{none}&
\mbox{$1\times$O8}&\mbox{$4\times$O7}\\
\hline
\end{array}
$$
Type B ({\it resp}. A) means that the involution preserves 
({\it resp}. reverses)
the orientation and thus can be used to define Type IIB ({\it resp}. IIA)
orientifolds.

Let us classify possible choices of the data $(B,{\mathcal L},\sA,c)$.
We parametrize the B-field as
$$
B\,=\,b\,\dd x\wedge \dd y,
$$
where we may take $b$ from the range $0\leq b<2\pi$.
The condition $[B+\inv^*B]\in H^2(T^2,2\pi \Z)$
requires $2b=\in 2\pi\Z$ (i.e. $b=0$ or $\pi$) for Type B 
but impose no constraint for Type A.
Let us first consider Type B with $b=0$ and Type A for which 
the twist connection must be flat and can be written as
$$
\sA=\alpha_x\dd x+\alpha_y\dd y.
$$
For the involutions (i) and (ii), we can turn off $\sA$ using 
a flat $\Lambda$ gauge transformation.
For (vi), we find that $\alpha_x$ and $\alpha_y$ must be $0$ or $\pi$ 
by the constraint that $\sA-\inv^*\sA=2\sA$ has trivial holonomy. 
For (iii) and (iv), a flat $\Lambda$ gauge transformation shifts $\sA$ by
$\Lambda+\inv^*\Lambda=2\lambda_y\dd y$ and hence $\alpha_y$ can be 
turned off. The constraint that $\sA-\inv^*\sA=2\alpha_x\dd x$ is trivial
requires $\alpha_x=0$ or $\pi$.
However, in (iv), the gauge transformation $\Lambda=2\pi y\dd x$ turns
off $\alpha_x=\pi$.
Thus, we may set $\sA=0$ in this case. On the other hand, in (iii),
$\alpha_x$ cannot be turned off by a $\Lambda$ gauge transformation.
For (v), a flat $\Lambda$ gauge transformation shifts $\sA$ by
$\Lambda+\inv^*\Lambda=(\lambda_x+\lambda_y)(\dd x+\dd y)$
and hence we may set $\alpha_y=0$ again.
The constraint that $\sA-\inv^*\sA=\alpha_x(\dd x-\dd y)$ has trivial holonomy
requires $\alpha_x=0$. Thus in this case we can turn of the twist connection,
$\sA=0$.

It remains to consider Type B with $b=\pi$. In this case,
$\dd\sA=2\pi\dd x\wedge\dd y$ and hence
${\mathcal L}$ is a complex line bundle
with first Chern class $-1$. In fact, with any choice of complex structure
of $T^2$ (there is one natural choice for a given metric), 
$({\mathcal L},\sA)$ can be regarded
as a holomorphic line bundle of degree $-1$, namely, 
${\mathcal O}(-p)$ for a point $p$ of $T^2$ 
--- the holomorphic line bundle that has a meromorphic section with 
a simple pole at $p$ and without zero.
We note that $({\mathcal L}\otimes\inv^*{\mathcal L}^*,
\sA-\inv^*\sA)\cong {\mathcal O}(-p+\inv(p))$ and that 
it is trivial if and only if $\inv(p)=p$.
For the involution (ii), there is no $\sA$ that obey the condition, since 
$\inv(p)\ne p$ for any point $p$.
For (i), the condition $\inv(p)=p$ is satisfied for any $p$
and hence any $\sA$ will do. In this case, however, 
${\mathcal O}(-p)$ for all $p$'s are related by
flat $\Lambda$ gauge transformations.
Therefore there is only one choice.
For (vi), $\inv(p)=p$ requires that
$p$ must be one of the four fixed points.

Let us discuss what types of orientifold planes are possible in each case.
If $\sA=0$, then the crosscap section $c$ is constant and hence
all O-planes (if there exist) are of the same type. 
If $\sA$ is a 2-torsion, i.e., if $\sA$ is flat and non-trivial 
but $2\sA$ is trivial,
a half of the O-planes are of the type O${}^-$ and the other
half is of the type O${}^+$. 
In the case (vi) with $b=\pi$, where 
$({\mathcal L},\sA)\cong {\mathcal O}(-p)$ for one of the four fixed 
points $p$,
the O7-plane at $p$ is of the opposite type compared to
the other three O7-planes.
We shall see this last point by an explicit construction below,
and also in Section~\ref{subsec:holo} applying a general argument
for holomorphic involutions.

For illustration, let us explicitly construct a twist connection
for the case (vi) with $b=\pi$.
Note that
$\dd \sA=2\pi \dd x\wedge \dd y$ is solved by
\beq
\sA=2\pi x\,\dd y.
\label{sA2}
\eeq
This determines a connection of a line bundle
${\mathcal L}$ over $T^2=\R^2/\Z^2$ which is
defined as the quotient of the trivial line bundle over
$\R^2$ with a global frame $\sigma(x,y)$, by the relations 
\beq
\sigma(x,y) \equiv \sigma(x+1,y)\e^{-2\pi i y}\equiv 
\sigma(x,y+1).
\label{relnL}
\eeq
The pull back connection $\inv^*\sA$ on $\inv^*{\mathcal L}$ has an expression 
$2\pi (-x)\,\dd (-y)$ with respect to
the pull-back frame $\inv^*\sigma(x,y)=([x,y],\sigma(-x,-y))$.
Note that this 1-form is exactly the same as (\ref{sA2}) 
and the frame $\inv^*\sigma(x,y)$ obeys exactly the same relations
as (\ref{relnL}).
Therefore the line bundle 
$\inv^*{\mathcal L}\otimes {\mathcal L}^{-1}$
has a global frame $u([x,y])=\inv^*\sigma(x,y)\otimes 
\sigma(x,y)^{-1}$ and 
the connection $\inv^*\sA-\sA$ 
is represented by 0 with respect to it.
That is, $u$ is a parallel section.
Thus, this $({\mathcal L},\sA)$ satisfies the condition for a twist 
connection.

Let us evaluate $u$ at the four fixed points, 
$p_1=[0,0]$, $p_0=[{1\over 2},0]$, $p_2=[0,{1\over 2}]$
and
$p_3=[{1\over 2},{1\over 2}]$:
\beqa
&&u(p_1)=(p_1,\sigma(0,0))\otimes \sigma(0,0)^{-1}=1,\nn\\
&&\mbox{$u(p_0)=(p_0,\sigma(-{1\over 2},0))\otimes \sigma({1\over 2},0)^{-1}
=(p_0,\sigma({1\over 2},0))\otimes \sigma({1\over 2},0)^{-1}=1$},\nn\\
&&\mbox{$u(p_2)=(p_2,\sigma(0,-{1\over 2}))\otimes\sigma(0,{1\over 2})^{-1}
=(p_2,\sigma(0,{1\over 2}))\otimes\sigma(0,{1\over 2})^{-1}=1$},\nn\\
&&\mbox{$u(p_3)
=(p_3,\sigma(-{1\over 2},-{1\over 2}))\otimes\sigma({1\over 2},{1\over 2})^{-1}
=(p_3,\sigma({1\over 2},{1\over 2})(-1))\otimes 
\sigma({1\over 2},{1\over 2})^{-1}
=-1$}.\nn
\eeqa
We have used the defining relations (\ref{relnL}) 
in the latter three lines, and also the
canonical isomorphism (\ref{canoiso}) for the evaluation.
We see that the value at $p_3$ is opposite to the
value at the other three points.
Since the crosscap section $c$ is proportional to $u$, 
the type of O7-plane at
$p_3$ is opposite to the type of the other three
O7-planes.

We now show that, for any choice of complex structure of $T^2$,
 this twist connection $\sA$
determines a holomorphic structure on ${\mathcal L}$ 
which is isomorphic to ${\mathcal O}(-p_3)$.
Let us take $z=-y+\tau x$ as a complex coordinate (${\rm Im}(\tau)>0$).
It is straightforward to see that 
$\sigma(x,y)^{-1}\e^{\pi i x^2}\vartheta_3(-y+\tau x,\tau)$,
where $\vartheta_3$ is Jacobi theta function
$$
\vartheta_3(z,\tau)=\sum_{n\in\Z}\e^{\pi i \tau n^2+2\pi i nz},
$$
is invariant under $(x,y)\to (x+1,y)$ and $(x,y+1)$.
Thus, it defines a global section of ${\mathcal L}^{-1}$.
We can also see that this section is holomorphic with respect to
the holomorphic structure determined by
the connection $-\sA$. 
Since $\vartheta_3(z,\tau)$ has a simple zero 
at $z={1\over 2}+{1\over 2}\tau$ (mod $\Z+\tau\Z$),
we find that $({\mathcal L}^{-1},-\sA)\cong {\mathcal O}(p_3)$, or 
equivalently,
$({\mathcal L},\sA)\cong {\mathcal O}(-p_3)$.

Other twist connections must differ from (\ref{sA2}) by a
2-torsion and hence are given by $\sA_0=2\pi x\dd y+\pi\dd x$,
$\sA_1=2\pi x\dd y+\pi(\dd x+\dd y)$
and $\sA_2=2\pi x\dd y+\pi\dd y$.
Repeating the above analysis, we find for the twist
$({\mathcal L},\sA_i)$ ($i=0,1,2$)
that the O-plane at $p_i$ is of the opposite type compared to the other three
O-planes, and also that $({\mathcal L},\sA_i)\cong {\mathcal O}(-p_i)$.

The classification is summarized in the table below:
\begin{center}
\begin{tabular}{|c||c|c|c|c|c|}
\hline
involution &\multicolumn{4}{c|}{(i)}&(ii)\\
\hline
$b$&\multicolumn{2}{c|}{$0$}&\multicolumn{2}{c|}{$\pi$}&$0$\\
\hline
$({\mathcal L},\sA)$&\multicolumn{2}{c|}{trivial}&
\multicolumn{2}{c|}{$c_1=-1$}
&trivial\\
\hline
O-plane&O9${}^-$&O9${}^+$&O9${}^-$&O9${}^+$&none\\
\hline
\end{tabular}

\begin{tabular}{|c||c|c|c|c|c|c|}
\hline
involution &\multicolumn{3}{c|}{(iii)}&(iv)&\multicolumn{2}{|c|}{(v)}\\
\hline
$b$&\multicolumn{3}{c|}{arbitrary}&arbitrary&\multicolumn{2}{c|}{arbitrary}\\
\hline
$({\mathcal L},\sA)$&\multicolumn{2}{c|}{trivial}&2-torsion
&trivial
&\multicolumn{2}{c|}{trivial}\\
\hline
O-plane&$2\cdot$O8${}^-$&$2\cdot$O8${}^+$&
O8${}^-$ \& O8${}^+$&none&O8${}^-$&O8${}^+$\\
\hline
\end{tabular}

\begin{tabular}{|c||c|c|c|c|c|}
\hline
involution &\multicolumn{5}{c|}{(vi)}\\
\hline
$b$&\multicolumn{3}{c|}{$0$}&\multicolumn{2}{c|}{$\pi$}\\
\hline
$({\mathcal L},\sA)$&\multicolumn{2}{c|}{trivial}&2-torsion&
\multicolumn{2}{c|}{${\mathcal O}(-p),\,\,\inv(p)=p$}\\
\hline
O-plane&$4\cdot$O7${}^-$&$4\cdot$O7${}^+$&
$2\cdot$O7${}^-$ \& $2\cdot$O7${}^+$&
O7${}^+$ \& $3\cdot$O7${}^-$&
O7${}^-$ \& $3\cdot$O7${}^+$\\
\hline
\end{tabular}
\end{center}

There are T-duality relations among them.
(i), (iii) and (vi) with trivial twist are obviously T-dual
to one another.
When T-duality is applied to (ii), $(x,y)\mapsto (x+{1\over 2},y)$,
in the $x$ direction, as in the
case of the circle, we find (iii) with 2-torsion twist.
The latter in turn is T-dual to (vi) with 2-torsion twist.
When T-duality is applied to (ii) in the $y$ direction, we find (iv).
Finally, (i) with $b=\pi$ and (vi) with $b=\pi$ are both T-dual
to (v).
To see this, let us take (\ref{sA2}) as the twist connection for (i) and (vi)
with $b=\pi$.
Since that expression is invariant under translations in $y$, we may perform
T-duality in the $y$ direction.
Since the orientifold action on the Wilson lines is
\beqa
\mbox{(i)}&&
a_x\dd x+a_y\dd y~\longmapsto
-(a_x\dd x+a_y\dd y)+2\pi x\dd y
=-a_x\dd x+(-a_y+2\pi x)\dd y,\nn\\
\mbox{(vi)}&&
a_x\dd x+a_y\dd y~\longmapsto
-(-a_x\dd x-a_y\dd y)+2\pi x\dd y
=a_x\dd x+(a_y+2\pi x)\dd y,\nn
\eeqa
the action on the T-dual coordinates
$(x,\tilde{y})=(x,a_y/2\pi)$ is
\beqa
\mbox{(i)}&& (x,\tilde{y})~\longmapsto~(x,-\tilde{y}+x),\nn\\
\mbox{(vi)}&& (x,\tilde{y})~\longmapsto~(-x,\tilde{y}+x).\nn
\eeqa
These actions are equivalent to the involution (v), $(x,y)\mapsto (y,x)$,
under the coordinate change
$$
(x_{\rm v},y_{\rm v})=(x_{\rm i}-\tilde{y}_{\rm i},\tilde{y}_{\rm i})
=(x_{\rm vi}+\tilde{y}_{\rm vi}, \tilde{y}_{\rm vi}).
$$
This shows that (v) is obtained from (i) and (vi) with $b=\pi$ by T-duality.
Note that the $\tilde{y}_{\rm vi}$ and $\tilde{y}_{\rm i}$
directions are respectively parallel and orthogonal to the fixed line
$x_{\rm v}=y_{\rm v}$. 
Thus, we may also say that (vi) with $b=\pi$
and (i) with $b=\pi$ are obtained from
(v) by T-duality in these two directions.

Let us comment on the structure group of the Chan-Paton bundle $E$
in Case (i) where the involution $\inv$ is the identity.
We suppose that $E$ is purely even and has rank $N$ (we know that
we must set $N=32$ for tadpole cancellation).
In the $b=0$ case, the twist is trivial and
the orientifold isomorphism defines a unitary map
$\indpU:E^*\to E$ such that $\indpU^t=\indpU$ or $-\indpU$. 
This reduces the structure group of $E$ from $U(N)$ 
to $G=O(N)$ or $USp(N)$ --- an orthonormal 
frame $\sigma$ of $E$ is $G$-admissible 
if $U$ maps the dual frame $\sigma^*$ to
$\sigma$ times the identity matrix ${\bf 1}_N$
or the symplectic matrix ${\bf J}_N$.
The condition
$A=\indpU(-A^t)\indpU^{-1}+i^{-1}\indpU\dd \indpU^{-1}$ says that
the connection $A$ preserves the reduction, i.e.,
$A$ can be regarded as an $O(N)$ or $USp(N)$ gauge field.
In the $b=\pi$ case, the twist is non-trivial,
$c_1({\mathcal L})=-1$, and
the orientifold isomorphism defines a unitary map
$\indpU:E^*\otimes {\mathcal L}\to E$ such that $\indpU^t=\indpU$
or $-\indpU$. 
This defines a principal bundle with the structure group
$$
G'=O(N)/\{\pm{\bf 1}_N\}\quad\mbox{or}\quad
USp(N)/\{\pm{\bf 1}_N\}
$$
--- a local section is given by an expression
$\sigma\otimes u^{-{1\over 2}}$ where $\sigma$ is an orthonormal frame
of $E$ and $u$ is a frame of ${\mathcal L}$ with unit length such that
$U$ maps $\sigma^*\otimes u$  to
$\sigma\cdot {\bf 1}_N$ or $\sigma\cdot{\bf J}_N$.
The condition $A=\indpU(-A^t+\sA)\indpU^{-1}+i^{-1}\indpU\dd \indpU^{-1}$
reads
$$
A'=\indpU(-A^{\prime \, t})\indpU^{-1}+i^{-1}\indpU\dd\indpU^{-1}\quad
\mbox{for}\quad A'=A-{1\over 2}\sA.
$$
It says that $A'$ defines a connection of the principal $G'$-bundle.
This $G'$-bundle does not lift to a $G$-bundle
since $c_1({\mathcal L})$ is odd and there is an obstruction to
define a square root ${\mathcal L}^{1\over 2}$.
We find that Case (i) with $b=\pi$ is a {\it
compacitification without a vector structure},
which is the assertion made earlier in \cite{SS,NVS} for
O9${}^-$.
T-duality to (vi) with $b=\pi$ was originally argued in \cite{NVS}.
T-duality to (v) was discussed more recently in \cite{BBBLW}.

\subsection{Higher Dimensional Torus}

Let us classify orientifolds on higher dimensional torus $T^n$, $n\geq 3$.
Instead of looking for orientifolds by all possible involutions, 
we just look for equivalence classes under T-duality and diffeomorphisms.
We use the coordinates $x^1,\ldots, x^n$ of periods $1$.

By T-duality, we can map any involution to the identity or a half-period shift.
Thus, we only have to consider such involutions.
Note that $B+\inv^*B=2B$ for such involutions and hence the B-field components
$B_{ij}$ must be $0$ or $\pi$.
Also, once $(\inv,B)$ is specified, 
all choices of allowed twist connection are equivalent up to flat
$\Lambda$ gauge transformations (as $\Lambda+\inv^*\Lambda=2\Lambda$).
Thus we only need to classify admissible data $(\inv,B)$,
i.e., those which admit a twist connection.
Since the analysis is straightforward, we just record the result
of classification:

If $(\inv',B')$ is admissible on $T^{n'}$ for $n'<n$,
then it determines an admissible data $(\inv,B)$ on $T^n$
--- we set $(\inv,B)=(\inv',B')$ on the first $n'$ coordinates 
and $(\inv,B)=({\rm id},0)$ for the remaining coordinates and components.
As we increase the dimension by one, exactly one new class of
admissible $(\inv, B)$ appears. 
The new class that appears for $T^n$ at even $n$ 
has $\inv=$ the identity and the B-field of maximal rank, say
$$
B=\sum_{i=1}^{n\over 2}\pi\,\dd x^{2i-1}\!\wedge\dd x^{2i}
=\pi \,\dd x^1\wedge \dd x^2
+\cdots +\pi\, \dd x^{n-1}\wedge \dd x^n.
$$
For this $(\inv, B)$, the two choices of crosscap section, $c$ and $-c$,
are inequivalent.
The new class that appears at odd $n$
has $\inv=$ a half-period shift, say, in the $x^n$ direction,
and a maximal rank B-field in the transverse directions, 
$B=\pi \dd x^1\wedge \dd x^2+\cdots +\pi \dd x^{n-2}\wedge \dd x^{n-1}$
for example.
For this $(\inv, B)$, the two choices of crosscap section are equivalent.

We can see this equally easily in the T-dual picture in which the involution
$\inv$ is the inversion $x\to -x$. 
Let us describe the new class of orientifolds that appears at even $n$ 
in this picture.
It has a maximal rank B-field, say
$B=\pi \dd x^1\wedge \dd x^2+\cdots +\pi \dd x^{n-1}\wedge \dd x^n$.
To describe the twist connection, we view the torus $T^n$ as the
product $T^2_{12}\times T^2_{34}\times \cdots\times T^2_{(n-1)n}$,
where $T^2_{ij}$ is the two-torus in the $x^i$-$x^j$ directions.
Then the twist connection is the sum
$\sA_{12}+\sA_{34}+\cdots+\sA_{(n-1)n}$,
where $\sA_{ij}$ is the connection on $T^2_{ij}$ of
 the type that appears in Case (vi) with $b=\pi$ 
in the previous subsection.
The $2^n$ O-planes can be labeled by 
$\vec{p}=(p_{12},p_{34},\ldots, p_{(n-1)n})$
where $p_{ij}$ is one of the four fixed points of the inversion of
$T^2_{ij}$. Let us denote the four fixed points by $p_{ij}^{(0)},p_{ij}^{(1)},
p_{ij}^{(2)}, p_{ij}^{(3)}$ 
and let us assume that $\sA_{ij}$ is such that $p_{ij}^{(3)}$
is the distinguished point, just as in the explicit construction on $T^2$.
Then, the O-plane type is classified according to whether
the number $n_{\vec{p}}$
of $p_{ij}=p^{(3)}_{ij}$ components
is even or odd. Let us count the number of O-plane with odd $n_{\vec{p}}$:
For $n_{\vec{p}}=1$, we have ${n\over 2}$ choices for the 
$p_{ij}=p^{(3)}_{ij}$ component and, for each of them, there are 
$3^{{n\over 2}-1}$ choices of the fixed points in other components.
Generalization to $n_{\vec{p}}\geq 3$ is obvious, and the total is
$$
\sum_{i\,{\rm odd}\atop 1\leq i\leq {n\over 2}}{{n\over 2}\choose i}
3^{{n\over 2}-i}=2^{n-1}-2^{{n\over 2}-1}.
$$
This number is $1,6,28,120$ for $n=2,4,6,8$.
Thus, the number of O${}^{\mp}$-planes
in this orientifold is as below (or the opposite):
$$
\begin{array}{|c||c|c|c|c|}
\hline
n&2&4&6&8\\
\hline
\mbox{O-plane}&\mbox{O7${}^+$ \& $3\cdot$O7${}^-$}
&\mbox{$6\cdot$O5${}^+$ \& $10\cdot$O5${}^-$}
&\mbox{$28\cdot$O3${}^+$ \& $36\cdot$O3${}^-$}
&\mbox{$120\cdot$O1${}^+$ \& $136\cdot$O1${}^-$}
\\
\hline
\end{array}
$$
The new class of orientifolds that appears at odd $n$ has maximal rank
B-field, say 
$B=\pi \dd x^1\wedge \dd x^2+\cdots +\pi \dd x^{n-2}\wedge \dd x^{n-1}$,
and $\sA=\sA_{12}+\cdots +\sA_{(n-2)(n-1)}+\pi\dd x^n$. It has equal number
of O-planes. The distribution of the O-planes at $x^n=0$ 
is as in the table above and the one at $x^n={1\over 2}$ is opposite to it.

The classification is summarized in the table below, 
in the T-duality frame in which $\inv$ is the inversion
and there are $2^n$ O$(9-n)$-planes.
{\small
$$
\begin{array}{|cccccccccc|}
\hline
0&1&2&3&4&5&6&7&8&9\\
\hline
(1,0)&(2,0)&(4,0)&(8,0)&(16,0)&(32,0)&(64,0)&(128,0)&(256,0)&(512,0)\\
&(1,1)&(2,2)&(4,4)&(8,8)&(16,16)&(32,32)&(64,64)&(128,128)&(256,256)\\
&&(3,1)_{{}_2}&(6,2)_{{}_2}&(12,4)_{{}_2}&(24,8)_{{}_2}&(48,16)_{{}_2}
&(96,32)_{{}_2}&(192,64)_{{}_2}&(384,128)_{{}_2}\\
&&&(4,4)_{{}_2}&(8,8)_{{}_2}&(16,16)_{{}_2}&(32,32)_{{}_2}&(64,64)_{{}_2}
&(128,128)_{{}_2}&(256,256)_{{}_2}\\
&&&&(10,6)_{{}_4}&(20,12)_{{}_4}&(40,24)_{{}_4}&(80,48)_{{}_4}
&(160,96)_{{}_4}&(320,192)_{{}_4}\\
&&&&&(16,16)_{{}_4}&(32,32)_{{}_4}&(64,64)_{{}_4}&(128,128)_{{}_4}
&(256,256)_{{}_4}\\
&&&&&&(36,28)_{{}_6}&(72,56)_{{}_6}&(144,112)_{{}_6}&(288,224)_{{}_6}\\
&&&&&&&(64,64)_{{}_6}&(128,128)_{{}_6}&(256,256)_{{}_6}\\
&&&&&&&&(136,120)_{{}_8}&(272,240)_{{}_8}\\
&&&&&&&&&(256,256)_{{}_8}\\
\hline
\end{array}
$$}
\noindent
$(a,b)_m$ is a theory in which
there are $a$ O${}^-$-planes and $b$ O${}^+$-planes
(or the opposite) and the B-field is of rank $m$. 
No subscript means $B=0$.

Toroidal compactifications of Type I with non-zero B-fields was originally
studied in \cite{BPS} and parts of this table were indeed constructed there. 
See also \cite{Bianchi}.
Classification of type distributions of O-planes
on $T^n/\Z_2$ had been given in \cite{triples,BGS} for the case $n=3$ 
and our result reproduces it.
The methods used in these references both look
computationally more involved for higher $n$ and classification
had not been carried out. 
In contrast,
our construction is very simple and quickly led us to the above result.

\subsection{Holomorphic Involutions}
\label{subsec:holo}

As the last class of examples, 
we consider Type II orientifold on $M\times \R^{10-2n}$,
where $M$ is an $n$-dimensional compact complex manifold,
by an involution $\inv$ which is holomorphic on $M$ 
and trivial on $\R^{10-2n}$.
We assume that the B-field is a $(1,1)$ form on $M$.
Then, a twist connection $\sA$ has a curvature without $(0,2)$-form component,
$\bartial_{\sA}^2=0$, 
and hence defines a holomorphic structure on ${\mathcal L}$.
The assumption is automatic if $H^{2,0}(M)=H^{0,2}(M)=0$, e.g., when 
$M$ is a simply connected Calabi-Yau three-fold, in which case
 any $B$ can be made into a $(1,1)$ form by a $\Lambda$
gauge transformation.
Note also that there is a unique twist $({\mathcal L},\sA)$
for any B-field such that $[B+\inv^*B]\in H^2(M,2\pi \Z)$
as long as $M$ is simply connected.

A parallel section of
$\inv^*{\mathcal L}\otimes {\mathcal L}^{-1}$ with respect to
$\inv^*\sA-\sA$ is of course holomorphic.
Conversely, a holomorphic section of 
$\inv^*{\mathcal L}\otimes {\mathcal L}^{-1}$
is necessarily parallel, since its ratio with a parallel section
must be a holomorphic function of a compact complex manifold
and must be a constant.
Thus, a holomorphic section $c$ of 
$\inv^*{\mathcal L}\otimes {\mathcal L}^{-1}$ qualifies as 
a crosscap section if it satisfies $\inv^*c\cdot c=1$.
It can be regarded as a linear map
$c:{\mathcal L}\to \inv^*{\mathcal L}$ and also there is a canonical
map $\inv:\inv^*{\mathcal L}\to {\mathcal L}$ over the map on the base 
$\inv:M\to M$. By composing the two, we find a lift
$\check{\inv}_c$ of $\inv$ to ${\mathcal L}$:
\beq
\begin{array}{cccccc}
\check{\inv}_c:\!\!&\!{\mathcal L}&\!\!\stackrel{c}{\longrightarrow}\!\!&
\inv^*{\mathcal L}&\!\!\stackrel{\inv}{\longrightarrow}\!\!&
{\mathcal L}\\
&\Big\downarrow&&\Big\downarrow&&\Big\downarrow\\
&M&\!\!\stackrel{{\rm id}}{\longrightarrow}\!\!&
M&\!\!\stackrel{\inv}{\longrightarrow}\!\!&M
\end{array}
\eeq
The condition $\inv^*c\cdot c=1$ is equivalent to the statement that
$\check{\inv}_c$ is an involution of ${\mathcal L}$,
\beq
\check{\inv}_c\circ \check{\inv}_c={\rm id}^{}_{\mathcal L}.
\eeq
The value of $c$ at a fixed point of $\inv$ is equal to the value of
$\check{\inv}_c$ at that point.
Thus, {\it a crosscap section may be regarded as
 a holomorphic and involutive lift of $\inv$ to ${\mathcal L}$},
and {\it the type of an O-plane is determined by its value
according to (\ref{Otype})}.

This observation is very useful to find the types of the O-planes.
For illustration, let us consider a particular Calabi-Yau manifold
with a class of holomorphic involutions which were studied in detail
in \cite{BHHW}.
We first introduce a toric variety $X$ 
realized as a symplectic quotient of $\C^6$
by $U(1)\times U(1)$ with the action
$$
x=(x_1,x_2,x_3,x_4,x_5,x_6)\longmapsto
x\cdot (g,h):=(hx_1,hx_2,gx_3,gx_4,gx_5,gh^{-2}x_6).
$$
The quotient is smooth if the parameters $r_1$ and $r_2$ 
for the moment map equations,
$|x_3|^2+|x_4|^2+|x_5|^2+|x_6|^2=r_1$ and
$|x_1|^2+|x_2|^2-2|x_6|^2=r_2$,
are both positive.
As a complex manifold, $X$ is the quotient of
the complement of $\Delta=\{x_1=x_2=0\}\cup \{x_3=x_4=x_5=x_6=0\}$ in $\C^6$
by the complexified group $\C^{\times}\times\C^{\times}$.
Our Calabi-Yau manifold $M$ is a hypersurface of $X$ defined by the
equation
$$
x_1^8x_6^4+x_2^8x_6^4+x_3^4+x_4^4+x_5^4=0.
$$
Let us denote by ${\mathcal O}(q_1,q_2)$ the holomorphic line bundle
over $X$ (and over $M$ by restriction)
whose total space is defined by the quotient of
$(\C^6-\Delta)\times\C$ by the
action
$
(x,v)\longmapsto (x\cdot (g,h), g^{q_1}h^{q_2}v)
$
of $\C^{\times}\times \C^{\times}$.
The first Chern classes of ${\mathcal O}(1,0)$ 
and ${\mathcal O}(0,1)$ form a basis of the cohomology lattice $H^2(M,\Z)$.
As the involutions $\inv$, we consider the sign flips of the
coordinates: $x_i\mapsto\epsilon_ix_i$ for $i=1,\ldots,5$ and 
$x_6\mapsto x_6$. 
Fixed points are found by solving the equation
\beq
\inv(x)=x\cdot (g,h).
\label{2pfxd}
\eeq
The solutions are listed below, together with $(g,h)$
needed for (\ref{2pfxd}):
\newcommand{\lw}[1]{\smash{\lower1.8ex\hbox{#1}}}
\newcommand{\lww}[1]{\smash{\lower4.8ex\hbox{#1}}}
\begin{center}
\begin{tabular}{|c||l|c|l|}
\hline
involution&solution&$(g,h)$&description\\
\hline
\hline
$(+\!+\!+\!+\!+)$&any $x$&$(1,1)$&the whole $M$\\
\hline
$(+\!+\!-\!+\!+)$&$x_3=0$&$(1,1)$&a divisor\\
\hline
\lw{$(+\!+\!-\!-\!+)$}&$x_3=x_4=0$&$(1,1)$&a curve ($C_9$)\\
\cline{2-4}&$x_5=x_6=0$&$(-1,1)$&four lines\\
\hline
\lw{$(+\!+\!-\!-\!-)$}&$x_3=x_4=x_5=0$&$(1,1)$&eight points\\
\cline{2-4}&$x_6=0$&$(-1,1)$&a divisor ($C_3\times \PP^1$)\\
\hline
\lw{$(+\!-\!+\!+\!+)$}&$x_2=0$&$(1,1)$&a divisor (K3)\\
\cline{2-4}&$x_1=0$&$(-1,1)$&a divisor (K3)\\
\hline
\lw{$(+\!-\!-\!+\!+)$}&$x_2=x_3=0$&$(1,1)$&a curve ($C_3$)\\
\cline{2-4}&$x_1=x_3=0$&$(1,-1)$&a curve ($C_3$)\\
\hline
\lww{$(+\!-\!-\!-\!+)$}&$x_2=x_3=x_4=0$&$(1,1)$&four points\\
\cline{2-4}&$x_1=x_3=x_4=0$&$(1,-1)$&four points\\
\cline{2-4}&$x_2=x_5=x_6=0$&$(-1,1)$&four points\\
\cline{2-4}&$x_1=x_5=x_6=0$&$(-1,-1)$&four points\\
\hline
\lw{$(+\!-\!-\!-\!-)$}&$x_2=x_6=0$&$(-1,1)$&a curve ($C_3$)\\
\cline{2-4}&$x_1=x_6=0$&$(-1,-1)$&a curve ($C_3$)\\
\hline
\end{tabular}
\end{center}
$C_g$ in the table stands for a curve of genus $g$.

Any of these involutions acts trivially on
the second cohomology group, as one 
can see by noting that the divisors defining the generating
line bundles
${\mathcal O}(1,0)$ and ${\mathcal O}(0,1)$
are invariant under the sign flips of the coordinates $x_i$.
In particular, 
$B=b_1 c_1({\mathcal O}(1,0))+b_2 c_1({\mathcal O}(0,1))$ satisfies the
condition $[B+\inv^*B]\in H^2(M,2\pi\Z)$ if and only if $2b_1,
2b_2\in 2\pi\Z$.
Thus, we may set $b_1=\pi q_1$ and $b_2=\pi q_2$ where $q_1,q_2=0$
or $1$. For this $B$, the twist is by a holomorphic line bundle
isomorphic to ${\mathcal O}(-q_1,-q_2)$.

As the lift of the involution $\inv$ to the line bundle
${\mathcal L}\cong {\mathcal O}(-q_1,-q_2)$, we may take
\beq
\check{\inv}[x,v]=[\inv(x),v].
\eeq
The value at a fixed point can be expressed in terms of the element
$(g,h)$ that realizes (\ref{2pfxd}):
$\check{\inv}[x,v]=[x\cdot (g,h),v]
=[x,g^{q_1}h^{q_2}v]$.
That is
\beq
\check{\inv}=g^{q_1}h^{q_2}\,\quad\mbox{at the fixed point obeying
(\ref{2pfxd}).}
\eeq
For this choice of lift, $\check{\inv}_c=\check{\inv}$,
the O-plane types are shown in the table below for each value
of $(b_1,b_2)$. 
For the opposite choice, $\check{\inv}_c=-\check{\inv}$, 
the types are all opposite to the table.
\begin{center}
\begin{tabular}{|c||c|c|c|c|}
\hline
involution&$(0,0)$&$(\pi,0)$&$(0,\pi)$&$(\pi,\pi)$\\
\hline
\hline
$(+\!+\!+\!+\!+)$&O9${}^-$&O9${}^-$&O9${}^-$&O9${}^-$\\
\hline
$(+\!+\!-\!+\!+)$&O7${}^-$&O7${}^-$&O7${}^-$&O7${}^-$\\
\hline
\lw{$(+\!+\!-\!-\!+)$}&O5${}^+(C_9)$ \&&O5${}^+(C_9)$ \&&O5${}^+(C_9)$ \&&
O5${}^+(C_9)$ \&\\
&$4\cdot$O5${}^+(\PP^1)$&$4\cdot$O5${}^-(\PP^1)$&$4\cdot$O5${}^+(\PP^1)$&
$4\cdot$O5${}^-(\PP^1)$\\
\hline
\lw{$(+\!+\!-\!-\!-)$}&$8\cdot$O3${}^+$ \&&$8\cdot$O3${}^+$ \&&
$8\cdot$O3${}^+$ \&&$8\cdot$O3${}^+$ \&\\
&O7${}^-(C_3\times \PP^1)$&O7${}^+(C_3\times \PP^1)$&
O7${}^-(C_3\times \PP^1)$&O7${}^+(C_3\times \PP^1)$\\
\hline
\lw{$(+\!-\!+\!+\!+)$}&O7${}^-$(K3)&O7${}^-$(K3)&O7${}^-$(K3)&
O7${}^-$(K3)\\
&O7${}^-$(K3)&O7${}^+$(K3)&O7${}^-$(K3)&
O7${}^+$(K3)\\
\hline
\lw{$(+\!-\!-\!+\!+)$}&O5${}^+(C_3)$&O5${}^+(C_3)$&O5${}^+(C_3)$&
O5${}^+(C_3)$\\
&O5${}^+(C_3)$&O5${}^+(C_3)$&O5${}^-(C_3)$&
O5${}^-(C_3)$\\
\hline
\lww{$(+\!-\!-\!-\!+)$}&$4\cdot$O3${}^+$&$4\cdot$O3${}^+$&$4\cdot$O3${}^+$&
$4\cdot$O3${}^+$\\
&$4\cdot$O3${}^+$&$4\cdot$O3${}^+$&$4\cdot$O3${}^-$&
$4\cdot$O3${}^-$\\
&$4\cdot$O3${}^+$&$4\cdot$O3${}^-$&$4\cdot$O3${}^+$&
$4\cdot$O3${}^-$\\
&$4\cdot$O3${}^+$&$4\cdot$O3${}^-$&$4\cdot$O3${}^-$&
$4\cdot$O3${}^+$\\
\hline
\lw{$(+\!-\!-\!-\!-)$}&O5${}^+(C_3)$&O5${}^-(C_3)$&O5${}^+(C_3)$&
O5${}^-(C_3)$\\
&O5${}^+(C_3)$&O5${}^-(C_3)$&O5${}^-(C_3)$&
O5${}^+(C_3)$\\
\hline
\end{tabular}
\end{center}
The result for the values $(b_1,b_2)=(0,\pi)$ and $(\pi,\pi)$
matches with the result obtained in \cite{BHHW}
based on continuation of RR-charges to the Gepner point and the 
tadpole cancellation condition there.
The same problem was studied in \cite{BrHe}.

As another application, let us revisit the problem of finding
the O-plane types in the orientifold of two-torus, for
Case (vi) with $b=\pi$.
In that case we found that the twist connection determines
the holomorphic line bundle ${\mathcal O}(-p)$ where $p$
is one of the four fixed points.
Let us choose a flat coordinate $z$ defined on a neighborhood
$U_0$ of the point $p$ such that $z(p)=0$. The inversion $\inv$ acts on
it as $z\mapsto -z$.
The line bundle ${\mathcal O}(-p)$ has a local frame $\sigma_0$
on $U_0$ and another frame $\sigma_1$ on a complement 
of $p$, $U_1=T^2-\{p\}$, which are related on the overlap 
by
\beq
\sigma_0(x)=\sigma_1(x)\cdot z(x),\qquad x\in U_0\cap U_1.
\label{defO-p}
\eeq
We may assume that $U_0$ is $\inv$-invariant.
As a lift of $\inv$, we can take
\beqa
&&\check{\inv}(\sigma_0(x)):=\sigma_0(\inv(x))\cdot (-1),\nn\\
&&\check{\inv}(\sigma_1(x)):=\sigma_1(\inv(x)).\nn
\eeqa
Note that the relation (\ref{defO-p}) is maintained
by a minus sign in one of the two equations.
We see that the value of $\check{\inv}$ is $-1$ at $p$ and
$+1$ at the other three fixed points.
This shows that the O7-plane at $p$ is of the opposite type
compared to the three other O7-planes.

\medskip

\section{Topology Of D-Branes}
\label{sec:K}

The goal of this section is to classify the configurations 
of the space filling D-branes
up to continuous deformations including tachyon condensation.

\subsection{Tachyons And Fredholm Operators}

Let $\Hi$ be a separable Hilbert space over $\R$, $\C$, or $\HH$.
One can introduce a topology in the set of bounded linear operators on $\Hi$,
and subsets therein, using the norm
$
|| f||:=\sup_{v\ne 0}|f(v)|/|v|.
$
Let $GL(\Hi)$ be the group of bounded operators of $\Hi$ with bounded inverses
and let $U(\Hi)$ be the subgroup consisting of unitary operators.
The most important fact for us is
{\bf Kuiper's theorem}\cite{Kuiper}:

{\it The groups $GL(\Hi)$ and $U(\Hi)$ are
 contractible to a point.}

\noindent
In particular, any vector bundle over a space $X$
with the fibre $\Hi$ and the structure group 
$GL(\Hi)$ or $U(\Hi)$ is trivializable.

For the real and quaternionic fields, $U(\Hi)$ may as well be denoted
by $O(\Hi)$ and $USp(\Hi)$ respectively. In what follows,
all of the three fields appear. To avoid confusion,
we shall put the field as the subscript, as $\Hi_{\R}$,
$\Hi_{\C}$ and $\Hi_{\HH}$.

\subsubsection{Type IIB --- $\Fr(\Hi_{\C})$}

In Type IIB string theory, D-brane/anti-D-brane pair 
is regarded as the vacuum without 
a D-brane if the tachyon defines a linear isomorphism between the 
Chan-Paton bundles supported by the branes and antibranes
\cite{SenT,WittenK}.
One way to motivate this from the worldsheet point of view
is to look at the long distance behaviour of
the boundary interaction (\ref{At}):
the tachyon enters as the potential term, ${1\over 2}T^2$,
and its positive values have the effect to contract the worldsheet boundary.
Thus, as long as the topology is concerned, we can freely add or remove
such trivial brane-antibrane pairs,
finite or infinite.
We may also consider finite deformation of the tachyon itself.
For example, even if the tachyon is not originally an isomorphism,
if we can make it into an isomorphism by a finite deformation,
the brane-antibrane system is continuously connected to the vacuum.

Let $(\gE,\gA,\gT)$ be a D9-brane configuration. 
Recall that $\gE=E^0\oplus E^1$ and that the tachyon $T$
determines a linear bundle map $T_{10}:E^0\to E^1$.
Let us add infinitely many trivial brane-antibrane pairs.
This is done, for example, by adding a (trivial) Hilbert bundle 
$\underline{\Hi}'_{\C}=X\times \Hi'_{\C}$ to both of $E^0$ and $E^1$
and extending $T_{10}$ by the identity of $\Hi'_{\C}$;
\beq
\begin{array}{ccc}
\underline{\Hi}'_{\C}&\stackrel{\underline{\rm id}}{\longlongrightarrow}&
\underline{\Hi}'_{\C}\\[-0.1cm]
\oplus&&\oplus\\[-0.1cm]
E^0&\stackrel{T_{10}}{\longlongrightarrow}&E^1.
\end{array}
\label{extt}
\eeq
By Kuiper's theorem the two vector bundles
can be trivialized,
$E^0\oplus\underline{\Hi}'_{\C}\cong \underline{\Hi}_{\C}\cong
E^1\oplus\underline{\Hi}'_{\C}$.
Then, (\ref{extt}) can be regarded as
an endomorphism of the trivial bundle $\underline{\Hi}_{\C}=X\times \Hi_{\C}$,
i.e., we have a function of $X$ with values 
in the operators of $\Hi_{\C}$, denoted again by $T_{10}(x)$.
The kernel and the cokernel of $T_{10}(x)$ are finite 
dimensional for any $x$, as they are bounded by the ranks of $E^0$ and $E^1$.
That is, $T_{10}(x)$ is a Fredholm operator. 
Thus, we obtained a continuous map
\beq
T_{10}:X\longrightarrow \Fr(\Hi_{\C}),
\label{tmap}
\eeq
where $\Fr(\Hi_{\C})$ is the space of Fredholm operators on $\Hi_{\C}$.
Continuous deformation of the original D9-brane configuration
$(E,A,T)$ corresponds to continuous deformation of the map (\ref{tmap}),
and vice versa.
Furthermore addition or subtraction of trivial brane-antibrane pairs
to or from $(E,A,T)$ results in the same map $T_{10}$ or at least
a map that can be connected by continuous deformation.
In this sense, the set of homotopy classes of the maps (\ref{tmap}),
which is denoted by
\beq
[X,\Fr(\Hi_{\C})],
\label{XFr}
\eeq
classifies the topology of D-branes.
The space $\Fr(\Hi_{\C})$ is closed under composition of operators. 
This induces the structure of a semi-group in the set (\ref{XFr}).

The spacetime for string theory
is non-compact in almost all cases and we typically impose some
conditions on the configurations of fields, branes, etc.
For example, we often assume translational invariance
in some of the dimensions, say the time plus a part of the space,
in order to describe static configurations of particles and branes. 
In such a case, we simply ignore such `irrelevant' dimensions. 
Also, if we have spatial infinities,
we usually impose certain boundary condition in order to achieve finite energy
or finite tension. In such a case, the map (\ref{tmap}) must obey
the respective boundary condition, or alternatively we 
take one point compactification of the relevant dimensions
and impose the condition at the infinity point.
In what follows, we shall often
assume that the `spacetime' $X$ is compact or compactified
for these reasons.

\subsubsection{Type IIA --- $\skFr(\Hi_{\C})$}
\label{subsub:IIA}

Let us next consider Type IIA string theory.
A D9-brane configuration is regarded as the vacuum 
if the tachyon defines 
an isomorphism of the Chan-Paton bundle to itself.
Note that any vector bundle admits an isomorphism --- the identity.
Therefore, if the spacetime $X$ is compact, any D9-brane
supporting a finite rank vector bundle is continuously connected to 
the vacuum since there is a finite deformation of the tachyon
to the identity, say
\beq
\ugT_t=t\, {\rm id}+(1-t)\ugT.
\label{homotopytoid}
\eeq
Thus, unlike in the Type IIB case, 
in order to have a non-trivial D-brane configuration on a compact space $X$, 
the vector bundle $\ugE$ must be of infinite rank to start with.
If $\ugE$ is indeed of infinite rank, it can be trivialized 
by Kuiper's theorem, 
and the tachyon $\ugT$ can be regarded as a continuous function
with values in self-adjoint operators on a Hilbert space $\Hi_{\C}$.
In order to have finite energy, we need 
$\Ker\, \ugT(x)\cong \Coker\, \ugT(x)$ to be finite dimensional
at any point $x$. Therefore $\ugT(x)$ is a Fredholm operator.
Furthermore, we may assume that the tachyon has infinitely many positive and 
infinitely many negative eigenvalues. To motivate this, 
suppose that $\ugT$ has only finitely many negative eigenvalues.
Then, (\ref{homotopytoid})
defines a homotopy between $\ugT$ and the identity operator, and hence
the brane is continuously deformable to the vacuum.
The similar homotopy works for those with finitely many positive eigenvalues
--- we just replace ${\rm id}$ by $-{\rm id}$.
Following the literature,
we denote by $\skFr(\Hi_{\C})$ the space of {\it skew-adjoint} Fredholm 
operators on $\Hi_{\C}$ which have infinitely many positive imaginary and 
infinitely many negative imaginary eigenvalues. 
Then, we have a map
\beq
\ugT:X\longrightarrow i^{-1}\skFr(\Hi_{\C}),
\eeq
and the set of homotopy classes of such maps
\beq
[X,\skFr(\Hi_{\C})]
\eeq
classifies the topology of D-branes.

For illustration, let us consider the case $X=S^1\times\R^9$ 
where we ignore the dependence in the $\R^9$ direction.
We use the coordinate $x$ of $S^1$ which have periodicity 
$x\equiv x+1$. We consider the tachyon $\ugT(x)$ whose eigenvalues
are
\beq
\lambda_n(x)=\lambda(x-x_0+n).
\label{D8IIA}
\eeq
where $n$ runs over all integers and $\lambda$ is a real number.
This represents a single BPS D8-brane at $x=x_0$.
Note that this operator $\ugT(x)$ is not bounded, but
the usual trick, $\ugT\to \ugT/\sqrt{1+\ugT^*\ugT}$, turns it into
a bounded operator.
In fact, $|T|=\infty$ is the natural value at the vacuum, in the
framework in which the tachyon $T$ appears in the boundary interaction
as (\ref{At}). Thus, we always assume this trick in order to put 
things in the context of Fredhom operators.
Another remark that has to be made here is that a single 
BPS D8-brane at a point of a circle violate the tadpole cancellation
condition and is inconsistent in the full string theory.
However, the tadpole condition can be ignored in the classical limit 
where the string coupling constant is set equal to zero. 
Alternatively, the above $\ugT(x)$ enters as a building block into
the D9-brane configuration for a BPS D$p$-brane at a point
of $S^1$, which has no tadpole problem for $p=0,2,4$.

\begin{figure}[htb]
\psfrag{x}{$x^1$}
\centerline{\includegraphics{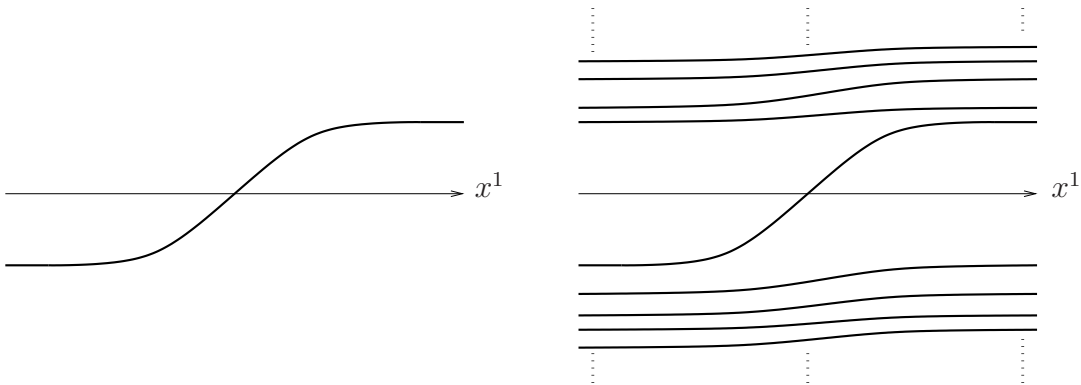}}
\caption{D9-brane configurations for a single D8-brane}
\label{fig:D8}
\end{figure}

If $X$ is non-compact, the configuration can be non-trivial
even when $\rank\,\ugE$ is finite. For example, a BPS D8-brane in $\R^{10}$
can be realized by a tachyon configuration on a rank one vector bundle.
In Section~\ref{subsec:ABS}, 
it is provided by the linear profile $\ugT(x^1)=x^1$. 
If we apply $\ugT\to \ugT/\sqrt{1+\ugT^*\ugT}$ to it,
we obtain a kink as shown in Figure~\ref{fig:D8} (left).
We see that the topology of the configuration is stable under finite 
deformation. For example, the zero point of the tachyon, i.e.,
the location of the BPS D8-brane, 
cannot be removed. Pairs of new zeroes may be created, 
but the number of zeroes with positive slope minus the number of zeroes
with negative slope is always 1.
In non-compact situations, we often partially compactify
the space by attaching one point to the relevant boundary directions.
Let us see whether or how the configuration can be extended 
to the infinity point in the present example. 
We compactify the real line $\R^1$ of $x^1$ to the circle $S^1$
by attaching one point at infinity.
The original kink does not extend to the infinity point
since the limiting values at the two boundaries, 
$x^1\to +\infty$ and $x^1\to -\infty$,
are different --- they even have opposite signs.
To cure this problem, we may add trivial D9-brane configurations,
i.e. those whose tachyons are everywhere non-zero.
We immediately notice
that infinitely many trivial D9-branes are required, in order 
for the spectrum to have the same limit 
as $x^1\to+\infty$ and $x^1\to -\infty$.
In the end, we have infinitely many positive and infinitely many
negative tachyon values, as shown in Figure~\ref{fig:D8} (right). 
(The resulting configuration 
is essentially the same as
the periodic configuration (\ref{D8IIA}) on the circle.)
We are automatically led to a map to $i^{-1}\skFr(\Hi_{\C})$
by the attempt to extend the configuration to the infinity point.

\subsubsection{Type II Orientifolds Without Twist --- $\Tc^k$}
\label{subsub:Tk}

Let us next discuss the topology of D9-brane configurations in
Type II orientifolds with trivial twist.
Before starting, let us collect some useful facts on linear algebra
of hermitian vector spaces and its duals.

\subsubsection*{\it Some Linear Algebras}

\newcommand{\cc}{\boldsymbol{\varsigma}}

We shall take the convention that
a hermitian inner product on a complex vector space
$V$, denoted by $(v,v')_V$, 
is antilinear in the left entry and linear in the right entry.
We define an antilinear map
\beq
h_V:V\longrightarrow V^*,\quad\mbox{by}\quad 
\langle h_V(v),v'\rangle=(v,v')_V.
\label{hV}
\eeq 
It is easy to prove that
$h_{V^*}=h_V^{-1}$, say, by choosing an orthonormal basis and 
its dual basis. For a linear map 
$f:V\to W$, we have
\beq
h_W\circ f\circ h_V^{-1}=(f^{\dag})^t=(f^t)^{\dag}.
\label{P1}
\eeq
This can be proved by straightforward computation;
$\langle h_Wfh_V^{-1}(v^*),w\rangle=(fh_V^{-1}(v^*),w)_W
=(h_V^{-1}(v^*),f^{\dag}(w))_V=\langle v^*,f^{\dag}(w)\rangle
=\langle (f^{\dag})^t(v^*),w\rangle$.
For a linear map $U:V^*\to V$ we define an antilinear map
from $V$ to itself by
\beq
\cc=U\circ h_V:V\stackrel{h_V}{\longrightarrow}V^*
\stackrel{U}{\longrightarrow} V.
\label{P2}
\eeq
Using the above results, we find the property
$\cc^2=Uh_VUh_V=Uh_VUh_{V^*}^{-1}=U(U^{\dag})^t$.
If $U$ is unitary, $U^{\dag}=U^{-1}$,
then we have
\beq
U=\pm U^t\,\,\Longrightarrow\,\, \cc^2=\pm \,{\rm id}_V.
\label{P3}
\eeq

If an antilinear map $\cc:V\to V$ obeys $\cc^2={\rm id}_V$, then 
$V$ has the structure of the complexification of a real vector space,
$V\cong V_{\R}\otimes \C$, and $\cc$ is the complex conjugation.
Indeed let us define a subspace $V_{\R}\subset V$ as a real vector space
by the set of vectors $v$ such that $\cc(v)=v$.
Then, any vector $v\in V$ can be written as ${\rm Re}(v)+i{\rm Im}(v)$
for ${\rm Re}(v)=(v+\cc(v))/2\in V_{\R}$ and
${\rm Im}(v)=(v-\cc(v))/2i\in V_{\R}$.

If an antilinear map $\cc:V\to V$ obeys $\cc^2=-{\rm id}_V$, then 
$V$ has the structure of a quaternionic vector space, $V=V_{\HH}$,
and $\cc$ is the multiplication by $j\in\HH$.
Indeed, we define the $i$, $j$, $k$ multiplications by
$i\cdot v=iv$, $j\cdot v=\cc(v)$, $k\cdot v=i\cc(v)$ respectively.
It is straightforward to check that these obey the quaternion 
algebra, $ij=-ji=k, k^2=-1$, etc.

We define the hermitian conjugate $f^{\dag}$ of an antilinear
map $f:V\to W$ by 
\beq
(f^{\dag}(w),v)_V=(f(v),w)_W.
\label{defhc}
\eeq
If $U$ is unitary, $\cc$ is also unitary
\beq
\cc^{\dag}=\cc^{-1},
\label{P4}
\eeq
i.e.,
$(\cc(v),\cc(v'))_V=(Uh_V(v),Uh_V(v'))_V=(h_V(v),h_V(v'))_{V^*}
=\langle v,h_V(v')\rangle=(v',v)_V$.

Vector spaces in the above remarks are implicitly assumed
to be finite dimensional.
Let $\Hi_{\C}$ be a complex separable Hilbert space. One can define
a map $h_{\Hi_{\C}}:\Hi_{\C}\to \Hom_{\C}(\Hi_{\C},\C)$ just as in
the finite dimensional cases (\ref{hV}). We define the subspace $\Hi_{\C}^*
\subset \Hom_{\C}(\Hi_{\C},\C)$ by the image of $h_{\Hi_{\C}}$.
Then, $\Hi_{\C}^*$ itself is naturally a separable complex Hilbert space.
Under this definition of the dual, 
all of the above remarks apply to Hilbert spaces as well. 
\footnote{In the standard terminology, our $\Hi_{\C}^*$ is 
nothing but the conjugate vector space
$\overline{\Hi}_{\C}$. The latter is equal to $\Hi_{\C}$ as a set,
but with the scalar multiplication rule modified by complex conjugation.
We would also like to warn the reader on possibly confusing use of notation:
We write $f^*$ for the complex conjugation of a linear map/operator $f$
on $\Hi_{\R}\otimes\C$.}

\subsubsection*{\it The Space With Involution \,$\Tc^k$}

Let us take a D9-brane configuration in a Type II orientifold on
a space $X$ with an involution $\inv:X\to X$.
In addition to the gauge field and the tachyon,
we have a unitary bundle map, $\indpU:\inv^*\gE^*\to \gE$ for Type IIB
and $\ugU:\inv^*\ugE^*\to\ugE$ for Type IIA, that obey certain conditions
depending on a mod $8$ integer $k$ (denoted by $[k]$ 
in Introduction) that is determined by the codimension and the 
type of the O-planes. 
By adding infinitely many empty branes if necessary, 
we can trivialize the resulting complex Hilbert bundle(s) by Kuiper's theorem.
In particular, we have families of Fredholm and unitary operators
over $X$,
\beqa
\mbox{$k$ even}:&&
T_{10}(x)\in \Fr(\Hi_{\C})
~~\mbox{and}~~
U(x):\Hi_{\C}^*\oplus \Hi_{\C}^*\to\Hi_{\C}\oplus\Hi_{\C},
\nn\\
\mbox{$k$ odd}:&&
\ugT(x)\in i^{-1}\skFr(\Hi_{\C})~~\mbox{and}~~
\ugU(x):\Hi_{\C}^*\to \Hi_{\C}.
\nn
\eeqa
These obey the following conditions from
(\ref{iiBcond})-(\ref{iiAcond}), where 
$
T\!=\!\left(\begin{array}{cc}
0&\!T_{10}^{\dag}\!\\
\!T_{10}\!&0
\end{array}\right)
$.
$$
\begin{array}{c|cl}
\hline
k&~~~~~~~~~~~~~\mbox{Conditions}\!\!\!\!\!\!\!\!&\\
\hline
\hline\\[-0.5cm]
0& ~  \mbox{$U$ even},~\, U=\inv^*U^t,&T=U\inv^*T^tU^{-1}
\\[0.1cm]
1& ~\ugU=\inv^*\ugU^t,&\ugT=-\ugU\inv^*\ugT^t\ugU^{-1}
\\[0.1cm]
2& ~\mbox{$U$ odd},~\, U=\inv^*U^t, &T=-U\inv^*T^tU^{-1}
\\[0.1cm]
3&~ \ugU=-\inv^*\ugU^t,&\ugT=\ugU\inv^*\ugT^t\ugU^{-1}
\\[0.1cm]
4&~ \mbox{$U$ even},~\, U=-\inv^*U^t,&T=U\inv^*T^tU^{-1}
\\[0.1cm]
5&~ \ugU=-\inv^*\ugU^t,&\ugT=-\ugU\inv^*\ugT^t\ugU^{-1}
\\[0.1cm]
6&~ \mbox{$U$ odd},~\, U=-\inv^*U^t,&T=-U\inv^*T^tU^{-1}
\\[0.1cm]
7&~ \ugU=\inv^*\ugU^t,&\ugT=\ugU\inv^*\ugT^t\ugU^{-1}
\\[0.15cm]
\hline
\end{array}
$$
Using again Kuiper's theorem, now for $\C$, $\R$ and $\HH$,
we can prove that there is a choice of trivialization 
of the Hilbert bundle(s)
such that $U(x)$ or $\ugU(x)$ is $x$-independent.
Since the argument is the same for $k$ even and $k$ odd cases,
we shall only spell out the proof in the latter cases.

The first step is to focus on the fixed point set $X^{\inv}$ 
of the involution. If $X^{\inv}$ is empty, this step is absent 
--- go to the next step.
On this set, the condition $\ugU=\pm \inv^*\ugU^t$ becomes
$\ugU=\pm \ugU^t$. Defining the family of antilinear operators $\cc(x)$
on $\Hi_{\C}$ as in (\ref{P2}),
we find using (\ref{P3}) that $\cc(x)^2=\pm 1$ for any $x$. In particular,
for each point $x\in X^{\inv}$,
$\cc(x)$ provides
$\Hi_{\C}$
with the structure of the complexification of a real Hilbert space
$\Hi'_{\R}(x)$
or of a quaternionic Hilbert space $\Hi'_{\HH}(x)$.
Thus, we find a Hilbert bundle 
$$
\bigcup_{x\in X^{\inv}}\Hi'_{\R}(x)\quad\mbox{or}\quad
\bigcup_{x\in X^{\inv}}\Hi'_{\HH}(x).
$$
Using Kuiper's theorem for $\R$ or $\HH$, we see that we can trivialize
it as the real or quaternionic Hilbert bundle.
Namely, we have found a frame in which the antilinear operator
$\cc(x)$ is independent of $x$. This in turn means that 
$\ugU(x)$ is independent of $x$.

The next step is to construct the frame over the open subset $X-X^{\inv}$.
We take a cell decomposition of $X$ which is $\inv$-compatible, 
i.e., the $\inv$ image of any cell is another cell
and any open cell is either inside or outside $X^{\inv}$.
The strategy is to construct the frame recursively with respect to
the dimension of the cells, so that $\ugU$
is a constant. Let us choose such a constant $\ugU_*$. It must satisfy
$\ugU_*=\pm\ugU_*^t$ if the condition for
$\ugU$ is $\ugU=\pm \inv^*\ugU^t$.
Recall that we have already made a choice of frame on the cells
inside $X^{\inv}$, on which we may assume 
$\ugU\equiv \ugU_*$.
Note that a frame change is given by a unitary operator $\ugM(x)$
that acts on $\ugU(x)$ by
$$
\ugU(x)\longmapsto \ugM(x)\ugU(x)\ugM(\inv(x))^t.
$$
Let us start with the 0-dimensional cells. For a 0-cell $x_i\in X-X^{\inv}$,
we choose unitary operators $\ugM(x_i)$ and $\ugM(\inv(x_i))$
such that $\ugM(x_i)\ugU(x_i)\ugM(\inv(x_i))^t=\ugU_*$.
Then at the mirror 0-cell, $x_{\inv(i)}=\inv(x_i)$,
we also find
\beqa
\lefteqn{\ugM(x_{\inv(i)})\ugU(x_{\inv(i)})\ugM(\inv(x_{\inv(i)}))^t
=\ugM(\inv(x_i))\ugU(\inv(x_i))\ugM(x_i)^t}\nn\\
&=&\pm \ugM(\inv(x_i))\ugU(x_i)^t\ugM(x_i)^t
=\pm (\ugM(x_i)\ugU(x_i)\ugM(\inv(x_i))^t)^t
=\pm \ugU_*^t=\ugU_*.
\nn
\eeqa
Thus, we are done with the 0-cells.
Next let us move on to 1-cells.
Take a 1-cell $\gamma$ which is not inside $X^{\inv}$. 
At the two end points of $\gamma$, say
$x_i$ and $x_j$, and their mirror points,
the frame changing operators $\ugM$ are already chosen.
We choose a path $\ugM(\gamma)$ in $U(\Hi_{\C})$
that connects $\ugM(x_i)$ and $\ugM(x_j)$. 
This is possible because the unitary group $U(\Hi_{\C})$
is connected. Then we choose
$\ugM$ along the mirror 1-cell $\inv(\gamma)$ so that
$\ugM\ugU\inv^*\ugM^t\equiv \ugU_*$ holds along $\gamma$.
Then it also holds along the mirror $\inv(\gamma)$
as we can show just as for the 0-cells.
Now we are done for 1-cells.
This recursive procedure never fails since
the unitary group $U(\Hi_{\C})$ is contractible (Kuiper's theorem).
This is what we wanted to show.

We can now assume that the antilinear operator,
$\cc=U\circ h_{\Hi_{\C}\oplus \Hi_{\C}}$ on $\Hi_{\C}\oplus \Hi_{\C}$
({\it resp}. $\cc=\ugU\circ h_{\Hi_{\C}}$ on $\Hi_{\C}$),
is constant over $X$. We also know that it is even/odd
and satisfies
$\cc^2=\pm 1$ if the o-isomorphism is even/odd and obeys
$\indpU=\pm \inv^*\indpU^t$ ({\it resp}. $\ugU=\pm\inv^*\ugU^t$). 
Using (\ref{P1}) and the hermiticity of
 the tachyon, we see that
 the condition on the tachyon reads as
\beq
\inv^*T=(-1)^{k\over 2}\cc\circ T\circ\cc^{-1}\quad
\left(\mbox{{\it resp}.}\quad
\inv^*\ugT=(-1)^{k+1\over 2} \cc\circ \ugT\circ\cc^{-1}\right).
\eeq
Let us see what it means for each $k\in\Z/8\Z$.

\subsubsection*{\underline{$k=0$}}

The antilinear operator $\cc$ on $\Hi_{\C}\oplus \Hi_{\C}$
is even and squares to $1$.
Therefore it introduces the structure of the complexification of a 
real Hilbert space $\Hi_{\R}$ 
in each of the first and the second Hilbert spaces,
and $\cc$ is simply the complex conjugation operator. 
The condition of the tachyon is therefore
$\inv^*T=T^*$.
In particular, $\urT:=T_{10}:\Hi_{\R}\otimes\C\to \Hi_{\R}\otimes\C$ 
also satisfies
$$
\inv^*\urT=\urT^*.
$$

\subsubsection*{\underline{$k=1$ ({\it resp}. $k=7$)}}

The antilinear operator $\cc$ on $\Hi_{\C}$
squares to $1$.
Thus $\Hi_{\C}$ has the structure of 
the complexification of a real Hilbert space
$\Hi_{\R}$ and $\cc$ is simply the complex conjugation operator.
The condition for $\urT:=\ugT:\Hi_{\R}\otimes \C\to \Hi_{\R}\otimes \C$ 
is therefore
$$
\inv^*\urT=-\urT^*\quad ({\it resp}. \quad \inv^*\urT=\urT^*).
$$

\subsubsection*{\underline{$k=2$ ({\it resp}. $k=6$)}}

The antilinear operator $\cc$ on $\Hi_{\C}\oplus \Hi_{\C}$
is odd and squares to $1$ ({\it resp}. $-1$).
Note that
$\cc T \cc^{-1}=\cc T\cc^{-1}=\cc T^{\dag}\cc^{\dag}=\cc (\cc T)^{\dag}$,
where we have used the hermiticity of $T$ and unitarity (\ref{P4})
of $\cc$. 
Thus, $T\to -\cc T\cc^{-1}$ does $\cc T\to 
-\cc^2(\cc T)^{\dag}=
-(\cc T)^{\dag}$
({\it resp}. $(\cc T)^{\dag}$).
Note that $\cc T$ is even and antilinear.
Let us denote by $\urT:\Hi_{\C}\to \Hi_{\C}$ its restriction 
to the first copy of $\Hi_{\C}$.
The condition on the tachyon reads for this antilinear operator as
$$
\inv^*\urT=-\urT^{\dag} \quad ({\it resp}. \quad \inv^*\urT=\urT^{\dag}).
$$

\subsubsection*{\underline{$k=3$ ({\it resp}. $k=5$)}}

The antilinear operator $\cc$ on $\Hi_{\C}$
squares to $-1$.
Thus $\Hi_{\C}$ has the structure of a quaternionic
Hilbert space, $\Hi_{\C}=\Hi_{\HH}$,
and $\cc$ is multiplication by $j\in \HH$.
The condition for $\urT:=\ugT:\Hi_{\HH}\to \Hi_{\HH}$
is therefore
$$
\inv^*\urT=j\urT j^{-1} \quad ({\it resp}. \quad \inv^*\urT=-j\urT j^{-1}).
$$

\subsubsection*{\underline{$k=4$}}

The antilinear operator $\cc$ on $\Hi_{\C}\oplus \Hi_{\C}$
is even and squares to $-1$.
Therefore it introduces the structure of a 
quaternionic Hilbert space $\Hi_{\HH}$
in each of the first and the second Hilbert spaces,
and $\cc$ is simply multiplication by $j\in \HH$. 
The condition of the tachyon is therefore
$\inv^*T=jTj^{-1}$.
In particular, $\urT:=T_{10}:\Hi_{\HH}\to \Hi_{\HH}$ 
also satisfies
$$
\inv^*\urT=j\urT j^{-1}.
$$

Let us introduce a space with an involution, 
$\Tc^k=\Tc^k(\Hi)$,
consisting of certain type of Fredholm operators 
on a Hilbert space $\Hi$ as follows:
$$
\begin{array}{c|ll}
\hline
~k~&~\mbox{Fredholm operators}&~\mbox{The involution}\\
\hline
\hline\\[-0.5cm]
0&~\urT:\Hi_{\R}\!\otimes\!\C\longrightarrow \Hi_{\R}\!\otimes\!\C,
~\mbox{$\C$-linear}&~\Inv:\urT\longmapsto \urT^*\\[0.1cm]
1&~\urT:\Hi_{\R}\!\otimes\!\C\longrightarrow \Hi_{\R}\!\otimes\!\C,
~\mbox{$\C$-linear, self-adjoint, ipin}&~\Inv:\urT\longmapsto -\urT^*\\[0.1cm]
2&~\urT:\Hi_{\C}\longrightarrow \Hi_{\C},~\mbox{$\C$-antilinear}&~
\Inv:\urT\longmapsto -\urT^{\dag}\\[0.1cm]
3&~\urT:\Hi_{\HH}\longrightarrow \Hi_{\HH},~
\mbox{$\C$-linear, self-adjoint, ipin}&~
\Inv:\urT\longmapsto j\urT j^{-1}\\[0.1cm]
4&~\urT:\Hi_{\HH}\longrightarrow \Hi_{\HH},~\mbox{$\C$-linear}&~
\Inv:\urT\longmapsto j\urT j^{-1}\\[0.1cm]
5&~\urT:\Hi_{\HH}\longrightarrow \Hi_{\HH},~
\mbox{$\C$-linear, self-adjoint, ipin}&~
\Inv:\urT\longmapsto -j\urT j^{-1}\\[0.1cm]
6&~\urT:\Hi_{\C}\longrightarrow \Hi_{\C},~\mbox{$\C$-antilinear}&~
\Inv:\urT\longmapsto \urT^{\dag}\\[0.1cm]
7&~\urT:\Hi_{\R}\!\otimes\!\C\longrightarrow \Hi_{\R}\!\otimes\!\C,
~\mbox{$\C$-linear, self-adjoint, ipin}&~
\Inv:\urT\longmapsto \urT^*\\[0.15cm]
\hline
\end{array}
$$
Here ``ipin'' means that the operator has infinitely many positive and 
infinitely many negative eigenvalues.
We have seen that the tachyon determines a map $\urT$
from $X$ to the space $\Tc^k$ which is equivariant with respect to
the involution $\inv$ on $X$ and the involution $\Inv$ on $\Tc^k$, that is,
$\urT(\inv(x))=\Inv(\urT(x))$.
The set of such $\Z_2$-equivarient maps modulo $\Z_2$-equivarient homotopies,
denoted by
\beq
[X,\Tc^k]_{\Z_2},
\eeq
classifies the topology of D-branes in the orientifold.

\subsection{Clifford Algebras}
\label{subsec:Clifford}

We shall show that the spaces with involution, $\Tc^k$,
can be characterized in terms of Clifford algebras.

Let $\Cl_n$ be the Clifford algebra over $\R$ generated by $\J_1,\ldots,\J_n$
which obey the relations $\{\J_i,\J_j\}=-2\delta_{i,j}$.
It is isomorphic to the following algebra:
$$
\begin{array}{c|ccccccccc}
\hline
n&~~~~0~~~~&~~~~1~~~~&~~~~2~~~~&~~~~3~~~~&~~~~4~~~~&~~~~5~~~~&
~~~~6~~~~&~~~~7~~~~\\
\hline
\Cl_n&\R&\C&\HH&\HH\oplus \HH&\HH(2)&\C(4)&\R(8)&\R(8)\oplus \R(8)\\
\hline
\end{array}
$$
${\bf k}(m)$ stands for the algebra of $m\times m$ matrices over a field 
${\bf k}$.
Also, $\Cl_{n+8}\cong \Cl_n(16)$ where the operation
${\mathcal A}\mapsto {\mathcal A}(16)$ 
replaces each simple factor ${\bf k}(m)$ of
an algebra ${\mathcal A}$ by ${\bf k}(16m)$. For example,
$\Cl_8\cong \R(16)$, $\Cl_{11}\cong \HH(16)\oplus \HH(16)$, etc. 
Irreducible representations of $\Cl_n$ are
${\bf k}^{m}$ for each simple factor ${\bf k}(m)$
and any representation consists of a sum of these.
We refer to the paper \cite{ABS} by Atiyah-Bott-Shapiro 
 for these facts.

For $k\geq 1$, let $\Hi_{\R}$ be a real Hilbert space with
a representation of $\Cl_{k-1}$ such that the generators $\J_i$
act as skew-adjoint operators and
each irreducible representation has infinite multiplicity.
Let $\Frst^k(\HRC)$ be the space of skew-adjoint
Fredholm operators $\rmA$ on $\HRC$
that satisfy the following conditions:
\begin{itemize}
\item[(i)] $\rmA$ anticommutes with $\J_1,\ldots, \J_{k-1}$.
\item[(ii)] For odd $k$, (i) implies that the operator
$$
w(\rmA)=\J_1\J_2\cdots \J_{k-1}\rmA
$$
commutes with $\J_1,\ldots, \J_{k-1}$ and $\rmA$.
It is self-adjoint for $k=-1$ mod 4 and skew-adjoint for $k=1$ mod 4.
The condition is that $w(\rmA)$ ({\it resp}. $i^{-1}w(\rmA)$)
has infinitely many positive and infinitely 
many negative eigenvalues.
\end{itemize}
Conditions (i) and (ii) are vacuous for $k=1$, and hence we have
$\Frst^1(\HRC)=\skFr(\HRC)$. 
For $k=0$, by definition, we put
$\Frst^0(\HRC):=\Fr(\HRC)$ for a real Hilbert space $\Hi_{\R}$.
An important point for us is that
{\it the complex conjugation, $\rmA\mapsto \rmA^*$, defines an
involution of the space $\Frst^k(\HRC)$.}
Indeed Fredholm property and the conditions (i), (ii) 
are invariant under the complex conjugation since
$\Ker\, \rmA^*=(\Ker\, \rmA)^*$, 
 $\J_i^*=\J_i$ and $w(\rmA^*)=w(\rmA)^*$.

The Hilbert space $\Hi_{\R}$ with a representation of $\Cl_{k-1}$
of the above type is unique up to isomorphisms.
Therefore we may simply write $\Frst^k$ for $\Frst^k(\HRC)$.
Also, because of the mod 8 periodicity of the Clifford
algebras and their representations, we have
$\Frst^{k+8}\cong \Frst^k$.

We claim that $\Tc^k$ can be identified 
with $\Frst^k$ 
as the space with involution.
In fact, we shall provide a bijection
\beq
\Frst^k(\HRC)~\stackrel{\cong}{\longrightarrow}~\Tc^k(\Hi)
\eeq
for a certain Hilbert space $\Hi$ related to $\Hi_{\R}$,
which is equivariant with respect to the complex conjugation of $\Frst^k(\HRC)$
and the involution $\Inv$ of $\Tc^k(\Hi)$.
The proof that the map indeed sends $\Frst^k(\HRC)$ to
$\Tc^k(\Hi)$, is $\Z_2$-equivariant and is bijective is
straightforward and is left as an exercise for the reader.

\subsubsection*{\underline{$k=0$}}

$\Frst^0(\HRC)$ and $\Tc^0(\HRC)$ are identical
as the space with involution.

\subsubsection*{\underline{$k=1$}}

The map is $\Frst^1(\HRC)\longto \Tc^1(\HRC)$
given by $\rmA\longmapsto \urT=i^{-1}\rmA$.

\subsubsection*{\underline{$k=2$}}

By $\Cl_1\cong \C$ (where $\J_1\leftrightarrow i$),
the Hilbert space $\Hi_{\R}$ itself
has the structure of a complex Hilbert space, $\Hi_{\R}=\Hi_{\C}$.
The map is
$$
\rmA=\rmA_1+i\rmA_2\in \Frst^2(\HRC)~\longmapsto~
\urT=\rmA_1+i\rmA_2\in \Tc^2(\Hi_{\C}),
$$
where $\rmA_1$ and $\rmA_2$ are the real and the imaginary part of $\rmA$
(we shall use this notation in what follows as well).

\subsubsection*{\underline{$k=3$}}

By $\Cl_2\cong \HH$ (say, $\J_1\leftrightarrow i$, $\J_2\leftrightarrow j$
and $\J_1\J_2\leftrightarrow k$),
the Hilbert space $\Hi_{\R}$
can be regarded as a quaternionic Hilbert space, $\Hi_{\R}=\Hi_{\HH}$.
The map is
$$
\rmA=\rmA_1+i\rmA_2\in \Frst^3(\HRC)~\longmapsto~
\urT=w(\rmA_1)+iw(\rmA_2)\in \Tc^3(\Hi_{\HH}).
$$

\subsubsection*{\underline{$k=4$}}

By $\Cl_3\cong \HH\oplus \HH$ (say, 
$\J_1\leftrightarrow (i,-i)$, $\J_2\leftrightarrow (j,-j)$
and $\J_3\leftrightarrow (k,-k)$)
and the assumption on the multiplicity
of the irreducible representations of the simple factors,
we may write $\Hi_{\R}=\Hi^+_{\HH}\oplus \Hi^-_{\HH}$, where
$\Hi^+_{\HH}$ ({\it resp}. $\Hi^-_{\HH}$)
is the subspace on which the first ({\it resp}. second) $\HH$ acts 
non-trivially. Note that any operator from $\Hi^+_{\HH}$ to $\Hi^-_{\HH}$
that anticommutes with $\J_i$'s is $\HH$-linear.
The map is
$$
\rmA=\rmA_1+i\rmA_2\in \Frst^4(\HRC)~\longmapsto~
\urT=\rmA_1+i\rmA_2\in \Tc^4(\Hi^+_{\HH},\Hi^-_{\HH}).
$$

\subsubsection*{\underline{$k=5$}}

By $\Cl_4\cong \HH(2)$, we may write $\Hi_{\R}
=\HH^2\otimes_{\HH}^{}\Hi_{\HH}$ for a quaternionic
Hilbert space $\Hi_{\HH}$. 
For an operator $f$ of $\Hi_{\R}$ that commutes with
$\Cl_4$, there is an $\HH$-linear operator
$\widehat{f}$ of $\Hi_{\HH}$ 
such that $f={\rm id}_{\HH^2}\otimes \widehat{f}$.
The map to
$$
\rmA=\rmA_1+i\rmA_2\in \Frst^5(\HRC)~\longmapsto~
\urT=i^{-1}\left(\widehat{w(\rmA_1)}+i\,\widehat{w(\rmA_2)}\right)
\in  \Tc^5(\Hi_{\HH}).
$$

\subsubsection*{\underline{$k=6$}}

By $\Cl_5\cong \C(4)$, we may write $\Hi_{\R}
=\C^4\otimes_{\C}^{}\Hi_{\C}$ for a complex
Hilbert space $\Hi_{\C}$. 
Also, the automorphism of $\Cl_5$ that flips the sign of
$\J_i$'s can be mapped to the automorphism
$\varphi\mapsto J\varphi^* J^{-1}$ of $\C(4)$ where $J$ is a real
antisymmetric $4\times 4$ matrix.
For an operator $f$ of $\Hi_{\R}$ that anticommutes with $\J_i$'s,
there is a $\C$-antilinear operator $\widehat{f}$ of
$\Hi_{\C}$ such that $f(v\otimes h)=J(v^*)\otimes \widehat{f}(h)$.
The map is
$$
\rmA=\rmA_1+i\rmA_2\in \Frst^6(\HRC)~\longmapsto~
\urT=\widehat{\rmA}_1+i\,\widehat{\rmA}_2
\in\Tc^6(\Hi_{\C}).
$$

\subsubsection*{\underline{$k=7$}}

By $\Cl_6\cong\R(8)$, we may write $\Hi_{\R}=\R^8\otimes_{\R}\Hi'_{\R}$
for a real Hilbert space $\Hi'_{\R}$.
For an operator $f$ of $\Hi_{\R}$ that commutes with
$\Cl_6$, there is an $\R$-linear operator $\widehat{f}$ of $\Hi'_{\R}$
such that $f={\rm id}_{\R^8}\otimes \widehat{f}$.
The map is
$$
\rmA=\rmA_1+i\rmA_2\in \Frst^7(\HRC)~\longmapsto~
\urT=\widehat{w(\rmA_1)}+i\,\widehat{w(\rmA_2)}
\in  \Tc^7(\Hi'_{\R}\!\otimes\!\C).
$$

\subsection{Fredholm Operators And K-Theory}
\label{subsec:K}

\newcommand{\sus}{\,{\rm S}}

Let $\K(X)$ be the Grothendieck group of the category of finite rank complex 
vector bundles over a compact space $X$.
It may be defined as the set of pairs $(E^0,E^1)$ of 
isomorphism classes of vector bundles
modulo the equivalence relations generated by
$(E^0,E^1)\sim (E^0\oplus F,E^1\oplus F)$.
It is a group under the sum $(E^0,E^1)+(F^0,F^1)
:=(E^0\oplus F^0,E^1\oplus F^1)$
 --- the zero is $(F,F)$, and the negative is $-(E^0,E^1)
=(E^1,E^0)$.
For a space with a base point, $X=(X,x_0)$,
we denote by $\rK(X)$ the subgroup of $\K(X)$ 
consisting of elements that restrict to zero at the base point $x_0$.
For a closed subspace $Y$ of a space $X$, we denote
by $X/Y$ the space with a base point
obtained from $X$ by contracting $Y$ to one point which becomes 
the base point.
Then we put $\K(X,Y):=\rK(X/Y)$.
For a space with a base point $X=(X,x_0)$, 
we define its reduced suspension by
$$
\sus X={I\times X\over (\partial I\times X)\cup (I\times x_0)}.
$$
$\sus^iX$ denotes the $i$-times operation of $\sus$ on $X$.
For a space $X$, we denote by $X^+$ the disjoint union of $X$ and a point
$*$ which is regarded as the base point of $X^+$.
It is easy to see that $\sus^iX^+=(I^i\times X)/(\partial I^i\times X)$.
We put
\beqa
&&\K^{-i}(X):=\rK(\sus^iX^+)=\K(I^i\times X,\partial I^i\times X),\nn\\
&&\K^{-i}(X,Y):=\rK(\sus^i(X/Y)).\nn
\eeqa

For a space $X$ with an involution $\inv:X\to X$, 
a {\it Real vector bundle} $E$ is a complex vector bundle with an antilinear
involution over $\inv$, i.e., 
the involution is an antilinear map
of the fibre over $x$ to the fibre over $\inv(x)$.
The Grothendieck group
of the category of finite rank Real vector bundles over
a compact space with involution, $X=(X,\inv)$, is denoted by
$\KR(X)$. We also have $\KR(X,Y)$
for an invariant closed subspace $Y$ of $X$.
We define $\KR^{-i}(X), \KR^{-i}(X,Y)$ as before
where we assume trivial action of the involution on the $I$ factors.
When the involution $\inv$ is the identity map of $X$,
a Real vector bundle $E$ is the complexification of
a real vector bundle on which the involution acts as the complex conjugation.
See the paragraph below (\ref{P3}).
Hence the category of Real vector bundles is identical to
the category of real vector bundles. In such a case, we
write $\KO^{-i}$ for $\KR^{-i}$.

These K-theory functors enjoy the property of generalized cohomology 
theory, such as the long exact sequence for a pair $Y\subset X$
and Mayer-Vietoris property.
They also obey Bott periodicity,
$\K^{-i}\cong \K^{-(i+2)}$
and $\KR^{-i}\cong \KR^{-(i+8)}$.

D-brane charges in Type I, Type IIB and Type IIA string theories
are classified by K-theories,
$\KO$, $\K$, and $\K^{-1}$ respectively, as
proposed and shown in \cite{MM,WittenK} (for I and IIB)
and in \cite{WittenK,Horava} (for IIA).
It was also proposed and argued in \cite{WittenK,H,BGH}
that D-brane charges in Type II orientifold with O$p^-$
({\it resp}. O$p^+$) planes only is classified 
by $\KR^{-(9-p)}$ ({\it resp}. $\KR^{-(5-p)}$).

\subsubsection{The Theorem of Atiyah and J\"anich}

\newcommand{\ind}{{\rm index}}
\newcommand{\urV}{{\rm V}}

The interpretation of $\K(X)$ as the lattice of D-brane charges
in Type IIB string theory on $X$ is very natural --- the pair $(E^0,E^1)$
consists of the Chan-Paton vector bundles supported 
on branes and antibranes, and
the relation $(E^0,E^1)\sim (E^0\oplus F,E^1\oplus F)$
corresponds to brane-antibrane creation and annihilation.
On the other hand, we have seen that the semi-group $[X,\Fr(\Hi_{\C})]$
classifies the topology of D-branes in the same theory.
This implies that the two classifying (semi-)groups,
$\K(X)$ and $[X,\Fr(\Hi_{\C})]$, must be equivalent in some way.

In fact, a direct link between the two (semi-)groups had been established
a long time ago (around 1964) by Atiyah and J\"anich \cite{Atiyah,Janich}:
{\it There is a natural isomorphism of semi-groups}
\beq
\ind:[X,\Fr(\Hi_{\CC})]~\longrightarrow~ \K(X).
\label{classK}
\eeq
For a map $\urT:X\to\Fr(\Hi_{\C})$, 
the kernel and the cokernel
of $\urT_x=\urT(x)$ are finite dimensional for any $x\in X$.
If their dimensions are constant, they form vector bundles, denoted by
$\Ker(\urT)$ and $\Coker(\urT)$.
Then, we put $\ind(\urT)=(\Ker(\urT),\Coker(\urT))\in \K(X)$.
In general, the dimensions of $\Ker\,\urT_x$ and $\Coker\, \urT_x$
may jump as $x$ varies while their difference, the index of $\urT_x$, 
stays the same.
In such a case, $\ind(\urT)$ is defined as follows \cite{Atiyah}.
First, for any point $x\in X$ we denote by $\urV_x$ the orthogonal 
complement of $\Ker\, \urT_x$ in $\Hi_{\C}$. One can show that
there is an open neighborhood $U_x$ of $x$ such that
$\Ker\,\urT_y\cap \urV_x=\{0\}$ for any $y\in U_x$ and that
the family of vector spaces $\Hi_{\C}/\urT_y(\urV_x)$ 
parametrized by $y\in U_x$ form a trivial vector bundle over $U_x$.
Since $X$ is compact, it can be covered by finite number of such 
open subsets, say $U_{x_i}$'s. 
Denote the intersection of all $\urV_{x_i}$'s by $\urV$.
It is a closed subspace of $\Hi_{\C}$ of finite codimension.
Then, we have $\Ker\, \urT_x\cap \urV=\{0\}$
for any $x\in X$ and the family of vector spaces 
$\Hi_{\C}/\urT_x(\urV)$ form a vector bundle over $X$, denoted by
$\underline{\Hi}_{\C}/\urT(\urV)$.
Then, we put
$\ind(\urT)=(\underline{\Hi}_{\C}/\urV,\underline{\Hi}_{\C}/\urT(\urV))$.
The index map is injective since the kernel can be shown to be
equal to $[X,GL(\Hi_{\C})]$, and that is one point by Kuiper's theorem.

The definition of the index map
is very natural from the tachyon condensation picture.
Also, we had already constructed the inverse map:
Given $(E^0,E^1)\in \K(X)$ we can find a family of Fredholm operators
over $X$ by (\ref{extt}) in which we may set $T_{10}=0$.
It is not difficult to see that the resulting family gives rise to
$(E^0,E^1)$ under the index map.
Thus, the index map is precisely what we expected as the relation between
the two (semi-)groups 
through D9-brane configurations in Type IIB string theory.

The same holds for D-branes in Type I string theory. The relevant map is
\beq
\ind:[X,\Fr(\Hi_{\R})]~\longmapsto ~ \KO(X).
\label{classKO}
\eeq

\subsubsection{The Atiyah-Singer Theorem}

The interpretation of $\K^{-1}(X)$ as the group of D-brane charges
in Type IIA string theory is less straightforward. 
In fact, it is best to go through $[X,\skFr(\Hi_{\C})]$ 
which we had already established in Section~\ref{subsub:IIA} as the set 
that classifies the topology of D-branes.
Again, the relevant mathematical fact for us had been obtained 
a long time ago (1969) by Atiyah-Singer \cite{ASFred}:
{\it There is a homotopy equivalence}
\beq
\alpha:\skFr(\Hi_{\C})~\longrightarrow ~\Omega\Fr(\Hi_{\C}),
\eeq
where $\Omega\Fr(\Hi_{\C})$ is the based loop space at
${\rm id}_{\Hi_{\C}}$, i.e., 
the space of maps $f:I\longrightarrow \Fr(\Hi_{\C})$
with the boundary condition $f(0)=f(1)={\rm id}_{\Hi_{\C}}$.
Note that $[X,\Omega\Fr(\Hi_{\C})]$ is identical to
the set $[\sus X^+,\Fr(\Hi_{\C})]_0$
of homotopy classes of maps from 
$\sus X^+=(I\times X)/(\partial I\times X)$ to 
$\Fr(\Hi_{\C})$ that send the base point
$[\partial I\times X]$ to the base point ${\rm id}_{\Hi_{\C}}$.
By the index map, the latter set
 is mapped bijectively onto $\rK(\sus X^+)=\K^{-1}(X)$.
Thus, we find a natural bijection
\beq
[X,\skFr(\Hi_{\C})]~\stackrel{\alpha}{\longrightarrow}~
[X,\Omega\Fr(\Hi_{\C})]=[\sus X^+,\Fr(\Hi_{\C})]_0
~\stackrel{\ind}{\longrightarrow}~ \K^{-1}(X).
\label{classK1}
\eeq
This interpretation of $\K^{-1}(X)$
as the classifying set of D-brane charges
was used in \cite{H} in the definition of T-duality map.
It was also revisited in \cite{Wittenoverv}.

The paper \cite{ASFred} also shows something that will be important
for us.
Let $\Hi_{\R}$ be the real Hilbert space with 
a $\Cl_{k-1}$ action as in the definition of $\Frst^k(\HRC)$.
We denote by $\Frst^k(\Hi_{\R})$ the space of skew-adjoint Fredholm 
operators on $\Hi_{\R}$ satisfying (i) and, for $k=-1$ mod $4$, (ii).
It can be regarded as the subspace of $\Frst^k(\HRC)$ 
consisting of real operators, 
i.e., the fixed point set of the complex conjugation on $\Frst^k(\HRC)$.
Let $\Hi_{\C}$ be the complex analog of $\Hi_{\R}$, which can
be realized, say, by $\Hi_{\C}=\HRC$.
We denote by $\Frst^k(\Hi_{\C})$ the space of skew-adjoint Fredholm 
operators on $\Hi_{\C}$ satisfying (i) and (ii).
When $\Hi_{\C}=\HRC$, it is simply the same space as
$\Frst^k(\HRC)$ in which we forget about the real structure.
Let $\Hi$ be such $\Hi_{\R}$ {\it resp}.\! $\Hi_{\C}$.
By the assumption on the multiplicity of the irreducible representations,
the representation of $\Cl_{k-1}$ on $\Hi$ extends to
a representation of $\Cl_{k+1}\supset \Cl_{k-1}$.
In particular, $\J_k$, regarded as an operator of $\Hi$,
belongs to $\Frst^k(\Hi)$ (condition (ii) is satisfied since
$w(\J_k)\J_{k+1}=-\J_{k+1}w(\J_k)$ for odd $k$), and is taken
as its base point. We put $\J_0={\rm id}_{\Hi}$ for $k=0$.
Let $\Omega\Frst^{k-1}(\Hi)$ be the based loop space at $\J_{k-1}$.
The result of \cite{ASFred} is that
{\it there is a homotopy equivalence}
\beq
\alpha:\Frst^k(\Hi)~\longrightarrow~\Omega\Frst^{k-1}(\Hi),
\label{ASth}
\eeq
given by
\beq
\alpha(\rmA)(t)=\left\{\begin{array}{ll}
\J_{k-1}\cos(2\pi t)+\rmA\sin(2\pi t)&0 \leq t\leq {1\over 2}\\[0.2cm]
\J_{k-1}\cos(2\pi (1-t))+\J_k\sin(2\pi(1-t))&{1\over 2}\leq t\leq 1.
\end{array}\right.
\label{alphadef}
\eeq
The base point $\J_k$ of
$\Frst^k(\Hi)$ is mapped to a loop that is contractible in
$\Frst^{k-1}(\Hi)$ to the constant loop at $\J_{k-1}$.
In particular, we have
$$
[X,\Frst^k(\Hi)]\,=\,[X^+,\Frst^k(\Hi)]_0\,\stackrel{\alpha_*}{\cong}\,
[X^+,\Omega\Frst^{k-1}(\Hi)]_0\,\cong \,
[\sus X^+,\Frst^{k-1}(\Hi)]_0
$$
where we used the 
standard relation
$[(X,x_0),\Omega(Y,y_0)]\cong [\sus(X,x_0),(Y,y_0)]$
in the last step.
Applying this repeatedly, and applying the index map at the end,
we find that $\Frst^k(\Hi)$ is
a classifying space for $\KO^{-k}$ {\it resp}.\! $\K^{-k}$:
\beqa
&&[X,\Frst^k(\Hi_{\R})\,] ~ \cong ~ \KO^{-k}(X),
\label{classKOk}\\
&&[X,\Frst^k(\Hi_{\C})\,] ~ \cong ~ \K^{-k}(X).
\label{classKk}
\eeqa

\subsubsection{$\Tc^k$ And $\KR^{-k}$}

We shall now show that $\Tc^k\cong \Frst^k$ is the classifying
$\Z_2$-space for $\KR^{-k}$. We start with the case $k=0$
where $\Frst^0(\HRC)=\Fr(\HRC)$. It is known \cite{Matumoto-Fred}
that {\it there is a natural isomorphism of semi-groups}
\beq
\ind:[X,\Fr(\HRC)]_{\Z_2}~\longrightarrow~ \KR(X).
\label{classKR}
\eeq

Let us first define this map. Note that the trivial bundle 
$X\times \HRC$ has a Real structure
$$
\widehat{\inv}:(x,h)~\longmapsto ~(\inv(x),h^*).
$$
For a $\Z_2$-map $\urT:X\to \Fr(\HRC)$, 
i.e., $\urT_{\inv(x)}=\urT_x^*$,
the map $\widehat{\inv}$ induces Real structures 
in the families of vector spaces,
$\Ker\,\urT_x$ and $\Coker\,\urT_x$.
If they define complex vector bundles,
then, $\ind(\urT)=(\Ker(T),\Coker(\urT))$ is an element of $\KR(X)$.
In general, we proceed as in the definition of (\ref{classK}).
We can find finite number of points $x_i$ with the neighborhoods
$U_{x_i}$ satisfying the condition as in the complex case
such that $U_{x_i}$'s and $U_{\inv(x_i)}$'s cover $X$.
Define $\urV$ to be the intersection of $\urV_{x_i}$'s
and $\urV_{\inv(x_i)}$'s.
Since $\overline{\urV}_{x_i}=\urV_{\inv(x_i)}$,
it is invariant under the complex conjugation, $\overline{\urV}=\urV$.
The trivial bundle $\underline{\Hi}_{\C}/\urV$ is then
a trivial Real bundle by $\widehat{\inv}$.
The same map $\widehat{\inv}$ introduces the structure of a Real bundle on
$\underline{\Hi}_{\C}/\urT(\urV)$ as well, since
$\urT_x(\urV)^*=\urT_x^*(\urV)=\urT_{\inv(x)}(\urV)$.
Then we put $\ind(\urT)=(\underline{\Hi}_{\C}/\urV,
\underline{\Hi}_{\C}/\urT(\urV))\in \KR(X)$.

A Real bundle pair $(E^0,E^1)$ over $X$
defines a $k=0$ D9-brane configuration with vanishing tachyon.
Applying the construction of Section~\ref{subsub:Tk} to it,
we find a $\Z_2$-map $X\to \Fr(\HRC)$ whose index is $(E^0,E^1)\in \KR(X)$.
This gives a right inverse to the index map.
Therefore, to see that they are bijections, we need to
show that the index map (\ref{classKR}) is injective.

The kernel of the map (\ref{classKR}) is the set $[X,GL(\HRC)]_{\Z_2}$
where the involution of $GL(\HRC)$ is the complex conjugation,
as one can show by adapting the argument used in \cite{Atiyah}.
We want to show that it consists of one point, i.e.,
 that any $\Z_2$-map $f:X\to GL(\HRC)$
can be deformed by a $\Z_2$-homotopy to the constant map to 
the identity element ${\bf 1}$.
This is true by the $\Z_2$-equivariant contractibility
of $GL(\HRC)$ \cite{Segal-Gcontr}.
For a space $X$ with a $\inv$-compatible cell decomposition,
we may also proceed as in Section~\ref{subsub:Tk} where we proved
that $\ugU(x)$ can be made constant.
First we focus on the fixed point set $X^{\inv}$.
On this set the $\Z_2$-map satisfies $f(x)=f(x)^*$, that is,
it is a map into $GL(\Hi_{\R})\subset GL(\HRC)$.
Using Kuiper's theorem for $\R$ we can find a homotopy from
$f|_{X^{\inv}}$ to the constant map to ${\bf 1}$.
Next, we extend it to a $\Z_2$-homotopy on $X-X^{\inv}$,
using a $\inv$-compatible cell decomposition.
This is possible thanks to Kuiper's theorem for $\C$.
This shows that the set $[X,GL(\HRC)]_{\Z_2}$ consists of one point,
thus proving injectivity and hence bijectivity of
the index map (\ref{classKR}).

Let us next proceed to higher $k$.
The map (\ref{ASth}) for $\Hi=\HRC$ is equivariant with respect to the
complex conjugation.
By the Atiyah-Singer theorem,
it is a homotopy equivalence of ordinary spaces
and also induces a homotopy equivalence between 
the subspaces of $\Z_2$-fixed points.
Then, is it a homotopy equivalence of $\Z_2$-spaces?
A question of this type had been asked in 
\cite{Bredon,Matumoto-GCW,James-Segal}:
Let $G$ be a compact group and let $f:Y\to Z$ be a $G$-equivariant map between
$G$-spaces. Suppose $f:Y^H\to Z^H$ is a homotopy 
equivalence for any closed subgroup $H\subset G$. Then,
is $f$ a homotopy equivalence of $G$-spaces? 
Affirmative answers of various levels were obtained
under various additional assumptions.
For us the following from Theorem (5.4) of \cite{Bredon}, Chapter II 
(see also \cite{Matumoto-GCW}) suffices:
If $X$ is a $G$-space with a $G$-compatible cell decomposition, 
  then $f$ induces a bijection $[X,Y]_G\cong [X,Z]_G$.
Applying this to our problem,
assuming that $X$ has a $\Z_2$-compatible cell decomposition,
we find the bijection in the middle,
$$
\begin{array}{ccc}
[\, X,\Frst^k(\HRC)\,]_{\Z_2}&&\\
\Vert&&\\
{[}\,X^+,\Frst^k(\HRC)\,]_{\Z_2,0}&\cong&
[\,X^+,\Omega\Frst^{k-1}(\HRC)\,]_{\Z_2,0}\\
&&\Vert\\
&&[\,\sus X^+,\Frst^{k-1}(\HRC)]_{\Z_2,0}.
\end{array}
$$
Applying this repeatedly and applying the index map
at the end, we find a bijection
\beq
[\,X,\,\Frst^k(\HRC)\,]_{\Z_2}~\cong~\KR^{-k}(X).
\eeq 
Namely, $\Tc^k\cong \Frst^k(\HRC)$ is a classifying 
$\Z_2$-space for $\KR^{-k}$.

Since we have seen that $[X,\Tc^k]_{\Z_2}$ classifies the topology of
D-branes in the Type II orientifold on $X$, so does $\KR^{-k}(X)$.
In particular, the proposal of \cite{H} is derived.

\subsection{Type II Orientifolds With Twists}
\label{subsec:twistK}

\newcommand{\whinv}{\widehat{\inv}}

Let us now consider Type II orientifolds with non-trivial twist,
in which the D9-branes have the structure 
as summarized in Section~\ref{subsec:summary}.
The goal is to describe the classification of the topology of
D-branes in terms of a certain kind of K-theory.

\subsubsection{Twisted Real Bundles}

Let $X$ be a compact space with an involution $\inv:X\to X$.
Let ${\mathcal L}$ be a complex line bundle over $X$
such that $\inv^*\overline{\mathcal L}\otimes {\mathcal L}$ is
topologically trivial.
We choose a trivialization 
$c:\inv^*\overline{\mathcal L}\otimes {\mathcal L}\to 
\underline{\C}$ such that $\inv^*c$ is equal to $c^*$,
the complex conjugate of $c$.
We shall call a space $X$ with such data, $\inv,{\mathcal L}$ and $c$, a
{\it twisted Real space}.
A {\it twisted Real vector bundle} over a twisted Real space
$(X,\inv,{\mathcal L},c)$
is a complex vector bundle $E$ over $X$ equipped with
an antilinear map $\whinv:E\to E\otimes {\mathcal L}^{-1}$ over
$\inv$ that squares to $c$. That is, we have an antilinear
map $\whinv_x:E_x\to (E\otimes {\mathcal L}^{-1})_{\inv(x)}$ for each $x\in X$
such that the composition
\beq
E_x\stackrel{\whinv_x}{\longrightarrow}
E_{\inv(x)}\otimes {\mathcal L}^{-1}_{\inv(x)}
\stackrel{\whinv_{\inv(x)}}{\longrightarrow}
E_x\otimes {\mathcal L}^{-1}_x\otimes \overline{\mathcal L}^{-1}_{\inv(x)}
\eeq
is equal to multiplication by 
$c(x)\in (\inv^*\overline{\mathcal L}\otimes{\mathcal L})^{-1}_x$.
An isomorphism between two twisted Real bundles, say,
from $(E_1,\whinv_1)$ to $(E_2,\whinv_2)$,
is provided by a linear map $f:E_1\to E_2$ such that
$\whinv_{2x}\circ f_x=f_{\inv(x)}\circ\whinv_{1x}$.
We denote by $\KR(X,\inv,{\mathcal L},c)$ or more simply by
$\KR(X;c)$ the Grothendieck group
of the category of finite rank twisted Real bundles
over the twisted Real space $(X,\inv,{\mathcal L},c)$.
For an ordinary Real space $(X,\inv)$, i.e.,
when ${\mathcal L}$ is the trivial bundle $X\times \C$
and if $c=1$, then a twisted Real bundle is an ordinary Real bundle
over $(X,\inv)$ and the group $\KR(X;c)$ is equal to the ordinary KR group
$\KR(X)$.
Also, we shall later define $\KR^{-i}(X,\inv,{\mathcal L},c)=\KR^{-i}(X;c)$
which agrees with $\KR^{-i}(X)$ for an ordinary Real space.

Let $(B,{\mathcal L},\sA,c)$ be the data for
a Type II orientifold on $(X,\inv)$.
Then, $(X,\inv,{\mathcal L},c)$ satisfies the condition for
a twisted Real space.
Indeed, the hermitian inner product on ${\mathcal L}$
yields an isomorphism
$\overline{\mathcal L}\cong {\mathcal L}^*(={\mathcal L}^{-1})$,
and we recover the conditions
$c:\inv^*\overline{\mathcal L}\otimes{\mathcal L}\cong \underline{\C}$
and $\inv^*c=c^*$ from the properties (ii) and (iii) of
$({\mathcal L},\sA,c)$ from Section~\ref{subsec:summary}.
Conversely, given a twisted Real space $(X,\inv,{\mathcal L},c)$
we can find a hermitian metric $h$ on ${\mathcal L}$
and a unitary connection $\sA$ of $({\mathcal L},h)$ 
such that the properties (ii) and (iii) 
from Section~\ref{subsec:summary} hold, where 
$c$ is regarded as a 
trivialization of $\inv^*{\mathcal L}^*\otimes {\mathcal L}$
via the isomorphism $\overline{\mathcal L}\cong{\mathcal L}^*$
provided by $h$.
Thus, the part $({\mathcal L},\sA,c)$ of the orientifold data
can be identified as the twisted Real structure $({\mathcal L},c)$.

Given a D-brane data, say $(E,A,T,U)$ for Type IIB orientifold 
$(X,\inv,B,{\mathcal L},\sA,c)$,
we can construct a twisted Real vector bundle over the corresponding
twisted Real space. Indeed, the antilinear map is defined by
\beq
\whinv_x:=U_{\inv(x)}\circ h_{E_x}:
E_x\stackrel{h_{E_x}}{\longrightarrow}
E_x^*\stackrel{U_{\inv(x)}}{\longrightarrow}
(E\otimes {\mathcal L}^{-1})_{\inv(x)}.
\label{whinvUh}
\eeq
That it squares to $c$ follows from the identity (\ref{P1}),
the unitarity of $U$, and the condition $U=c\cdot \inv^*U^t$.
Conversely, given a twisted Real bundle $(E,\whinv)$ over
a twisted Real space $(X,\inv,{\mathcal L},c)$,
we can find the $(E,U)$-part of a D-brane data for the 
corresponding orientifold.
That is, we can find a
hermitian metric $h_E$ on $E$ and a unitary map 
$U:\inv^*E^*\otimes {\mathcal L}\to E$ that gives $\whinv$ by
(\ref{whinvUh}).
Indeed, if the anti-linear map $\whinv_x$ is represented by a matrix 
$M(\inv x)$ with respect to a frame of $E$, its image frame of $\inv^*E$
and a unitary frame of $\inv^*{\mathcal L}^{-1}$, then, we put 
$h_E=2(1+MM^{\dag})^{-1}$ with respect to the same frame of $E$, and define $U$
via (\ref{whinvUh}). Then, one can show that $U$ is indeed unitary
(and obeys $U=c\cdot\inv^*U^t$).
Furthermore, there is an isomorphism between twisted Real bundles
 if and only if there is an isomorphism between the corresponding
$(E,U)$'s. Here we say that
$(E_1,U_1)$ is isomorphic to $(E_2,U_2)$ if there is a unitary map
$f:E_1\to E_2$ such that (c.f. (\ref{isomDO}))
\beq
f\circ U_1\circ \inv^*f^t=U_2.
\label{isomDOv2}
\eeq
In what follows, we shall also refer to the $(E,U)$ part of the D-brane data 
a (twisted) Real bundle.

As the counterpart of (\ref{gtra}), there are gauge transformations of 
the twisted Real structure
\beq
({\mathcal L},\,c)~\longrightarrow~
({\mathcal L}\otimes L\otimes\inv^*\overline{L}^{-1},\,c)\quad\mbox{and}\quad
({\mathcal L},\,\lambda\cdot\inv^*\lambda^*\cdot c),
\label{gtratw}
\eeq
where $L$ is a complex line bundle and $\lambda$ is a $\C^{\times}$-valued
function with $\lambda^*$ being its complex conjugate.
The transformations 
\beq
(E,\,\widehat{\inv})~\longrightarrow~
(E\otimes L,\,\widehat{\inv})\quad
\mbox{and}\quad (E,\,\inv^*\lambda\cdot \widehat{\inv})
\eeq
send twisted Real bundles over the original twisted Real space
to those over the transformed spaces.
These in particular
determine isomorphisms of the group $\KR(X,\inv,{\mathcal L},c)$
to the ones corresponding to the transformed data (\ref{gtratw}).

We claim that the D-brane charges in the Type II orientifold
with data $(X,\inv,B,{\mathcal L},\sA,c)$
are classified by
the Grothendieck groups, $\KR^{-i}(X;c)=\KR^{-i}(X,\inv,{\mathcal L},c)$. 
To be precise,
\beq
\begin{array}{rcl}
({\rm B}_+)&\!\! :& \KR(X;c)\cong \KR^{-4}(X;-c),\\[0.2cm]
({\rm A}_-)&\!\! :& \KR^{-1}(X;c)\cong \KR^{-5}(X;-c),\\[0.2cm]
({\rm B}_-)&\!\! :& \KR^{-2}(X;c)\cong \KR^{-6}(X;-c),\\[0.2cm]
({\rm A}_+)&\!\! :& \KR^{-7}(X;c)\cong \KR^{-3}(X;-c).
\end{array}
\label{class4}
\eeq
Recall that $({\rm B}_{\pm})$ and $({\rm A}_{\pm})$ label the distinction
concerning whether $\inv$ is orientation preserving or not (i.e. IIB or IIA)
and whether the lift $\invS$ to Majorana spinors squares
to $1$ or $-1$.
The isomorphism between K-theory groups
\beq
\KR^{-i}(X;c)\,\cong\,\KR^{-i-4}(X;-c)
\label{Bott4}
\eeq
is the Bott periodicity in this context.

In Case $({\rm B}_+)$, it is easy to understand
that the group $\KR(X;c)$ appears as the classification of the
D-brane charges.
In this case,
the Chan-Paton bundle for D9-branes is graded, $E=E^0\oplus E^1$, and
the o-isomorphism is even, $U={\rm diag}(U_0,U_1)$.
Then, we have a pair of twisted Real bundles, $((E^0,U_0),(E^1,U_1))$,
which represents an element of the group $\KR(X;c)$ as long as
$E$ is of finite rank.
The classification in the other three cases can also be
guessed by taking the decompactification limit and matching
with the classification in the untwisted cases
--- observe that (\ref{class4}) reduces to the
result of the previous subsection when the twist $({\mathcal L},\sA)$
is trivial so that $c\equiv 1$ or $-1$.
In what follows, we shall give a derivation of the claim (\ref{class4}).

\subsubsection{The Hyperbolic Bundle}
\label{subsub:hyperbolic}

As before, we add infinitely many empty branes if necessary
so that we have a Hilbert bundle as the Chan-Paton bundle
on which the tachyon acts as a Fredholm operator.
The first step in the classification is to establish that there is
no freedom in the choice of underlying Real bundle.
In general, a {\it hyperbolic bundle} is the twisted Real bundle
$(H_F,U_F)$ for some complex vector bundle $F$ where
\beq
\begin{array}{l}
H_F~=~F\,\,\oplus\,\, 
(\inv^*F^*\otimes {\mathcal L}),\\[0.2cm]
U_F~=~\left(\begin{array}{cc}
0&{\rm id}^{}_F\otimes c\\
{\rm id}^{}_{\inv^*F^*\otimes {\mathcal L}}&0
\end{array}\right).
\end{array}
\label{hyperbolic}
\eeq
We will show that
any twisted Real Hilbert bundle $(E,U)$ is isomorphic
to a hyperbolic bundle $(H_F,U_F)$ for some Hilbert bundle $F$.
Note that $F$ is necessarily trivial,
$F\cong \underline{\Hi}_{\C}$, by Kuiper's theorem.

When $E$ is graded and $U$ is odd (Case $({\rm B}_-)$),
$E=E^0\oplus E^1$,
$$
U=\left(\begin{array}{cc}
0&U_{01}\\
U_{10}&0
\end{array}\right),
\quad
\begin{array}{ll}
U_{01}:\inv^*E^{1*}\otimes {\mathcal L}\stackrel{\cong}{\longrightarrow} E^0,\\
U_{10}:\inv^*E^{0*}\otimes {\mathcal L}\stackrel{\cong}{\longrightarrow} E^1,
\end{array}
$$
$(E,U)$ is indeed isomorphic to the hyperbolic bundle
$(H_{E^0},U_{E^0})$.
For Case $({\rm B}_+)$, we need to show that both 
of $(E^0,U_0)$ and $(E^1,U_1)$ are isomorphic to hyperbolic bundles.
The vector bundle $\ugE$ in Type IIA cases $({\rm A}_{\pm})$ is ungraded.
Thus, the main task is 
to show the assertion when $E$ is ungraded.

To this end, we introduce the notion of ``Lagrangian subbundles''.
A subbundle $V\subset E$ is a {\it Lagrangian} subbundle of a
twisted Real bundle $(E,U)$ when
the following is an exact sequence of vector bundles
\beq
0\,\,\to\,\,
V\,\,\stackrel{i}{\longrightarrow}\,\,
E\,\,\stackrel{\inv^*i^t\circ U^{-1}}{\longrightarrow}\,\,
\inv^*V^*\otimes{\mathcal L}\,\,\to \,\,0.
\label{Lagrangian0}
\eeq
Note that $F\subset H_F$ is a Lagrangian subbundle of $(H_F,U_F)$.
We shall show that  (i) a twisted Real bundle having a Lagrangian
subbundle $V$ is isomorphic to the hyperbolic bundle $(H_V,U_V)$, and 
(ii) a Lagrangian subbundle always exists
in a twisted Real Hilbert bundle.

\subsubsection*{\it Proof Of (i)}

 Suppose $V\subset E$ is a Lagrangian subbundle of $(E,U)$.
Using the hermitian inner products on the vector bundles involved,
we can find a splitting of the exact sequence (\ref{Lagrangian0}).
I.e., an exact sequence in the opposite direction
$$
0\,\,\leftarrow\,\,
V\,\,\stackrel{t}{\longleftarrow}\,\,
E\,\,\stackrel{s}{\longleftarrow}\,\,
\inv^*V^*\otimes{\mathcal L}\,\,\leftarrow\,\,0,
$$
such that 
$t\circ i={\rm id}_V$,
$j\circ s={\rm id}_{\inv^*V^*\otimes {\mathcal L}}$
and
$i\circ t+s\circ j={\rm id}_E$, for $j=\inv^*i^t\circ U^{-1}$.
Then, we have an isomorphism of vector bundles
$(i,s):V\oplus (\inv^*V^*\otimes {\mathcal L})\to E$.
This defines an isomorphism of twisted Real bundles,
$(H_V,U_V)\cong (E,U)$, in the sense of 
(\ref{isomDOv2}). That is,
$$
(i,s)\left(\begin{array}{cc}
0&{\rm id}_V\otimes c\\
{\rm id}_{\inv^*V^*\otimes{\mathcal L}}&0
\end{array}\right)
\left(\begin{array}{c}
\inv^*i^t\\
\inv^*s^t
\end{array}\right)\,\,=\,\, U.
$$
This can be shown using the condition
$U=c\cdot \inv^*U^t$, the unitarity of $U$ 
and the defining properties of the splitting.

\subsubsection*{\it Proof Of (ii)}

\newcommand{\bilin}{\boldsymbol{\beta}}

Let $(E,U)$ be a twisted Real Hilbert bundle.
The first step is to find a Lagrangian subbundle over the fixed point set
$X^{\inv}$ on which $c$ can be canonically 
identified as a number which is either $+1$ or $-1$ in each component.
On this set, $U$ defines a non-degenerate bilinear form
$\bilin:E\times E\to {\mathcal L}$ 
which is symmetric or antisymmetric depending on
$c\equiv +1$ or $c\equiv -1$.
Then, ``Lagrangian'' is in the usual sense 
--- a subbundle of $E$ is Lagrangian if it is equal to its own
orthocomplement with respect to $\bilin$.
We take a cell decomposition of $X^{\inv}$ which is fine enough
so that, over each cell, 
${\mathcal L}$ is trivialized and $\bilin$ takes values in $\C$.
We try to construct a Lagrangian subbundle
recursively with respect to the dimension of the cells.
The question is whether a Lagrangian subbundle over the boundary of 
a cell can be extended to the interior.
The answer is yes because the space of all Lagrangian subspaces
in $(\Hi_{\C},\bilin)$ (the Lagrangian Grassmannian $\Lambda(\Hi_{\C},\bilin)$)
is contractible, as we show below.
Here $\Hi_{\C}$ is a complex Hilbert space and $\bilin$ is
a symmetric or antisymmetric bilinear form 
which is compatible with the inner product (i.e. the map
$\Hi_{\C}^*\to \Hi_{\C}$ defined by $\bilin$ is unitary).

\noindent
\underline{$\bilin$ symmetric}~~In this case, $\Hi_{\C}$ is
a complexification of a real Hilbert space $\Hi_{\R}$
on which $\bilin$ agrees with the inner product.
Let us take an orthogonal complex structure $J$ of $\Hi_{\R}$. 
For example, take an orthonormal basis
$\{e_n\}_{n=1}^{\infty}\subset \Hi_{\R}$ and put $J(e_{2m-1})=e_{2m}$
and $J(e_{2m})=-e_{2m-1}$ for all $m$. 
Then, we have a Lagrangian subspace $V_J\subset \Hi_{\C}=
\Hi_{\R}\otimes_{\R}\C$
consisting of antiholomorphic vectors,
$v\in V_{J}$ $\Leftrightarrow$ $J(v)=-iv$.
Conversely any Lagrangian $V\subset \Hi_{\C}$ determines an orthogonal
complex structure $J_V$ of $\Hi_{\R}$.
To see this, take an orthonormal basis $\{v_m\}_{m=1}^{\infty}\subset V$
with respect to the hermitian inner product, and put
$e_{2m-1}=\sqrt{2}{\rm Re}(v_m)$ and $e_{2m}=\sqrt{2}{\rm Im}(v_m)$.
These vectors $e_n$ define an orthonormal basis of $\Hi_{\R}$.
The complex structure $J_V$ is obtained by applying the above construction 
to this basis $\{e_n\}_{n=1}^{\infty}$.
It is straightforward to see that $J\mapsto V_J$ and $V\mapsto J_V$
are inverse to each other.
Thus, we find
$$
\Lambda(\Hi_{\C},\bilin)~\cong ~ O(\Hi_{\R})/U(\Hi_{\R},J_0)
$$
where $U(\Hi_{\R},J_0)$ is the subgroup of $O(\Hi_{\R})$ that commutes with
a fixed complex structure $J_0$ and is isomorphic to
the group $U(V_{J_0})$ of unitary operators of the Hilbert space $V_{J_0}$.
Both $O(\Hi_{\R})$ and $U(V_{J_0})$ are contractible by Kuiper's theorem
and hence so is the Grassmannian.

\noindent
\underline{$\bilin$ antisymmetric}~~
In this case, $\Hi_{\C}$ has the structure of a quaternionic Hilbert space
$\Hi_{\HH}$ (with $\HH$ acting from the right, say), such that
the quaternionic and hermitian inner products are related by
$(v,w)_{\HH}=(v,w)_{\C}-j\bilin(v,w)$.
Let us take an orthonormal basis $\{e_n\}_{n=1}^{\infty}$ of $\Hi_{\HH}$. 
Then, it spans over $\C$ a Lagrangian subspace of $(\Hi_{\C},\bilin)$.
Conversely, given a Lagrangian subspace $V$ of $(\Hi_{\C},\bilin)$
choose an orthonormal basis of $V$ with respect to the hermitian inner product
$(\,,\,)_{\C}$. Then it is an orthonormal basis of $\Hi_{\HH}$.
Therefore, we have a one to one correspondence between
Lagrangian subspaces of $(\Hi_{\C},\bilin)$ and
orthonormal bases of $\Hi_{\HH}$ up to complex unitary base changes.
This shows
$$
\Lambda(\Hi_{\C},\bilin)~\cong~ USp(\Hi_{\HH})/U(V_0)
$$
where $V_0$ is a fixed Lagrangian subspace of $(\Hi_{\C},\bilin)$.
Both $USp(\Hi_{\HH})$ and $U(V_0)$ are contractible by Kuiper's theorem
and hence so is the Grassmannian.

Having constructed a Lagrangian subbundle on each component of the fixed 
point set $X^{\inv}$, the next task is to extend it over the entire space 
$X$. We do it by recursive construction using
a $\Z_2$-compatible cell decomposition, as in Section~\ref{subsub:Tk}.
By a moment of thought, we see that there is no obstruction in
the recursive process.
This establishes the existence of a Lagrangian subbundle of
$(E,U)$.

\subsubsection{The Classification}

Since there is no freedom in the underlying twisted Real bundle,
the focus of the classification of D-branes 
is that of the tachyon configurations.
Let us write down the condition for the tachyons,
$T$ (for Type IIB) or $\ugT$ (for Type IIA), which is a Fredholm operator
at each point $x\in X$.

\noindent
$({\rm B}_+)$~~
We can take $(E^0,U_0)=(E^1,U_1)=(H_{\Hi_{\C}},U_{\Hi_{\C}})$, and
the condition $T=U\inv^*T^tU^{-1}$ reads
for $T_{10}:H_{\Hi_{\C}}\to H_{\Hi_{\C}}$ as
\beq
T_{10}=U_{\Hi_{\C}}\circ\inv^*T_{10}^{\dag t}\circ U_{\Hi_{\C}}^{-1}.
\label{condB+}
\eeq

\noindent
$({\rm B}_-)$~~
 We can take $(E,U)=(H_{\Hi_{\C}},U_{\Hi_{\C}})$
and the condition is $T=-U_{\Hi_{\C}}\inv^*T^t U_{\Hi_{\C}}^{-1}$.
In particular it reads as
\beq
T_{10}=-c^{-1}\cdot \inv^*T_{10}^t.
\label{condB-}
\eeq

\noindent
$({\rm A}_{\pm})$~~
We can take $(\ugE,\ugU)=(H_{\Hi_{\C}},U_{\Hi_{\C}})$
and the condition $\ugT=\pm\ugU\inv^*\ugT^t\ugU^{-1}$ reads for
$
\ugT=\left(\begin{array}{cc}
A&B\\
C&D
\end{array}\right)
$
as
\beq
A=\pm\, \inv^*D^t,\quad
B=\pm \, c\cdot \inv^*B^t.
\label{condApm}
\eeq
Note that $A=A^{\dag}, C=B^{\dag}, D=D^{\dag}$
by the hermiticity of $\ugT$. We recall that, at each point $x$,
$\ugT(x)$ must
have infinitely many positive and infinitely many negative eigenvalues.

These conditions can be recast into another form using Clifford algebras.
Let $\Fr(H_{\Hi_{\C}})=\Frst^0(H_{\Hi_{\C}})$ 
be the bundle of Fredholm operators on $H_{\Hi_{\C}}$ 
--- the fibre at $x\in X$ is the space of Fredholm operators 
on the Hilbert space $H_{\Hi_{\C}}|_x$. 
For $k\geq 1$, 
let $\Frst^k(H_{\Hi_{\C}})$ be the bundle of {\it skew-adjoint}
Fredholm operators of $H_{\Hi_{\C}}$ satisfying
the conditions (i) and (ii) as in Section~\ref{subsec:Clifford}.
Here we assume that the Clifford algebra $\Cl_{k-1}$ acts on
$H_{\Hi_{\C}}=\underline{\Hi}_{\C}\oplus 
(\underline{\Hi}_{\C}^*\otimes{\mathcal L})$ as
\beq
\J_i=\left(\begin{array}{cc}
J_i&0\\
0&-J_i^t\otimes {\rm id}
\end{array}\right),\qquad i-1,\ldots, k-1,
\eeq
where $J_i$ are skew-adjoint $\C$-linear operators on $\Hi_{\C}$
that determine a complex representation of $\Cl_{k-1}$ on $\Hi_{\C}$.
We assume that each irreducible representation $\Cl_{k-1}$ occurs
in $\Hi_{\C}$ with infinite multiplicity.
These $\J_i$'s are chosen so that they commute with 
the twisted Real structure $\widehat{\inv}$ determined by $U_{\Hi_{\C}}$,
$$
\widehat{\inv}\circ\J_i=\J_i\circ \widehat{\inv}.
$$
It follows that the conjugation 
$\rmA\longmapsto\widehat{\inv}\circ \rmA\circ\widehat{\inv}^{-1}$
defines an involution of $\Frst^k(H_{\Hi_{\C}})$ over 
the one $\inv$ on the base $X$.

The assumption on the multiplicity of
$\Cl_{k-1}$ representations in $\Hi_{\C}$ is vacuous when $(k-1)$ is even
since there is only one irreducible representation.
When $(k-1)$ is odd, there are
the two representations of $\Cl_{k-1}$ 
distinguished by the value of the center $J_1\cdots J_{k-1}$. Since
$$
(-J_1^t)\cdots(-J_{k-1}^t)=(-1)^{k(k-1)\over 2}(J_1\cdots J_{k-1})^t,
$$
the representations determined by $J_i$'s and $-J_i^t$'s 
are the same for $(k-1)=3$ mod 4 and are opposite for $(k-1)=1$ mod 4.
This means that the assumption on the multiplicity in $\Hi_{\C}$
is unnecessary when $(k-1)=1$ mod $4$.
To see this, suppose that only one of the two representations occurs 
in $\Hi_{\C}$.
Then, $\underline{\Hi}_{\C}^*\otimes{\mathcal L}$ consists of the other 
representation.
Let us take a decomposition $\Hi_{\C}=\Hi_1\oplus \Hi_2$,
where $\Hi_1$ and $\Hi_2$ are both infinite dimensional and invariant under
$\Cl_{k-1}$.
We choose a trivialization $\underline{\Hi}_2^*\otimes{\mathcal L}\cong
\underline{\Hi}'_2$ and put $\Hi_{\C}'=\Hi_1\oplus \Hi_2'$.
Then, $(H_{\Hi_{\C}},U_{\Hi_{\C}})$ is isomorphic to 
$(H_{\Hi'_{\C}},U_{\Hi'_{\C}})$ as a twisted Real bundle
and each of the two representations occurs in $\Hi_{\C}'$ 
with infinite multiplicity.
This does not happen when $(k-1)=3$ mod $4$ --- the assumption on
the multiplicity in $\Hi_{\C}$ is necessary.

When the twist ${\mathcal L}$ is trivial and $c=1$,
$\Frst^k(H_{\Hi_{\C}})$ is the trivial bundle with fibre
$\Frst^k(\Hi_{\R}\otimes\C)$ as defined in Section~\ref{subsec:Clifford}.
To see this, we choose a complex structure $J$ of
$\Hi_{\R}$ that commutes with the $\Cl_{k-1}$ generators.
Then, we identify $\Hi_{\C}$ resp. $\Hi_{\C}^*\cong\overline{\Hi}_{\C}$
as the complex subspaces $\Hi_{\R}^{1,0}$ resp. $\Hi_{\R}^{0,1}$
of $\Hi_{\R}\otimes \C$ consisting of vectors satisfying $Jv=iv$
resp. $Jv=-iv$, so that $H_{\Hi_{\C}}=\underline{\Hi_{\R}\otimes\C}$.
Under this identification, 
the conjugation by $\widehat{\inv}$ is equal to the complex conjugation.

We shall consider sections of $\Frst^k(H_{\Hi_{\C}})$
that are equivariant with respect to the involution $\inv$ on $X$
and the conjugation by $\widehat{\inv}$, i.e.,
\beq
\inv^*\rmA=\widehat{\inv}\circ \rmA\circ \widehat{\inv}^{-1},
\label{condFrT}
\eeq
or equivalently,
$\rmA=U_{\Hi_{\C}}\circ\inv^*\rmA^{\dag t}\circ U_{\Hi_{\C}}^{-1}$.
We denote by $\Gamma(X,\Frst^k(H_{\Hi_{\C}}))^{}_{\Z_2(c)}$
the space of such sections.
Now we can state the main result on the classification:
Tachyon configurations are in one to one correspondence with 
equivariant sections of Fredholm bundles of the following types
\beq
\begin{array}{rcl}
({\rm B}_+)&\!\! :&\Gamma(X,\Fr(H_{\Hi_{\C}}))^{}_{\Z_2(c)},\\[0.2cm]
({\rm A}_-)&\!\! :&\Gamma(X,\Frst^1(H_{\Hi_{\C}}))^{}_{\Z_2(c)},\\[0.2cm]
({\rm B}_-)&\!\! :&\Gamma(X,\Frst^2(H_{\Hi_{\C}}))^{}_{\Z_2(c)},\\[0.2cm]
({\rm A}_+)&\!\! :&\Gamma(X,\Frst^3(H_{\Hi_{\C}}))^{}_{\Z_2(-c)}.
\end{array}
\label{classFr4}
\eeq
For Case $({\rm B}_+)$, the correspondence is obviously $T_{10}=\rmA$
--- the condition (\ref{condB+}) is nothing but (\ref{condFrT}).
To see the rest, let us write down the condition (\ref{condFrT})
for the skew adjoint operator $\rmA$ written as $\left(\begin{array}{cc}
\alpha&\beta\\
\gamma&\delta
\end{array}\right)$ according to the decomposition
$H_{{\Hi}_{\C}}=\underline{\Hi}_{\C}\oplus
(\underline{\Hi}_{\C}^*\otimes{\mathcal L})$,
\beq
\alpha=-\inv^*\delta^{t},\quad
\beta=-c\cdot \inv^*\beta^{t}.
\label{condskad}
\eeq
Note that $\alpha=-\alpha^{\dag},\gamma=-\beta^{\dag}$ and
$\delta=-\delta^{\dag}$ by the skew-adjointness of $\rmA$.

\noindent
$({\rm A}_-)$~~ The correspondence is given by $\ugT=i^{-1}\rmA$.
The condition (\ref{condApm}) agrees with (\ref{condskad})
and the conditions concerning the eigenvalues also match.

\noindent
$({\rm B}_-)$~~ The two complex representations of $\Cl_1$ 
are $J_1=i$ and $J_1=-i$. By the remark above for the case $(k-1)=1$ mod $4$,
we may assume that the operator $\J_1$ on $H_{\Hi_{\C}}$ is given by
$$
\J_1=\left(\begin{array}{cc}
i&0\\
0&-i
\end{array}\right).
$$
Anticommutativity $\rmA\J_1=-\J_1 \rmA$ requires $\alpha=\delta=0$
and the condition for $\beta$ agrees with the one for
$T_{01}=T_{10}^{\dag}$. Thus, we find a correspondence
$T=i^{-1}\rmA$.

\noindent
$({\rm A}_+)$~~ We may take
$$
J_1=\left(\begin{array}{cc}
0&-1\\
1&0
\end{array}\right),\quad
J_2=\left(\begin{array}{cc}
0&i\\
i&0
\end{array}\right)
$$ 
for a decomposition $\Hi_{\C}=\Hi_1\oplus \Hi_1$.
Anticommutativity $\rmA\J_i=-\J_i \rmA$ ($i=1,2$) requires 
$$
\alpha=\left(\begin{array}{cc}
\alpha_1&0\\
0&-\alpha_1
\end{array}\right),\quad
\beta=\left(\begin{array}{cc}
0&\beta_1\\
\beta_1&0
\end{array}\right),\quad
\gamma=\left(\begin{array}{cc}
0&\gamma_1\\
\gamma_1&0
\end{array}\right),\quad
\delta=\left(\begin{array}{cc}
\delta_1&0\\
0&-\delta_1
\end{array}\right).
$$
The condition (\ref{condskad}) takes the same form,
$\alpha_1=-\inv^*\delta_1^t$ and $\beta_1=-c\cdot\inv^*\beta_1^t$.
Let us compute $w(\rmA)=\J_1\J_2\rmA$:
$$
w(\rmA)=-i\left(\begin{array}{cccc}
\alpha_1&&&\beta_1\\
&\alpha_1&-\beta_1&\\
&-\gamma_1&-\delta_1&\\
\gamma_1&&&-\delta_1
\end{array}\right)
\simeq
-i\left(\begin{array}{cccc}
\alpha_1&\beta_1&&\\
\gamma_1&-\delta_1&&\\
&&\alpha_1&\beta_1\\
&&\gamma_1&-\delta_1
\end{array}\right),
$$
where we made a unitary basis change in the second equality.
If we put
$$
\ugT=-i\left(\begin{array}{cc}
\alpha_1&\beta_1\\
\gamma_1&-\delta_1
\end{array}\right),
$$
the conditions on $\ugT:H_{\Hi_1}\to H_{\Hi_1}$ 
agree with those on $\rmA\in \Gamma(X,\Frst^3(H_{\Hi_{\C}}))^{}_{\Z_2(-c)}$, 
including Fredholmness as well as the condition on the eigenvalues.

\subsubsection*{\it Bott Periodicity}

We next show the periodicity
\beq
\Gamma(X,\Frst^k(H_{\Hi_{\C}}))^{}_{\Z_2(c)}\cong 
\Gamma(X,\Frst^{k+4}(H_{\Hi'_{\C}}))^{}_{\Z_2(-c)},
\label{BottFr4}
\eeq
which in particular leads to mod 8 periodicity.
The key is the isomorphism $\Cl_{k+3}\cong \Cl_{k-1}\otimes\Cl_4$
\beq
e_i~\longleftrightarrow~ \left\{\begin{array}{ll}
e_i\otimes e_1e_2e_3e_4&i=1,\ldots, k-1,\\
1\otimes e_{i-(k-1)}&i=k,\ldots,k+3.
\end{array}\right.
\eeq
Suppose $\Hi_{\C}$ is the Hilbert space with an admissible $\Cl_{k-1}$
representation, i.e., each irreducible 
representation appears with infinite multiplicity.
Then, a Hilbert space with an admissible $\Cl_{k+3}$ representation
is obtained by $\Hi'_{\C}=\Hi_{\C}\otimes V_4$ where 
$V^{}_4\cong \C^4$ is the (unique) irreducible representation of $\Cl_4$.
The isomorphism (\ref{BottFr4}) is given by
\beq
\rmA=\left(\begin{array}{cc}
\alpha&\beta\\
\gamma&\delta
\end{array}\right)~\longleftrightarrow~
\rmA'=\left(\begin{array}{cc}
\alpha'&\beta'\\
\gamma'&\delta'
\end{array}\right)
=\left(\begin{array}{cc}
\alpha\otimes J_{1234}&\beta\otimes u\\
\gamma\otimes u^{\dag}&\delta\otimes J_{1234}^t
\end{array}\right),
\label{BottFr4'}
\eeq
where $J_{1234}:V_4^{}\to V_4^{}$ is defined by $J_1J_2J_3J_4$
 and $u:V_4^*\to V_4^{}$ is such that $J_iu=uJ^t_i$ for $i=1,2,3,4$.
It is straightforward to check that $\Cl_{k-1}$ anticommutativity of
$\rmA$ corresponds to $\Cl_{k+3}$ anticommutativity of $\rmA'$.
Repeating the computation in Section~\ref{subsec:bfPa}, we find
$$
u^t=-u,
$$
from which it follows that
$$
\beta=-c\cdot\inv^*\beta^t~\Longleftrightarrow~
\beta'=c\cdot \inv^*\beta^{\prime t}.
$$
Since $\alpha=-\inv^*\delta^t$ is obviously equivalent to
$\alpha'=-\inv^*\delta^{\prime t}$, we find that 
(\ref{BottFr4'}) indeed gives rise to the isomorphism
(\ref{BottFr4}).

\subsubsection*{\it K-Theory}

Let $[X,\Frst^k(H_{\Hi_{\C}})]^{}_{\Z_2(c)}$ be the set of connected components
of the space $\Gamma(X,\Frst^k(H_{\Hi_{\C}}))^{}_{\Z_2(c)}$.
For $k=0$ there is a bijection
\beq
{\rm index}:[X,\Fr(H_{\Hi_{\C}})]^{}_{\Z_2(c)}~\longrightarrow~
\KR(X;c).
\eeq
It is defined as follows.
Let $\urT$ be an equivariant section of $\Fr(H_{\Hi_{\C}})$, i.e., 
$\widehat{\inv}_x\circ \urT_x=\urT_{\inv(x)}\circ \widehat{\inv}_x$.
We see that $\widehat{\inv}_x$ maps $\Ker\,\urT_x$ to
$\Ker\,\urT_{\inv(x)}\otimes {\mathcal L}^{-1}_{\inv(x)}$,
and the same holds on $\Ker\,\urT^{\dag}$.
Thus, if $\Ker\,\urT$ and $\Coker\,\urT$ have constant ranks, the
pair $(\Ker\,\urT,\Coker\,\urT)$ determines an element of $\KR(X;c)$. 
In general, we can find a subspace 
$\urV\subset \Hi_{\C}$ of finite codimension
such that  $H_{\Hi_{\C}}/\urT(H_{\urV})$ defines a finite rank twisted Real 
vector bundle, and we put 
${\rm index}(\urT)=(H_{\Hi_{\C}/\urV},H_{\Hi_{\C}}/\urT(H_\urV))\in \KR(X;c)$.
The proof that it is a bijection is similar to the earlier cases and 
is omitted here.

For $k\geq 1$, we may put {\it by definition}
\beq
\KR^{-k}(X;c)=[X,\Frst^k(H_{\Hi_{\C}})]^{}_{\Z_2(c)}.
\label{def1KR-k}
\eeq
Then, the classification (\ref{class4}) and the periodicity (\ref{Bott4})
follow from (\ref{classFr4}) and (\ref{BottFr4}).
Also, when the twist ${\mathcal L}$
is trivial and $c=1$, $\KR^{-k}(X;c)$ agrees with $\KR^{-k}(X)$.

Alternatively, we may seek for a definition within the
category of {\it finite rank} twisted Real bundles.
Suspension cannot be used in general,
since the twist ${\mathcal L}$ may be non-trivial along
a subspace that is to be contracted. Here we quote an alternative definition
of the group ${\rm K}(X,Y)$ as the Grotherndieck group of the category
of the pair $(E,\psi)$ of a finite rank vector bundle on
$X$ which is trivial over $Y$ and a trivialization 
$\psi:E|_Y\to Y\times\C^r$ over $Y$.
We would like to define $\KR(X,Y;c)$ in the similar way.
The question is when do we say that a twisted Real bundle is
{\it trivial}?
Here we propose to say that $(E,U)$ is trivial
when it is isomorphic to the hyperbolic bundle $(H_{\C^r},U_{\C^r})$ for
the trivial vector bundle $\underline{\C}^r$ of arbitrary rank. 
This is so that the index map
$$
{\rm index}:
[(X,Y),(\Fr(H_{\Hi_{\C}}),{\rm id})]^{}_{\Z_2(c)}
~\longrightarrow~\KR(X,Y;c)
$$
becomes a bijection, where
the domain stands for the set of connected components of the space
of equivariant sections of $\Fr(H_{\Hi_{\C}})$ which is the identity
over $Y$.
Then, we define
$\KR^{-k}(X;c)$ by $\KR(I^k\times X,\partial I^k\times X;c)$,
where we extend $({\mathcal L},c)$ uniformly in the $I^k$ direction.

Do the two definitions agree?
The key to this question is whether there is a homotopy equivalence 
like (\ref{ASth}) also in the twisted case.
We first note that we can take $\J_k$, obtained by extension of
Clifford algebra action on $\Hi_{\C}$, 
as the base section of the bundle $\Frst^k(H_{\Hi_{\C}})$
(we put $\J_0={\rm id}$ for $k=0$).
Then, the formula (\ref{alphadef}) defines a map
\beq
[(X,Y),(\Frst^k(H_{\Hi_{\C}}),\J_k)]^{}_{\Z_2(c)}
\stackrel{\alpha_*}{\longrightarrow}
[(I\times X,(I\times Y)\cup (\partial I\times X)),
(\Frst^{k-1}(H_{\Hi_{\C}}),\J_{k-1})]^{}_{\Z_2(c)}.
\eeq
We claim without proof that it is a bijection.
If that is indeed the case, using it iteratively, we find
that the two definitions agree,
$$
[X,\Frst^k(H_{\Hi_{\C}})]^{}_{\Z_2(c)}
~\cong~
[(I^k\times X,\partial I^k\times X),(\Fr(H_{\Hi_{\C}}),{\rm id})]^{}_{\Z_2(c)}.
$$

\medskip

\section{${\mathcal N}=1$ Supersymmetry}
\label{sec:SUSY}

Type II orientifold on a Calabi-Yau three-fold with D-branes
is one way to obtain ${\mathcal N}=1$ supersymmetric
theories in $3+1$ dimensions with non-zero Newton's constant.
In this section, we shall study the structure of Chan-Paton factors
for space-filling D-branes in Type IIB orientifolds by 
holomorphic involutions, with a focus on
${\mathcal N}=1$ supersymmetry and categorical description.

\newcommand{\cone}{{\rm Cone}}

\subsection{${\mathcal N}=2$ Worldsheet Supersymmetry}

To start with, let us focus on the supersymmetric sigma model 
on the internal space $M$
which we take for now to be an $n$-dimensional K\"ahler manifold.
It has an extended ${\mathcal N}=(2,2)$ supersymmetry.
We are interested in D-branes that preserve a diagonal
${\mathcal N}=2_B$ subalgebra with $U(1)$ R-symmetry
which acts on the worldsheet fields as
\beqa
{\mathcal N}=2_B:&\!\!\!\!
\left\{
\begin{array}{ll}
\delta x^i=\epsilon\psi^i,&\delta x^{\bi}=-\bepsilon\psi^{\bi},\\
\delta \psi^i=-2i\bepsilon \dot{x}^i,&\delta\psi^{\bi}=2i\epsilon\dot{x}^{\bi}
\end{array}\right.
\label{N=2B}
\\
U(1)_R:&(x^{\mu},\psi^i,\psi^{\bi})\mapsto 
(x^{\mu},\e^{-i\alpha}\psi^i,\e^{i\alpha}\psi^{\bi}).
\eeqa
We shall summarize the condition and properties of boundary interactions
with this extended supersymmetry. We refer the reader to \cite{HHP}
for more detail.

\subsubsection*{\it The Conditions}

The condition for the D9-brane configuration $(\gE,\gA,\gT)$ to preserve
the symmetry is as follows:
$\gE$ has a $\Z$-grading that reduces modulo $2$ to the original 
$\Z_2$-grading,
the gauge field $\gA$ has degree $0$
and has a $(1,1)$-form curvature, and the tachyon
$\gT$ can be written as a sum $\gT=iQ-iQ^{\dag}$
where $Q$ has degree $1$, is holomorphic $D_{\bi}Q=0$, and squares to
zero $Q^2=0$. Such a data defines a complex of holomorphic vector bundles.
$$
\cdots\longrightarrow {\mathcal E}^i\stackrel{Q}{\longrightarrow} 
{\mathcal E}^{i+1}\stackrel{Q}{\longrightarrow} {\mathcal E}^{i+2}
\longrightarrow\cdots
$$
For the purpose of our discussion, it is appropriate to generalize
the boundary interaction ${\mathcal A}$ so that it has higher powers 
of the fermions $\psi$.
${\mathcal N}=1$ supersymmetry requires the form
\beq
{\mathcal A}_t=-\dot{x}^{\mu}{\partial \over\partial\psi^{\mu}}{\mathcal T}
+{i\over 2}\psi^{\mu}{\partial\over\partial x^{\mu}}{\mathcal T}
+{1\over 2}{\mathcal T}^2
\label{At2}
\eeq
where ${\mathcal T}$ depends both on $x$ and $\psi$.
The condition of ${\mathcal N}=2_B$ supersymmetry with $U(1)$ R-symmetry
is that $E$ has a $\Z$-grading and
${\mathcal T}=i{\mathcal Q}-i{\mathcal Q}^{\dag}$ where
\beqa
&&\mbox{${\mathcal Q}$ has degree 1},
\label{deg1c}\\
&&{\partial \over\partial \psi^i}{\mathcal Q}=0,
\label{psindep}\\
&&\psi^{\bi}{\partial \over\partial x^{\bi}}{\mathcal Q}+{\mathcal Q}^2=0.
\label{MCeq1}
\eeqa
By the first two conditions, we may write
\beq
{\mathcal Q}=\cQ^{(0)}+\cQ^{(1)}+\cQ^{(2)}+\cdots+\cQ^{(n)},
\qquad
\cQ^{(p)}={1\over p!}\psi^{\bi_1}\cdots\psi^{\bi_p}
\cQ_{\bi_1...\bi_p}(x),
\label{Qexpa}
\eeq
where $\cQ_{\bi_1...\bi_p}$ has degree $(1-p)$, i.e.,
it maps $E^i$ to $E^{i+1-p}$.
For the two term case ${\mathcal Q}=\cQ^{(0)}+\cQ^{(1)}$,
we find, after rewriting $\cQ^{(0)}=Q$ and $\cQ^{(1)}=i\psi^{\bi}A_{\bi}$,
that the boundary interaction (\ref{At2})
agrees with the one (\ref{At}) for $(E,A,T)$
where $A=A_{\bi}\dd x^{\bi}+A_{\bi}^{\dag}\dd x^i$
and $T=iQ-iQ^{\dag}$.
Also, the last equation (\ref{MCeq1}) splits to $Q^2=0$, $D_{\bi}Q=0$ and
$F_{\bi\bj}=0$, which are indeed the condition quoted above
for $(E,A,T)$.
In general, (\ref{MCeq1}) splits to $(n+1)$ equations
starting from $\cQ^{(0)\,2}=0$.

\subsubsection*{\it D-term Deformations And Brane-Antibrane Annihilation}

Our primary interest is the infra-red limit of the boundary interaction
${\mathcal A}_t$. There are two types of operations that do not change the 
low energy behaviour. One is the boundary D-term deformation
--- deformation of ${\mathcal A}_t$ by the terms of the form 
${\bf Q}\overline{\bf Q}\epsilon$ and $\overline{\bf Q}{\bf Q}\epsilon'$
for some expressions $\epsilon$, $\epsilon'$ of $x$, $\psi$ and its derivatives,
where
\beqa
&&
i{\bf Q}\lambda=i{\bf Q}_{x,\psi}\lambda
+{\mathcal Q}\lambda-(-1)^{|\lambda|}\lambda{\mathcal Q},
\\
&&
i\overline{\bf Q}\lambda=i\overline{\bf Q}_{x,\psi}\lambda
-{\mathcal Q}^{\dag}\lambda+(-1)^{|\lambda|}\lambda{\mathcal Q}^{\dag}.
\eeqa
${\bf Q}_{x,\psi}$ etc are the supersymmetry variations 
of $x^{\mu},\psi^{\mu}$, i.e.,
$\delta=i\epsilon \overline{\bf Q}_{x,\psi}-i\bepsilon {\bf Q}_{x,\psi}$
in (\ref{N=2B}).
Note that any deformation of ${\mathcal Q}$ leads to 
$\delta{\mathcal A}_t={i\over 2}{\bf Q}\delta{\mathcal Q}^{\dag}
-{i\over 2}\overline{\bf Q}\delta{\mathcal Q}$.
In particular, deformation of ${\mathcal Q}$ of the form
$\delta{\mathcal Q}=i{\bf Q}\beta$
is a boundary D-term deformation.
Such a $\beta$ must be of degree 0 and be independent of $\psi^i$
(but it can depend on $\psi^{\bi}$),
in order for the deformed ${\mathcal Q}$ to satisfy (\ref{deg1c}) and
(\ref{psindep}). The deformation then takes the following form
under which (\ref{MCeq1}) is manifestly invariant:
\beq
\delta{\mathcal Q}
=\psi^{\bi}{\partial\over\partial x^{\bi}}\beta+[{\mathcal Q},\beta ].
\eeq
The other operation is brane-antibrane annihilation,
which is to discard a part of D-brane that is empty in the infra-red limit.
Note that the boundary interaction ${\mathcal A}$ includes the potential term
\beq
V={1\over 2}\{\cQ^{(0)},\cQ^{(0)\dag}\}
\eeq
where $\cQ^{(0)}$ is the leading term of ${\mathcal Q}$.
When all of its eigenvalues are positive everywhere on $M$, then
the boundary has no degree of freedom at low enough energies.
Thus, such a brane is empty in the infra-red limit. 
Below, we will propose a more general characterization of empty branes
in terms of homological algebra.

\subsubsection*{\it Boundary Chiral Ring}

The cohomology classes of boundary NS vertex operators with respect to
the supercharge ${\bf Q}$ form a ring, the boundary chiral ring.
It is protected from renormalization and is isomorphic to
the ring of boundary chiral primary fields in the infra-red superconformal 
field theory. 
Let us describe it in the zero mode approximation.

Under the replacement $\psi^{\bi}\to\dd x^{\bi}$,
$\cQ$ can be regarded as a differential form with values in endomorphisms
of $E$, with $\cQ^{(p)}\in \Omega^{0,p}(M,Hom^{1-p}(E,E))$,
and the condition (\ref{MCeq1}) can be written as 
\beq
\bartial{\mathcal Q}+{\mathcal Q}^2=0.
\label{condcalQ}
\eeq
Let us consider the open string between
D-branes $\cB_1=(E_1,\cQ_1)$ and $\cB_2=(E_2,\cQ_2)$ 
satisfying the conditions above.
In the zero mode approximation, NS vertex operators of 
canonical\footnote{``Canonical R-charge'' is the na\"ive, ultra-violet 
R-charge. It may not be the same the as
the actual R-charge in the infra-red conformal field theory.
The terminology is after ``canonical dimension''.}
R-charge $i$ are represented by differential forms in 
\beq
\mathscr{C}^i(\cB_1,\cB_2):=
\bigoplus_{p+q=i}\Omega^{0,p}(M,Hom^q(E_1,E_2)),
\label{Ci}
\eeq
and the supercharge ${\bf Q}$ is represented by the Dolbeault type
operator
\beq
i{\bf Q}\phi =\bartial\phi+{\mathcal Q}_2\,\phi
-(-1)^{|\phi|}\phi\,{\mathcal Q}_1.
\label{Qzero}
\eeq
We denote the spaces of $\bfQ$-closed elements and $\bfQ$-cohomology classes
by $\mathscr{Z}^i(\cB_1,\cB_2)$ and
$\mathscr{H}^i(\cB_1,\cB_2)$ respectively.
For three D-branes $\cB_1$, $\cB_2$ and $\cB_3$, we have a product
\beq
\mathscr{H}^i(\cB_1,\cB_2)\times
\mathscr{H}^j(\cB_2,\cB_3)\longrightarrow
\mathscr{H}^{i+j}(\cB_1,\cB_3)
\label{Hprod}
\eeq
induced from the obvious product in $\mathscr{C}^{{}^{\cdot}}_{}(\cB_a,\cB_b)$.
This is the boundary chiral ring.

It is important to keep in mind 
that elements of $Hom^q(E,F)$ with $q$ odd are given odd statistics 
here ---
for example, they anticommute with $\dd x^{\bi}$'s. This is assumed in
the products discussed above, including the ones in (\ref{condcalQ})
and (\ref{Qzero}).
However, we may work with a formulation in which
no such statistics is given to odd homomorphisms.
This is done, for example, by placing all $\psi^{\bi}$'s
on the left of homomorphisms before identifying them as
differential forms.
One advantage of this choice is that $\psi^{\bi}\partial_{\bi}\phi$ can be 
identified with $\bartial\phi$ without a sign.
The product in this formulation, denoted by ``$\wedge$'',
is related to the graded product used above by
\beq
\varphi_1\varphi_2=(-1)^{q_1p_2}\varphi_1\wedge \varphi_2,
\label{2prods}
\eeq
for $\varphi_1\in \Omega^{0,p_1}(M,Hom^{q_1}(F,G))$ and
$\varphi_2\in \Omega^{0,p_2}(M,Hom^{q_2}(E,F))$.
This remark is particularly useful when we discuss shifts of grading,
which we do next.

\subsubsection*{\it Shifts Of Gradings}

The ``shift by one to the left'', 
$\cB=(E,{\mathcal Q})\mapsto\cB[1]=(E[1],\cQ[1])$, 
is defined by
\beq
E[1]^i=E^{i+1},\quad
\cQ[1]^{(p)}=(-1)^{p+1}\cQ^{(p)}.
\label{defshift}
\eeq
The conditions (\ref{deg1c})-(\ref{MCeq1}) are preserved under this operation. 
For an integer $i$,
we write $\cB[i]$ for the $i$-times shift of $\cB$.
The space $\mathscr{H}^j(\cB_1,\cB_2)$ 
is isomorphic to
$\mathscr{H}^{j-i}(\cB_1,\cB_2[i])$ and also to 
$\mathscr{H}^{j-i}(\cB_1[-i],\cB_2)$. This follows from
the isomorphisms of complexes
\beq
\begin{array}{ccccc}
\!\!\mathscr{C}(\cB_1,\cB_2)\!\!&\cong&
\!\!\mathscr{C}(\cB_1,\cB_2[i])[-i]\!\!&\cong&
\!\!\mathscr{C}(\cB_1[-i],\cB_2)[-i]\!\!\\[0.2cm]
\phi^{(p)}&\longleftrightarrow & (-1)^{ip}\phi^{(p)}
&\longleftrightarrow & (-1)^{i(j-i)}\phi^{(p)},
\end{array}
\label{ismoo}
\eeq
where $\phi^{(p)}$ is the $p$-form
component of a degree $j$ element $\phi\in \mathscr{C}^j(\cB_1,\cB_2)$.
The shift of the complexes, $\mathscr{C}\mapsto \mathscr{C}[-i]$,
is the standard one; $\mathscr{C}[-i]^j=\mathscr{C}^{j-i}$
and ${\bf Q}[-i]=(-1)^i{\bf Q}$.

To understand the significance of the signs,
let us check $\mathscr{C}(\cB_1,\cB_2)\cong \mathscr{C}(\cB_1[-i],\cB_2)[-i]$.
For $\phi=\sum_{p=0}^n\phi^{(p)}\in \mathscr{C}^j(\cB_1,\cB_2)$,
we have
\beqa
i({\bf Q}\phi)^{(p)}&=&
\bartial \phi^{(p-1)}
+\sum_{k+l=p}\left(
\cQ_2^{(k)} \phi^{(l)}-(-1)^j\phi^{(l)}\cQ_1^{(k)}\right)\nn\\
&=&\bartial \phi^{(p-1)}
+\sum_{k+l=p}\left(
(-1)^{(1-k)l}\cQ_2^{(k)} \wedge\phi^{(l)}
-(-1)^j(-1)^{(j-l)k}\phi^{(l)}\wedge\cQ_1^{(k)}\right).
\nn
\eeqa
If we regard $\phi$ as an element of $\mathscr{C}^{j-i}(\cB_1[-i],\cB_2)$,
where $\cQ_1[-i]^{(k)}=(-1)^{i(k+1)}\cQ_1^{(k)}$ by definition (\ref{defshift}), 
we have
\beqa
i({\bf Q}\phi)^{(p)}&=&
\bartial \phi^{(p-1)}
+\sum_{k+l=p}\left(
\cQ_2^{(k)} \phi^{(l)}-(-1)^{j-i}\phi^{(l)}\cQ_1[-i]^{(k)}\right)\nn\\
&&\!\!\!\!\!\!\!\!\!\!\!\!\!\!\!\!\!\!\!\!\!\!\!\!\!\!\!\!\!\!\!\!\!\!\!\!
\!\!\!\!\!
=~\bartial \phi^{(p-1)}
+\sum_{k+l=p}\left(
(-1)^{(1-k)l}\cQ_2^{(k)} \wedge\phi^{(l)}
-(-1)^{j-i}(-1)^{(j-i-l)k}\phi^{(l)}\wedge (-1)^{i(k+1)}\cQ_1^{(k)}\right).
\nn
\eeqa
The two indeed agree.
This demonstrates the necessity of the sign 
$(-1)^{p+1}$ in the shift (\ref{defshift}).
The sign $(-1)^{i(j-i)}$ in (\ref{ismoo}) is just to guarantee that
the sign in ${\bf Q}[-i]=(-1)^i{\bf Q}$ is reproduced correctly.
The other relation $\mathscr{C}(\cB_1,\cB_2)\cong
\mathscr{C}(\cB_1,\cB_2[i])[-i]$ can be shown in the same way,
in which the sign $(-1)^{ip}$ in (\ref{ismoo}) plays a more important r\^ole.
The apparent asymmetry in the r\^ole of the signs originates from
the choice (\ref{2prods}) of convention.
Note also that there is an ambiguity in the choice of signs in (\ref{ismoo})
--- they can be modified by factors depending only on $i$.

Using the two isomorphisms in (\ref{ismoo}) repeatedly, 
we find an isomorphism of complexes
\beq
\begin{array}{c}
[i]:\mathscr{C}(\cB_1,\cB_2)~\stackrel{\cong}{\longrightarrow}~
 \mathscr{C}(\cB_1[i],\cB_2[i])\\[0.2cm]
~~~~~~~~~~~~~~~~~~~~~~~~\,
\phi^{(p)}~~~~\longmapsto ~~~~\phi[i]^{(p)}=(-1)^{i(|\phi|+p)}\phi^{(p)}.
\end{array}
\label{shiftmor}
\eeq
The sign $(-1)^{i(|\phi|+p)}$
is uniquely determined by the condition that the shift 
preserves the composition rule,
$(\varphi_1\varphi_2)[i]=\varphi_1[i]\varphi_2[i]$,
and by $[i]\circ [j]=[i+j]$.
A useful way to write the shift (\ref{defshift})
and (\ref{shiftmor}) is
\beq
\cQ[1]=-\cQ|_{\psi\to -\psi}=\sigma \cQ\sigma,\qquad
\phi[1]=(-1)^{|\phi|}\phi|_{\psi\to-\psi}=\sigma_2\phi\sigma_1.
\label{shiftcon}
\eeq

\subsubsection*{\it Cone}

For an element 
$\phi\in\mathscr{Z}^0(\cB_1,\cB_2)$, its cone
$\cone(\phi)=(E_{\phi},\cQ_{\phi})$ is defined by
\beq
\begin{array}{l}
E_{\phi}\,\,=\,\,\,E_1[1]\,\,\oplus \,\,E_2,\\[0.2cm]
\cQ_{\phi}\,\,=\left(\begin{array}{cc}
\cQ_1[1]&0\\
\phi&\cQ_2
\end{array}\right).
\end{array}
\label{cone}
\eeq
$\phi$ in the expression for $\cQ_{\phi}$ is regarded as an element of
$\mathscr{Z}^1(\cB_1[1],\cB_2)$ by (\ref{ismoo}).
It is straightforward to check that it satisfies the condition
$\bartial \cQ_{\phi}+\cQ_{\phi}^2=0$.

\subsubsection*{\it Zero Objects}

A D-brane $\cB$ is said to be a {\it zero object}
if the cohomology space vanish 
$$
\mathscr{H}(\cB,\cB)=0.
$$
It follows that $\mathscr{H}(\cB,\cB')=\mathscr{H}(\cB',\cB)=0$
for any brane $\cB'$.
For example, a complex of vector bundles is a zero object if and only if 
its cohomology sheaves all vanish, that is, it is an exact complex.
Note that for a bounded exact complex
$({\mathcal E}^{\cdot},Q)$ the boundary potential
 $V={1\over 2}\{Q,Q^{\dag}\}$ is positive definite everywhere,
and therefore the corresponding D-brane is empty in the infra-red limit.
This motivates us to propose that the D-brane corresponding to a zero
object is infra-red empty.

\subsubsection*{\it Quasi-Isomorphisms}

An element $s\in \mathscr{Z}^0(\cB_1,\cB_2)$
is said to be a {\it quasi-isomorphism}  if it represents
an isomorphism in $\mathscr{H}^0(\cB_1,\cB_2)$,
that is, if it has an inverse --- 
an element $s^{-1}\in\mathscr{Z}^0(\cB_2,\cB_1)$
such that $ss^{-1}\simeq {\rm id}_{E_2}$ and $s^{-1}s\simeq {\rm id}_{E_1}$
where $\simeq$ means equality modulo $\bfQ$-exact terms. 
Multiplication by a quasi-isomorphism $s$
induces the isomorphisms 
$$
\mathscr{H}^i(\cB_3,\cB_1)
\stackrel{\cong}{\longrightarrow}
\mathscr{H}^i(\cB_3,\cB_2),\quad
\mathscr{H}^i(\cB_2,\cB_3)
\stackrel{\cong}{\longrightarrow}
\mathscr{H}^i(\cB_1,\cB_3),
$$
for any $\cB_3$.
We may also define a {\it quasi-isomorphism of degree} $j$
in an analogous way --- an element 
$s\in \mathscr{Z}^j(\cB_1,\cB_2)$ with an ``inverse''
$s^{-1}\in \mathscr{Z}^{-j}(\cB_1,\cB_2)$ in the same sense as above.
By (\ref{ismoo}), it can be regarded as a quasi-isomorphism 
(of degree 0) from $\cB_1$ to $\cB_2[j]$, or from
$\cB_1[-j]$ to $\cB_2$.

For example, let us consider D-branes given by
complexes of vector bundles, ${\mathcal E}_1^{\cdot}$ and 
${\mathcal E}_2^{\cdot}$. A cochain map 
$s^{\cdot}:{\mathcal E}_1^{\cdot}\to{\mathcal E}_2^{\cdot}$
can be regarded an element of
$\mathscr{Z}^0({\mathcal E}_1^{\cdot},{\mathcal E}_2^{\cdot})$.
It is a quasi-isomorphism if it induces an isomorphism of the 
cohomology sheaves at each degree. In fact, this is the traditional meaning
of quasi-isomorphism.

An important fact is: {\it 
A ${\bf Q}$-closed element
 is a quasi-isomorphism if and only if its cone
is a zero object.} The proof of ``only if'' part is a straightforward 
computation.
``If'' part can be seen by noting that the identity 
of the cone must be ${\bf Q}$-exact.

If there is a quasi-isomorphism from $\cB_1$ to $\cB_2$,
there is a chain of D-term deformations and brane-antibrane annihilation 
that connects the two D-branes.
This was shown in \cite{HHP} for quasi-isomorphisms in the traditional
sense, but the derivation there goes through for quasi-isomorphisms
in the present sense as well,
provided that the cone, a zero object, 
indeed corresponds to an empty brane which can be annihilated.
In particular, the two D-branes flow to the same fixed point
in the infra-red limit.
In this sense {\it a quasi-isomorphism yields an isomorphism between 
the low energy D-branes.}

\subsubsection*{\it Ramond Ground States}

\newcommand{\phiR}{\uphi_{{}_{\rm R}}}

Cohomological description is possible also for Ramond ground states which,
in a superstring theory on $M\times \R^D$ ($D+2n=10$), give rise to
massless fermions in the spacetime $\R^D$.
Let us consider the Ramond sector of the open string between two D-branes
${\mathcal B}_i=(E_i,\cQ_i)$, $i=1,2$.
To be specific, we choose the $(-+)$ spin structure of the open string.
In the zero mode approximation, wavefunctions
are represented by spinors with values in $Hom(E_1,E_2)$, on which 
$\psi_0^i$ and $\psi_0^{\bi}$ act as the Gamma matrices.
On a K\"ahler manifold $M$, the spin bundle is isomorphic to
$\sqrt{K}\otimes \bigwedge\overline{T}^*_M$ where 
$\sqrt{K}$ is a line bundle which squares to the canonical bundle
$K=\det(T^*_M)$. The choice of $\sqrt{K}$ corresponds to the choice
of the spin structure of $M$. Thus, zero mode wavefunctions are
elements of
\beq
\mathscr{C}^{i}_{\rm R}(\cB_1,\cB_2)
:=\bigoplus_{p-{n\over 2}+q=i}\Omega^{0,p}(M,\sqrt{K}\otimes
Hom^q(E_1,E_2)),
\label{Rzer}
\eeq
where $i$ shows the canonical R-charge.
The supercharge ${\bf Q}$ is represented by
\beq
i{\bf Q}\phiR=\bartial\phiR+\cQ_2\phiR-(-1)^{|\phiR|}\phiR\!\left(
-i\cQ_1\bigl|_{\psi\to i\psi}\right),
\label{QRamond}
\eeq
where $|\phiR|$ is the canonical R-charge plus ${n\over 2}$.
We have the factor $-i\cQ_1|_{\psi\to i\psi}$ rather than $\cQ_1$ 
in the last term, because the left boundary has the opposite orientation,
as in the case of ${\mathcal N}=1$ supersymmetry (\ref{Q1}).
The spaces of ${\bf Q}$-cohomology classes, which correspond to 
Ramond ground states, are denoted by $\mathscr{H}_{\rm R}^i(\cB_1,\cB_2)$.
Note that the chiral ring naturally acts on them,
\beq
\mathscr{H}^i(\cB_1,\cB_2)\times
\mathscr{H}_{\rm R}^j(\cB_2,\cB_3)\times
\mathscr{H}^k(\cB_3,\cB_4)\longrightarrow
\mathscr{H}_{\rm R}^{i+j+k}(\cB_1,\cB_4).
\eeq
The left action is the na\"ive one and the right action is via 
${\psi\to i\psi}$.

\subsection{O-Isomorphisms From Quasi-Isomorphisms}
\label{subsec:qois}

\newcommand{\bfs}{{\bf s}}

Let us consider the Type II orientifold
of $M\times\R^D$ with respect to a holomorphic involution
$\inv$ of $M$ and a twist $({\mathcal L},\sA)$ which is holomorphic,
$\bartial_{\sA}^2=0$. 
The parity transform (\ref{pagen}) of the ${\mathcal N}=2_B$ preserving D-brane
$\cB=(E,\cQ)$ reads
\beq
\cQ~\longmapsto~\Ptr(\cQ)=\left.\spst\inv^*\cQ^T
\right|_{\psi\to\spst\psi}+i\psi^{\bi}\sA_{\bi}.
\label{defPfunct1}
\eeq
The image bundle can be given a $\Z$-grading compatible with 
the $\Z_2$,
\beq
\Ptr(E)^i=\inv^*(E^{-i})^*\otimes {\mathcal L},
\label{defPfunctgr}
\eeq
with respect to which $\Ptr(\cQ)$ has degree $1$.
$\Ptr(Q)$ is obviously independent of $\psi^i$. It also
obeys (\ref{MCeq1}): 
$\psi^{\bi}\partial_{\bi}\cQ+\cQ^2=0$ implies
$\psi^{\bi}\partial_{\bi}\cQ^T-(\cQ^T)^2=0$ which, after $\psi\to\spst\psi$,
yields
$
\psi^{\bi}\partial_{\bi}\left(\spst\cQ^T|_{\psi\to\spst\psi}
\right)+\left(\spst\cQ^T|_{\psi\to\spst\psi}\right)^2
=0.
$
This relation is preserved by the pull back by the holomorphic map $\inv$
and also by the shift by the holomorphic connection $i\psi^{\bi}\sA_{\bi}$.
Thus, the image brane $\Ptr(\cB)$ preserves
${\mathcal N}=2_B$ supersymmetry with $U(1)$ R-symmetry.

To define the parity operator on open string states,
an o-isomorphism must be specified.
Let $\gU:\Ptr(E)\to E$ be a unitary o-isomorphism of a D-brane 
given by a complex of vector bundles, $\cQ=Q+i\psi^{\bi}A_{\bi}$.
The condition that it maps the boundary interaction for
$\Ptr(\cB)$ to the one for $\cB$,
(\ref{Tco}) and (\ref{Aco}), can be written as
\beq
\bfQ\gU=0\quad\mbox{and}\quad
\bfQ^{\dag}\gU=0
\eeq
for $\gU$ regarded as an element of $\mathscr{C}(\Ptr(\cB),\cB)$.
Here $\bfQ^{\dag}$ is defined in the analogous way to (\ref{Qzero}).
In particular, the parity operator defined via $\gU$ preserves
${\mathcal N}=2_B$ supersymmetry,
and also the $U(1)$ R-symmetry if $\gU$ has a definite degree.

Such an example can be obtained from a ``holomorphic o-isomorphism'',
i.e., an isomorphism
between a complex of holomorphic vector bundles
and its parity image,
$\gU_{\it hol}:(\Ptr({\mathcal E}),\Ptr(Q))\to({\mathcal E},Q)$, which
satisfies (\ref{condforU}).
Indeed, one can construct a hermitian metric of
${\mathcal E}$ with respect to which $\gU_{\it hol}$ is unitary:
We first construct it on the $\inv$-fixed point set $M^{\inv}$. 
This is possible essentially due to the fact that $G_{\C}/G$ 
is contractible, for $G=O(N)$ or $USp(N)$.
We then extend it to the entire $M$, recursively using a $\inv$-compatible
cell decomposition (as in analogous constructions in Section~\ref{sec:K}).
Because of the contractibility of $G_{\C}/G$, for $G$ as above and $U(N)$, 
the metric is unique up to continuous deformation.
Being holomorphic and unitary,
$\gU_{\it hol}$ sends the hermitian connection of $\Ptr({\mathcal E})$
to that of ${\mathcal E}$.
In this way we obtain a unitary o-isomorphism.

What we are really interested in, however, are
the infra-red fixed points of the boundary interactions
and the orientifold action on them.
It is enough that we have an isomorphism between the infra-red limit
of the brane and its parity image.
In the above example of complex of holomorphic vector bundles,
a unitary o-isomorphism of course does the job, but that is not necessary.
For example, we could have made a ``wrong'' choice of the fibre metric so that
$\gU_{\it hol}$ fails to be unitary, in which case
${\bf Q}\gU_{\it hol}=0$ holds but ${\bf Q}^{\dag}\gU_{\it hol}=0$ fails.
Such a brane is connected to the one with a unitary o-isomorphism
by a continuous deformation of the metric, and that induces an isomorphism
in the infra-red limit. Therefore, we have an infra-red isomorphism 
between the brane and its parity image.
More generally, as we discussed,
a quasi-isomorphism induces an isomorphism in the
infra-red superconformal field theory. 
Therefore, it is enough that we have a ``quasi-o-isomorphism'',
i.e., a quasi-isomorphism between the brane and its parity image.

It is rare that we explicitly know the infra-red isomorphism induced from a
quasi-isomorphism.
Therefore, a quasi-o-isomorphism is usually helpless for
writing down the parity operator on {\it all} open string states.
However, it does help us write down the parity operator 
on the chiral sector, as we now describe.

First, the parity transform of NS vertex operators
in the zero mode approximation,
$\Ptr:\mathscr{C}(\cB_1,\cB_2)\to \mathscr{C}(\Ptr(\cB_2),\Ptr(\cB_1))$,
is given by
\beq
\Ptr(\phi)=\left.\inv^*\phi^T\right|_{\psi\to\spst\psi}.
\label{defPfunct2}
\eeq
Unlike in (\ref{parmap})
we do not need the parallel transport factor $h_{\sA}$ in the zero mode sector.
It has the property
\beqa
\Ptr(i\bfQ\phi)&=&
\left.\inv^*\left(\bartial\phi+\cQ_2\phi-(-1)^{|\phi|}\phi\cQ_1\right)^T
\right|_{\psi\to\spst\psi}\nn\\
&=&\left.\inv^*\left(\bartial\phi^T+(-1)^{|\phi|}\phi^T\cQ_2^T
-\cQ_1^T\phi^T\right)\right|_{\psi\to\spst\psi}
\nn\\
&=&\spst\bartial\Ptr(\phi)-\spst(-1)^{|\phi|}\Ptr(\phi)\Ptr(\cQ_2)
+\spst\Ptr(\cQ_1)\Ptr(\phi)
=\spst i\bfQ\Ptr(\phi).\label{PtrQ}
\eeqa
Note also that
\beq
\Ptr(\phi_1\phi_2)=(-1)^{|\phi_1||\phi_2|}\Ptr(\phi_2)\Ptr(\phi_1),
\label{functoriality}
\eeq
for $\phi_1\in\mathscr{C}(\cB_2,\cB_3)$ and 
$\phi_2\in\mathscr{C}(\cB_1,\cB_2)$.

Let $\cB$ be a D-brane and let
$\bfs:\Ptr(\cB)\to\cB$ be a quasi-isomorphism, i.e. an element
$\bfs\in\mathscr{Z}(\Ptr(\cB),\cB)$ with an ``inverse''
$\bfs^{-1}\in\mathscr{Z}(\cB,\Ptr(\cB))$ 
in the sense explained earlier.
Let us put
\beq
\Po(\phi)= \bfs\,\Ptr(\phi)\,\bfs^{-1}(-1)^{|\bfs||\phi|}.
\label{defPch}
\eeq
The property (\ref{PtrQ}) leads to the commutation relation
\beq
\Po\,\bfQ=\spst\,\bfQ\,\Po.
\label{PQQP}
\eeq
In particular, $\Po$ maps ${\bf Q}$-closed {\it resp}.\! exact elements 
to  ${\bf Q}$-closed {\it resp}.\! exact elements.
Also, if $\phi$ is ${\bf Q}$-closed,
shifts of $\bfs$ and $\bfs^{-1}$ by $\bfQ$-exact terms
only affect $\Po(\phi)$ by $\bfQ$-exact terms and hence do not affect
its ${\bf Q}$-cohomology class.
Thus, (\ref{defPch}) defines a parity operator
$\Po:\mathscr{H}(\cB,\cB)\to \mathscr{H}(\cB,\cB)$.
This represents the parity action on chiral primary fields
in the infra-red superconformal field theory,
as $\bfQ$ is expected to flow to the superconformal generator
$G_{-{1\over 2}}$.
The relation (\ref{PQQP}) is compatible with this expectation
since $G_{-{1\over 2}}$ satisfies 
$\Po G_{-{1\over 2}}\Po^{-1}=\spst G_{-{1\over 2}}$.

The condition for $\Po^2=(-1)^F$ can be found in the same way as in
Section~\ref{subsec:o-isom}. By definition,
$\Po^2(\phi)=\bfs \,\Ptr(\bfs^{-1})\,\Ptr^2(\phi)\,
\Ptr(\bfs)\,\bfs^{-1}(-1)^{|\bfs|}$.
Using $\Ptr(\bfs^{-1})(-1)^{|\bfs|}\simeq \Ptr(\bfs)^{-1}$ and
\beq
\Ptr^2(\phi)=\phi^{TT}|_{\psi\to -\psi}
=\imath\phi\imath^{-1}|_{\psi\to -\psi}
=(-1)^{|\phi|}\imath\sigma^{-1}
\phi\sigma\imath^{-1},
\label{Ptrsq}
\eeq
we find $\Po^2(\phi)\simeq (-1)^{|\phi|}{\bf t}\,
\phi\,{\bf t}^{-1}$
for ${\bf t}=\bfs \,\Ptr(\bfs)^{-1}\imath\sigma^{-1}$.
The condition is thus
\beq
\bfs~\simeq ~(\sigma\imath^{-1}\otimes \gc)\,\Ptr(\bfs),
\label{condfors}
\eeq
for some nowhere vanishing holomorphic section $\gc$ of 
$\inv^*{\mathcal L}\otimes {\mathcal L}^*$. 
Note that, for such a $\gc$,
the factor $(\sigma\imath^{-1}\otimes \gc)$ is a (quasi-)isomorphism
from $\Ptr^2(\cB)$ back to $\cB$. 
For a pair of D-branes with quasi-o-isomorphisms,
say $(\cB_1,\bfs_1)$ and $(\cB_2,\bfs_2)$,
the parity operator $\Po:\mathscr{H}(\cB_1,\cB_2)\to\mathscr{H}(\cB_2,\cB_1)$
is defined by 
$\Po(\phi)=\bfs_1\,\Ptr(\phi)\,\bfs_2^{-1}(-1)^{|\phi||\bfs_2|}$.
The consistency condition $\Po^2=(-1)^F$ requires that
$\bfs_1$ and $\bfs_2$ both satisfy (\ref{condfors}) with a common $\gc$.

The condition (\ref{condfors}) is the quasi-o-isomorphism version of 
the condition (\ref{condforU}) for unitary o-isomorphisms.
The consistency condition requires that the section $\gc$ in (\ref{condfors}) 
must be equal to the one for unitary o-isomorphisms,
i.e., the crosscap section, as the notation already implies.
In particular, $\gc$ and the mod 2 degree of $\bfs$
are correlated with the types and the dimensions of O-planes as 
 (\ref{basic2}) and (\ref{statU}).

\subsubsection*{\it Parity On Ramond Ground States}

Let us describe the parity action on Ramond ground states.
For this purpose, 
we must manifest the dependence on the phase $\spst=\mp i$, say, 
by the subscript, $\Ptr_{(\spst)}$, $\bfs_{(\spst)}$ etc.
Following the discussion in Section~\ref{subsec:Ramond}
we propose that the quasi-o-isomorphisms for the two phases are related by
\beq
\bfs_{(i)}=\kappa\,\bfs_{(-i)}\sigma^T.
\label{relsism}
\eeq
Indeed, under this relation
 $\bfs_{(i)}$ is ${\bf Q}$-closed as an element of
$\mathscr{C}(\Ptr_{(i)}(\cB),\cB)$ if and only if
$\bfs_{(-i)}$ is ${\bf Q}$-closed as an element of 
$\mathscr{C}(\Ptr_{(-i)}(\cB),\cB)$.
That can be shown using $\Ptr_{(i)}(\cQ)=\sigma^T\Ptr_{(-i)}(\cQ)\sigma^T$
which holds since $\cQ$ is odd, $\sigma\cQ|_{\psi\to -\psi}\sigma=-\cQ$.
Let $(\cB_i,\bfs_i)$ be D-branes with quasi-o-isomorphisms, for $i=1,2$.
The parity $\wtPo$ is represented on the zero mode Ramond sector
sector (\ref{Rzer}) by
\beq
\wtPo(\phiR)=\bfs_{2(i)}\inv^*\!\phiR^T
\left(\bfs^{-1}_{1(-i)}|_{\psi\to i\psi}\right)(-1)^{|\phiR||\bfs_1|}
\eeq
It is straightforward to check that it commutes with the supercharge
${\bf Q}$,
\beq
\wtPo\,{\bf Q}={\bf Q}\,\wtPo.
\label{wtPQ}
\eeq
In particular, $\wtPo$ acts on the ${\bf Q}$-cohomology classes.
This represents the parity action on the Ramond ground states.
The relation (\ref{wtPQ}) is compatible with the expectation that
${\bf Q}$ flows to the superconformal zero mode $G_0$
which commutes with $\wtPo$.

\subsubsection{The Degree Of O-Isomorphisms}
\label{subsub:Odeg}

Let $\cB$ be a D-brane with a quasi-o-isomorphism $\bfs:\Ptr(\cB)\to \cB$
of a certain degree. 
Let us see if the brane shifted by one to the left,
$\cB[1]$, defined by (\ref{defshift}), also has a quasi-o-isomorphism.
Note that we have a quasi-isomorphism 
$\bfs[1]:\Ptr(\cB)[1]\to\cB[1]$ of the same degree as $\bfs$
(see (\ref{shiftmor}) for the definition).
But we want $\Ptr(\cB[1])$ in the place of $\Ptr(\cB)[1]$.
How are they related?
The former has
$$
\Ptr(E[1])^i=\inv^*(E[1]^{-i})^*\otimes {\mathcal L}
=\inv^*(E^{-i+1})^*\otimes{\mathcal L},
$$
and
$\Ptr(\cQ[1])
=\inv^*(\cQ[1])^{T_{E[1]}}|_{\psi\to\spst\psi}+i\psi^{\bi}\sA_{\bi}$.
Here we have denoted the transpose by $(-)^{T_{E[1]}}$ to emphasize that
it is with respect to the shifted degree. The relation to
the one before the twist can be found from (\ref{grshift}):
$f^{T_{E[1]}}=\sigma_{E^*}f^{T_E}\sigma_{E^*}$ where 
$\sigma_{E^*}=\sigma_E^{T_E}=\sigma^T$.
Note also that $\cQ[1]=\sigma\cQ\sigma$ by (\ref{shiftcon}).
Thus, we find that $(\cQ[1])^{T_{E[1]}}$ is equal to $\cQ^{T_E}=\cQ^T$ 
and hence that
$\Ptr(\cQ[1])=\Ptr(Q)$ in the end.
On the other hand, $\Ptr(\cB)[1]$ has
$$
\Ptr(E)[1]^i=\Ptr(E)^{i+1}=\inv^*(E^{-i-1})^*\otimes {\mathcal L},
$$
and $\Ptr(\cQ)[1]=\sigma^T\Ptr(Q)\sigma^T$.
Thus, we have a degree $-2$ isomorphism
\beq
\sigma^T:\Ptr(\cB[1])\longrightarrow\Ptr(\cB)[1].
\label{sigmaf}
\eeq
Composing this with $\bfs[1]$, we obtain a quasi-isomorphism 
\beq
\bfs[1]\circ\sigma^T:\Ptr(\cB[1])\longrightarrow\cB[1]
\label{barbfs}
\eeq
of degree $|\bfs|-2$.
It is straightforward to see that it 
satisfies the condition (\ref{condfors}) for the same $\gc$ as $\bfs$. 
Thus, it is a quasi-o-isomorphism of $\cB[1]$.
Note that $\cB[1]$ may be interpreted as the antibrane of $\cB$. 
Then $\bfs\to \bfs[1]\circ\sigma^T$ is compatible with the general formula
(\ref{antiU}) that relates the unitary o-isomorphisms between
a brane and its antibrane.
As in that case, in order to maintain the relation (\ref{relsism}),
we need to choose an opposite sign for opposite $\spst$,
say, $\overline{\bfs}_{(\pm i)}=\pm s[1]_{(\pm i)}\circ\sigma^T$.
Applying this repeatedly, we find that $\cB[i]$ has 
a quasi-o-isomorphism of degree $|\bfs|-2i$.

The shift by even integers, $\cB\mapsto \cB[2i]$, which results
in $|\bfs|\to |\bfs|-4i$, has no effect on 
the resulting D-brane in the string theory.
Thus, we can always bring the degree of the quasi-o-isomorphism
into one of the two values in a window of length four.
For example, we may always assume 
$|\bfs|=0$ or $2$ ({\it resp}.\! $|\bfs|=\pm 1$)
if $|\bfs|$ is even ({\it resp}.\! odd).
It is also true that shift by odd integers, say $\cB\mapsto\cB[1]$,
does not change the infra-red boundary interaction itself. 
However, $\cB[1]$ must be distinguished
from $\cB$ in string theory --- it is the antibrane of $\cB$, 
which yields the opposite 
GSO projection for the open string with other branes.
For example, the orientifold projection condition
is opposite in the Ramond sector. 
To summarize, all branes are classified into two classes,
distinguished by the degree of quasi-o-isomorphisms mod 4.
We shall discuss the significance of this
in Section~\ref{subsec:sf} from the view point of spacetime supersymmetry.

The degree of o-isomorphisms may be traded into the degree of
the parity transform. Suppose we have a D-brane $\cB$ with a 
quasi-o-isomorphism $\bfs:\Ptr(\cB)\to \cB$ of degree $|\bfs|=-r_o$.
Using (\ref{ismoo}), we may regard it as a quasi-isomorphism
of degree $0$ from $\Ptr(\cB)[r_o]$ to $\cB$.
This motivates us to change the definition of the parity transform
to 
\beq
\Ptr_{r_o}(\cB)=\Ptr(\cB)[r_o].
\eeq 
Accordingly, we change the definition of the parity transform 
of vertex operators
to $\Ptr_{r_o}(\phi)=\Ptr(\phi)[r_o]$.
In this formulation, the parity operator is defined by 
\beq
\Po(\phi)=\bfs\,\Ptr_{r_o}(\phi)\,\bfs^{-1}.
\eeq
There is no sign factor $(-1)^{|\bfs||\phi|}$
since $\bfs$ is regarded to have degree $0$ here. 
Using (\ref{ismoo}) and (\ref{shiftmor}) along with (\ref{2prods}),
one can check that this parity operator is identical to the one 
in (\ref{defPch}).
The condition (\ref{condfors}) for $\Po^2=(-1)^F$ translates in this formulation 
to
\beq
\bfs~\simeq~\left\{\begin{array}{ll}
(\sigma\imath^{-1}\otimes\gc)\Ptr_{r_o}(\bfs)&\mbox{$r_o$ even},\\[0.2cm]
-(\imath^{-1}\otimes\gc)\Ptr_{r_o}(\bfs)&\mbox{$r_o$ odd}.
\end{array}
\right.
\label{condforss}
\eeq
The precise sign on the right hand side is important
if we want $\gc$ here to be identical to the crosscap section. 
If we are interested only in branes within one of the two classes 
mentioned above,
we may fix $r_o$ and restrict our attention 
to those with degree zero quasi-o-isomorphisms.

\subsubsection{An Example}
\label{ex:qois}

For illustration, let us present 
an example in which there is a quasi-o-isomorphism but not
a unitary o-isomorphism.
It is in Type IIB orientifold on $T^2\times\R^8$
by $\inv=$ the inversion of $T^2$, with
all four O7-planes of type O${}^-$
(for this, we must take $B=0$, the trivial twist, and $\gc\equiv\spst$). 
In this case, the degree of quasi-o-isomorphisms is odd.
Alternatively, we may take the parity transform to be
$\Ptr_{r_o}$ for $r_o=\pm 1$, say, and
seek for branes with degree $0$ quasi-o-isomorphisms.
Let $p\in T^2$ be one of the four fixed points.
The holomorphic line bundle ${\mathcal O}(p)$ has a section $\vartheta_p$
(a theta function) that vanishes at $p$. The brane we consider is given by 
the following complex
\beq
\cB:~~~~~~~~~
{\mathcal O}~
\stackrel{\vartheta_p}{\longrightarrow}~
\,\underline{\!\!{\mathcal O}(p)\!\!}\,~
\label{D7}
\eeq
where the underline shows the position of the degree $0$ component.
Its image by the transform $\Ptr_{1}$ is isomorphic to 
the following complex
\beq
\Ptr_{1}(\cB)':~~~
{\mathcal O}(-p)\,
\stackrel{\spst\vartheta_p}{\longrightarrow}~~
\underline{{\mathcal O}}~~~~~~~~~~~
\label{imD7}
\eeq
via $\inv^*{\mathcal O}(p)^*\cong {\mathcal O}(-p)$
and $\inv^*{\mathcal O}^*\cong{\mathcal O}$ which we assume in what follows. 
Both (\ref{D7}) and (\ref{imD7}) represent a D7-brane at $p$ and hence
the two must be isomorphic.\footnote{A single D7-brane at an O7${}^-$ is 
actually inconsistent \cite{HIS,triples}.
However, at the classical level,
we should be able to construct such a configuration.
Alternatively, we may consider a single D3-brane at one of the four
O3${}^-$-planes in orientifold of $T^2\times \R^8$
by an involution which flips the sign of four coordinates of $\R^8$
in addition to the two of $T^2$. We also note here that
we do not (need to) respect other constraints such as those of the type
discussed in \cite{Uranga}.}
However, one cannot find a holomorphic
isomorphism between them --- the 
two vector bundles, ${\mathcal O}(p)\oplus{\mathcal O}$
and ${\mathcal O}\oplus {\mathcal O}(-p)$, are simply not isomorphic.
But there is a quasi-isomorphism.
An element $\bfs\in \mathscr{C}^0(\Ptr_{1}(\cB)',\cB)$ given by
\beq
\begin{array}{ccc}
\,{\mathcal O} \!\!\!&\stackrel{\vartheta_p}{\longrightarrow}& 
\!\!\!{\mathcal O}(p)\!\!\!\\
\!\!\!\scriptstyle{s^{(0)}_{-1}}\Big\uparrow &
\,\,\nwarrow^{\!\!s^{(1)}_0}  \!\! &
\Big\uparrow\scriptstyle{s^{(0)}_0}\!\!\!\!\!\!\!\\
\!\!\!\!{\mathcal O}(-p)\!\!\!\!
 &\stackrel{\spst\vartheta_p}{\longrightarrow}& \!{\mathcal O}
\end{array}
\eeq
is ${\bf Q}$-closed if $s^{(0)}_{-1}=\spst s^{(0)}_0$ and 
$\bartial s^{(0)}=\vartheta_p s^{(1)}_0$. A solution is determined 
for any choice of $s^{(0)}_0$ which is holomorphic 
in a neighborhood of the point $p$ at which $\vartheta_p$ vanishes.
This is a quasi-isomorphism, i.e.,
there is an inverse $\bfs^{-1}\in\mathscr{C}^0(\cB,\Ptr_{1}(\cB)')$,
$\bfs\bfs^{-1}\simeq {\rm id}_{\cB}$ and
$\bfs^{-1}\bfs\simeq {\rm id}_{\Ptr_{1}(\cB)'}$,
as long as $s^{(0)}_0$ is non-vanishing at $p$ (note that such an
$s^{(0)}_0$ cannot be globally holomorphic).
The condition (\ref{condforss}) for $r_o=1$ reads
$$
(-\spst\inv^*s^{(0)t}_{-1},-\inv^*s^{(1)t}_0,\spst\inv^*s^{(0)t}_{0})
\simeq (s^{(0)}_0,s^{(1)}_0,s^{(0)}_{-1}),
$$
and is satisfied for a suitable choice of $s^{(0)}_0$.
Thus, $\cB$ has a quasi-o-isomorphism.

Existence of such an example raises a question concerning
our classification of topology in Section~\ref{sec:K} in which
we assumed that all branes have unitary o-isomorphisms.
Does it miss some of the branes?
In fact, there would be no problem if each brane has a representative,
if not itself, which does admit a unitary o-isomorphism.
Let us examine such a possibility for the example (\ref{D7}).


Unfortunately,
no representative of (\ref{D7}) admits a unitary o-isomorphism
if its Chan-Paton bundle is of finite rank.
To see this, suppose
$\cB_F=(F,\cQ_F)$, with $F$ of finite rank, has a unitary o-isomorphism.
Then, we have isomorphisms of vector bundles,
$F^{\rm ev}\cong\inv^*F^{{\rm od}*}$ and
$F^{\rm od}\cong\inv^*F^{{\rm ev}*}$.
It then follows that $F^{\rm ev}$ and $F^{\rm od}$
have the same rank and their first Chern classes
are related by $c_1(F^{\rm ev})=\inv^*c_1(F^{{\rm od}*})
=-c_1(F^{\rm od})$, where we have used the fact 
that $\inv^*$ is the identity on second cohomology classes.
In particular, the Chern character of the brane is
${\rm ch}(\cB_F)=c_1(F^{\rm ev})-c_1(F^{\rm od})
=2c_1(F^{\rm ev})$. 
On the other hand, if $\cB_F$ and $\cB$ are quasi-isomorphic,
their Chern characters must also agree.
However, 
${\rm ch}(\cB)$ is an integral generator of ${\rm H}^2(T^2,\Z)$
and cannot be equal to ${\rm ch}(\cB_F)\in 2{\rm H}^2(T^2,\Z)$.

However,
there is an infinite rank representative
which admits a unitary o-isomorphism.
It is given by a complex
$$
\cB_{\infty}:~~E^{-1}\stackrel{Q}{\longrightarrow}\underline{E^0}
$$
where the vector bundle $E^i$ is the quotient of the trivial bundle
over the complex plane $\C$ 
with fibre $\oplus_{n,m\in\Z}\C|n,m\rangle_{{}_i}$
by the equivalence relation
$$
(z,|n,m\rangle_{{}_i})\sim (z+1,|n-1,m\rangle_{{}_i})\sim
(z+\tau,|n,m-1\rangle_{{}_i}),
$$
and $Q$ is given by
\beq
Q:\,(z,|n,m\rangle_{{}_{-1}})~\longmapsto~(z,|n,m\rangle_{{}_0})
\cdot (z+n+\tau m).
\eeq
Note that $Q$ vanishes exactly at one point $[z]=[0]\in T^2$
on a rank one subbundle.
That is, 
it represents a D7-brane at $[0]$.
In particular, it represents the same brane as (\ref{D7}), 
if we identify $p$ as the point $[0]$.
Let us next describe the orientifold image $\Ptr_{1}(\cB_{\infty})$.
The Chan-Paton bundle consists of degree $-1$ component
$\inv^*E^{0*}$ and degree $0$ component $\inv^*E^{-1*}$.
The fibre of $\inv^*E^{i*}$ at $[z]\in T^2$ is
spanned by $\inv^*(z,{}_{{}_i}\!\langle n,m|)
:=([z],(-z,{}_{{}_i}\!\langle n,m|))$,
where $\{(z,{}_{{}_i}\!\langle n,m|)\}_{n,m}$ form the dual frame to
$\{(z,|n,m\rangle_{{}_i})\}_{n,m}$. The tachyon configuration is
obtained from
$$
\Ptr_{1}(Q):\,\inv^*(z,{}_{{}_0}\!\langle n,m|)~\longmapsto
\inv^*(z,{}_{{}_{-1}}\!\langle n,m|)\cdot \spst(z-n-\tau m).
$$
There is a unitary o-isomorphism $\gU:\Ptr_{1}(\cB_{\infty})\to \cB_{\infty}$
given by
\beqa
&&
\inv^*(z,{}_{{}_0}\!\langle n,m|)~\longmapsto~ 
\spst (z,|-n,-m\rangle_{{}_{-1}}),
\nn\\
&&
\inv^*(z,{}_{{}_{-1}}\!\langle n,m|)~\longmapsto~
 (z,|-n,-m\rangle_{{}_0}).
\nn
\eeqa
It is straightforward to check that it sends $\Ptr_{1}(Q)$ to $Q$
and satisfies the condition (\ref{condforss}) for $r_o=1$.
The brane $(\cB_{\infty},\gU)$ is isomorphic to $(\cB,\bfs)$
as D-branes in the orientifold, i.e., there is
a quasi-isomorphism $f\in \mathscr{C}^0(\cB_{\infty},\cB)$ such that
\beq
f\circ \gU\circ \Ptr_{1}(f)\simeq \bfs,
\label{isomD7}
\eeq
which is the quasi-isomorphism version of (\ref{isomDO}).
To construct such $f$, let us choose local frames of ${\mathcal O}$
and ${\mathcal O}(p)$ in the neighborhood $|z|<3\epsilon$
of $p=[0]$ with respect to which we have $\vartheta_p=z$.
Let us put $f=0$ on $(z,|n,m\rangle_{{}_i})$ if $|z+n+\tau m|> 2\epsilon$
and
\beqa
&&f:(z,|0,0\rangle_{{}_{-1}})\in E^{-1} \longmapsto \rho \in 
\Omega^0({\mathcal O}),
\nn\\
&&f:(z,|0,0\rangle_{{}_0})\in E^0\longmapsto 
\left(\rho,{1\over z}\bartial \rho\right)\in \Omega^0({\mathcal O}(p))
\oplus \Omega^{0,1}({\mathcal O}),
\nn
\eeqa
where $\rho$ is a smeared step function which is constantly $1$
for $|z|<\epsilon$ and zero outside $|z|<2\epsilon$.
Then $f$ is ${\bf Q}$-closed. And it satisfies (\ref{isomD7}) if
$s_0^{(0)}=\rho\cdot\inv^*\rho$.

This exercise illustrates that even if a brane does not admit a unitary
o-isomorphism, there can be another representative that admits one.
However, we have not given a proof of general existence.
We leave this question open in this paper.

\subsection{The Spectral Flow}
\label{subsec:sf}

\newcommand{\dBRST}{\delta_{{}_{\rm BRST}}}

When the first Chern class of $M$ vanishes, $c_1(M)=0$,
the non-linear sigma model on $M$ admits the topological B-twist.
We shall use this to define a one to one correspondence between 
Ramond ground states and chiral ring elements, 
known as the {\it spectral flow} \cite{LVW},
and study the spacetime supersymmetry that results from it.

The topological B-twist turns the spinors $\psi_{\pm}^{\bi}$ and
$\psi_{\pm}^i$ into scalars and one-forms respectively
\cite{Wittentop}.
The scalars are denoted by $\eta^{\bi}$ and $\theta_i$
and the one-forms are written as $\rho^i$.
The spinor supercharge ${\bf Q}$ becomes a scalar, with the 
transformation rule 
\beq
\begin{array}{c}
\delta x^i=0,\quad \delta x^{\bi}=-\bepsilon \eta^{\bi}\\
\delta \rho^i=-2i\bepsilon \dd x^i, \quad \delta \eta^{\bi}=0,\quad 
\delta\theta_i=0,
\end{array}
\label{BRST}
\eeq
for $\delta=-i\bepsilon {\bf Q}$.
In the absence of spinors, there is no need to choose spin structure, and
in particular, there is no $(+)$ versus $(-)$ distinction in
the boundary condition for a D-brane.
For the space-filling brane with data $(E,\cQ)$,
we impose the boundary condition $\theta_i=0$ and $\rho^i_n=0$
compatible with (\ref{BRST}), and
make the replacement $\psi^{\bi}\to\eta^{\bi}$ and
$\psi^i\to\rho^i_{\tau}$
inside the boundary interaction ${\mathcal A}_{\tau}$.

\begin{figure}[htb]
\psfrag{NS}{NS vertex operator}
\psfrag{R}{R-sector state}
\psfrag{twist}{Twist}
\centerline{\includegraphics{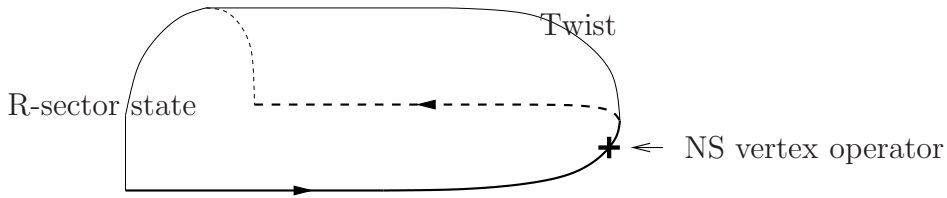}}
\caption{The spectral flow}
\label{fig:sfr}
\end{figure}

On a flat region of the worldsheet, the twisted model is indistinguishable
from the original ${\mathcal N}=2$ theory,
provided we choose an appropriate boundary condition in the latter.
For example, at the boundary of the upper-half plane, 
vertex operators of the twisted model correspond to
NS vertex operators in the untwisted model.
As another example, the twisted model on a flat strip corresponds to
the Ramond sector of the untwisted model.
These two flat geometries can be connected by
the `quarter-sphere' diagram as shown in Fig.~\ref{fig:sfr}.
In the twisted model, the scalar supercharge ${\bf Q}$ 
is conserved and 
has a closed one form as its current, even in the curved region.
The diagram therefore determines a one to one correspondence between
${\bf Q}$-cohomology classes of vertex operators and 
${\bf Q}$-cohomology classes of states, i.e.
between chiral ring elements and Ramond ground states.
This is the spectral flow.

\newcommand{\spf}{\Scr{U}}

To be precise, 
there is a minor difference between the twisted and untwisted models
on the flat strip. 
It is in the boundary interaction ${\mathcal A}_t$ on the left.
Because the time runs in the opposite direction on the left boundary,
we make a replacement
$\psi^{\bi}\to i\psi^{\bi}$ and $\psi^i\to i\psi^i$ inside ${\mathcal A}_t$
in the untwisted model
(here we take the $(-+)$ spin structure ---
see Section~\ref{subsec:Ramond}).
In the twisted model, on the other hand,
we do $\eta^{\bi}\to \eta^{\bi}$ and $\rho^i_t\to -\rho^i_t$.
The difference is 
$\psi^{\bi}\to i\psi^{\bi}$ and $\psi^i\to -i\psi^i$, which is
nothing but the R-symmetry transformation with phase $i$.
At this point, we introduce the representation of the R-symmetry on 
the Chan-Paton bundle, $R_{\cB}(\lambda):E\to E$,
which acts as multiplication by $\lambda^i$ on the degree $i$ 
component $E^i$.
The condition that $\cQ$ has degree 1 can be expressed as
$R_{\cB}(\lambda)\cQ|_{\psi\to\lambda\psi}R_{\cB}(\lambda)^{-1}=\lambda \cQ$.
In particular, conjugation 
${\mathcal A}_t\to R_{\cB}(\lambda)^{-1}{\mathcal A}_t R_{\cB}(\lambda)$
has the effect of the transformation $\psi^{\bi}\to\lambda\psi^{\bi}$ and 
$\psi^i\to \lambda^{-1}\psi^i$ inside ${\mathcal A}_t$.

Based on this consideration, we consider the following map
of NS vertex operators to Ramond sector states in the zero mode
approximation, which sends $\mathscr{C}^i(\cB_1,\cB_2)$ to
$\mathscr{C}_{\rm R}^{i-{n\over 2}}(\cB_1,\cB_2)$,
\beq
\spf\,:\,\phi~\longmapsto~\sqrt{\mit\Omega}\otimes\phi\circ R_{\cB_1}(i).
\label{mapf}
\eeq
To be precise, this is a map from $(++)$ to $(-+)$.
Here $\sqrt{\mit\Omega}$ is a holomorphic trivialization of
the line bundle $\sqrt{K}$ (we are choosing a spin structure of $M$
that admits it).
It can be regarded as a square root
of a holomorphic volume form $\mit\Omega$ of $M$
which exists when $c_1(M)=0$.
The presence of $\sqrt{\mit\Omega}$ in (\ref{mapf})
is consistent with the definition of
the topological B-model measure 
under which the worldsheet of Euler number $\chi$ comes with
the factor $\mit\Omega^{\chi}$; the sphere has $\chi=2$ and hence the
quarter-sphere may be assigned $\chi={1\over 2}$.
The map (\ref{mapf}) indeed commutes with the supercharge,
$$
{\bf Q}\circ \spf=\spf\circ {\bf Q},
$$
where ${\bf Q}$ on the left {\it resp}.\! right hand side is
defined by  (\ref{QRamond}) {\it resp}.\! (\ref{Qzero}).
In particular, it determines a one to one correspondence between
chiral ring elements and Ramond ground states.

\newcommand{\phS}{\varrho_{\rm S}}

Using this map let us compare the parity
$\wtPo$ on Ramond ground states
and the parity $\Po$ on chiral ring elements.
Note first that, under the holomorphic involution $\inv$,
the holomorphic volume form $\mit\Omega$ is invariant 
up to a sign,
$\inv^*\mit\Omega=\pm \mit\Omega$, 
and hence its square root $\sqrt{\mit\Omega}$ is 
invariant up to a phase, $\inv^*\sqrt{\mit\Omega}
=\phS\sqrt{\mit\Omega}$ with
$\phS^2=\pm 1$.
By a straightforward computation, we find, for a brane with
a quasi-o-isomorphism $(\cB,\bfs)$
\beq
\wtPo(\spf (\phi))
=\kappa\epsilon_{|\bfs|}i^{|\phi|}
\spf(\Po(\phi)),
\label{relPwtP}
\eeq
where $\epsilon_{|\bfs|}:=\phS i^{-|\bfs|}$.
The choice of the phase $\phS$ is a part of the choice of the lift
$\invS$ discussed in Section~\ref{subsec:O-plane}.
Since we consider the trivial involution on $\R^D$, we must have
$\phS^2=\pm 1$ in Case $({\rm B}_{\pm})$.
Recall also that
$|\bfs|$ is even in Case $({\rm B}_+)$ and odd in Case $({\rm B}_-)$.
Therefore, the phase
$\epsilon_{|\bfs|}$ is a sign.

The formula (\ref{relPwtP}) exhibits the spacetime supersymmetry.
For concreteness let us consider the case $D=4$ and $n=3$, i.e.,
compactification on a Calabi-Yau three-fold $M$ down to four dimensions.
We consider a brane $\cB$ such that chiral ring elements flow
to chiral primary fields with
the actual R-charge very close to the canonical R-charge.
The states corresponding to massless fermions are of the form 
$\phiR\otimes |u,k\rangle_{{}_{\rm R}}$ where 
$\phiR$ is a Ramond ground state for the sigma model on $M$
and $|u,k\rangle_{{}_{\rm R}}$ is a
$4d$ massless spinor state.
Note that $\wtPo$ does 
$\psi_0^{\mu}\to \psi_0^{\mu}$ for the $\R^4$ directions $\mu$
in the $(-+)$ spin structure. 
Since all states $|u,k\rangle_{{}_{\rm R}}$ are related by multiplication 
of such $\psi_0^{\mu}$'s, we may choose to define $\wtPo$ to 
act trivially on $|u,k\rangle_{{}_{\rm R}}$.
For this choice, the phase $\kappa$ is a sign, and also, 
the orientifold projection condition 
takes the form 
$$
\wtPo(\phiR)=\phiR.
$$
On the other hand, massless vectors correspond, in the 0-picture, to
states of the form
$\phi_0\otimes (\zeta\cdot\alpha_{-1}+\cdots)|k\rangle_{{}_{\rm NS}}$
where $\phi_0$ is the identity times the Chan-Paton factor,
and nearly massless scalars correspond to the states of the form
$\phi_1\otimes k\cdot\psi_{-{1\over 2}}|k\rangle_{{}_{\rm NS}}$
where $\phi_1$ is a chiral primary state of the R-charge close to
$|\phi_1|=1$.
Recall that the parity $\Po$ acts on the states
$(\zeta\cdot\alpha_{-1}+\cdots)|k\rangle_{{}_{\rm NS}}$ and
$k\cdot\psi_{-{1\over 2}}|k\rangle_{{}_{\rm NS}}$ by multiplication
by $(-1)$ and $(-i)$ respectively in the $(++)$ spin structure.
Thus, orientifold projection condition on these light bosons takes the form
$$
\Po(\phi_0)=-\phi_0\quad\mbox{and}\quad \Po(\phi_1)=i\phi_1.
$$
Inserting these to (\ref{relPwtP}), we find
$\wtPo(\spf(\phi_0))=-\kappa\epsilon_{|\bfs|}\spf(\phi_0)$
and
$\wtPo(\spf(\phi_1))=-\kappa\epsilon_{|\bfs|}\spf(\phi_1)$.
When $\kappa\epsilon_{|\bfs|}=-1$,
these are nothing but the orientifold projection condition for
the corresponding massless fermions. 
For this choice of $\kappa$, the unprojected degrees of freedom form
the vector and chiral multiplets of a possibly broken
${\mathcal N}=1$ supersymmetry.
For the other choice, $\kappa\epsilon_{|\bfs|}=1$,
the orientifold projection conditions
 for the bosons and fermions do not match
--- the branes and the orientifold preserve completely different
supersymmetries.
Of course, in that case, we may replace the brane $\cB$ by its antibrane, say
$\cB[1]$. If we do so, the quasi-o-isomorphism has a shifted
degree, $|\bfs|-2$, as we have discussed in Section~\ref{subsub:Odeg}. 
The shift $|\bfs|\to |\bfs|-2$ flips the sign of
$\epsilon_{|\bfs|}=\phS i^{-|\bfs|}$, and indeed we recover
the condition $\kappa\epsilon_{|\bfs|-2}=-1$.

This discussion shows the significance of
the division of branes into two classes by the mod 4 degree of 
quasi-o-isomorphisms (Section~\ref{subsub:Odeg}).
For a given $\kappa$,  branes from only one class is compatible with
the spacetime supersymmetry that is preserved by the orientifold
with orientation $\kappa$. 
I.e., there is a one to one correspondence between bosons and fermions.
Note that the supersymmetry may be broken ---
scalars may be massive or tachyonic while fermions are always massless.
On the other hand, the other class of branes have no chance;
there is not even a correspondence between bosons and fermions.
If we replace $\kappa$ by $-\kappa$, the r\^oles of the two classes
are exchanged.

\subsection{A Categorical Description}
\label{subsec:category}

\newcommand{\mO}{\mathfrak{O}}
\newcommand{\shift}{{\tt T}}
\newcommand{\DM}{\mathfrak{D}(M)}
\newcommand{\Pf}{{\tt P}}
\newcommand{\idf}{{\tt Id}}

D-branes in a theory of oriented strings
form a category $\cC$; objects are D-brane data,
morphisms for a pair of objects are open string states, 
and the composition of morphisms represents
gluing of two open strings into one.
The categorical description may also be extended to D-branes in orientifolds
\cite{HW}.
In the most basic form, 
it goes as follows.

An orientifold transform of D-branes can be represented as
a contravariant functor $\Pf:\mathcal{C}\to \mathcal{C}$.
We require that the square $\Pf^2=\Pf\circ\Pf$ 
is isomorphic to the identity functor of $\mathcal{C}$, 
and we choose such an isomorphism
\beq
\mathfrak{c}:\Pf^2~\stackrel{\cong}{\longrightarrow}~
\idf_{\cC}^{}.
\label{cisom}
\eeq
That is, to each object $X\in\cC$ is assigned an isomorphism
$\mathfrak{c}_X\in \Hom^{}_{\mathcal{C}}(\Pf^2(X),X)$
in such a way that the following diagram commutes
for each morphism $f\in\Hom^{}_{\mathcal{C}}(X,Y)$,
\beq\begin{CD}
\Pf^2(X) @>\mathfrak{c}_X>> X\\
@V\Pf^2(f)VV  @VVfV\\
\Pf^2(Y) @>\mathfrak{c}_Y>> Y
\end{CD}
\label{commd}
\eeq
We further require that it satisfies
\beq
\mathfrak{c}_{\Pf(X)}\cdot\Pf(\mathfrak{c}_X)
={\rm id}_{\Pf(X)}.
\label{condforcis}
\eeq
Then, we may define the category of D-branes in the orientifold,
$\mO=\mO(\cC;\Pf,\mathfrak{c})$, as follows.
An object is a brane with an o-isomorphism, i.e., a pair $(X,s)$ of
an object $X\in\mathcal{C}$ and an isomorphism
$s\in \Hom^{}_{\mathcal{C}}(\Pf(X),X)$ satisfying
the condition
\beq
s=\mathfrak{c}_X\cdot \Pf(s).
\label{condforsgen}
\eeq
The space of morphisms $\Hom^{}_\mO((X,s),(Y,t))$ is
the space $\Hom^{}_{\mathcal{C}}(X,Y)$. It is equipped with the operator
$P:\Hom^{}_{\mathcal{C}}(X,Y)\to \Hom^{}_{\mathcal{C}}(Y,X)$
defined by $P(f)=s\cdot \Pf(f)\cdot t^{-1}$.
This operator is an involution, $P^2={\rm id}$, thanks to
the condition (\ref{condforsgen})
and the commutativity of the diagram (\ref{commd}).
An isomorphism from $(X,s)$ to $(Y,t)$ is an isomorphism
$f\in \Hom_{\cC}(X,Y)$ such that
\beq
f\cdot s \cdot \Pf(f)=t.
\label{isomgen}
\eeq

Some remarks are in order.\\
(i)~ 
If the category is graded, i.e., if the space of morphisms
has a $\Z_2$ or $\Z$ grading, it is natural to assume that
$\Pf$ is a contravariant functor in the graded sense, 
i.e., there is a sign on the product of morphisms,
$\Pf(f\cdot g)=(-1)^{|f||g|}\Pf(g)\cdot \Pf(f)$.
Furthermore, we may need a sign in the definition of the parity operators as
$P(f)=s\cdot \Pf(f)\cdot t^{-1}(-1)^{|f||t|}$, if the
isomorphism $t$ has a non-zero degree.
\\
(ii)~
The property (\ref{condforcis}) is required for existence of
isomorphisms obeying (\ref{condforsgen}) --- use (\ref{condforsgen})
in itself.
\\
(iii)~
The structure is partly motivated by the results of the present paper
and in fact can be used to summarize them as discussed below.
For example, the isomorphism condition (\ref{isomgen})
is motivated by (\ref{isomDO}).
\\
(iv)~
A pair $(\Pf,\mathfrak{c})$ satisfying (\ref{cisom})
and (\ref{condforcis}) is known as a
{\it duality} of the category ${\mathcal C}$.
See, for example, \cite{QSS}.


\newcommand{\Ptrr}{\Ptr_{\!\!{}_{r_o}}}

As far as the chiral ring sector is concerned,
the structure found in this section can be summarized in this language, 
although the presence of worldsheet spinors requires
a minor modification.
We take the category $\cC=\DM$, which has 
D-brane data $\cB=(E,\cQ)$ as objects, 
cohomology classes in $\mathscr{H}(\cB_1,\cB_2)$ as morphisms between objects,
and the product (\ref{Hprod}) as the composition of morphisms.
The parity functor is the transform $\Ptr$ given by 
(\ref{defPfunct1}) and (\ref{defPfunctgr}). 
It depends on the choice of the phase $\spst$ 
(i.e. of the boundary spin structure), $\Ptr=\Ptr_{(\spst)}$.
Also, its square $\mathscr{P}^2$ is not isomorphic to the identity
but to the functor $(-1)^F$ which acts as the identity
on objects but as the $\Z_2$-grading on morphisms.
Indeed, given a crosscap section $\gc$, the assignment
$$
\cB=(E,\cQ)\,\longmapsto\,\mathbb{c}_{\cB}
=\sigma^{}_E\imath_E^{-1}\otimes \gc
$$
provides an isomorphism
$$
\mathbb{c}:\mathscr{P}^2~\stackrel{\cong}{\longrightarrow}~
(-1)^F.
$$
That is, the diagram (\ref{commd}) commutes if 
the vertical arrow $f$ on the right is replaced by $(-1)^{|f|}f$.
This was indeed seen in (\ref{Ptrsq}).
The condition for o-isomorphisms (\ref{condforsgen})
appears in (\ref{condfors}).

If we restrict our attention to D-branes compatible with the spacetime
supersymmetry preserved by the orientifold, 
we may assume that all o-isomorphisms have the same
degree. Alternatively, we may take a shifted parity functor
\beq
\Ptr_{r_o}=\shift^{r_o}\circ \Ptr,
\eeq
and restrict our attention to branes with o-isomorphisms of degree zero.
$\shift^j$ is the shift functor that acts as
$\cB\mapsto \cB[j]$ on objects and  as $\phi\mapsto \phi[j]$ on morphisms.
The relevant isomorphism $\Ptr_{r_o}^2\cong (-1)^F$ 
is provided by
\beq
\mathbb{c}_{r_o,(E,{\cQ})}
=(-1)^{r_o}\sigma_E^{r_o+1}\imath_E^{-1}\otimes\gc.
\eeq
See (\ref{condforss}).
We shall denote the category of such D-branes by
$\mathbb{O}(\DM,\Ptr_{r_o},\mathbb{c}_{r_o})$.
Translating the discussion in Section~\ref{subsub:Odeg},
given an o-isomorphism $\bfs:\Ptr_{r_o}(\cB)\to \cB$ of degree zero,
we have an o-isomorphism
$\bfs[1]\circ\sigma^T:\Ptr_{r_o+2}(\cB[1])\to\cB[1]$ again of degree zero.
This defines an equivalence
\beq
\mathbb{O}(\DM,\Ptr_{r_o},\mathbb{c}_{r_o})
~\cong ~ 
\mathbb{O}(\DM,\Ptr_{r_o+2},\mathbb{c}_{r_o+2})
\label{2perII}
\eeq
between the categories of D-branes compatible with the opposite
orientation of the orientifold.


\subsubsection*{\it Digression:  General Type II Orientifolds}

We make a brief digression to discuss whether the categorical description 
is possible for the entire sector of more general 
Type II orientifolds, as those considered in Section~\ref{sec:main}.
For this we need to define the product of open string states, as it
is used in the condition like (\ref{condforsgen}) as well as 
in the definition of the parity operator
where the isomorphism $\mathfrak{c}_{X}$ 
and o-isomorphisms are regarded as open string states.
One possibility is to use the $*$-product in open string field theory, but
this may not be useful in the current status where
even the identity element is represented by a complicated state.
As an alternative, we take the product of boundary vertex operators, which
has to be taken rather informally as it
depends on the insertion points of the operators.
A parity functor $\Ptr$ is given by (\ref{pagen}) and (\ref{parmap}).
For a brane $\cB=(E,A,T)$,
an isomorphism $\Ptr^2(\cB)\to \cB$ is provided by the
``vertex operator'' $\sigma_E\imath_E^{-1}\otimes \gc$.
It appears in the following form in the path-integral weight:
$$
W_{\cB}(\tau_f,\tau)\circ
(\sigma_E\imath_E^{-1}\otimes\gc)\bigl|_{\tau}\circ
W_{\Ptr^2(\cB)}(\tau,\tau_i),
$$
where $W_{\cB}(\tau_f,\tau_i)$ is the boundary interaction 
(\ref{WilsonE}) for the brane $\cB$. 
It is independent of the insertion point $\tau$
by the fact that $\gc$ is flat with respect to
$\inv^*\sA-\sA$.
Likewise, an o-isomorphism $\gU:\Ptr(\cB)\to\cB$ may be regarded as
a ``vertex operator'' which appears as
$$
W_{\cB}(\tau_f,\tau)\circ \gU\bigl|_{\tau}\circ 
W_{\Ptr(\cB)}(\tau,\tau_i).
$$
Again, it is independent of the insertion point $\tau$ 
thanks to the equations (\ref{Tco})-(\ref{Aco}).
By the presence of Ramond sector states, 
we need to consider both of the two boundary spin structures.
This is unlike in the discussion of the chiral ring, 
which is a part of the Neveu-Schwarz sector, where
we were able to work with a fixed one.
We have functors that flip the boundary spin 
structure, $(-1)^{F_R}$ and $(-1)^{F_L}$, which we discussed
in Section~\ref{subsec:Ramond}.
They obey the relations of the form
$(-1)^{F_L}\circ(-1)^{F_R}=(-1)^{F_L}\circ(-1)^{F_R}=(-1)^F$ and
$(-1)^{F_R}\circ \Ptr\cong \Ptr\circ (-1)^{F_L}$. 
The hom space has two $\Z_2$-gradings. One is the usual
worldsheet $\Z_2$-grading, and the other is of spacetime nature
--- Neveu-Schwarz states are even and Ramond states are odd.
Accordingly, we have
$$
\Ptr(\Phi\cdot \Psi)=(-1)^{|\Phi||\Psi|
+|\Phi|_{\rm st}|\Psi|_{\rm st}}
\Ptr(\Psi)\cdot \Ptr(\Phi).
$$
The spacetime sign factor has appeared in (\ref{extram}) for example.

\subsubsection*{\it Topological B-Model}

As we discussed earlier,
the topological model has no worldsheet spinors and hence there is no need of
choosing spin structure nor summing over the choices (i.e. no GSO projection).
In particular, there is a unique boundary condition for a D-brane
and a unique parity transformation for an orientifold
--- there is no $(+)$ versus $(-)$ nor
$\Omega$ versus $(-1)^{F_R}\Omega$, etc.
Vertex operators are always of ``NS'' type and the states
are always of ``Ramond'' type, and the parity must square to
the identity, not to $(-1)^F$.
This indicates that the categorical description in the basic form
applies without modification.

Let us first determine the parity transform of the worldsheet fields.
The guiding principle is to preserve the scalar supersymmetry (\ref{BRST}).
We notice that $\eta^{\bi}$ is the partner of $x^{\bi}$
and $\dd x^i$ is the partner of $\rho^i$.
This implies that the parity transform $\Omega$ which preserves
the symmetry (\ref{BRST}) is
\beq
x\to x\circ\Omega,\quad
\eta^{\bi}\to\eta^{\bi}\circ\Omega,\quad
\theta_i\to -\theta_i\circ \Omega,\quad
\rho^i\to\Omega^*\rho^i.
\eeq
The boundary interaction includes the terms
$$
i{\mathcal A}_{\tau}=\cdots
-{1\over 2}\rho_{\tau}^i\partial_i\cQ
+{1\over 2}\eta^{\bi}\partial_{\bi}\cQ^{\dag}
+{1\over 2}\{\cQ,\cQ^{\dag}\},
$$
where
$\cQ=\cQ(x,\eta)$ and $\cQ^{\dag}=\cQ^{\dag}(x,\rho_{\tau})$.
Under the parity $\Omega$, which reverses the orientation of the boundary,
composed with the transpose,
the interaction transforms as
$$
i{\mathcal A}_{\tau}~\longmapsto~\left(\cdots
-{1\over 2}(-\rho^i_{\tau})\partial_i\cQ
+{1\over 2}\eta^{\bi}\partial_{\bi}\cQ^{\dag}|_{\rho_{\tau}\to-\rho_{\tau}}
+{1\over 2}\{\cQ,\cQ^{\dag}\}|_{\rho_{\tau}\to-\rho_{\tau}}\right)^T\circ\Omega.
$$
We see that the effect of the parity transform is
\beq
\cQ\longmapsto -\cQ^T,
\qquad
\cQ^{\dag}\longmapsto (\cQ^{\dag})^T|_{\rho_{\tau}\to -\rho_{\tau}}.
\eeq
This is consistent with the
hermiticity relation in the theory before the topological twist
between the components of
$\cQ$ and $\cQ^{\dag}$,
since $(f^{\dag})^T=(-1)^{|f|}(f^T)^{\dag}$.

Combined with the involution $\inv:M\to M$ and the twist by
$({\mathcal L},\sA)$, we find a parity functor $\Pf_0:\DM\to \DM$,
\beq
\begin{array}{rcl}
E&\longmapsto&\Pf_0(E)=\inv^*E^*\otimes {\mathcal L},\\[0.2cm]
\cQ&\longmapsto&\Pf_0(\cQ)=-\inv^*\cQ^T+i\eta^{\bi}\sA_{\bi},\\[0.2cm]
\phi\in \mathscr{C}(\cB_1,\cB_2)&\longmapsto&
\Pf_0(\phi)=\inv^*\phi^T\in\mathscr{C}(\Pf_0(\cB_2),\Pf_0(\cB_1)). 
\end{array}
\label{defPtop}
\eeq
We may also consider the shifted versions, $\Pf_{r_o}=\shift^{r_o}\circ \Pf_0$.
The square $\Pf_{r_o}^2$ is isomorphic to
the identity functor by
\beq
\mathfrak{c}_{r_o\,(E,\cQ)}
= \sigma_E^{r_o}\imath_E^{-1}\otimes c_{{}_{\rm top}}
\label{defctop}
\eeq
where $c_{{}_{\rm top}}$ is a holomorphic section of 
$\inv^*{\mathcal L}\otimes {\mathcal L}^*$
with the property $c_{{}_{\rm top}}\cdot\inv^*c_{{}_{\rm top}}=1$
for (\ref{condforcis}).
This gives us an orientifold category consisting of
pairs $(\cB,s)$ where $s$ is of degree zero and obeys 
$
s\simeq \mathfrak{c}_{r_o\,\cB}\cdot
\Pf_{r_o}(s).
$
The resulting parity operator, $P_{{}_{\rm top}}$, squares to the identity,
$P_{{}_{{}^{\rm top}}}^2={\rm id}$.
The shift functor $\shift$ yields an equivalence of categories
\beq
\mO(\DM,\Pf_{r_o},\mathfrak{c}_{r_o})~\cong ~
\mO(\DM,\Pf_{r_o+2},-\mathfrak{c}_{r_o+2}),
\label{4pertop}
\eeq
if we take the common $c_{{}_{\rm top}}$ for both
$\mathfrak{c}_{r_o}$ and $\mathfrak{c}_{r_o+2}$.
Note the appearance of a minus sign for the isomorphism 
$\mathfrak{c}$, in contrast with (\ref{2perII}).
That means that the section $c_{{}_{\rm top}}$ 
does not simply reflect the type of the O-planes in the corresponding 
Type II model. We now see a more explicit relation.

\subsubsection*{\it Relation To Type II}

Let us see how the information of Type II orinetifold can be recovered from
the D-brane category for topological orientifold.
To this end, let us compare the parity transform of 
the fields at the boundary:
\beqa
\mbox{Type II}:&&\psi^\bi\to\spst\psi^\bi\circ\Omega,
\quad\psi^i\to\spst\psi^i\circ\Omega,
\nn\\
\mbox{topological}:&&\,\eta^{\bi}\to\eta^{\bi}\circ\Omega,\quad~~
\rho_{\tau}^i\to-\rho_{\tau}^i\circ \Omega.
\nn
\eeqa
We see that the two are related by the R-symmetry transform,
$\psi^{\bi}\to \spst^{-1}\psi^{\bi}$ and
$\psi^i\to\spst\psi^i$. This prompts us to look into the representation 
$R_{\cB}(\lambda)$ of the R-symmetry on the Chan-Paton bundle.
Note that $R_{\shift^{-r_o}(\cB)}(\lambda)=\lambda^{-r_o}R_{\cB}(\lambda)$ and
$R_{\Ptr(\cB)}(\lambda)=R_{\Pf_0(\cB)}(\lambda)=R_{\cB}(\lambda)^{-T}$.
Comparison of (\ref{defPfunct1}) and (\ref{defPfunct2}) on the one hand
and (\ref{defPtop}) on the other shows the relation
\beqa
&&R_{\Ptr(\cB)}(\spst)\Ptr(\cQ)R_{\Ptr(\cB)}(\spst)^{-1}
=\Pf_0(\cQ),\nn\\
&&R_{\Ptr(\cB_1)}(\spst)\Ptr(\phi)R_{\Ptr(\cB_2)}(\spst)^{-1}
=\spst^{|\phi|}\Pf_0(\phi).
\nn
\eeqa
The same holds for $\Ptr_{r_o}$ versus $\Pf_{r_o}$.
The o-isomorphisms
in the Type II and topological theories can also be related.
Let $\bfs$ and $s$ be o-isomorphisms of
an object $\cB$ in the two categories based on
$(\Ptr_{r_o},\mathbb{c}_{r_o})$
and $(\Pf_{r_o},\mathfrak{c}_{r_o})$.
If we put
\beq
\bfs\propto s\,R_{\Ptr_{r_o}(\cB)}(\spst),
\label{sstop}
\eeq
then $\bfs$ and $s$ obey the condition 
(\ref{condforsgen}) in the respective categories at the same time.
(\ref{sstop}) is consistent with the relation
(\ref{relsism}) between $\bfs_{(i)}$ and $\bfs_{(-i)}$
since $R_{\Ptr_{r_o}(\cB)}(-1)\propto\sigma_{\cB}^T$.
The resulting parity operators in the two theories are simply related by
\beq
\Po(\phi)=\spst^{|\phi|}P_{{}_{\rm top}}(\phi).
\label{PPtoprel}
\eeq
Combining this with (\ref{relPwtP}), or by a direct comparison,
we also find the relation to the parity operator
on the Ramond ground states,
\beq
\wtPo=\kappa\epsilon_{-r_o} \spf\circ
P_{{}_{\rm top}}\circ \spf^{-1}.
\eeq
We also find from (\ref{sstop}) the precise relation between
the crosscap section $\gc$ in the Type II
theory and the topological counterpart $c_{{}_{\rm top}}$,
\beq
\gc=\spst^{-r_o}c_{{}_{\rm top}}.
\label{cvsctop}
\eeq
This relation reproduces the fact  (\ref{4pertop})
 that $c_{{}_{\rm top}}$ changes by a sign under the shift 
$r_o\to r_o+2$ (since $\gc$ is invariant).

\subsubsection*{\it Triangles}

We shall make a comment on the category $\cC=\DM$ and the parity functor
$\Pf_0$ defined in (\ref{defPtop}) or its shifts $\Pf_j=\shift^j\Pf$.
Here and in what follows, we take only degree zero morphisms
unless otherwise stated, i.e.,
we redefine the morphism space as $\Hom_{\cC}(X,Y)=\mathscr{H}^0(X,Y)$.
For a morphism $u\in \Hom^{}_{\cC}(X,Y)$, we have a sequence of objects 
and morphisms,
\beq
X\,\,\stackrel{u}{\longrightarrow}\,\,
Y\,\,\stackrel{{0\choose 1}}{\longrightarrow}\,\,
\cone(u)\,\,\stackrel{(1,0)}{\longrightarrow}\,\,
\shift(X),
\label{triangle1}
\eeq
A {\it triangle} is
a sequence of the form
$A\stackrel{a}{\to} B\stackrel{b}{\to} C\stackrel{c}{\to}\shift(A)$
which is isomorphic to the above for some $u:X\to Y$.
The functor $\shift$ and the set of triangles 
obey a certain set of axioms and make the category $\cC$
a {\it triangulated category}. We refer the reader to \cite{RD} for details.
One of the axioms is the rotation axiom: if 
$X\stackrel{u}{\to}Y\stackrel{v}{\to}Z\stackrel{w}{\to}\shift(X)$
is a triangle, then 
$Y\stackrel{v}{\to}Z\stackrel{w}{\to}\shift(X)
\stackrel{-\shift(u)}{\longrightarrow}
\shift(Y)$
is also a triangle. It is very important to be careful about the minus sign
on the last arrow.
For example, under the same assumption,
$\shift(X)\stackrel{\shift(u)}{\to}\shift(Y)\stackrel{\shift(v)}{\to}
\shift(Z)\stackrel{\shift(w)}{\to}\shift^2(X)$
is {\it not} always a triangle but is so when a minus sign is placed
at each arrow, or alternatively, a minus sign at one of the three arrows
---  we can flip the sign of two arrows by a 
change of basis (which is an isomorphism).

Given a triangle
$X\stackrel{u}{\to}Y\stackrel{v}{\to}Z\stackrel{w}{\to}\shift(X)$,
its image under the parity $\Pf=\Pf_j$, more precisely,
\beq
\Pf\shift(X)\,\,\stackrel{\Pf(w)}{\longrightarrow}\,\,
\Pf(Z)\,\,\stackrel{\Pf(v)}{\longrightarrow}\,\,
\Pf(Y)\,\,\stackrel{\sigma_{\Pf(X)}\circ\Pf(u)}{\longrightarrow}\,\,
\shift\Pf\shift(X)
\eeq
is also a triangle, for any $j\in\Z$. 
Note that $\sigma_{\Pf(X)}$ defines an isomorphism
$\Pf(X)\cong \shift\Pf\shift(X)$.
We shall call such a category with duality $(\cC,\Pf,\mathfrak{c})$
a {\it triangulated category with duality}.
Here $\mathfrak{c}=\mathfrak{c}_j$ for $\Pf=\Pf_j$
defined in (\ref{defctop}).


The category $\DM$ is equivalent as a triangulated category
to the full subcategory
of the bounded derived category of sheaves of 
${\mathcal O}_M$ modules consisting of complexes with coherent cohomology 
sheaves \cite{BonRos,Block}.
If $M$ is algebraic, that is equivalent to the derived category of coherent 
sheaves on $M$. 
For the case $\inv={\rm id}_M^{}$, the parity functor
$\Pf_0$ is essentially the same as the duality functor 
$\R{\it hom}(-,{\mathcal L})$.

\subsection{Some Binding/Decay Channels}

Let us discuss possible channels of bound state formation or decay.
We put $(\cC,\Pf,\mathfrak{c})=(\DM,\Pf_j,\mathfrak{c}_j)$ for some $j$.

\subsubsection*{\it Invariant Cones}

The superposition of a brane and its orientifold image gives rise to
an invariant brane. Indeed,
for any object $L\in \cC$, the direct sum $H_L=\Pf(L)\oplus L$
has an o-isomorphism  $s_L:\Pf^2(L)\oplus \Pf(L)\to \Pf(L)\oplus L$,
\beq
s_L=\left(\begin{array}{cc}
0&{\rm id}_{\Pf(L)}\\
\mathfrak{c}_L&0
\end{array}\right).
\label{sL}
\eeq
Note that it satisfies the condition (\ref{condforsgen}),
$$
\mathfrak{c}_{H_L}\Pf(s_L)=\left(\begin{array}{cc}
\mathfrak{c}_{\Pf(L)}&0\\
0&\mathfrak{c}_L
\end{array}\right)
\left(\begin{array}{cc}
0&\Pf(c_L)\\
\Pf({\rm id}_{\Pf(L)})&0
\end{array}\right)
=\left(\begin{array}{cc}
0&\mathfrak{c}_{\Pf(L)}\Pf(\mathfrak{c}_L)\\
\mathfrak{c}_L{\rm id}_{\Pf^2(L)}&0
\end{array}\right)
=s_L,
$$
where we used (\ref{condforcis}).
An invariant object of this form is called {\it hyperbolic}.
This construction works if
$(\cC,\Pf,\mathfrak{c})$ is a general category with duality.

When the brane and its orientifold image are bound together
by an open string state, does it form an invariant brane?
More specifically, does the cone of a morphism $u:\shift^{-1}\Pf(L)\to L$ 
admits an o-isomorphism?
As the simplest candidate, let us see if $s_L$ in (\ref{sL})
can serve as an o-isomorphism of the cone $C=\cone(u)$.
It is enough to check whether it is ${\bf Q}$-closed.
Note that
$$
\cQ_C=\left(\begin{array}{cc}
\cQ_{\Pf(L)}&0\\
u&\cQ_L
\end{array}\right),\qquad
\cQ_{\Pf(C)}=\left(\begin{array}{cc}
\cQ_{\Pf^2(L)}&-\Pf(u)\\
0&\cQ_{\Pf(L)}
\end{array}\right)
$$
where $u$ here is regarded as a degree $1$ map $\Pf(L)\to L$.
The condition for ${\bf Q}s_L=0$ is
\beq
u+\mathfrak{c}_L\cdot\Pf(u)=0.
\label{iccond}
\eeq
Under this condition, $(C,s_L)$ is an invariant object.
We shall call it an {\it invariant cone} 
of $u\in\mathscr{Z}^1(\shift^{-1}\Pf(L),L)$ 
satisfying (\ref{iccond}),
and denote it by $\cone(u,L)$.
When the actual R-charge of $u$ is smaller than $1$,
$L$ and $\Pf(L)$ are bound together to form 
the invariant cone $\cone(u,L)$.
When the charge is greater than $1$, the invariant cone
splits to $L$ and $\Pf(L)$.

The construction can be extended to 
a general triangulated category with duality.
For this purpose, we first rewrite (\ref{iccond}) 
as a condition for $u$ regarded as a degree zero morphism 
$u:\shift^{-1}\Pf(L)\to L$. It reads
\beq
u=\mathfrak{c}_L\cdot\sigma_{\Pf^2(L)}\cdot\shift^{-1}\Pf(u),
\label{iccond2}
\eeq
where $\sigma_{\Pf^2(L)}$ is regarded as an isomorphism 
$\shift^{-1}\Pf\shift^{-1}\Pf(L)\to\Pf^2(L)$.
The morphism $u$ extends to a triangle
$\shift^{-1}\Pf(L)\stackrel{u}{\to}L\stackrel{v}{\to}C\stackrel{w}{\to}\Pf(L)$.
Applying $\Pf$ to it and rotating once, we have another triangle
which appears as the top line in the diagram below.
It is important that no sign is needed on the arrows.
\beq\begin{CD}
\shift^{-1}\Pf(L) @>\sigma_{\Pf^2(L)}\shift^{-1}\Pf(u)>> 
\Pf^2(L) @>\Pf(w)>> \Pf(C) @>\Pf(v)>> \Pf(L)\\
@V{\rm id}VV  @V\mathfrak{c}_L VV @V\exists\varphi VV @VV {\rm id}V\\
\shift^{-1}\Pf(L) @>u>>
L @>v>>
C @>w>>
\Pf(L)
\end{CD}
\label{ocone}
\eeq
The left square commutes because of (\ref{iccond2}). 
Then, by one of 
the axioms of triangulated category, there exists a morphism,
denoted $\varphi$ in the diagram,
such that the remaining two squares also commute.
Applying the functor $\Pf$ to the two squares, and using
the fact that $\mathfrak{c}$ is an isomorphism of $\Pf^2$ to $\idf_{\cC}^{}$
obeying (\ref{condforcis}),
we find that the diagram still commutes even if $\varphi$ is replaced by
$\mathfrak{c}_C\Pf(\varphi)$, and hence also by
the average $\varphi'={1\over 2}(\varphi+\mathfrak{c}_C\Pf(\varphi))$.
Note that $\mathfrak{c}_C\Pf(\mathfrak{c}_C\Pf(\varphi))
=\mathfrak{c}_C\Pf^2(\varphi)\Pf(\mathfrak{c}_C)
=\varphi\mathfrak{c}_{\Pf(C)}\Pf(\mathfrak{c}_C)=\varphi$.
This means $\varphi'=\mathfrak{c}_C\Pf(\varphi')$.
Therefore, we may assume from the beginning that $\varphi$ obeys 
this equation, replacing it by the average $\varphi'$ if necessary.
The fact that the identities and $\mathfrak{c}_L$ are isomorphisms means that
$\varphi$ is also an isomorphism. That is, $\varphi:\Pf(C)\to C$
is an o-isomorphism! One can also show that $(C,\varphi)$ is unique up to 
an isomorphism in $\mO(\cC,\Pf,\mathfrak{c})$.
We may also call it the invariant cone of
$u:\shift^{-1}\Pf(L)\to L$ obeying (\ref{iccond2}) and denote it by
$\cone(u,L)$.
This procedure is taken from P. Balmer's work
\cite{Balmer} in which it is called the ``symmetric cone construction''.
An earlier source is M. Knebusch's work \cite{Knebusch} 
on the category $\mathfrak{Bil}(M)$ of 
holomorphic vector bundles with symmetric bilinear forms over $M$. 
See the review \cite{Balmerintro} and a survey in \cite{handbookK} by Balmer.
``Invariant objects'' here are called ``bilinear space'' by Knebusch
and ``symmetric space'' by Balmer.

An invariant object of this type is called {\it metabolic}.
The object $L$ is called the {\it Lagrangian} of the invariant
cone $\cone(u,L)$ because it fits into a triangle that includes
$$
L\,\,\stackrel{v}{\longrightarrow}\,\,
C\,\,\stackrel{\Pf(v)\circ\varphi^{-1}}{\longrightarrow}\,\,
\Pf(L).
$$
Compare this with the exact sequence (\ref{Lagrangian0})
that defines Lagrangian subbundle of a twisted Real bundle.
In the present language, the assertions (i) and (ii) in
Section~\ref{subsub:hyperbolic} are stated as (i) any metabolic object
is hyperbolic in the category of topological twisted Real bundles,
 and (ii) any object is metabolic 
(and hence hyperbolic by (i)) in the category of topological 
twisted Real Hilbert bundles.
In general, and in particular for
$(\cC,\Pf,\mathfrak{c})=(\DM,\Pf_j,\mathfrak{c}_j)$, neither is true:
there are metabolic but non-hyperbolic objects
and there are non-metabolic objects.
See the survey by Balmer in \cite{handbookK} and a reference therein
for examples on an elliptic curve.

\subsubsection*{\it Binding Invariant Objects}

For two invariant branes,
$(X,s)$ and $(Y,t)$, their direct sum $(X\oplus Y,s\oplus t)$
is obviously an invariant brane.
Let us see if they can be bound together
by the cone construction.
For $f:\shift^{-1}X\to Y$, its cone $Z=\cone(f)$ and its
orientifold image have $\cQ$-profiles
$$
\cQ_Z=\left(\begin{array}{cc}
\cQ_X&0\\
f&\cQ_Y
\end{array}\right),\qquad
\cQ_{\Pf(Z)}=\left(\begin{array}{cc}
\cQ_{\Pf(X)}&-\Pf(f)\\
0&\cQ_{\Pf(Y)}
\end{array}\right).
$$
For a trial o-isomorphism
$
\varphi=\left(\begin{array}{cc}
s+{\mit\Delta} s&a\\
b&t+{\mit\Delta} t
\end{array}\right),
$
the condition for ${\bf Q}\varphi=0$ is
\beqa
&{\bf Q}{\mit\Delta} s=0,\quad
(s+{\mit\Delta} s)\Pf(f)+i{\bf Q}a=0,\nn\\
&f(s+{\mit\Delta} s)+i{\bf Q}b=0,\quad
fu+v\Pf(f)+i{\bf Q}{\mit\Delta} t=0,
\nn
\eeqa
and the condition (\ref{condforsgen}) is 
${\mit\Delta} s=\mathfrak{c}_X\Pf({\mit\Delta} s)$,
${\mit\Delta} t=\mathfrak{c}_Y\Pf({\mit\Delta} t)$,
$a=\mathfrak{c}_X\Pf(b)$ and 
$b=\mathfrak{c}_Y\Pf(a)$.
If we assume ${\mit\Delta} s=0$, or if $s+{\mit\Delta} s$ is still a 
quasi-isomorphism, we must conclude that $f$ is ${\bf Q}$-exact.
It then follows that the resulting object
is isomorphic to the direct sum.
Indeed, if $f={\bf Q}g$ (and ${\mit\Delta} s=0$ for simplicity), 
we may take $a=-s\Pf(g)$, $b=-gs$ and ${\mit\Delta} t=gs\Pf(g)$
as the solution, for which $(Z,\varphi)$ is isomorphic to 
$(X\oplus Y,s\oplus t)$ by
$$
\left(\begin{array}{cc}
s&-s\Pf(g)\\
-gs&t+gs\Pf(g)
\end{array}\right)
=\left(\begin{array}{cc}
{\rm id}_X&0\\
-g&{\rm id}_Y
\end{array}\right)
\left(\begin{array}{cc}
s&0\\
0&t
\end{array}\right)
\left(\begin{array}{cc}
{\rm id}_{\Pf(X)}&-\Pf(g)\\
0&{\rm id}_{\Pf(Y)}
\end{array}\right).
$$
The only way out would be to find ${\mit\Delta}s$ such that
$s+{\mit\Delta}s$ is no longer a quasi-isomorphism and that $\varphi$
is a quasi-isomorphism. However, there is no general way to find such
a ${\mit\Delta}s$. 
Thus, the cone construction does not lead to anything new in general.

This of course does not mean that it is impossible to bind 
two invariant branes together.
A rather trivial example is to bind two hyperbolic
objects $(H_{L_1},s_{L_1})$ and $(H_{L_2},s_{L_2})$
into another hyperbolic object $(H_L,s_L)$
where $L$ is the cone of a morphism $f:\shift^{-1}L_1\to L_2$. 
Note that $H_L$ is not of the form of a cone since
the binding arrows go in both ways --- $f$ goes from $H_{L_1}$ to $H_{L_2}$
while $-\Pf(f)$ goes oppositely.
A slightly less trivial example is obtained by replacing 
the hyperbolic objects by invariant cones in this construction.
A yet another example is to bind an invariant object $(X,s)$
and a hyperbolic object $(H_L,s_L)$. 
Let us try the following $\cQ$-profile and o-isomorphism
$$
\cQ=\left(\begin{array}{ccc}
\cQ_X&w&\\
&\cQ_{\Pf(L)}&\\
v&u&\cQ_L
\end{array}\right),\qquad
\varphi=\left(\begin{array}{ccc}
s&&\\
&&{\rm id}_{\Pf(L)}\\
&\mathfrak{c}_L&
\end{array}\right),
$$
for $v\in\mathscr{C}^1(X,L)$, $w\in \mathscr{C}^1(\Pf(L),X)$
and $u\in\mathscr{C}^1(\Pf(L),L)$. The condition for this to determine
an invariant brane is
\beqa
&{\bf Q}v=0,\quad {\bf Q}w=0,\quad
vw+i{\bf Q}u=0,\nn\\
&w+s\Pf(v)=0,\quad u+\mathfrak{c}_L\Pf(u)=0.\nn
\eeqa
This channel may correspond to a ``tertiary vertex'' in flow 
trees for orientiholes \cite{OH}.

\subsection{K-Theory, Revisited}

In Section~\ref{sec:K}, we discussed the classification of the topology
of D-brane configurations in terms of K-theory, where
the machinery of Grothendieck group is applied to the categories of 
topological vector bundles (with additional structures).
We may also apply it to the type of categories discussed 
in the present Section.

The Grothendieck group \cite{Grothendieck}
of a triangulated category $\cC$, denoted by $\K(\cC)$,
is 
the free Abelian group of isomorphism classes of objects of
$\cC$ divided by the relation $[X]-[Y]+[Z]=0$ for each triangle
$X\stackrel{u}{\to} Y\stackrel{u}{\to}Z\stackrel{w}{\to}\shift(X)$.
For the category $\cC=\DM$, it is equal to the Grothendieck group
$\K_{\omega}(M)$ of holomorphic vector bundles over $M$.
There is a forgetful map to the topological K-theory 
\beq
f:\K_{\omega}(M)\longrightarrow\K(M).
\label{forgetK}
\eeq
In general, it is neither injective nor surjective.

The Grothendieck group of a category with duality is referred to 
as the {\it Hermitian K-theory}, or the Grothendieck-Witt group, and
comes with another version, called the Witt group. They have origins
in surgery theory and the theory of symmetric or antisymmetric bilinear forms.
The theory is introduced and developed by Wall, Novikov, Karoubi, and 
many people.
See \cite{MilHuse,handbookK,Novikov,Karoubi,Mishchenko}, for example.
The theory for the category $\mathfrak{Bil}(M)$ was developed by Knebusch
\cite{Knebusch}.
Relevant for us is the case of triangulated categories 
with duality \cite{Balmer,Walter}, which we now describe.

The Grothendieck-Witt group (or the Hermitian K-theory) 
of a triangulated category with duality
$(\cC,\Pf,\mathfrak{c})$, denoted by 
$\GW(\cC,\Pf,\mathfrak{c})$, is the free Abelian group 
of isomorphism classes of objects of $\mO(\cC,\Pf,\mathfrak{c})$
divided by the relations $[(X,s)]+[(Y,t)]=[(X\oplus Y),(s\oplus t)]$
and $[\cone(u,L)]=[(H_L,s_L)]$.
The class of an invariant cone is identified with the underlying 
hyperbolic object.
The Witt group of $(\cC,\Pf,\mathfrak{c})$, denoted by 
$\Witt(\cC,\Pf,\mathfrak{c})$, is obtained by 
dividing further by the relation $[\cone(u,L)]=0$.
Only ``truly invariant'' objects are non-zero in this group.
There is an exact sequence of groups
\beq
\K(\cC)\,\,\longrightarrow\,\,
\GW(\cC,\Pf,\mathfrak{c})\,\,\longrightarrow\,\,
\Witt(\cC,\Pf,\mathfrak{c})\,\,\to\,\,0.
\eeq
The first map sends $[X]$ to $[(H_X,s_X)]$.

This can be applied to our category with dualities
$(\DM,\Pf_j,\mathfrak{c}_j)$.
Let us regard them as a series where we use a fixed section $c_{{}_{\rm top}}$
to define $\mathfrak{c}_j$. It is 4-periodic by (\ref{4pertop}).
If $(\cC,\Pf,\mathfrak{c})$ is one of them, we define
$\GW^i(\cC,\Pf,\mathfrak{c})$ and $\Witt^i(\cC,\Pf,\mathfrak{c})$
as the Grotheidieck-Witt and Witt groups for the duality 
at the ``$i$-steps ahead''. For example,
$\GW^i(\DM,\Pf_0,\mathfrak{c}_0)=\GW(\DM,\Pf_i,\mathfrak{c}_i)$.
Note that (\ref{4pertop}) implies that
$\GW^{i+2}(\cC,\Pf,\mathfrak{c})\cong \GW^i(\cC,\Pf,-\mathfrak{c})$
and similarly for the Witt group. 
These series of groups are of course 4-periodic.
Here we would like to quote from \cite{Balmer} 
a cohomological interpretation of the Witt groups $\Witt^i$.
First, note that $\mathfrak{c}_j$ and $\mathfrak{c}_{j-1}$ are related by
$\mathfrak{c}_{j-1 X}=\mathfrak{c}_{j X}\cdot\sigma_{\Pf_j^2(X)}$
where $\sigma_{\Pf_j^2(X)}$ is regarded as an isomorphism 
$\Pf_{j-1}^2(X)\to \Pf_j^2(X)$.
Then, for the case $(\Pf,\mathfrak{c})=(\Pf_j,\mathfrak{c}_j)$
the condition (\ref{iccond2}) can be written as 
$$
u=\mathfrak{c}_{j-1\, L}\cdot\Pf_{j-1}(u)
$$
for $u$ regarded as a degree 0 map $u:\Pf_{j-1}(L)\to L$.
Thus, the invariant cone construction is to construct from 
a brane with an ``o-morphism'' for the duality $(\Pf_{j-1},\mathfrak{c}_{j-1})$
a brane with an o-isomorphism of the next duality $(\Pf_j,\mathfrak{c}_j)$.
The image of this map consists of metabolic objects for 
$(\Pf_j,\mathfrak{c}_j)$. The kernel
of the next map consists of brane with o-morphisms for
$(\Pf_j,\mathfrak{c}_j)$ whose cones are trivial.
Note that the cone of a morphism is trivial if and only if the morphism is
an isomorphism. Therefore, the kernel consists of
branes with o-isomorphisms, i.e., invariant branes.
Therefore, cohomology classes are
invariant objects modulo metabolic ones, which are nothing but
elements of the Witt group
$\Witt(\DM,\Pf_j,\mathfrak{c}_j)=\Witt^i(\DM,\Pf_0,\mathfrak{c}_0)$.

Let us consider D-branes in Type II orientifold on $M\times \R^D$
with data $(B,{\mathcal L},\sA,c)$. We assume that everything non-trivial
occurs in the $M$-component. Then, the relevant K-group is
$\GW(\DM,\Pf_{r_o},\mathfrak{c}_{r_o})$ where $r_o$ is even for $({\rm B}_+)$
and odd for $({\rm B}_-)$. Here
$\mathfrak{c}_{r_o}$ or $c_{{}_{\rm top}}$ is related to the crosscap section 
$\gc$ by (\ref{cvsctop}), i.e., $\gc=\spst^{-r_o}c_{{}_{\rm top}}$.
At this point, we recall the crosscap section $c$ introduced 
in Section~\ref{subsec:summary},
$\gc=c$ for $({\rm B}_+)$
and $\gc=\spst c$ for $({\rm B}_-)$,
which is used in many places including
Section~\ref{subsec:twistK} for K-theory classification of topology.
If we identify $c_{{}_{\rm top}}$ with $c$, i.e., if we put
$$
\mathfrak{c}_{r_o \,\cB}=\sigma_E^{r_o}\imath_E^{-1}\otimes c,
$$
then, we must take $r_o=0$ for $({\rm B}_+)$
and $r_o=-1$ for $({\rm B}_-)$.
That is, the relevant Hermitian K-theory is
\beq
\begin{array}{ll}
({\rm B}_+)\,\,:&
\GW^0(\DM,\Pf_0,\mathfrak{c}_0)\cong \GW^2(\DM,\Pf_0,-\mathfrak{c}_0),
\\[0.2cm]
({\rm B}_-)\,\,:&
\GW^{-1}(\DM,\Pf_0,\mathfrak{c}_0)\cong \GW^1(\DM,\Pf_0,-\mathfrak{c}_0),
\end{array}
\eeq
Compare this with the topological classification (\ref{class4}).
This implies that the group
$\GW^{-i}(\DM,\Pf_0,\mathfrak{c}_0)$ is related to
$\KR^{-2i}(M,c)$ in a way similar to the relation between the
algebraic and topological K-theories. We might have a forgetful map
between them like (\ref{forgetK}), but that requires us to find
a representative with a holomorphic o-isomorphism
for each invariant object with a quasi-o-isomorphism.
See the discussion in the example~\ref{ex:qois}.

K-theory discussed in this subsection and the topological K-theory
discussed in Section~\ref{sec:K} carry different information,
as shown by the fact that the forgetful map (\ref{forgetK})
is in general neither injective nor surjective.
We may ask whether the category (with duality) also knows about
the topological K-theory. The answer is known to be {\it no} in general,
as long as we view the category only as the triangulated category. However,
in many cases it comes from a differential graded category.
For our example of D-branes on $M$, 
as the space of morphisms we may take the whole ${\bf Q}$-complex 
$\mathscr{C}(\cB_1,\cB_2)$, with (\ref{Ci}) as the $i$-th component,
instead of the $0$-th cohomology space.
Alternatively (and equivalently \cite{BonRos,Block}), 
we may take the differential graded category of perfect complexes on $M$.
Then, it is known that, as long as $M$ is a (smooth) projective variety,
the topological K-theory of the underlying topological space,
$\K(M)$, can be recovered from the differential graded category. 
(This was shown in \cite{Toen} based on ``Semi-topological K-theory''
by Friedlander and Walker \cite{FWstK}).
It is an interesting question if that can be extended to 
the differential graded category with duality.

\medskip

\section*{Acknowledgement}

We would like to thank B.~Acharya, M.~Aganagic, A.~Bondal, 
I.~Brunner, R.~Buchweitz,
R.~Douglas, D.~Freed, K.~Fukaya, E.~Getzler, S.~Hellerman,
M.~Herbst, M.~Hopkins, 
P.~Horja, K.~Hosomichi,
M.~Kapranov, M.~Karoubi, the late Lev Kofman,
D.~Krefl, M.~Kontsevich, J.~Martin, E.~Meinrenken,
G.~Mikhalkin, R.~Myers, Y-G.~Oh, H.~Ohta,
Y.~Okawa, K.~Ono, D.~Page, E.~Poppitz, R.~Rabadan, S.~Ryu,
K.~Saito, G.~Segal, I.~Singer, J.~Solomon, 
S.~Sugimoto, T.~Takayanagi,
B.~To\"en, A.~Tsuchiya, C.~Vafa, J.~Walcher and J.-Y.~Welschinger
for enjoyable conversations, useful discussions, 
collaboration in related subjects,
illuminating instructions and encouragement at difficult stages.

DG wants to thank the Department of Physics, the University of Toronto, 
and IPMU, the University of Tokyo, where the work was done.
KH would like to thank the town of Strasbourg where he realized 
the importance of graded duality during his visit in May 2005.
He would also like to thank other towns and/or institutions
for the hospitality during his visits where important progress was made,
in particular,
(K)ITP, Santa Barbara (1999, 2003, 2005),
RIMS/YITP, Kyoto University, Kyoto (2004),
Fields Institute, Toronto (2005),
Paris (2006), DESY, Hamburg (2008),
CIRM, Luminy (2009), and Seoul (2010).

Parts of this work were supported by 
NSF-PHY 9514797 and DOE-DE-AC03-76SF00098  at U.C. Berkeley/LBL;
NSF-DMS 9709694 and 0074329 at Harvard University;
NSF-PHY 0070928 at IAS, Princeton;
NSERC, the Alfred P. Sloan Foundation and Connaught Foundation 
at the University of Toronto; 
JSPS Grant-in-Aid for Scientific Research No. 21340109
and WPI Initiative, MEXT, Japan at IPMU, the University of Tokyo.




\bigskip
\noindent
{\it E-mail addresses}:\\
\texttt{dfgao@wipm.ac.cn},
\texttt{kentaro.hori@ipmu.jp}

\end{document}